\documentclass[12pt]{article}

\usepackage{graphicx,psfrag,epsf}
\usepackage{enumerate}
\usepackage{natbib}
\usepackage{url} 
\usepackage{calc}
\usepackage{array}
\usepackage{booktabs}

\usepackage{amsthm, amsmath, amsfonts, amssymb}
\usepackage{bigints}
\usepackage{commath}
\usepackage{upgreek}
\usepackage{mathrsfs}
\usepackage{bbm}
\usepackage{paralist}

\allowdisplaybreaks

\usepackage{dsfont}
\usepackage{mathtools}
\usepackage{makeidx}
\usepackage{enumitem}
\usepackage{tabularx}
\usepackage{rotating}
\usepackage{url}
\usepackage{hyperref}
\usepackage{float}
\usepackage{lscape} 
\usepackage{setspace}
\usepackage{booktabs}
\usepackage{colortbl}
\usepackage{xcolor}
\usepackage{color}
\usepackage{dcolumn}
\usepackage{eurosym}
\usepackage{subcaption}
\usepackage{comment}

\usepackage{accents}
\newlength{\dhatheight}
\newcommand{\doublehat}[1]{%
  \settoheight{\dhatheight}{\ensuremath{\widehat{#1}}}%
  \addtolength{\dhatheight}{-0.35ex}
  \widehat{%
    \vphantom{\rule{0pt}{\dhatheight}}
    \smash{\widehat{#1}}%
  }%
}
\usepackage{xr}

\setcitestyle{authoryear,round}
\hypersetup{colorlinks=true, citecolor=blue, linkcolor=blue}

\newcolumntype{d}[1]{D{.}{.}{#1}}  

\newcommand\tab[1][1cm]{\hspace*{#1}}

\newtheorem{theorem}{Theorem}
\theoremstyle{plain}
\newtheorem{assumption}{Assumption}
\newtheorem{prop}{Proposition}

\newtheorem{lemma}[theorem]{Lemma}

\newtheorem{remark}{Remark}

\newcommand{\bra}[1]{\left( #1 \right)}
\newcommand{\sbra}[1]{\left[ #1 \right]}
\newcommand{\cbra}[1]{\left\{ #1 \right\}}
\newcommand{\vbra}[1]{\left| #1 \right|}
\newcommand{\vvbra}[1]{\left\| #1 \right\|}

\DeclareMathOperator{\tr}{tr}
\DeclareMathOperator{\trace}{\tr}
\DeclareMathOperator{\vect}{vec}
\DeclareMathOperator{\vecth}{vech}

\DeclareMathOperator*{\argmax}{arg\,max}

\def\pconv{\xrightarrow{\mathbb{P}}}
\def\asconv{\xrightarrow{a.s.}}
\def\easconv{\xrightarrow{e.a.s.}}

\def\dIndicator{\mathds{1}}

\DeclareMathOperator{\diag}{diag}

\def\trans{^{\top}}
\def\inv{^{-1}}

\def\dA{\boldsymbol{A}}

\def\dD{\boldsymbol{D}}
\def\dI{\boldsymbol{I}}
\def\dK{\boldsymbol{K}}

\def\dV{\boldsymbol{V}}
\def\dZ{\boldsymbol{Z}}

\def\dPi{\boldsymbol{\Pi}}

\def\dDt{\boldsymbol{D}_t}

\def\du{\boldsymbol{u}}
\def\dut{\boldsymbol{u}_t}

\def\dhathatut{\doublehat{\boldsymbol{u}}_t}

\def\dS{\boldsymbol{S}}
\def\dSt{\boldsymbol{S}_t}
\def\dG{\boldsymbol{G}}

\def\dv{\boldsymbol{v}}
\def\dvt{\boldsymbol{v}_t}
\def\dvtp{\boldsymbol{v}_{t+1}}

\def\dz{\boldsymbol{z}}

\def\depsi{\boldsymbol{\epsilon}}

\def\dzeroN{\boldsymbol{0}_N}
\def\dIdentityN{\boldsymbol{I}_N}

\def\dzeroN{\boldsymbol{0}_N}

\def\dzeror{\boldsymbol{0}_r}
\def\dIdentityN{\boldsymbol{I}_N}
\def\dIdentityr{\boldsymbol{I}_r}

\def\dx{\boldsymbol{x}}
\def\dxt{\boldsymbol{x}_t}

\def\dxi{\boldsymbol{e}}
\def\dxit{\boldsymbol{e}_t}

\def\Cittithetaz{{C}_{i,t|t-1}(\dtheta_0)}

\def\hatCittithetamle{\widehat{C}_{i,t|t-1}(\dthetamle)}

\def\Pittithetaz{{P}_{i,t|t-1}(\dtheta_0)}

\def\hatPittithetamle{\widehat{P}_{i,t|t-1}(\dthetamle)}

\def\dF{\boldsymbol{F}}
\def\dFt{\boldsymbol{F}_t}

\def\dFtp{\boldsymbol{F}_{t+1}}
\def\dhatF{\widehat{\boldsymbol{F}}}
\def\dhatFt{\widehat{\boldsymbol{F}}_t}

\def\dhathatFtp{\doublehat{\boldsymbol{F}}_{t+1}}

\def\dbarF{\bar{\boldsymbol{F}}}
\def\dbarFt{\bar{\boldsymbol{F}}_t}

\def\dmu{\boldsymbol{\mu}}
\def\dmutpt{\boldsymbol{\mu}_{t+1|t}}
\def\dmutti{\boldsymbol{\mu}_{t|t-1}}

\def\dhatmu{\widehat{\boldsymbol{\mu}}}
\def\dhatmutpt{\widehat{\boldsymbol{\mu}}_{t+1|t}}
\def\dhatmutti{\widehat{\boldsymbol{\mu}}_{t|t-1}}

\def\dhathatmu{\doublehat{\boldsymbol{\mu}}}
\def\dhathatmutpt{\doublehat{\boldsymbol{\mu}}_{t+1|t}}
\def\dhathatmutti{\doublehat{\boldsymbol{\mu}}_{t|t-1}}

\def\dhathatmuthetatti{\doublehat{\boldsymbol{\mu}}_{t|t-1}(\dtheta)}

\def\dhathatmuthetamletti{\doublehat{\boldsymbol{\mu}}_{t|t-1}(\dthetamle)}

\def\dSigma{\boldsymbol{\Sigma}}

\def\dSigmatti{\boldsymbol{\Sigma}_{t|t-1}}

\def\dhatSigmatti{\widehat{\boldsymbol{\Sigma}}_{t|t-1}}

\def\dhathatSigmatti{\doublehat{\boldsymbol{\Sigma}}_{t|t-1}}

\def\dhathatSigmathetatti{\doublehat{\boldsymbol{\Sigma}}_{t|t-1}(\dtheta)}

\def\dhathatSigmathetamletti{\doublehat{\boldsymbol{\Sigma}}_{t|t-1}(\dthetamle)}

\def\dsigma{\boldsymbol{\sigma}^2}
\def\dsigmatpt{\boldsymbol{\sigma}^2_{t+1|t}}
\def\dsigmatti{\boldsymbol{\sigma}^2_{t|t-1}}

\def\dhatsigma{\widehat{\boldsymbol{\sigma}}^2}
\def\dhatsigmatpt{\widehat{\boldsymbol{\sigma}}^2_{t+1|t}}
\def\dhatsigmatti{\widehat{\boldsymbol{\sigma}}^2_{t|t-1}}

\def\dhathatsigma{\doublehat{\boldsymbol{\sigma}}^2}
\def\dhathatsigmatpt{\doublehat{\boldsymbol{\sigma}}^2_{t+1|t}}
\def\dhathatsigmatti{\doublehat{\boldsymbol{\sigma}}^2_{t|t-1}}

\def\dhathatsigmathetamletti{\doublehat{\boldsymbol{\sigma}}^2_{t|t-1}(\dthetamle)}

\def\dhathatsigmathetatti{\doublehat{\boldsymbol{\sigma}}^2_{t|t-1}(\dtheta)}

\def\dsigmaotti{{\sigma}^2_{1,t|t-1}}

\def\dsigmartti{{\sigma}^2_{r,t|t-1}}

\def\dpsi{\boldsymbol{\psi}}
\def\dpsitpt{\boldsymbol{\psi}_{t+1|t}}
\def\dpsitti{\boldsymbol{\psi}_{t|t-1}}

\def\dbarpsi{\bar{\boldsymbol{\psi}}}
\def\dbarpsitpt{\bar{\boldsymbol{\psi}}_{t+1|t}}
\def\dbarpsitti{\bar{\boldsymbol{\psi}}_{t|t-1}}

\def\dhatpsi{\widehat{\boldsymbol{\psi}}}

\def\dhatpsitti{\widehat{\boldsymbol{\psi}}_{t|t-1}}

\def\dpsithetatpt{\boldsymbol{\psi}_{t+1|t}(\dtheta)}
\def\dpsithetatti{\boldsymbol{\psi}_{t|t-1}(\dtheta)}

\def\dhatpsi{\widehat{\boldsymbol{\psi}}}
\def\dhatpsithetatpt{\widehat{\boldsymbol{\psi}}_{t+1|t}(\dtheta)}
\def\dhatpsithetatti{\widehat{\boldsymbol{\psi}}_{t|t-1}(\dtheta)}

\def\dhathatpsi{\doublehat{\boldsymbol{\psi}}}
\def\dhathatpsitti{\doublehat{\boldsymbol{\psi}}_{t|t-1}}
\def\dhathatpsithetatpt{\doublehat{\boldsymbol{\psi}}_{t+1|t}(\dtheta)}
\def\dhathatpsithetatti{\doublehat{\boldsymbol{\psi}}_{t|t-1}(\dtheta)}

\def\dtildepsi{\widetilde{\boldsymbol{\psi}}}

\def\dtildepsitti{\widetilde{\boldsymbol{\psi}}_{t|t-1}}

\def\dtildepsithetatpt{\widetilde{\boldsymbol{\psi}}_{t+1|t}(\dtheta)}
\def\dtildepsithetatti{\widetilde{\boldsymbol{\psi}}_{t|t-1}(\dtheta)}

\def\dsqrtSigmatti{\boldsymbol{\Sigma}^{1/2}_{t|t-1}}

\def\dW{\boldsymbol{W}}

\def\dtheta{\boldsymbol{\theta}}
\def\dvartheta{\boldsymbol{\vartheta}}
\def\dthetamle{\widehat{\boldsymbol{\theta}}}

\def\limitcov{\boldsymbol{\mathcal{I}}}

\def\dTheta{\boldsymbol{\Theta}}

\def\ddelta{\boldsymbol{\delta}}

\def\dPhi{\boldsymbol{\Phi}}
\def\dK{\boldsymbol{K}}
\def\dUpsilon{\boldsymbol{\Upsilon}}
\def\dOmega{\boldsymbol{\Omega}}

\def\dLambda{\boldsymbol{\Lambda}}

\def\dH{\boldsymbol{H}}
\def\dhatH{\widehat{\boldsymbol{H}}}

\def\Ee{\mathbb{E}}

\def\Nn{\mathbb{N}}
\def\Pp{\mathbb{P}}
\def\Rr{\mathbb{R}}
\def\Vv{\mathbb{V}}
\def\Zz{\mathbb{Z}}

\def\calK{\boldsymbol{\mathcal{K}}}
\def\calC{\boldsymbol{\mathcal{C}}}
\def\calX{\mathcal{X}}
\def\calY{\mathcal{Y}}

\def\calF{\mathcal{F}}
\def\calH{\mathcal{H}}
\def\calN{\mathcal{N}}
\def\calZ{\mathcal{Z}}

\def\calP{\boldsymbol{\mathcal{P}}}

\def\calV{\boldsymbol{\mathcal{V}}}
\def\calW{\boldsymbol{\mathcal{W}}}

\begin{document}

\def\spacingset#1{\renewcommand{\baselinestretch}%
{#1}\small\normalsize} \spacingset{1}


{
  \title{\bf 
  High-Dimensional Approximate Factor Models with Score-Driven Dynamics: Estimation and Inference
  }
   \author{
  Matteo Barigozzi$^{a}$
   and  
  Enzo D'Innocenzo$^{a,}$\thanks{
   Corresponding author; e-mail: \href{mailto:enzo.dinnocenzo2@unibo.it}{enzo.dinnocenzo2@unibo.it}}
\\[1cm]
\normalsize $^a$ Department of Economics, Alma Mater Studiorum  Università di Bologna}
  \maketitle
} 


\spacingset{1.45} 

\begin{abstract}
We propose a  factor model for high-dimensional time series where  the dynamics of the latent factors is generated by a multivariate score-driven model, thus allowing to model nonlinearities and heavy tails in both the conditional mean and the conditional variance. Estimation is in two steps: first the factors are extracted via Principal Components  and then the parameters of the score-driven model are estimated via Quasi Maximum Likelihood based on the estimated factors. Since factors are not identified, we first establish set-consistency and asymptotic normality for the estimated parameters governing their dynamics, as both the number of time series and the sample size diverge to infinity. Then, we construct valid asymptotic prediction intervals, which do not depend on the chosen identifying constraints for the factors. Numerical results confirm the goodness of our estimator.
\end{abstract}

\section{Introduction}
We propose a class of  factor models designed for analyzing high-dimensional datasets of time series that display nonlinear dynamics and potentially heavy tails. TTo this end, we assume that the latent-factor dynamics are nonlinear and generated by the multivariate score-driven framework introduced by \citet{Creal2013} and \citet{Harvey2013}. Our model framework is designed to flexibly accommodate both high-dimensional conditional mean \citep{Gasperoni2023,DInnocenzo2023} and conditional covariance models \citep{Creal2011,Buccheri2021}.

The factor model we consider has three main features. First, it is approximate in the sense that it is characterized by the assumption of weakly cross-sectionally correlated idiosyncratic components, a realistic feature of high-dimensional panels of time series. Second, since the factors are serially dependent, we explicitly model their possibly nonlinear time evolution. Third, although we allow for serially correlated idiosyncratic components, we leave their dynamics unspecified. Our modeling strategy is consistent with most existing literature on factor models for high-dimensional time series \citep[see][and references therein]{Stock2016}. The last aspect is what distinguishes our work from existing score-driven factor models, which, however, are studied only under very restrictive assumptions \citep{Artemova2023}.


Estimation is as follows. First, we extract latent factors using Principal Component (PC) analysis \citep{Bai2003}. Then, we apply the score-driven methodology to model the time-varying parameters of the latent factors, estimating the relevant parameters via Quasi Maximum Likelihood (QMLE) \citep{Blasques2022}. This two-step strategy yields flexible high-dimensional nonlinear time series models that are both statistically tractable and computationally efficient.



It is well known that, in an approximate factor model, the factor-driven common component can be disentangled from the idiosyncratic component only as the cross-sectional dimension $N$ diverges \citep{BarigozziHallin2026}. Hence our asymptotic theory is developed under the double asymptotic framework $N,T\to\infty$, where $T$ is the sample size. Furthermore, since the factors are in general not identified, we must develop our theory by explicitly mapping the effect of this indeterminacy on all considered estimators. This task is far from trivial because the estimators of the parameters governing the dynamics of the factors have not a closed form expression.

We make three theoretical contributions. First, we formally derive conditions for strict stationarity, ergodicity, and filter invertibility of our general nonlinear score-driven high-dimensional approximate dynamic factor model in the case of Gaussian and heavy-tailed conditional location (mean) and conditional covariance models. 
Second, to address the aforementioned factor-identification issue, we develop an asymptotic theory by adapting the framework of \cite{Gallant1988} and \cite{Potscher1997} for set-identified nonlinear dynamic models, allowing us to establish low-level conditions based on the model's primitive structure, such as contraction, smoothness and domination conditions. 
This allows us to derive appropriate convergence rates, ensuring the consistency and asymptotic normality of the proposed QMLE of the parameters driving the score-driven model for the latent factors.
Third, we construct  asymptotically valid prediction intervals for the filtered process which do not depend on any specific identification condition one might impose on the factors.


Our approach aligns with the recent work of \cite{Blasques2023}, who introduced the class of quasi score-driven models where the dynamic equation is decoupled from the log-likelihood function while preserving the robustness properties of traditional score-driven models. In this sense, our model can be viewed as a quasi score-driven model, as PC can be formulated within a Gaussian latent variable framework, though the driving force of the score-driven model remains non-Gaussian. 
Furthermore, PC factor estimation has the same consistency rate as QML estimation \citep{Bai2016,BarigozziLuciani2019}. Alternatives to PCs requiring fewer moment constraints are briefly discussed and studied numerically \citep{he2020elliptic,FL22, barigozzi2024tail}.
 


\vspace{-.5cm}
\paragraph{Related literature.} 
 So far, to the best of our knowledge, only two studies develop a complete QML theory for score-driven factor models. While \citet{Artemova2023} focuses on estimation, \citet{Buccheri2024} also addresses the factor-identification problem. Unlike our approach, however, both studies keep the cross-sectional dimension fixed and therefore consider exact factor models with cross-sectionally uncorrelated idiosyncratic components. Moreover, their asymptotic theories rely on high-level conditions concerning the convergence of the log-likelihood function and its derivatives. Extending the approach of \citet{Artemova2023} to a high-dimensional setting is far from trivial, because high-dimensional maximization creates difficulties concerning both the theoretical existence of the maximizers and the numerical convergence of the optimization algorithms employed.

Our work is the first to provide a joint $N,T$ asymptotic theory for high-dimensional factor generalized autoregressive conditional heteroskedasticity (GARCH) models These models, dating back to \cite{Diebold1989}, \cite{Engle1990}, and \citet{HarveyRuizSentana1992}, were developed to address two major challenges: the exponential growth in the number of parameters as dimensionality increases and the algebraic constraints required to ensure the positive definiteness of the conditional covariance matrix.  To the best of our knowledge no theory has ever been developed for high-dimensional approximate factor GARCH models. While only a few works address the related estimation theory in a fixed  dimensional exact factor model setting \citep{Sentana2001,Weide2002,Vrontos2003,Fiorentini2004,SentanaCalzolariFiorentini2008}, other works extend the model to a high-dimensional setting, but provide either partial theoretical results, or just numerical results \citep{Lanne2007,FanWangYao2008,Zhang2009,LiGaoLiYao2016,TruciosEtAl2022}.

Finally, our theory generalizes to the nonlinear dynamics setting the works on factor augmented regressions, where the dynamics of the factors is assumed to be generated by a linear multivariate autoregressive model. Two steps estimators of such models, having PCA as first step, are studied by \citet{bai2006confidence}, \citet{forni2009opening}, and \citet{doz2011two}.

The rest of the paper is organized as follows. Section \ref{sec_model} outlines the model and its main features. Section \ref{sec_estimator} describes estimation. Section \ref{sec:asymptotic_theory} contains all the asymptotic properties of the estimators and the required assumptions to derive them. Section \ref{sec:asymptotic_theory_J} studies the extension to the joint mean-variance model. Section \ref{sec:MCsim} contains simulation results. Additional details on the model, the relation between the score-driven parameters and the factors, all proofs, and more numerical results are in the Appendix.

\vspace{-.5cm}
\paragraph{Basic notation.} For a vector $\dv$, we denote its Euclidean norm by $\Vert\dv\Vert$. For a matrix $\dA$, we denote its spectral norm by $\Vert\dA\Vert$ and its determinant by $\vert\dA\vert$. For a random vector $\boldsymbol x$, we denote its $L_p$-norm by $\Vert\boldsymbol x\Vert_p=(\Ee[\Vert\boldsymbol x\Vert^p])^{1/p}$, for $1\leq p<\infty$. Convergence in probability is denoted by $\pconv$, almost-sure convergence by $\asconv$, exponentially fast almost-sure convergence by $\easconv$, and convergence in distribution by $\Rightarrow$. Boundedness in probability and almost surely are denoted by $O_{\Pp}$ and $O_{a.s.}$, respectively.
\section{Model}
\label{sec_model}

We consider a real $N$-valued weakly stationary time series $\cbra{ \dxt }_{t\in\Zz}$, such that $\dxt = \bra{x_{1t} \cdots x_{Nt}}\trans \in \Rr^{N}$, for all $ t\in \Zz$, and, without loss of generality, let $\dxt$ be zero-mean. 
We consider the following model
\begin{align}
\label{eq:Model_xt}
\dxt &= \dLambda \dFt + \dxit, \tab t\in\Zz,  \\
\label{eq:Model_yt}
\dFt &= \boldsymbol q\bra{\dpsitti, \dvt}, 
\end{align}
where $\dLambda \in \Rr^{N\times r}$ is the matrix of factor loadings, $\cbra{\dFt}_{t\in\Zz}$ is the $r$-dimensional process of latent factors with
$\dFt \in\Rr^r$, for all $ t\in \Zz$, and $r \ll N$ being the number of factors, $\cbra{\dxit}_{t\in\Zz}$ is the $N$-dimensional process of idiosyncratic components
with $\dxit\in\Rr^N$, for all $t\in\Zz$, which are allowed to be weakly serially and cross-correlated, 
%
$\boldsymbol q\bra{\,\cdot\,, \cdot}:\mathbb R^r \times \mathbb R^r \to\mathbb R^r$ is a link function which determines the (possibly) nonlinear relation between a time-varying parameter $\{\dpsitti\}_{t\in\Zz}$, and a noise process $\cbra{\dvt}_{t\in\Zz}$, which jointly generate the factor process. 

In particular, $\dpsitti$ is a measurable function of the sigma-field $\calF_{t-1} = \sigma \cbra{\dF_{t-1}, \dF_{t-2}, \dots}$ generated by $\cbra{\dF_{t}}_{t\in\Zz}$, hence  the subscript $t|t-1$, following a first-order auto-regressive equation of the form
\begin{equation}
\label{eq:SD_Model_psit}
\dpsitpt = \ddelta + \dPhi \dpsitti + \dK \dut, \tab t\in\Zz,
\end{equation}
where $\dut$ is the score vector with respect to $\dpsitti$ of the quasi log-likelihood contribution of $\dF_t$, conditional on $\calF_{t-1}$, possibly scaled by a proper matrix, e.g., the conditional Information matrix. The scaled-score process, $\cbra{\dut}_{t\in\Zz}$, then plays the role of an innovation process for $\{\dpsitti\}_{t\in\Zz}$. In  \eqref{eq:SD_Model_psit}, $\ddelta$ is an $r$-dimensional vector of intercepts, whereas $\dPhi$ and $\dK$ are $r\times r$ matrices of static parameters which drive the dynamics of the unobservable process $\{\dpsitti\}_{t\in\Zz}$.  We denote as $\dtheta$ the vector containing the values of the parameters governing \eqref{eq:SD_Model_psit},
which contains $(\ddelta^\prime, \vect\dPhi^\prime, \vect\dK^\prime)^\prime$ plus any parameter specific to the chosen distribution. 
More precisely, let $\ell_{\dF}(\dF_t;\dtheta)$ be the conditional quasi log-likelihood of $\dF_t$, where, for simplicity of notation, we leave implicit the conditioning on $\calF_{t-1}$. 
Then, the rescaled score vector is given by
$$
\dut=\dS_t\inv\frac{\partial \ell_{\dF}(\dF_t;\dtheta)}{\partial \dpsitti}=\dS_t\inv\frac{\partial \ell_{\dF}(\boldsymbol q(\dpsitti,\dvt);\dtheta)}{\partial \dpsitti} , \tab t\in\Zz,
$$
where $\dS_t$ is a scaling $r\times r$ matrix and in the second step we used  \eqref{eq:Model_yt}  to highlight the dependence on the time-varying parameter, while, for simplicity of notation, we leave implicit the dependence of $\dpsitti$ and $\dut$ on $\dtheta$.


Clearly, depending on the considered conditional distribution we obtain a different recurrence equation for the unobservable process $\{\dpsitti\}_{t\in\Zz}$ and a different parameter vector $\dtheta$. Popular examples that satisfy the relation in \eqref{eq:Model_yt}-\eqref{eq:SD_Model_psit} are the classical VAR and multivariate GARCH models and score-driven models for multivariate time-varying locations or covariance matrices. In this paper, we consider in detail the four cases listed below.

\vspace{-.5cm}
\paragraph{Time-varying locations.}
In this case, we suppose that the non-stochastic link function, $\boldsymbol q\bra{\,\cdot\,, \dvt}$, generates the process $\cbra{\dFt}_{t\in\Zz}$ with a linear relation of the following form
\begin{equation}
\label{eq:SD_Model_location}
\dFt =\boldsymbol q\bra{\dpsitti, \dvt}= \dmutti + \dvt,
\end{equation}
so that in this case $\dpsitti \equiv \dmutti\in\Rr^{r}$ is the vector of time-varying conditional locations.  
In particular,  \eqref{eq:SD_Model_psit} assumes the following form (recall that we assumed $\dFt$ to have zero mean)
\begin{equation}
\label{eq:SD_Model_mut}
\dmutpt = \dPhi \dmutti + \dK \dut.
\end{equation}

\noindent\textit{Gaussian updates.} We assume that $\dvt \sim \calN\bra{\dzeror, \dSigma}$, 
i.e., $\cbra{\dvt}_{t\in \Zz}$ is Gaussian with $r$-dimensional zero location vector $\dzeror$ and $\dSigma\in\Rr^{r\times r}$ positive definite covariance matrix. The corresponding conditional quasi log-likelihood contribution is
\begin{align}
\label{eq:log_likelihood_gaussian_locations}
\ell_{\dF}\bra{\dF_t ; \dtheta}
=&\,
- \frac{r}{2}\log \bra{2\pi}
-\frac{1}{2}\log| \dSigma |
 - 
\frac{1}{2} 
\bra{\dFt - \dmutti}\trans 
\dSigma\inv 
\bra{\dFt - \dmutti}
,
\end{align}
and it is straightforward to show that, when we set $\dS_t=\dSigma\inv$,
$
\dut =  \dFt -\dmutti.
$
From \eqref{eq:SD_Model_mut} and \eqref{eq:log_likelihood_gaussian_locations} we see that the unknown parameter vector is given by $\dtheta = (\vecth\dSigma\trans, \vect\dPhi\trans, \vect\dK\trans)\trans$, which has dimension $d = r(r+1)/2 + 2 r^2$.

\noindent\textit{Student's {t} updates.}
Following  \cite{DInnocenzo2023} we assume that $\dvt\sim t_\nu\bra{\dzeror, \dSigma}$, i.e., $\cbra{\dvt}_{t\in\Zz}$ follows a multivariate Student's \emph{t} distribution with $\nu>0$ degrees of freedom, an $r$-dimensional zero location vector $\dzeror$  and $\dSigma\in\Rr^{r\times r}$ positive definite scale matrix. 
The conditional log-likelihood of the factors is:
\begin{align}
\label{eq:log_likelihood_student_locations}
\ell_{\dF}\bra{\dF_t ; \dtheta}
=&\,
\log\Gamma \bra{ \frac{\nu + r}{2} } - \log\Gamma \bra{ \frac{\nu}{2} } - \frac{r}{2}\log \bra{\pi \nu}
-\frac{1}{2}\log| \dSigma |
\nonumber\\
& - 
\frac{\nu + r}{2} 
\log\bra{
1 +
\frac{\bra{\dFt - \dmutti}\trans 
\dSigma\inv 
\bra{\dFt - \dmutti}}{\nu}
}.
\end{align}
Thus,
the scaled-score vector assumes the following form
$
\dut = w_t \big( \dFt -\dmutti\big)$, with $w_t=\big\{1 + ( \dFt -\dmutti)\trans \dSigma\inv ( \dFt -\dmutti)/ \nu\big\}^{-1}.
$
From \eqref{eq:SD_Model_mut} and \eqref{eq:log_likelihood_student_locations} we see that the unknown parameter vector is  given by $\dtheta = (\nu, \vecth\dSigma\trans, \vect\dPhi\trans, \vect\dK\trans)\trans$,  which has dimension
 $d=1 + r(r+1)/2 + 2 r^2$. When $\nu\rightarrow\infty$, the model with Student's \emph{t} score updates reduces to the Gaussian ones.

\vspace{-.5cm}
\paragraph{Time-varying variances.}
In this case, we suppose that the non-stochastic link function, $\boldsymbol q\bra{\,\cdot\,, \dvt}$, generates the process $\cbra{\dFt}_{t\in\Zz}$ with a multiplicative relation of the following form
\begin{equation}
\label{eq:SD_Model_covariance}
\dFt =\boldsymbol q\bra{\dpsitti, \dvt}= \dsqrtSigmatti\dvt, 
\end{equation}
such that $\dSigmatti\in\Rr^{r\times r}$ is its associated time-varying conditional covariance matrix assumed to be positive definite, and with $\dsqrtSigmatti$ we denote its square root.
To model the full covariance matrix, we assume that the time-varying conditional covariance matrix has a diagonal structure, such that $\dSigmatti = \diag(\dsigmatti)$, with $\dsigmatti=(\dsigmaotti, \dots, \dsigmartti)\trans$ \citep{Sentana2001}. Thus, in this case $\dpsitti \equiv \dsigmatti
\in\mathbb R^r$ is the time-varying vector of conditional variances. 
Then, \eqref{eq:SD_Model_psit} specified for $ \dsigmatti$ becomes
\begin{equation}
\label{eq:SD_Model_sigmat}
\dsigmatpt = \ddelta + \dPhi \dsigmatti + \dK \dut.
\end{equation}
Furthermore, consider the $r^2\times r$ matrix $\dW_r =
 \sum_{i=1}^r\bra{\mathbf{e}_i \otimes  \mathbf{E}_i}$, where, for $i = 1, \dots, r$, $\mathbf{E}_i \in \Rr^{r\times r}$ with all zero entries except for the identity in $(i, i)$,  and $\mathbf{e}_i\in\Rr^r$  is a vector with all zeros except for the $i$-th position. Notice also that $\dW_r\trans\dW_r=\dIdentityr$. Then,
$
\vect \dSigmatti =\dW_r \dsigmatti$  and $\dsigmatti= \dW_r\trans \vect \dSigmatti. 
$
Notice that while it is clear that we can always derive $\dsigmatti$ from the full matrix $\dSigmatti$,  in our setting the vice versa is also possible since we assume the matrix $\dSigmatti$ to be diagonal for each $t$. The recursion in \eqref{eq:SD_Model_sigmat} can also be extended to incorporate leverage effects as detailed in Appendix \ref{app:leverage_effects}.

\noindent\textit{Gaussian updates.} Analogously to the previous dynamic location, if we assume that $\dvt\sim \calN\bra{\dzeror, \dIdentityr}$, i.e., $\cbra{\dvt}_{t\in\Zz}$ follows a multivariate standard Gaussian distribution,
the conditional log-likelihood of the factors is:
\begin{align}
\label{eq:log_likelihood_gaussian_covariances}
\ell_{\dF}\bra{\dF_t ; \dtheta}
=&\,
- \frac{r}{2}\log \bra{2\pi}
 - 
\frac{1}{2} 
\bra{
\log| \dSigmatti |
+\dFt\trans 
\dSigmatti\inv 
\dFt
}
,
\end{align}
and the scaled-score vector assumes the following form
$
\dut = 
\dW_r\trans
\bra{ \bra{\dFt \otimes \dFt} - \vect \dSigmatti}.
$
From \eqref{eq:SD_Model_sigmat} and \eqref{eq:log_likelihood_gaussian_covariances} we see that the unknown parameter vector is  given by $\dtheta = (\ddelta\trans, \vect\dPhi\trans, \vect\dK\trans)\trans$,  which has dimension
 $d= r + 2 r^2$.

\noindent\textit{Student's {t} updates.}
Here, the time-varying conditional covariance matrix is driven by the score of the multivariate Student's \emph{t} density, see \cite{Creal2011}. In particular, we assume that $\dvt\sim t_\nu\bra{\dzeror, \dIdentityr}$, i.e.,  $\cbra{\dvt}_{t\in\Zz}$ follows a multivariate standard Student's \emph{t} distribution with $\nu>2$ degrees of freedom, an $r$-dimensional zero mean vector $\dzeror$, and $\dIdentityr\in\Rr^{r\times r}$ as scale matrix.
The conditional log-likelihood of the factors is:
\begin{align}
\label{eq:log_likelihood_student_covariances}
\ell_{\dF}\bra{\dF_t ; \dtheta}
=&\,
\log\Gamma \bra{ \frac{\nu + r}{2} } - \log\Gamma \bra{ \frac{\nu}{2} } - \frac{r}{2}\log \bra{\pi (\nu-2)}
\nonumber\\
& -
\frac{1}{2}
\bra{
\log| \dSigmatti |
+
\bra{\nu + r} 
\log\bra{
1 +
\frac{\dFt\trans 
\dSigmatti\inv 
\dFt }{\nu-2}
}}.
\end{align}
Following the proof of \citet[Theorem 1]{Creal2011} we set (see Appendix \ref{app:creal} for more details)
\begin{align}
\dut&=\dW_r\trans \bra{
 \frac{\nu + r }{\nu - 2 + \dFt\trans\dSigmatti\inv\dFt} \bra{\dFt \otimes \dFt} - \vect \dSigmatti
}.\nonumber
\end{align}
This choice greatly simplifies calculations and is in line with the discussion by \cite{Opschoor2018}. From \eqref{eq:SD_Model_sigmat} and \eqref{eq:log_likelihood_student_covariances} the unknown parameter vector is  given by
 $\dtheta = (\nu, \ddelta\trans, \vect\dPhi\trans, \vect\dK\trans)\trans$,  which has dimension
 $d=1 + r + 2 r^2$. When $\nu\rightarrow\infty$, the model with Student's \emph{t} score updates reduces to the Gaussian ones.

\color{black}
\section{Estimation}
\label{sec_estimator}

\paragraph{Factors.} Consider the $N\times N$ sample covariance matrix $\widehat{\boldsymbol{\Gamma}}_x=T^{-1}\sum_{t=1}^T \dxt\dxt\trans$, with  normalized eigenvectors corresponding to its $r$ largest eigenvalues collected in the columns of the $N\times r$ matrix $\widehat{\boldsymbol P}$. Then, the PC estimators are given by 
\begin{align}
\widehat{\boldsymbol \Lambda}=\sqrt N\widehat{\boldsymbol P},\label{eq:loadPCA}\\
\widehat{\dF}_t=
 \frac{\widehat{\boldsymbol P} \trans \dxt}{\sqrt N}.\label{eq:factor_lin_proj}
\end{align}
Notice that in this case $N^{-1}\widehat{\boldsymbol \Lambda}\trans \widehat{\boldsymbol \Lambda}=\dIdentityr$ and $T^{-1}\sum_{t=1}^T \dhatF_t \dhatF_t\trans $ is an $r\times r$ diagonal matrix containing the eigenvalues of $\widehat{\boldsymbol{\Gamma}}_x$ divided by $N$. This is a standard formulation of PC estimators used, e.g., by \citet{stock2002forecasting} and differs from the one in \citet{Bai2003} just by a different rescaling of the estimated loadings. Alternatives to PCs are discussed in the next section.

\vspace{-.2cm}
\begin{remark}
\upshape{
We notice that a Student's $t$ with $\nu=4+\delta$, for some $\delta>0$, degrees of freedom would satisfy our assumptions for PC to be consistent but still allow for quite heavy tails. In fact, as shown in \citet[][Proposition 1]{Gasperoni2023}, for a standard $r$-variate Student's \emph{t} the convergence rate to the Gaussian slows with the dimension $r$, and a conservative bound for Gaussian-like integration constant is $\nu\simeq 30r^2$. Therefore, a moderate $r$ and a $\nu$ slightly larger than $4$ make the Student's \emph{t} behave closer to Gaussian for covariance/eigenspace accuracy, while preserving tail-robustness. Hence, the PC approach is likely to allow us to still capture heavy-tailed reality. This is confirmed by the numerical results in Section \ref{sec:MCsim}.
}
\end{remark}
%
%
%
\vspace{-.5cm}
\paragraph{Score-driven model parameters.}
 For observed factors $\dF=(\dF_1\cdots \dF_T)\trans$ it is clear that we can estimate the unknown vector of parameters in $\dtheta$ by 
maximizing the quasi $\log$-likelihood  
$\sum_{t=1}^T\ell_{\dF}\bra{\dF_t ; \dtheta}$ with respect to $\dtheta$.
However, 
$\dF_t$, $\dmutti$, and  $\dsigmatti$ are not observable. Still, given the first estimation step, we can replace $\dFt$ with $\dhatFt$ for $t=1,\dots,T$, which, in turn, can be employed to obtain the score-driven filtered paths of $\dhathatmuthetatti$ and $\dhathatSigmathetatti$ or $\dhathatsigmathetatti$ for any given $\dtheta$.
This will give rise to the approximate and feasible quasi $\log$-likelihood function $\sum_{t=1}^T \doublehat{\ell}_{\dF}(\dhatF_t ; \dtheta)$, which is a function of the estimated factors $\dhatFt$, of the score-driven filters $\dhathatmuthetatti$ and $\dhathatsigmathetatti$, for $t=1,\dots,T$, and of $\dtheta$.

Thus, we consider the QMLE obtained by solving the following problem:
\begin{align}
\label{eq:MLE}
\dthetamle 
=
\argmax_{\dtheta \in \dTheta\subset \mathbb R^d}
\sum_{t=1}^T
\doublehat{\ell}_{\dF}\bra{\dhatF_t ; \dtheta}.
\end{align}
In general the true factors $\dFt$ are identified only up to left-multiplication by an invertible $r\times r$ matrix. Therefore, any derived asymptotic property must take the effect of such indeterminacy fully into consideration. This discussed in detail below. Here we just note that, while QMLE is classical in both the correctly specified or misspecified score-driven literature (see, for instance, \citealp{Blasques2023}), the presence of estimated factors makes our overall model {only \emph{set-identified}} at the first stage, according to the taxonomy of \cite{Gallant1988} and \cite{Potscher1997}.

\vspace{-.2cm}
\begin{remark}
\upshape{
In principle, the loadings could be estimated via QML jointly with the score-driven parameters. Besides the need to avoid computational issues discussed in the Introduction, our choice has also two theoretical motivations, which hold at least when considering a Gaussian likelihood. First, \citet{BarigozziLuciani2019} prove that, for large $N$, the QMLE in case of serially correlated factors is asymptotically equivalent to the QMLE of the case in which factors follow a dynamic parametric model. 
Second, \citet{Bai2016} prove that, for large $N$, the QMLE of the loadings has the same convergence rate as the PCA estimator.}
\end{remark}

\section{Asymptotic theory}
\label{sec:asymptotic_theory}


Hereafter, the factor model in \eqref{eq:Model_xt} is defined for an infinite dimensional process $\{x_{it},\, i\in\mathbb N,\, t\in\mathbb Z\}$, and the properties and related assumptions are defined for the $N$-dimensional sub-process $\{x_{it},\, i=1\ldots, N,\, t\in\mathbb Z\}$ either holding for all $N\in\mathbb N$ or in the limit $N\to\infty$. The properties of the estimators are then derived for a given $NT$-dimensional realization  $\{x_{it},\, i=1\ldots, N,\, t=1,\ldots, T\}$ and hold in the limit $N,T\to\infty$. 

Throughout this section, for any $i\in\mathbb N$, we denote with $\boldsymbol \lambda_{0i}$ the $r$-dimensional vector of true factor loadings of $\{x_{it}\}_{t\in\Zz}$, with $\boldsymbol \lambda_{0i}\in \Rr^{r}$. Then, for any $N\in\Nn$, we denote with
$\dLambda_0=(\boldsymbol \lambda_{01}\cdots \boldsymbol \lambda_{0N})\trans$ the true $N\times r$ matrix of loadings, with $\dLambda_0\in \Rr^{N\times r}$.
We also denote with
$\dtheta_0$ the  $d$-dimensional vector of true score-driven parameters, with $\dtheta_0\in\dTheta\subset \Rr^d$. 
The true conditional mean or variance processes are then generically indicated as $\{\dpsitti\bra{\dtheta_0}\}_{t\in \Zz}$, such that $\dpsitti\bra{\dtheta_0}\in\calP\subset \Rr^r$ for all $t\in\Zz$, 
and with either $\dpsitti\bra{\dtheta_0}=\dmutti\bra{\dtheta_0}$ or  $\dpsitti\bra{\dtheta_0}=\dsigmatti\bra{\dtheta_0}$, as defined in Section \ref{sec_model}. The process of true factors is denoted as $\{\boldsymbol F_t\}_{t\in\Zz}$, such that $\dF_t\in \Rr^r$ for all $t\in\Zz$, and the true rescaled score, i.e., computed when using the true factors and parameters $\dtheta_0$, is denoted as $\{\du_t\}_{t\in\Zz}$, such that $\du_t\in \Rr^r$ for all $t\in\Zz$.

Hence, the model in \eqref{eq:Model_xt}-\eqref{eq:SD_Model_psit} now reads:
\begin{align}
&x_{it} = \boldsymbol \lambda_{0i}\trans\boldsymbol F_t+e_{it},\qquad i\in\Nn,\quad t\in \Zz,\label{eq:Model_xt0}\\
&\dFt = \boldsymbol q\bra{\dpsitti\bra{\dtheta_0}, \dvt},\label{eq:Model_yt0}\\
&\dpsitpt(\dtheta_0) =\ddelta_0 + \dPhi_0 \dpsitti(\dtheta_0) + \dK_0 \dut(\dtheta_0).\label{eq:SD_Model_psit0}
\end{align}
Furthermore, a generic trajectory of the conditional mean or variance is denoted as $\{\dpsitti\}_{t\in \Zz}$ such that $\dpsitti\in\calP\subset \Rr^r$ for all $t\in\Zz$. Although this distinction implies that we should introduce a new notation for generic factors generated as $\dFt= \boldsymbol q(\dpsitti, \dvt)$, and the corresponding rescaled score, we refrain from doing so, since, in the following, the only case in which $\{\dF_t\}_{t\in\Zz}$ indicates a generic factor process is when we consider derivatives of the log-likelihood or of the rescaled score with respect to $\dF_t$.


%


\subsection{Asymptotic properties of the estimated factors and their loadings}
\label{susection:asy_theory}
We start by making the following assumptions on the noise process $\cbra{\dvt}_{t\in\Zz}$ and the link function $\boldsymbol q(\,\cdot\,,\cdot)$.
\begin{assumption}
\label{ass:noise}
\begin{inparaenum}
\item[(i)] The process $\cbra{\dvt}_{t\in\Zz}$ is geometrically $\beta$-mixing. In particular, letting $\beta_{\dv}(h)$ denote its $\beta$-mixing coefficients, there exist a finite positive real $C_{\beta}$ and $0<\rho_{\beta}<1$ such that $\beta_{\dv}(h)\leq C_{\beta}\rho_{\beta}^{h}$ for every $h\geq1$. Moreover, $\Ee[\dvt] = \dzeror$ and $\Ee[\dvt\dvt\trans]$ is finite and positive definite for all $t\in \Zz$.
\item[(ii)] The distribution of $\cbra{\dvt}_{t\in\Zz}$ has a density strictly positive in a neighbourhood of $0$ with respect to the Lebesgue measure on $\Rr^r$.
\item [(iii)] The link function
$\boldsymbol q:\Rr^r\times\Rr^r\rightarrow\Rr^r$ is non-stochastic, Borel measurable and continuously differentiable in its first argument.
Moreover, for the same $p>2$ as in Assumptions \ref{ass:pram_SE} and \ref{ass:contraction_cond}, there exist Borel measurable functions $B_t:\Rr^r\times\Rr^r\rightarrow[0,\infty)$ such that, for all $\dz,\dz^\star\in\Rr^r$,
$
\vvbra{
\boldsymbol q\bra{\dz,\dvt}
-
\boldsymbol q\bra{\dz^\star,\dvt}
}
\leq
B_t\bra{\dz,\dz^\star}
\vvbra{
\dz-\dz^\star
}$,
and $\sup_{t\in\Zz}\sup_{\dz,\dz^\star\in\Rr^r} \vvbra{B_t\bra{\dz,\dz^\star} }_{\frac{2p}{p-2}}\leq B$, for some finite positive real $B$ independent of $t$.
%
%
\end{inparaenum}
\end{assumption}
Part $(i)$ imposes centering and non-degeneracy of the innovations and controls their temporal dependence. Geometric $\beta$-mixing is used together with the near-epoch dependence results below in the finite-window empirical-process argument that yields the uniform root-$T$ rate in Proposition \ref{lemma:ULLN}. Since every $\beta$-mixing process is also $\alpha$-mixing, it also provides the mixing condition required by the near-epoch dependent central limit theorem underlying Theorem \ref{thm:asy_norm}. Part $(ii)$ rules out degenerate innovation distributions and guarantees sufficient local support for the score-driven process. Finally, part $(iii)$ ensures that perturbations of the filtered parameter process translate into controlled perturbations of the factor process. This Lipschitz property will be used in Proposition \ref{prop:rate_psit_decomp}, in Lemma \ref{lemma:NED}, and consequently in the likelihood approximation of Proposition \ref{lemma:ULLN}.

For the factor process $\{\dFt\}_{t\in\Zz}$ and the true  stochastic parameter process $\{\dpsitti\bra{\dtheta_0}\}_{t\in\Zz}$ we assume the following.

\begin{assumption}
\label{ass:pram_SE}
\begin{inparaenum}
\item[(i)] The process $\cbra{\dFt}_{t\in\Zz}$ with $\dFt=\dzeror$ for $t\le 0$, converges as $t\to\infty$ to a unique strictly stationary and ergodic process $\{\dbarFt\}_{t\in\Zz}$, and is such that $\Ee[\dFt] = \dzeror$ and $\Ee[\dFt\dFt\trans]=\boldsymbol\Gamma_F$ finite and positive definite.
\item[(ii)] The process $\{\dpsitti(\dtheta_0)\}_{t\in\Zz}$ solving \eqref{eq:SD_Model_psit0} when using $\cbra{\dFt}_{t\in\Zz}$ and initialized with some random variable $\dpsi_{1|0}\in \boldsymbol{\mathcal{P}}$ with $\Vert\dpsi_{1|0}\Vert_p \leq C_0$ for $p > 2$ and some finite positive real $C_0$, converges as $t\to\infty$ to a unique strictly stationary and ergodic $\{\dbarpsitti(\dtheta_0)\}_{t\in\Zz}$ solution of \eqref{eq:SD_Model_psit0} computed when using $\{\dbarFt\}_{t\in\Zz}$.
\item [(iii)] For all $t \in \Zz$, $\Ee[\Vert \dFt\Vert^4]\le Q_F$ for some finite positive real $Q_F$ independent of $t$.
\end{inparaenum}
\end{assumption}

Parts $(i)$ and $(ii)$ are standard in score-driven models and can be substituted with appropriate conditions on the parameters of the score-driven recursion in \eqref{eq:SD_Model_psit} \citep{DInnocenzo2023}.  Part $(iii)$ is standard in PC literature \citep{Bai2003}.

\vspace{-.2cm}
\begin{remark}\label{rem:birk}
\upshape{
Under Assumptions \ref{ass:noise}-\ref{ass:pram_SE} we obtain that $\Vert T^{-1}\sum_{t=1}^T \dFt\dFt\trans - \boldsymbol\Gamma_F\Vert = O_{\Pp}(T^{-1/2})$ as $T\rightarrow\infty$, which directly follows from an application of the ergodic theorem for real-valued sequences of random variables of \cite{Birkhoff1931}. }
\end{remark}
\vspace{-.2cm}
For the loadings, we make the following assumption. 
\begin{assumption}
\label{ass:factor_identifiability}
As ${N\to\infty}$, $\Vert N^{-1}{\dLambda_0\trans\dLambda_0}- \boldsymbol \Sigma_\Lambda\Vert = O_{\Pp}(N^{-1/2})$ for some $r\times r$  finite and positive definite matrix $\dSigma_\Lambda$, and, for all $i\in\mathbb N$, $\Vert\boldsymbol{\lambda}_{0i}\Vert\le M_\Lambda$ for some finite positive real $M_\Lambda$ independent of $i$. 
\end{assumption}

Note that the rate $\sqrt N$ is often not given in the literature but it is reasonable to assume it in absence of any specific ordering of the cross-sectional items 
\citep{BarigozziHallin2026}. 
This setting corresponds to considering only strongly pervasive factors.


For the idiosyncratic process $\cbra{\dxit}_{t\in\Zz}$, we make the following assumptions.
\begin{assumption}
\label{ass:idiosyncratic}
\begin{inparaenum}
	\item [(i)] For all $i\in\mathbb N$ and $t\in\mathbb Z$, $\Ee[e_{it}]= 0$ and $\sigma_i^2=\Ee[e_{it}^2]$ such that $\sigma_i^2\in[C_e ,C_e^\prime]$ for some finite positive reals $C_e$ and $C_e^\prime$ independent of  $i$ and $t$. 
	\item [(ii)] For all $i,j\in\mathbb N$,  $t\in\mathbb Z$, and  $k\in\mathbb Z$, $\vert \Ee[e_{it}e_{j,t-k}]\vert\le \rho^{\vert k\vert} M_{ij}$, where $\rho$ and $M_{ij}$ are finite positive reals independent of $t$ such that $0\le \rho <1$, $M_{ii}=\sigma_i^2$, $\sum_{j=1,i\ne j}^N M_{ij}\le M_e$, and $\sum_{i=1, j\ne i}^N M_{ij}\le M_e$ for some finite positive real $M_{e}$ independent of  $i$, $j$, and $N$. 
	\item [(iii)] For any fixed $t\in\mathbb Z$, as $N\to\infty$,
	$
	N^{-1/2}\sum_{i=1}^N\boldsymbol{\lambda}_{0i}e_{it}
	\Rightarrow
	\calN\bra{
	\boldsymbol 0,
    \boldsymbol\Pi_{t,t}
	}
	$,
	with
	$\boldsymbol\Pi_{s,u}
	=
	\lim_{N\to\infty}N^{-1}\sum_{i=1}^N\sum_{j=1}^N
	\boldsymbol{\lambda}_{0i}\boldsymbol{\lambda}_{0j}\trans
	\Ee[e_{is}e_{ju}]$.
\item [(iv)]  For all $i=1,\ldots, N$,  $t=1,\ldots, T$, and  $N,T\in\mathbb N$, $\Ee[e_{it}^4]\le Q_e$ for some finite positive real $Q_e$ independent of $i$ and $t$.
\item [(v)] For all $s,u=1,\ldots,T$ and $N,T\in\mathbb N$,
$
\Ee\big[
\big|
N^{-1/2}
\sum_{i=1}^{N}
\cbra{
e_{iu}e_{is}
-
\Ee\sbra{e_{iu}e_{is}}
}
\big |^{2}
\big]
\leq K_e,
$
for some finite positive real $K_e$ independent of $s$, $u$, $N$, and $T$.
 \end{inparaenum}
\end{assumption}

Regarding the relation between idiosyncratic components and factors, we assume the following.
\begin{assumption}
\label{ass:idiosyncraticY}
\begin{inparaenum}
	\item [(i)] The processes $\{\dFt\}_{t\in\Zz}$ and  $\{e_{it}\}_{t\in\Zz}$ are independent for all $i\in\Nn$.
	\item [(ii)]  For all  $t=1,\ldots, T$ and $N,T\in\mathbb N$,
	$
	\Ee\big[\big\Vert(NT)^{-1/2}\sum_{s=1}^T\sum_{k=1}^N \dF_s\big(e_{ks}e_{kt}-\Ee[e_{ks}e_{kt}]\big)
	\big\Vert^2
	\big]\le K_{e,1},
	$
	for some finite positive real $K_{e,1}$ independent of $t$, $N$, and $T$.
	\item [(iii)] For all $N,T\in\mathbb N$,
	$
	\Ee\big[\big\Vert
	(NT)^{-1/2}
	\sum_{t=1}^T\sum_{k=1}^N\dF_t \boldsymbol{\lambda}_{0k}\trans e_{kt}
	\big\Vert^2
	\big]\le K_{e,2},
	$
	for some finite positive real $K_{e,2}$ independent of $N$ and $T$.
	\item [(iv)] For any $i\in\Nn$, as $T\to\infty$,
$
T^{-1/2}\sum_{t=1}^T \dF_te_{it} \Rightarrow \calN\bra{\dzeror, \boldsymbol \Xi_i}
$, with $\boldsymbol \Xi_i= \lim_{T\to\infty}T^{-1}\sum_{t=1}^T\sum_{s=1}^T \Ee[\dF_t\dF_s\trans]\Ee[e_{it}e_{is}]$.
\item [(v)] The number of factors $r$ is independent of $N$ and the $r$ eigenvalues of $ \boldsymbol{\Gamma}_F\boldsymbol{\Sigma}_\Lambda$ are distinct, where $\boldsymbol{\Gamma}_F$ and $\boldsymbol{\Sigma}_\Lambda$ are defined in Assumptions \ref{ass:pram_SE}(i) and \ref{ass:factor_identifiability}, respectively.
 \end{inparaenum}
\end{assumption}

As shown in Appendix \ref{appendix_proofs}, the Assumptions \ref{ass:noise}--\ref{ass:idiosyncraticY} nest all assumptions in \citet{Bai2003}, hence we can easily adapt the proofs therein despite using here the definition of PC estimators by \citet{stock2002forecasting}.
The estimated loadings and factors are then consistently retrieving the true ones up to an invertible linear transformation.

\begin{prop}
\label{prop:rate_yt_decomp}
Under Assumptions \ref{ass:noise}--\ref{ass:idiosyncraticY}, for any $i=1,\ldots, N$ and $t=1,\ldots, T$, there exists an $r\times r$ matrix $\widehat \dH$, depending on $N$ and $T$, such that, 
$\dH_0={\mathbb P}\text{-}\lim_{N,T\to\infty} \widehat\dH$ is $r\times r$ finite and positive definite and,
as $N,T\to\infty$: 
\begin{inparaenum}
\item [(a)]
$\Vert \widehat \dF_t- \widehat\dH \dFt \Vert = O_{\mathbb P}\big(N^{-1/2}+T^{-1}\big)$;
\item [(b)]  if $\sqrt N/ T\to 0$, 
$
\sqrt N\big(\widehat \dF_t- \widehat\dH \dFt\big) \Rightarrow \calN\big(\boldsymbol 0_r, \dH_0\dSigma_\Lambda\inv\boldsymbol \Pi_{t,t} \dSigma_\Lambda\inv \dH_0\trans\big),
$
where $\boldsymbol \Pi_{t,t}$ is defined in Assumption \ref{ass:idiosyncratic}(iii);
\item [(c)] $\Vert\widehat{\boldsymbol {\lambda}}_i- \widehat{\dH}^{-1\top} \boldsymbol {\lambda}_{0i} \Vert= O_{\mathbb P}\big(N^{-1}+T^{-1/2}\big)$;
\item [(d)] if $\sqrt T/ N\to 0$,
$
\sqrt T\big(\widehat{\boldsymbol\lambda}_i - \widehat\dH^{-1\top}{\boldsymbol\lambda}_{0i}\big)
\Rightarrow\calN\big(\boldsymbol 0_r, \dH_0^{-1\top} \boldsymbol\Gamma_F\inv \boldsymbol \Xi_i   \boldsymbol\Gamma_F\inv \dH_0^{-1}\big),
$
where $\boldsymbol \Xi_i$ is defined in Assumption \ref{ass:idiosyncraticY}(iv);
\item [(e)]  $\Vert \widehat \dF_t- \dH_0 \dFt \Vert = O_{\mathbb P}\big(N^{-1/2}+T^{-1/2}\big)$
and $\Vert\widehat{\boldsymbol {\lambda}}_i- {\dH}^{-1\top}_0 \boldsymbol {\lambda}_{0i} \Vert=  O_{\mathbb P}\big(N^{-1/2}+T^{-1/2}\big)$.
\end{inparaenum}
\end{prop}

While parts (a)-(d) are standard, part (e) simply reflects the fact that by Remark \ref{rem:birk} and Assumption~\ref{ass:factor_identifiability}, $\Vert \widehat{\dH}-\dH_0\Vert=  O_{\mathbb P}\big(N^{-1/2}+T^{-1/2}\big)$, a result needed for proving the following results.

\vspace{-.2cm}
\begin{remark}
\upshape{
At least three other alternatives to PCs are possible which do not require  finite 4th moments. First, we could use
PCs based on the spatial Kendall’s tau matrix \citep{FanLiuWang2018,he2020elliptic}, which require no moment conditions and have the same consistency rates as in Proposition \ref{prop:rate_yt_decomp}. However, this approach is computationally very intensive and its consistency is proved only in the case of serially independent data following a multivariate elliptical distribution. In Appendix \ref{app:kendall} we show numerical results when using this methodology. Second, we could use diversified projections \citep{FL22}, which are $\sqrt N$-consistent for the factors and require moment conditions only for the idiosyncratic components. However, their efficacy is based on considering more projections than the number of factors and it is not clear how to choose the related weighted schemes. 
Last, we could use robust PCs, which are based on thresholding the data and require the existence of at least 2nd moments \citep{barigozzi2024tail}. For a fixed threshold this approach has the same consistency rates as in Proposition \ref{prop:rate_yt_decomp}. However, a threshold pre-selection step can be highly computationally demanding.
%
}
\end{remark}

\subsection{Asymptotic properties of estimated time-varying mean and variance}
\label{subsec:asy_prop_mean_var}

We introduce the following contraction and moment conditions for the scaled-score process $\cbra{\dut}_{t\in\Zz}$ and its derivatives. Recall that, at any given $t\in \Zz$, $\dut$ is a function of $\boldsymbol F_t$ and therefore also of $\dpsitti$.

\begin{assumption}
\label{ass:contraction_cond_I}
\begin{inparaenum}
 \item [(i)] $\dTheta$ is a compact subset of  $\Rr^d$.
\item [(ii)] The process $\cbra{\dut(\dtheta)}_{t\in\Zz}$ is such that\linebreak
$
\Ee [
\log^+
\sup_{\dtheta \in\dTheta}
\Vert 
\dut(\dtheta)
\Vert ] \leq K_{\du},
$ for some finite positive real $K_{\du}$ independent of $t$, and where  $\log^+(x) = \max\cbra{0, \log(x)}$. 
\item [(iii)] Let
$
X_{t,k}(\dtheta) = 
\sup_{\dpsi \in {\calP}}
\big\Vert
\prod_{j=1}^k
\boldsymbol{X}_{t-j+1}(\dtheta)
\big\Vert$
 with
$
\boldsymbol{X}_{t}(\dtheta) =
\dPhi + \dK \frac{\partial \dut(\dtheta)}{\partial\dpsitti\trans}\big|_{\dpsitti=\dpsitti(\dtheta)}$
then, for some $k\geq 1$,
$
\Ee [
\log
\sup_{\dtheta \in \dTheta} 
\vert
X_{t,k}(\dtheta)
\vert
 ]
 < 0$.
\end{inparaenum}
\end{assumption}

\begin{assumption}
\label{ass:contraction_cond}
\begin{inparaenum}
\item [(i)]  For some $k\geq 1$,
$
\sup_{\dtheta\in\dTheta}
\vbra{
X_{t,k}(\dtheta)}
\leq \calX_{k}$, almost surely, for finite positive real $\calX_{k}<1$ independent of $t$, with $X_{t,k}(\dtheta)$ defined in Assumption \ref{ass:contraction_cond_I}.
\item [(ii)] Let,
$
Y_{t}(\dtheta) =
\sup_{\dpsi \in \boldsymbol{\calP}}
\big\Vert\boldsymbol{Y}_{t}(\dtheta) \big\Vert
$
with
$\boldsymbol{Y}_{t}(\dtheta) =
\dK \frac{\partial \dut(\dtheta)}{\partial\dF_t\trans}\big|_{\dpsitti=\dpsitti(\dtheta)},
$
then,  for some $k\geq 1$,\footnote{The notation for $\boldsymbol{Y}_{t}(\dtheta)$ has to be intended that the derivative is with respect to a generic $\dF_t=\boldsymbol q(\dpsitti,\dv_t)$ and then computed in $\dF_t=\boldsymbol q(\dpsitti(\dtheta),\dv_t)$ for a given $\dtheta\in\dTheta$.}
\[
\max\cbra{
\vvbra{
\sum_{j=1}^{k}
\sup_{\dtheta\in\dTheta}
\vbra{{Y}_{t-j+1}(\dtheta)}
\vvbra{\dbarF_{t-j+1} - \dF_{t-j+1}}
}_p,
\vvbra{
\sum_{j=1}^{k}
\sup_{\dtheta\in\dTheta}
\vbra{{Y}_{t-j+1}(\dtheta)}
}_p
} \leq \calY_k,
\]
for finite positive real $\calY_k < \infty$ independent of $t$, and $p> 2$ with $\dbarFt$ defined in Assumption \ref{ass:pram_SE}.
\end{inparaenum}
\end{assumption}

We then establish a preliminary result concerning the approximation of the unobserved stochastic parameter process. Let $\big\{\dbarpsitti(\dtheta)\big\}_{t\in\Zz}$ denote the unique stationary and ergodic solution of \eqref{eq:SD_Model_psit0}, evaluated at some generic $\dtheta\in\dTheta$ and driven by the transformed stationary and ergodic factors $\big\{\dH_0\dbarFt\big\}_{t\in\Zz}$. Let $\big\{\dtildepsithetatti\big\}_{t\in\Nn}$ denote the filtered sequence driven by $\big\{\dH_0\dbarFt\big\}_{t\in\Zz}$ initialized at the fixed value $\dpsi^\star_{1|0}$. Moreover, let $\big\{\dhatpsithetatti\}_{t=1}^T$ be the filter driven by $\big\{\widehat{\dH}\dFt\big\}_{t=1}^T$ and initialized at $\dpsi^\star_{1|0}$, and let $\big\{\dhathatpsithetatti\big\}_{t=1}^T$ be the feasible filter driven by $\big\{\dhatFt\big\}_{t=1}^T$ and initialized at $\dpsi_{1|0}$.

We consider the decomposition
\begin{align}
\label{eq:psi_decomposition}
&\dhathatpsithetatti-\dbarpsitti(\dtheta)
\nonumber
\\
={}&
\bra{\dhathatpsithetatti-\dhatpsithetatti}
+
\bra{\dhatpsithetatti-\dtildepsithetatti}
+
\bra{\dtildepsithetatti-\dbarpsitti(\dtheta)}.
\end{align}
The first term reflects the factor-estimation error $\dhatFt-\widehat{\dH}\dFt$ together with the possibly different filter initializations. The second term is generated by
$
\widehat{\dH}\dFt-\dH_0\dbarFt
=
\big(\widehat{\dH}-\dH_0\big)\dFt
+
\dH_0\big(\dFt-\dbarFt\big),
$
and therefore captures both the transformation-estimation error and the convergence of the factor process $\big\{\dFt\big\}_{t\in\Nn}$ to its unique stationary solution $\big\{\dbarFt\big\}_{t\in\Zz}$. The third term contains only the effect of the filter initialization, since both processes are driven by the same stationary factor sequence.
The asymptotic behavior of these terms is given in the following proposition.

\begin{prop}
\label{prop:rate_psit_decomp}
Under Assumptions \ref{ass:noise}--\ref{ass:contraction_cond}, let $k\geq1$ and $0<\calX_k<1$ be as in Assumption \ref{ass:contraction_cond}. Then, for some $\varrho_k$ such that $\calX_k<\varrho_k<1$ and any fixed initial conditions $\dpsi_{1|0},\dpsi^\star_{1|0}\in\boldsymbol{\calP}$, for any $t\geq1$, as $N,T\rightarrow\infty$:
\begin{inparaenum}
\item[(a)] 
 $\sup_{\dtheta\in\dTheta}\big\Vert{\dhathatpsithetatpt-\dhatpsithetatpt}\big\Vert
=
O_{\Pp}\big({N^{-1/2}+T^{-1}}\big)
+
O_{a.s.}\big({\calX_k^{\lfloor t/k\rfloor}}\big)$;
\item[(b)] 
$\sup_{\dtheta\in\dTheta}\big\Vert{\dhatpsithetatti-\dtildepsithetatti}\big\Vert
\!=\!
O_{\Pp}\big(N^{-1/2}+T^{-1/2}\big)
+
O_{a.s.}\big({\varrho_k^{\lfloor t/k\rfloor}}\big)\!\!$;
\item[(c)] 
 $\sup_{\dtheta\in\dTheta}\big\Vert{\dtildepsithetatti-\dbarpsitti(\dtheta)}\big\Vert$
$\!=\!
O_{a.s.}\big({\varrho_k^{\lfloor t/k\rfloor}}\big)$;
\item[(d)] 
 the process
$\big\{\dbarpsitti(\dtheta_0)\big\}_{t\in\Nn}$ is near-epoch dependent of
size $-1$ on the $\beta$-mixing process $\cbra{\dvt}_{t\in\Zz}$;
\item[(e)]  
 $
\sup_{\dtheta\in\dTheta}
\big\Vert{
\dhathatpsithetatpt
-
\dbarpsitpt(\dtheta)
}\big\Vert
=O_{\Pp}\big(N^{-1/2}+T^{-1/2}\big)
+
O_{a.s.}\big({
\calX_k^{\lfloor t/k\rfloor}
+
\varrho_k^{\lfloor t/k\rfloor}
}\big).
$
\end{inparaenum}
\end{prop}

In the proof, the first two terms are controlled by combining the factor convergence rates in Proposition \ref{prop:rate_yt_decomp} with the contraction and moment conditions in Assumptions \ref{ass:contraction_cond_I} and \ref{ass:contraction_cond}. The last term is controlled through the invertibility of the filter by an application of Theorem 3.1 of \citet{Bougerol1993}. 

\vspace{-.2cm}
\begin{remark}
\label{remark:rate_psi_decomp}
\upshape{
An important implication of Proposition \ref{prop:rate_psit_decomp} is that the feasible score-driven filter inherits the full sample dependence generated by the first-step PC estimation. Indeed, although its dependence on $T$ is suppressed in the notation, each estimated factor $\dhatFt$ depends on $\widehat\dLambda$, which is computed using the entire sample $\cbra{\dxt}_{t=1}^{T}$. Consequently, each element of $\big\{\dhathatpsithetatti\big\}_{t=1}^{T}$ depends, through the recursive filter and the PC plug-in estimator, on the whole sample. The feasible filter must therefore be treated as a triangular array rather than as an ordinary stationary time series process. The rates in Proposition \ref{prop:rate_psit_decomp} reflect this structure. The terms depending on $N$ and $T$ account for the estimation of the factors and of their transformation, whereas the geometrically decreasing terms account for the different filter initializations. Filter contraction controls the propagation of these errors through the recursion, but it does not remove the common full sample dependence induced by the PC estimator.
}
\end{remark}

\subsection{Set-consistency of estimated score-driven parameters}
\label{subsec:consistency_QMLE}
The main difference between classical score-driven models treated in the literature
and our model is that, in the former, the process entering the score recursion is observed without error,
whereas in the latter the factor process $\cbra{\dFt}_{t\in\Zz}$ is latent. Consequently, while in classical score-driven models the parameter vector $\dtheta\in\dTheta$ governing the score update can be point identified, 
in the present setting, since the latent factor process 
is replaced by the estimated factors
$\big\{\dhatFt\big\}_{t=1}^{T}$, the factors are identified only up to an $r\times r$ invertible linear transformation, 
therefore, the population criterion therefore identifies an orbit rather
than a single parameter vector. This means that the parameter vector is set identified, with the required separation outside the identified
set imposed below through the regularity of the relevant level sets.

 The relevant consistency concept is therefore set-consistency. Formally, consider the general linear group of degree $r$,
$
GL\bra{r}
=
\cbra{
\dH\in\Rr^{r\times r}:
\vbra{\dH}\neq0
},
$
and, for every admissible $\dH\in GL\bra{r}$, let
$\boldsymbol r_{\dH}$ denote the invertible reparameterization such
that, whenever $\dtheta$ governs the dynamics of $\dFt$,
$\boldsymbol r_{\dH}\bra{\dtheta}$ governs the dynamics of
$\dH\dFt$. We refer to $\dH$ as the alignment matrix and to
$\boldsymbol r_{\dH}$ as the alignment mapping. For every fixed
$\dH$, the mapping and its inverse are linear in $\dtheta$, although
generally nonlinear in $\dH$ (see Appendix
\ref{appendix_parameter_identifications} for their explicit
expressions in the time-varying location and variance specifications).
We say that an estimator $\dthetamle$  is set-consistent for $\dtheta_0$ if
$
\inf_{\dvartheta\in\dTheta_{\dH}}
\big\Vert{
\dthetamle-\dvartheta
}\big\Vert
=
o_{\Pp}(1), 
$
where
$\dTheta_{\dH}
=
\cbra{
\boldsymbol r_{\dH}(\dtheta_0)\in\dTheta:
\dH\in GL\bra{r}
}.
$

To establish consistency, we require standard smoothness and domination conditions on the infeasible conditional quasi log-likelihood
$
\ell_{\dF}\bra{\dFt;\dtheta}
$
and its derivatives.

\begin{assumption}
\label{ass:smooth_domination_likelihood}
\begin{inparaenum}
\item [(i)] The process $\cbra{\ell_{\dF}\bra{\dFt ; \dtheta}}_{t\in\Nn}$ is almost surely Lipschitz-$L_p$ on $\dTheta$. 
\item [(ii)] For $\ell_{\dF}\bra{\dFt ; \dtheta}$ it holds that $\sup_{\dtheta\in\dTheta}\vbra{ \ell_{\dF}\bra{\dFt ; \dtheta} } \leq D_{1,t}$ where $\sup_{t\geq 1}\Ee[D_{1,t}^p]<\infty$ for $p >  2$.
\end{inparaenum}
\end{assumption}


\begin{assumption}
\label{ass:moment_score}
\begin{inparaenum}
\item[(i)] For any $t\ge 1$, $\ell_{\dF}\bra{\dFt ; \dtheta}$ is continuously differentiable with respect to $\dpsitti\in \calP$ and $\dFt\in\Rr^r$. 
%
%
\item[(ii)] Let 
$\nabla^{\dpsi} \ell_{\dF}(\dFt; \dtheta)
=
\frac{\partial \ell_{\dF}(\dFt; \dtheta)}{\partial \dpsitti}\big\vert _{\dpsitti=\dpsitti(\dtheta)}$ and  
$\nabla^{\dF} \ell_{\dF}(\dFt; \dtheta)
=
\frac{\partial \ell_{\dF}(\dFt; \dtheta)}{\partial \dF_t}\big\vert _{\dpsitti=\dpsitti(\dtheta)}
$,
then,
$\sup_{\dtheta\in\dTheta}\vbra{ \nabla^{\dpsi} \ell_{\dF}\bra{\dFt ; \dtheta} } \!\leq D_{2,\dpsitti}\!\!$ and $\sup_{\dtheta\in\dTheta}\vbra{ \nabla^{\dF} \ell_{\dF}\bra{\dFt ; \dtheta} } \!\leq D_{2,\dFt}$ where $\sup_{t\geq 1}\Ee[D_{2,\dpsitti}^q]\!<~\infty$ and $\sup_{t\geq 1}\Ee[D_{2,\dFt}^q]\!<\infty$ for $q= 2p/(p-2)$.~\item[(iii)] The processes $\cbra{\nabla^{\dpsi}\ell_{\dF}\bra{\dFt;\dtheta}}_{t\in\Nn}$ and $\cbra{\nabla^{\dF}\ell_{\dF}\bra{\dFt;\dtheta}}_{t\in\Nn}$ are almost surely locally Lipschitz with respect to $\dpsitti\in \calP$ and $\dFt\in\Rr^r$, uniformly over $\dtheta\in\dTheta$, and their Lipschitz coefficients have uniformly bounded moments of order $q=2p/(p-2)$.
\end{inparaenum}
\end{assumption}

Hereafter, we refer to
$
T^{-1}\sum_{t=1}^T
\doublehat{\ell}_{\dF}
\big(
\dhatF_t;
\boldsymbol r_{\widehat\dH}(\dtheta)
\big)
+
\log\big\vert \widehat\dH\big\vert 
$
as the aligned feasible criterion. Assumptions \ref{ass:smooth_domination_likelihood} and \ref{ass:moment_score} allow us to compare, uniformly over $\dTheta$, this criterion with the stationary infeasible criterion
$
T^{-1}\sum_{t=1}^T\ell_{\dF}\bra{\dbarF_t;\dtheta}
$
and its population counterpart. These assumptions rely on the almost sure Lipschitz continuity and domination concepts of \citet[][Definitions 3.5 and 3.16]{Gallant1988}.

The triangular array structure discussed in Remark \ref{remark:rate_psi_decomp} implies that the feasible quasi log-likelihood cannot be treated directly as the sample average of a stationary time-series process. In particular, the pointwise filter rates in Proposition \ref{prop:rate_psit_decomp} are not, by themselves, sufficient to obtain the rate of convergence of the time averaged criterion. The dependence generated by the PC plug-in estimator must instead be retained throughout the likelihood expansion.

Accordingly, we decompose the difference between the aligned feasible criterion and its population counterpart by separating the effect of factor estimation and its propagation through the score-driven filter from the effects of filter initialization, convergence to stationarity, and sampling variation. The factor estimation component is studied directly at the level of the time averaged likelihood, thereby preserving the averaging structure of the PC estimation error. The remaining components are controlled using filter contraction and the near-epoch dependence of the stationary factor and filter processes (see also Remark \ref{rem:detailProp3}). This delivers the uniform convergence rate stated in the following proposition.

\begin{prop}
\label{lemma:ULLN}
Under Assumptions \ref{ass:noise}--\ref{ass:moment_score}, with $p>2$, as $N,T\rightarrow\infty$,
\[
\sup_{\dtheta\in\dTheta}
\vbra{
\frac{1}{T}\sum_{t=1}^T
\doublehat{\ell}_{\dF}
\bra{
\dhatF_t;
\boldsymbol r_{\widehat\dH}(\dtheta)
}
+
\log\vbra{\widehat\dH}
-
\Ee\sbra{
\ell_{\dF}\bra{\dbarF_t;\dtheta}
}
}
=O_{\Pp}\bra{\frac 1N+\frac 1{\sqrt T}},
\]
where $\dbarF$ is defined in Assumption \ref{ass:pram_SE}(i).
\end{prop}
\begin{remark}
\upshape{
The term $\log\big\vert{\widehat\dH}\big\vert$ that appears in the statement of Proposition \ref{lemma:ULLN} is independent of $\dtheta$ and therefore has no effect on the maximizer of the feasible criterion. Its inclusion serves only to account for the change of variables induced by the mapping $\boldsymbol r_{\widehat\dH}(\dtheta)$.  
}
\end{remark}
\vspace{-.5cm}
\begin{remark}\label{rem:detailProp3}
\upshape{
Although the pointwise PC estimation error is of order
$
O_{\Pp}\big(
N^{-1/2}+T^{-1}
\big),
$
the feasible criterion in Proposition \ref{lemma:ULLN} does not inherit the pointwise cross-sectional rate $N^{-1/2}$. As discussed in Remark \ref{remark:rate_psi_decomp}, the filter
$\big\{\dhathatpsithetatti\big\}_{t=1}^{T}$ has a triangular array structure because each estimated factor depends on the entire sample through the PC estimator. The proof of Proposition \ref{lemma:ULLN} therefore studies the factor estimation error directly within the time averaged likelihood, without bounding each likelihood contribution separately. This retains the cancellations in the weighted time average of the PC estimation errors and gives
$
O_{\Pp}\big((NT)^{-1/2}\big)
$
for the first order component. The remaining factor estimation error enters the likelihood quadratically and is therefore of order
$
O_{\Pp}\big(
N^{-1}+T^{-1}
\big).
$
Combining these terms with the $T^{-1/2}$ sampling variation of the stationary criterion gives
$
O_{\Pp}\big(
T^{-1/2}
+
(NT)^{-1/2}
+
N^{-1}
+
T^{-1}
\big)
=
O_{\Pp}\big(
N^{-1}+T^{-1/2}
\big).
$
Hence, the term $N^{-1}$ in Proposition \ref{lemma:ULLN} arises from the time averaging of the first order factor estimation error and from the quadratic order of the corresponding remainder. It does not represent a faster \linebreak pointwise convergence rate for either the estimated factors or the feasible filter.
}
\end{remark}

\vspace{-.2cm}
We then assume a regularity condition of the level-sets (\citealp[Definition 4.1]{Potscher1997}).

\begin{assumption}
\label{ass:regular_level_set}
\begin{inparaenum}
\item[(i)] The compact parameter space $\dTheta\subset\Rr^d$ satisfies $\vbra{\dK}\neq0$.
\item[(ii)] Let
$
K_T
=
\sup_{\dtheta\in\dTheta}
\Ee\big[
\ell_{\dF}\bra{\dbarFt;\dtheta}
\big]$, then $K_T<\infty$ for every $T\in\Nn$.
\item[(iii)] 
let $\boldsymbol \vartheta\in\dTheta_{\dH}$ so that, by definition, there exists a matrix $\dH\in GL(r)$ such that $\boldsymbol {\vartheta}=\boldsymbol {r}_{\dH}(\dtheta_0)$, then,
$\Ee\big[
\ell_{\dF}\bra{\dbarFt;\boldsymbol \vartheta}
\big]
\geq K_T$;
%
\item[(iv)] The sequence of level sets is regular, that is, for every
$\varepsilon>0$,
\begin{align*}
\liminf_{T\rightarrow\infty}
\bra{
\inf_{\dtheta\in\dTheta_{\dH}}
\Ee\sbra{
\ell_{\dF}\bra{\dbarFt;\dtheta}
}
-
\sup_{{\dtheta\in\dTheta:
\inf_{\dvartheta\in\dTheta_{\dH}}
\vvbra{
\dtheta-\dvartheta
}
\geq\varepsilon}}
\Ee\sbra{
\ell_{\dF}\bra{\dbarFt;\dtheta}
}
}
>
0.
\end{align*}
\end{inparaenum}
\end{assumption}

Given Proposition \ref{lemma:ULLN} and the separation condition in Assumption \ref{ass:regular_level_set}, we can establish the set
consistency of the QMLE $\dthetamle$. 



\begin{theorem}
\label{thm:consistency}
Under Assumptions \ref{ass:noise}--\ref{ass:regular_level_set}, with $p>2$, for every fixed initialization $\dpsi_{1|0}\in\boldsymbol{\mathcal P}$, as $N,T\rightarrow\infty$,
$
\inf_{\dvartheta\in\dTheta_{\dH}}
\big\Vert
\dthetamle-\dvartheta
\big\Vert
=
o_{\Pp}(1).
$
Moreover, for $\dH=\dH_0$, with $\dH_0$ defined in Proposition \ref{prop:rate_yt_decomp}, as $N,T\rightarrow\infty$,
$
\big\Vert\dthetamle
-
\boldsymbol r_{\dH_0}
\bra{
\dtheta_0
}\big\Vert
=
o_{\Pp}(1)$,
and $
\big\Vert\boldsymbol r_{\widehat\dH}^{-1}
\big(
\dthetamle
\big)
-
\dtheta_0\big\Vert
=
o_{\Pp}(1).$ 
\end{theorem}

Theorem \ref{thm:consistency} establishes both the set-consistency of $\dthetamle$ with respect to the orbit $\dTheta_{\dH}$ and, after fixing the limiting factor orientation $\dH_0$, its point consistency for the corresponding representative $\boldsymbol r_{\dH_0}\bra{\dtheta_0}$. The last statement, which holds since $\widehat\dH\pconv\dH_0$ as $N,T\rightarrow\infty$, is useful for proving asymptotic normality.

\vspace{-.2cm}
\begin{remark}
\upshape{
The assumptions of Theorem \ref{thm:consistency} are sufficient for consistency, but not for deriving an exact convergence rate, which requires the local score and Hessian conditions introduced below. Nevertheless, Proposition \ref{lemma:ULLN} suggests the conjectured rate
$
\big\Vert
\dthetamle
-
\boldsymbol r_{\widehat\dH}
\bra{
\dtheta_0
}
\big\Vert
=
O_{\Pp}\big(N^{-1}+ T^{-1/2}\big).
$
Theorem \ref{thm:asy_norm} confirms this rate when the required additional assumptions are introduced.
}
\end{remark}

\subsection{Asymptotic normality of score-driven parameters}


Because $\big\Vert \widehat\dH-\dH_0\big\Vert=O_{}\big(N^{-1/2}+T^{-1/2}\big)$, we cannot use the first statement of Theorem \ref{thm:consistency} to prove asymptotic normality. To this end, it is more useful the last statement of Theorem \ref{thm:consistency}  which provides point-consistency of the aligned estimator which depends explicitly on $\widehat\dH$. 
%
In particular, for obtaining asymptotic normality of $\boldsymbol r_{\widehat\dH}^{-1}\big(\dthetamle\big)$, both the score vector $\nabla^{\dtheta}\ell_{\dF}\bra{\dFt;\dtheta}$ and the Hessian matrix $\nabla^{\dtheta\dtheta}\ell_{\dF}\bra{\dFt;\dtheta}$ need to satisfy appropriate smoothness and domination conditions. These conditions are required to establish the uniform convergence of the score and Hessian processes to their stationary and ergodic counterparts over $\dTheta$ (see Lemmas \ref{lemma:rate_psit_first_derivs} and \ref{lemma:rate_psit_second_derivs} for their explicit expressions). They rely on the same concepts employed in Assumptions \ref{ass:smooth_domination_likelihood} and \ref{ass:moment_score}. The required conditions are stated below \citep[][Lemma~8.1]{Potscher1997}.

\begin{assumption}
\label{ass:smooth_domination_score_Hessian}
\begin{inparaenum}
\item[(i)] For any $t\geq1$,  $\ell_{\dF}\bra{\dFt;\dtheta}$ is twice continuously differentiable with respect to $\dtheta$, and the processes
$
\cbra{
\nabla^{\dtheta}
\ell_{\dF}\bra{\dFt;\dtheta}
}_{t\in\Nn}
$ and $
\cbra{
\nabla^{\dtheta\dtheta}
\ell_{\dF}\bra{\dFt;\dtheta}
}_{t\in\Nn}
$
are almost surely Lipschitz-$L_p$ on $\dTheta$, for $p>2$.
 \item[(ii)] For $\nabla^{\dtheta}\ell_{\dF}\bra{\dFt;\dtheta}$ and $\nabla^{\dtheta\dtheta} \ell_{\dF}\bra{\dFt;\dtheta}$ it holds that $\sup_{\dtheta\in\dTheta}\vvbra{\nabla^{\dtheta}\ell_{\dF}\bra{\dFt;\dtheta}}\le D_{2,\dtheta,t}$ and \linebreak $\sup_{\dtheta\in\dTheta} \vvbra{\nabla^{\dtheta\dtheta}\ell_{\dF}\bra{\dFt;\dtheta}}\leq D_{3,t}$ where $\sup_{t\geq1}\Ee\sbra{D_{2,\dtheta,t}^{p}}<\infty$ and $\sup_{t\geq1}\Ee[D_{3,t}^{q}]<\infty$ for \linebreak $q=2p/(p-2)$. 
\item [(iii)] The processes $\cbra{\nabla^{\dtheta} \ell_{\dF}(\dFt; \dtheta)}_{t\in\Nn}$ and $\cbra{\nabla^{\dtheta\dtheta} \ell_{\dF}(\dFt; \dtheta)}_{t\in\Nn}$ are almost surely locally Lipschitz with respect to $\dpsitti\in \calP$ and $\dFt\in\Rr^r$, uniformly over $\dtheta\in\dTheta$, and their Lipschitz coefficients have uniformly bounded moments of order $q=2p/(p-2)$.
\item [(iv)] For any $p > 2$,
$
\vvbra{
\sup_{\dtheta\in\dTheta}
\sup_{\dpsitti \in \calP}
\Big\Vert{\frac{\partial\vect{\boldsymbol{X}_{t}(\dtheta) }}{\partial\dpsitti\trans} }\Big\Vert }_p \leq K_X^\prime,
$ and
$
\vvbra{
\sup_{\dtheta \in\dTheta}
\Big\Vert{
\frac{\partial\vect{\boldsymbol{X}_{t}\bra{\dtheta}} }{\partial\dF_t\trans}
}\Big\Vert }_p\leq~K_X^{\prime\prime},
$
for some finite positive reals $K_X^\prime$ and $K_X^{\prime\prime}$ independent of $t$, 
with $\boldsymbol{X}_{t}(\dtheta)$ defined in Assumption \ref{ass:contraction_cond_I}.
\item [(v)] Let $\dDt(\dtheta) = \frac{\partial \dpsitpt(\dtheta) }{\partial \dtheta\trans}$,
then, for any $p> 2$,
$ \vvbra{ \sup_{\dtheta\in\dTheta}\vvbra{\dDt(\dtheta)} }_p\leq K_D^{\prime}$,
$
\vvbra{ \sup_{\dtheta\in\dTheta}\sup_{\dpsitti\in\calP}\Big\Vert{\frac{\partial \vect{\dDt(\dtheta)}}{\partial \dpsitti\trans} }\Big\Vert }_p\leq K_D^{\prime\prime}
$, and
$\vvbra{
\sup_{\dtheta\in\dTheta}
\Big\Vert{\frac{\partial \vect{\dDt(\dtheta)}}{\partial \dF_t\trans} }\Big\Vert }_p\leq K_D^{\prime\prime\prime}
,
$
for some finite positive reals $K_D^\prime$, $K_D^{\prime\prime}$, and $K_D^{\prime\prime\prime}$ independent of $t$.
\item [(vi)] For any $p> 2$,
$
\vvbra{
\sup_{\dtheta\in\dTheta}\Big\Vert{\frac{\partial\vect}{\partial\dpsi\trans}\bra{\frac{\partial \vect {\boldsymbol{X}_{t}(\dtheta)}}{\partial \dtheta\trans}  }  }\Big\Vert}_p \leq C_X^\prime
$
and
$
\vvbra{
\sup_{\dtheta \in\dTheta}
\Big\Vert{
\frac{\partial\vect}{\partial\dF\trans}\bra{\frac{\partial \vect{\boldsymbol{X}_{t}(\dtheta)}}{\partial \dtheta\trans}  } 
}\Big\Vert}_p
\leq C_X^{\prime\prime}
,
$
for some finite positive reals $C_X^\prime$ and $C_X^{\prime\prime}$ independent of $t$.
\item [(vii)] For any $p> 2$,
$
\vvbra{ \sup_{\dtheta\in\dTheta}\Big\Vert{\frac{\partial \vect{\dDt(\dtheta)}}{\partial \dtheta\trans} }\Big\Vert}_p\leq C_D^{\prime}
$,\linebreak
$
\vvbra{
\sup_{\dtheta\in\dTheta}\Big\Vert{\frac{\partial \vect}{\partial \dpsi\trans}\bra{\frac{\partial \vect {\dDt(\dtheta)}}{\partial \dtheta\trans}} }\Big\Vert}_p\leq C_D^{\prime\prime}
$, and
$
\vvbra{
 \sup_{\dtheta\in\dTheta}
\Big\Vert{\frac{\partial \vect}{\partial \dF\trans}\bra{\frac{\partial \vect{\dDt(\dtheta)}}{\partial \dtheta\trans} } }\Big\Vert }_p\leq C_D^{\prime\prime\prime}
$,
for some finite positive reals  $C_D^\prime$, $C_D^{\prime\prime}$, and $C_D^{\prime\prime\prime}$  independent of $t$.
\end{inparaenum}
\end{assumption}

The next assumptions are standard in the QML estimation literature. 
\begin{assumption}
\label{ass:uniformly_pos_def}
The $d\times d$ matrix
$\limitcov\bra{\dtheta_0}
=
-\Ee\big[{
\nabla^{\dtheta\dtheta}
\ell_{\dF}\bra{\dbarF_t;\dtheta_0}
}\big]$
is positive definite.
\end{assumption}



\begin{assumption}
\label{ass:truePar}
$\dtheta_0$ lies in the interior of the (compact) parameter space $\dTheta$. 
\end{assumption}

Asymptotic normality then holds but the QMLE is not centered around the true value but around a nonlinear transformation thereof, which depends on $\widehat{\dH}$ accounting for the fact that the estimated factors do not identify the true ones. We therefore first derive the asymptotic distribution of the aligned estimator $\boldsymbol r_{\widehat\dH}^{-1}\big(\dthetamle\big)$ and subsequently map the result back to $\dthetamle$.

\begin{theorem}
\label{thm:asy_norm}
Under Assumptions \ref{ass:noise}--\ref{ass:truePar}, with
$p>2$, for every fixed initial condition $\dpsi_{1|0}\in\calP$,  and if ${\sqrt T}/{N} \rightarrow0$, as $N,T\rightarrow\infty$,
$
\sqrt{T}
\bra{
\dthetamle
-
\boldsymbol r_{\widehat{\dH}}
\bra{
\dtheta_0
}
}
\Rightarrow
\calN
\bra{
\boldsymbol 0_d,
\limitcov_{\dH_0}\inv(\dtheta_0)
\dV_{\dH_0}(\dtheta_0)
\limitcov_{\dH_0}\inv(\dtheta_0)
},
$
with
$
\limitcov_{\dH_0}(\dtheta_0)
=
{\boldsymbol R}_{\dH_0}\trans
\limitcov(\dtheta_0)
{\boldsymbol R}_{\dH_0},
$
$
\dV_{\dH_0}(\dtheta_0)
=
{\boldsymbol R}_{\dH_0}\trans
\dV(\dtheta_0)
{\boldsymbol R}_{\dH_0},$
and
$
{\boldsymbol R}_{\dH_0}
=
\frac{\partial \boldsymbol r_{\dH_0}^{-1}(\dtheta)}{\partial\dtheta\trans}\mid_{\dtheta=\dtheta_0}
,
$
and where
$
\limitcov(\dtheta_0)
=
\lim_{T\rightarrow\infty}
-
T^{-1}
\Ee\big[{
\nabla^{\dtheta\dtheta}
\ell_{\dF}
\bra{
\dbarF;
\dtheta_0
}
}\big],
$
$
\dV(\dtheta_0)
=
\lim_{T\rightarrow\infty}
T^{-1}
\Ee\big[{
\nabla^{\dtheta}
\ell_{\dF}
\bra{
\dbarF;
\dtheta_0
}
\nabla^{\dtheta}
\ell_{\dF}
\bra{
\dbarF;
\dtheta_0
}^{\trans}
}\big].
$
\end{theorem}

This result implies the consistency rate $O_\Pp(\max(N^{-1},T^{-1/2}))$ for the QMLE suggested by Proposition \ref{lemma:ULLN}. The constraint $\sqrt T/N\to 0$ is standard in factor models literature \citep{Bai2003} and it allows use to neglect the error made when estimating the factors, thus treating them as observed, hence, the classical $\sqrt T$ consistency rate for $\dthetamle$. Nevertheless, the above result still depend on $\widehat\dH$, hence, in general, it cannot be directly used for making inference.

\subsection{Asymptotic prediction intervals}\label{sec:API}


We derive the approximation error between the feasible filtered predictor of the conditional location $\hatCittithetamle  = \widehat{\boldsymbol\lambda}_i\trans \dhathatmuthetamletti$ and its unobserved infeasible counterpart $\Cittithetaz = \boldsymbol\lambda_{0i}\trans \dmutti(\dtheta_0)$. 
Similarly, 
we derive the approximation error between the feasible filtered predictor of the conditional variance $\hatPittithetamle  = \widehat{\boldsymbol\lambda}_{i}\trans \dhathatSigmathetamletti\widehat{\boldsymbol\lambda}_{i}$ and its unobserved infeasible counterpart $\Pittithetaz = {\boldsymbol\lambda}_{0i}\trans \dSigmatti(\dtheta_0){\boldsymbol\lambda}_{0i}$. We start with a consistency result.

\begin{theorem}
\label{thm:filter_approx_error_cons}
Under Assumptions \ref{ass:noise}--\ref{ass:truePar}, with $p>2$, for every fixed initial condition $\dpsi_{1|0}\in\calP$, for any $i=1,\ldots,N$ and
$t=1,\ldots,T$, as $N,T\rightarrow\infty$: 
\begin{inparaenum}
\item [(a)]
$
\big\vert{
\hatCittithetamle
-
\Cittithetaz
}\big\vert
=O_{\Pp}\big(N^{-1/2}+T^{-1/2}\big)
+
O_{a.s.}\big({
\calX_k^{\lfloor t/k\rfloor}
}\big);
$
\item [(b)]
$
\big\vert{
\hatPittithetamle
-
\Pittithetaz
}\big\vert
=O_{\Pp}\big(N^{-1/2}+T^{-1/2}\big)
+
O_{a.s.}\big({
\calX_k^{\lfloor t/k\rfloor}
}\big).
$
\end{inparaenum}
\end{theorem}


The rate coincides with the PCA rate of the estimated common component $\widehat{\boldsymbol\lambda}_i\trans\widehat{\dF}_t$ which follows directly from Proposition \ref{prop:rate_yt_decomp}  \citep[Theorem 3]{Bai2003}, but for an additional term which is due to initial conditions.
To control for this term we  introduce the following assumption.

\begin{assumption}
\label{ass:initial_conditions}
The initial condition $\dmu_{1|0}$ and $\dF_1$ are fixed and known, with $\dmu_{1|0} = \dF_1 = \boldsymbol 0_r$.
The initial condition $\dsigma_{1|0}$ and $\dF_1$ are fixed and known, with $\dsigma_{1|0} = \dW_r\trans\bra{\dF_1\otimes\dF_1} =  \bra{\boldsymbol I_r - \dPhi_0}\inv \ddelta_0$.
\end{assumption}

Clearly, Assumption \ref{ass:initial_conditions} is stronger than  the asymptotic negligibility of initial conditions established in Proposition \ref{prop:rate_psit_decomp}. Similar assumptions have been considered in the early univariate score-driven literature, such as \cite{Harvey2013} and \cite{HarveyLuati2014}, as well as by \cite{Patton2019} and \cite{Catania2023} for their non-Lipschitz score-driven filters. More recently, \cite{Hafner2023} and \cite{DInnocenzo2024} have also adopted similar assumptions for their multivariate score-driven specifications.

In this context, Proposition \ref{prop:rate_psit_decomp} can be further refined. Specifically, the estimation error of the factors determines the rate of convergence for the time-varying mean and variance.

\begin{prop}
\label{prop:filter_approx_error_cons_II}
Under Assumptions \ref{ass:noise}--\ref{ass:initial_conditions}, for
any $t=1,\ldots,T$, as $N,T\rightarrow\infty$:\\
\begin{inparaenum}
\item [(a)]
$
\sup_{\dtheta\in\dTheta}
\big\Vert{
\widehat\dH^{-1}
\dhathatmutti
\bra{
\boldsymbol r_{\widehat\dH}(\dtheta)
}
-
\dmutti(\dtheta)
}\big\Vert
=O_{\Pp}\big( N^{-1/2}+T^{-1}\big)
$;\\
\item [(b)]
$
\sup_{\dtheta\in\dTheta}
\big\Vert{
\big({
\widehat\dH^{-1}
\otimes
\widehat\dH^{-1}
}\big)
\dW_r
\dhathatsigmatti
\bra{
\boldsymbol r_{\widehat\dH}(\dtheta)
}
-
\dW_r
\dsigmatti(\dtheta)
}\big\Vert
=O_{\Pp}\big( N^{-1/2}+T^{-1}\big).
$
\end{inparaenum}
\end{prop}

\vspace{-.2cm}
\begin{remark}
\upshape{
The different rates in Propositions \ref{prop:rate_psit_decomp} and
\ref{prop:filter_approx_error_cons_II} are due to their different comparison targets. Proposition \ref{prop:rate_psit_decomp} compares the feasible filter with its stationary counterpart and therefore includes the $T^{-1/2}$ error arising from $\widehat\dH-\dH_0$, as well as the initialization effect. Proposition \ref{prop:filter_approx_error_cons_II}, instead, compares filters expressed under the same estimated transformation $\widehat\dH$, while Assumption \ref{ass:initial_conditions} removes the initialization effect. The remaining perturbation is therefore generated only by
$
\dhatFt-\widehat\dH\dFt,
$
whose rate is
$
O_{\Pp}\big(
N^{-1/2}+T^{-1}
\big).
$
Filter contraction preserves this rate, explaining the absence of a $T^{-1/2}$ term in Proposition \ref{prop:filter_approx_error_cons_II}.
}
\end{remark}

We then prove asymptotic normality of the common-component predictors.

\begin{theorem}
\label{thm:filter_approx_error_asy_norm}
Under Assumptions \ref{ass:noise}--\ref{ass:initial_conditions}, with $p>2$, for any $i=1,\dots,N$ and $t=1,\ldots, T$,  and if $T/ N\to 0$, as $N,T\rightarrow \infty$,
\begin{align}
&\sqrt{T}\frac{\hatCittithetamle -\Cittithetaz}{\sqrt{\boldsymbol a_{i,t|t-1}\trans(\dtheta_0) \dOmega(\dtheta_0) \boldsymbol a_{i,t|t-1}(\dtheta_0)}}\Big\vert\calF_{t-1}
\Rightarrow
\calN\bra{0, 1},\;\;\sqrt{T}\frac{
\hatPittithetamle
-
\Pittithetaz
}
{\sqrt{\boldsymbol b_{i,t|t-1}\trans(\dtheta_0) \dOmega(\dtheta_0) \boldsymbol b_{i,t|t-1}(\dtheta_0)}}
\Big\vert\calF_{t-1}
\Rightarrow
\calN\bra{0, 1},\nonumber
\end{align}
where
$
\boldsymbol a_{i,t|t-1}\trans(\dtheta_0)
=
\sbra {\dmutti(\dtheta_0)\trans \;\big\vert\; \boldsymbol \lambda_{0i}\trans \nabla^{\dtheta}\dmutti(\dtheta_0)
}$ and\\
$\boldsymbol b_{i,t|t-1}\trans(\dtheta_0) =
\sbra{
\dsigmatti(\dtheta_0)\trans
 \dW_r\trans
\bra{\boldsymbol I_r \otimes \boldsymbol\lambda_{0i} + \boldsymbol\lambda_{0i} \otimes \boldsymbol I_r}
\;\big\vert\; 
\bra{
{\boldsymbol\lambda}_{0i}
\otimes
 {\boldsymbol\lambda}_{0i}
 }
 \trans\dW_r
 \nabla^{\dtheta}\dsigmatti(\dtheta_0)
 }$
are $(r+d)$-dimensional row vectors 
and
\begin{align*}
\dOmega(\dtheta_0)
=
\begin{bmatrix}
\boldsymbol{\Gamma}_F^{-1} \boldsymbol \Xi_i \boldsymbol{\Gamma}_F^{-1} & \boldsymbol 0_{r\times d}\\
 	\boldsymbol 0_{d\times r}				&		\limitcov\inv(\dtheta_0)\dV(\dtheta_0)\limitcov\inv(\dtheta_0)
\end{bmatrix}
,
\end{align*}
is an $(r+d)\times (r+d)$ matrix with 
$\boldsymbol{\Gamma}_F$ defined in Assumption \ref{ass:pram_SE},
$\boldsymbol \Xi_i$ defined in Assumption \ref{ass:idiosyncraticY},
$\limitcov(\dtheta_0)$ and $\dV(\dtheta_0)$ defined in Theorem \ref{thm:asy_norm}.
\end{theorem}
 
Although the matrix $\widehat\dH$ appears in Proposition \ref{prop:filter_approx_error_cons_II}, it does not play any role in the last result. Indeed, the estimated loading vectors transform contravariantly to the filtered factor moments, so that the $\widehat\dH$ cancels from the common component predictors.
Consequently, the limiting distributions and their feasible estimators can be expressed entirely in terms of common-component quantities which do not depend on $\widehat\dH$.

\vspace{-.2cm}
\begin{remark}
\upshape{
The condition $T/N\rightarrow0$ in Theorem \ref{thm:filter_approx_error_asy_norm} ensures that the first-stage factor-estimation error and its propagation through the filter are negligible under the root-$T$ normalization. Indeed, Proposition \ref{prop:filter_approx_error_cons_II} gives this component the rate $ O_{\Pp}(N^{-1/2}+T^{-1})$. 
The restriction $T/N\rightarrow0$, therefore, allows the limiting distribution to be determined only by the loading estimation and QMLE score components.
For the Gaussian specifications considered below, the affine structure of the score updates makes it possible to retain the first-stage factor estimation component explicitly.  For the Student's $t$ specifications, instead, the score updates contain nonlinear random weights depending jointly on the estimated factors and filtered parameters. The corresponding first-stage component does not admit the same explicit linear representation. }
\end{remark}

\vspace{-.2cm}
For the Gaussian score updates, we can improve on Theorem  \ref{thm:filter_approx_error_asy_norm} and remove the restriction on the relative rates of $N$ and $T$. Indeed, the recursively propagated effect of the score updates can be written as a linear function of the PC estimation errors, so that its cross-sectional Gaussian limit can be combined with the time series limits of the loading estimator and the QMLE, yielding a variance containing both $N^{-1}$ and $T^{-1}$ components and resulting in the following Central Limit Theorem.

\begin{theorem}
\label{thm:CLT_Gaussian_models}
For the conditional location and variance models, suppose that the filters are based on the Gaussian quasi log-likelihoods in
\eqref{eq:log_likelihood_gaussian_locations} and \eqref{eq:log_likelihood_gaussian_covariances}, respectively. Then
under Assumptions \ref{ass:noise}--\ref{ass:initial_conditions}, with $p>2$, for any $i=1,\ldots,N$ and $t=1,\ldots, T$, as $N,T\rightarrow\infty$, 
\begin{align}
&
\frac{
\hatCittithetamle-\Cittithetaz
}{
\sqrt{
\frac{1}{N}
\boldsymbol\lambda_{0i}\trans
\calV_t(\dtheta_0)
\boldsymbol\lambda_{0i}
+
\frac{1}{T}
\boldsymbol a_{i,t|t-1}\trans(\dtheta_0)
\dOmega(\dtheta_0)
\boldsymbol a_{i,t|t-1}(\dtheta_0)
}
}
\Bigg\vert\calF_{t-1}
\Rightarrow
\calN\bra{0,1},
\nonumber
\\
&
\frac{
\hatPittithetamle-\Pittithetaz
}{
\sqrt{
\frac{1}{N}
\bra{
\boldsymbol\lambda_{0i}
\otimes
\boldsymbol\lambda_{0i}
}\trans
\calW_t(\dtheta_0)
\bra{
\boldsymbol\lambda_{0i}
\otimes
\boldsymbol\lambda_{0i}
}
+
\frac{1}{T}
\boldsymbol b_{i,t|t-1}\trans(\dtheta_0)
\dOmega(\dtheta_0)
\boldsymbol b_{i,t|t-1}(\dtheta_0)
}
}
\Bigg\vert\calF_{t-1}
\Rightarrow
\calN\bra{0,1},
\nonumber
\end{align}
where
\begin{align*}
\calV_t(\dtheta_0)
={}&
\sum_{k=0}^{t-1}
\sum_{h=0}^{t-1}
\bra{\dPhi_0-\dK_0}^{k}
\dK_0
\dSigma_{\Lambda}^{-1}
\dPi_{t-k-1,t-h-1}
\dSigma_{\Lambda}^{-1}
\dK_0\trans
\bra{\dPhi_0-\dK_0}^{h\top},
\\
\calW_t(\dtheta_0)
={}&
\sum_{k=0}^{t-1}
\sum_{h=0}^{t-1}
\bra{
\dW_r
\bra{\dPhi_0-\dK_0}
\dW_r\trans
}^{k}
\dW_r
\dK_0
\dW_r\trans
\\
&\times
\bra{
\boldsymbol I_r\otimes\dF_{t-k-1}
+
\dF_{t-k-1}\otimes\boldsymbol I_r
}
\dSigma_{\Lambda}^{-1}
\dPi_{t-k-1,t-h-1}
\dSigma_{\Lambda}^{-1}
\bra{
\boldsymbol I_r\otimes\dF_{t-h-1}
+
\dF_{t-h-1}\otimes\boldsymbol I_r
}^{\trans}
\\
&\times \dW_r
\dK_0\trans
\dW_r\trans
\bra{
\dW_r
\bra{\dPhi_0-\dK_0}
\dW_r\trans
}^{h\top}.
\end{align*}
The matrices $\calV_t(\dtheta_0)$ and $\calW_t(\dtheta_0)$ have dimensions $r\times r$ and $r^2\times r^2$, respectively. Moreover, $\dSigma_{\Lambda}$ and $\dPi_{s,u}$ are defined in Assumptions \ref{ass:factor_identifiability} and \ref{ass:idiosyncratic}, respectively, 
and $\boldsymbol a_{i,t|t-1}(\dtheta_0)$, $\boldsymbol b_{i,t|t-1}(\dtheta_0)$, and $\dOmega(\dtheta_0)$ are defined as in Theorem \ref{thm:filter_approx_error_asy_norm}.
\end{theorem}
\vspace{-.5cm}
\paragraph{Feasible covariance estimation.}
All the asymptotic variances in Theorems \ref{thm:filter_approx_error_asy_norm} and \ref{thm:CLT_Gaussian_models} are feasible. Let $\widehat e_{it}=x_{it}-\widehat{\boldsymbol\lambda}_{i}\trans\dhatF_t$ and, for a generic zero-mean vector process $\cbra{\boldsymbol g_t}_{t=1}^{T}$, define the Bartlett HAC estimator
$
\operatorname{HAC}_{M_T}\bra{\{\boldsymbol g_t\}_{t=1}^T}
=
\frac{1}{T}\sum_{t=1}^{T}\boldsymbol g_t\boldsymbol g_t\trans
+
\sum_{h=1}^{M_T}
\big(
1-\frac{h}{M_T+1}
\big)
\frac{1}{T}
\sum_{t=h+1}^{T}
\bra{
\boldsymbol g_t\boldsymbol g_{t-h}\trans
+
\boldsymbol g_{t-h}\boldsymbol g_t\trans
},
$
where we set $M_T=\lfloor T^{1/4}\rfloor$. Then, let
$
\widehat{\boldsymbol\Gamma}_F
=
T^{-1}\sum_{t=1}^{T}\dhatF_t\dhatF_t\trans
$,
$
\widehat{\boldsymbol\Xi}_i
=
\operatorname{HAC}_{M_T}\big(\big\{\dhatF_t\widehat e_{it}\big\}_{t=1}^{T}\big)
$,
$
\widehat{\limitcov}
=
-T^{-1}\nabla^{\dtheta\dtheta}\doublehat{\ell}_{\dF}\bra{\dhatF;\dthetamle}
$,
and
$
\widehat{\dV}
=
\operatorname{HAC}_{M_T}
\big(
\big\{
\nabla^{\dtheta}
\doublehat{\ell}_{\dF}
\big(
\dhatF_t;\dthetamle
\big)
\big\}_{t=1}^{T}
\big)
$ \citep{NeweyWest1987}. These quantities provide the estimator
\[
\widehat{\dOmega}
=
\begin{bmatrix}
\widehat{\boldsymbol\Gamma}_F^{-1}
\widehat{\boldsymbol\Xi}_i
\widehat{\boldsymbol\Gamma}_F^{-1}
&
\boldsymbol 0_{r\times d}
\\
\boldsymbol 0_{d\times r}
&
\widehat{\limitcov}^{-1}
\widehat{\dV}
\widehat{\limitcov}^{-1}
\end{bmatrix}.
\]
For Theorem \ref{thm:CLT_Gaussian_models}, we additionally estimate $\dSigma_\Lambda$ by $\widehat{\dSigma}_\Lambda=N^{-1}\widehat\dLambda\trans\widehat\dLambda$. Then, to accommodate weak cross-sectional dependence, let $\widehat{\boldsymbol\eta}_{i,t}=\widehat{\boldsymbol\lambda}_i\widehat e_{it}$ and define $\widehat d_{ij}=\vbra{1/\widehat\rho_{ij}}-1$, with $\widehat\rho_{ij}$ denoting the sample correlation between $\cbra{\widehat e_{it}}_{t=1}^{T}$ and $\cbra{\widehat e_{jt}}_{t=1}^{T}$. Following the AV-SHAC approach of \citet{Kim2022}, we use
\[
\widehat{\dPi}_{s,u}
=
\frac{1}{N}
\sum_{i=1}^{N}
\sum_{j=1}^{N}
\kappa\bra{
\frac{\widehat d_{ij}}{b_N}
}
\widehat{\boldsymbol\eta}_{i,s}
\widehat{\boldsymbol\eta}_{j,u}\trans,
\qquad
\kappa(x)
=
\bra{1-\vbra{x}}
\dIndicator_{\cbra{\vbra{x}\leq1}},
\]
where the cross-sectional bandwidth $b_N$ may be selected using the cluster wild-bootstrap procedure proposed by \citet{Kim2022}. Under covariance stationarity, estimation precision can be improved by averaging $\widehat{\dPi}_{s,u}$ over all time pairs sharing the same lag $s-u$. At lag zero, this yields the original contemporaneous AV-SHAC estimator. The matrices $\widehat{\calV}_t$ and $\widehat{\calW}_t$ are obtained by substituting the feasible quantities into their definitions in Theorem \ref{thm:CLT_Gaussian_models}. Importantly, both summations are retained, so that all matrices $\widehat{\dPi}_{t-k-1,t-h-1}$, including those associated with $k\neq h$, enter the estimated asymptotic variance. The vectors $\widehat{\boldsymbol a}_{i,t|t-1}$ and $\widehat{\boldsymbol b}_{i,t|t-1}$ are constructed analogously by replacing the unknown parameters, loadings, filters, and filter derivatives with their feasible estimated counterparts. Although the intermediate matrices are expressed in the factor coordinates selected by the PC estimator, the resulting scalar quadratic forms are invariant to the alignment of factors and loadings. Hence, neither $\widehat\dH$ nor $\dH_0$ plays any role, and all the asymptotic variances are directly computable in empirical applications.

\section{Joint time-varying conditional location-variance models}
\label{sec:asymptotic_theory_J}

The time-varying location and variance models defined in Section \ref{sec_model} can be combined, so that the full multivariate dynamic model generated by the non-stochastic link function $\boldsymbol q\bra{\,\cdot\,,\dvt}$ in \eqref{eq:Model_yt} assumes the form
\begin{equation}
\label{eq:SD_full_conditional}
\dFt
=
\boldsymbol q\bra{\dpsitti,\dvt}
=
\dmutti+\dSigmatti^{1/2}\dvt,
\qquad t\in\Zz,
\end{equation}
where $\dpsitti\equiv\big(\dmutti\trans,{\dsigmatti}\trans\big)\trans$. The noise process $\cbra{\dvt}_{t\in\Zz}$ satisfies Assumption \ref{ass:noise}, with $\Ee\sbra{\dvt\dvt\trans}=\boldsymbol I_r$. The conditional location and variance processes in
\eqref{eq:SD_full_conditional} are defined by
\begin{align}
\label{eq:SD_full_conditional_location}
\dmutpt
&=
\dPhi^\mu\dmutti
+
\dK^\mu\bra{\dFt-\dmutti},
\\
\label{eq:SD_full_conditional_residuals}
\depsi_t
&=
\dFt-\dmutti,
\\
\label{eq:SD_full_conditional_variance1}
\vect\dSigmatti
&=
\dW_r\dsigmatti,
\\
\label{eq:SD_full_conditional_variance2}
\dsigmatpt
&=
\ddelta^\sigma
+
\dPhi^\sigma\dsigmatti
+
\bra{
\dK^\sigma
+
\dUpsilon^\sigma\boldsymbol I\bra{\dFt}
}
\dW_r\trans
\bra{
\depsi_t\otimes\depsi_t
-
\vect\dSigmatti
},
\end{align}
where $\dI\bra{\dFt} = \diag\big(\dIndicator_{\cbra{F_{1t} \leq 0}},\ldots, \dIndicator_{\cbra{F_{rt} \leq 0}}\big)$. The main difference between the joint model in \eqref{eq:SD_full_conditional}--\eqref{eq:SD_full_conditional_variance2} and the time-varying conditional variance model in \eqref{eq:SD_Model_sigmat} is that the conditional variance process $\big\{\dsigmatti\big\}_{t\in\Zz}$ now depends on the unobserved conditional
location process $\big\{\dmutti\big\}_{t\in\Zz}$ through the residuals $\cbra{\depsi_t}_{t\in\Zz}$. Consequently, even after replacing the latent factors $\cbra{\dFt}_{t\in\Zz}$ with their first-step estimators $\big\{\dhatFt\big\}_{t=1}^T$, the variance filter cannot be evaluated without first obtaining a filtered path for the conditional location process. While the first-step PC estimator and its asymptotic properties remain unchanged, the score-driven estimation stage and its associated asymptotic theory must therefore be adapted to account for the additional estimation error generated by the filtered conditional location process.

\vspace{-.2cm}
\begin{remark}
\upshape{
The leverage indicator in \eqref{eq:SD_full_conditional_variance2} is defined in terms of $\dFt$, rather than the centered residual $\depsi_t(\dtheta^\mu)$, so that, conditional on the factor path, it depends neither on $\dtheta^\mu$ and $\dtheta^\sigma$ nor on the filtered location process $\dmutti(\dtheta^\mu)$. It therefore enters the conditional variance recursion as an exogenous measurable input and does not introduce an additional discontinuity in the stochastic recurrence equation. This preserves the differentiability and contraction structure required to apply the contraction and near-epoch-dependence arguments developed in Section \ref{sec:asymptotic_theory}.
}
\end{remark}
\vspace{-.6cm}
\paragraph{Sequential QML estimation.}
To overcome the issue discussed above, we propose a sequential QML estimation strategy. After retrieving the factors $\big\{\dhatF_t\big\}_{t=1}^T$ from the first-step, we first estimate the parameters governing the conditional location dynamics in \eqref{eq:SD_full_conditional_location} by means of a quasi-likelihood (QL) criterion constructed under a working homoskedastic assumption. More precisely, we use the classical quasi-Gaussian likelihood for the factors given in \eqref{eq:log_likelihood_gaussian_locations}, but we treat the conditional covariance matrix as fixed and equal to the identity matrix, such that $\dSigma$ is replaced by $\dIdentityr$. 

Hence, at the first-stage, for a generic parameter vector  $\dtheta^\mu = \big(\vect \dPhi^{\mu\top}, \vect\dK^{\mu\top}\big)\trans$ such that $\dtheta^\mu \in \dTheta^\mu\subset \Rr^{2r^2}$, we define
\begin{align}
\label{eq:MLE_mu_full_model}
\dthetamle^{\,\mu} 
=&
\argmax_{\dtheta^\mu \in \dTheta^\mu \subset \mathbb R^{2r^2}}
\sum_{t=1}^T
\doublehat{\ell}_{\dmu,\dF}\bra{\dhatF_t ; \dtheta^\mu},
\\
\label{eq:QLE_mu_full_model}
\doublehat{\ell}_{\dmu,\dF}\bra{\dhatF_t ; \dtheta^\mu}
=&
-\frac{1}{2}\bra{\dhatFt - \dhathatmutti(\dtheta^\mu)}\trans
\bra{\dhatFt - \dhathatmutti(\dtheta^\mu)},
\end{align}
up to an irrelevant additive constant, and where $\big\{\dhathatmutti(\dtheta^\mu)\big\}_{t=1}^T$ is generated recursively from the conditional location equation in \eqref{eq:SD_full_conditional_location} initialized at some arbitrary $\dmu_{1|0}$ and substituting the latent factors $\cbra{\dFt}_{t\in\Zz}$ with their first-step PC estimators $\big\{\dhatFt\big\}_{t=1}^T$. This first-stage criterion is not intended to provide a full specification of the joint conditional density of the factors, but rather to deliver a feasible estimator of the conditional location dynamics and, consequently, a filtered path for the unobservable process $\big\{\dmu_{t|t-1}\big\}_{t\in\mathbb{Z}}$.

Given the estimator $\dthetamle^{\,\mu}$, we then construct the filtered conditional location path $\big\{\dhathatmutti(\dthetamle^{\,\mu})\big\}_{t=1}^T$ and define the corresponding centered residuals
\begin{equation}
\label{eq:SD_resid_full_model}
\doublehat{\depsi}_t(\dthetamle^{\,\mu})
=
\dhatF_t - \dhathatmutti(\dthetamle^{\,\mu}),
\qquad t=1,\dots,T,
\end{equation}
which are the approximated version of the unobserved stationary residuals $\bar{\depsi}_t\bra{\dtheta_0^\mu}=\dbarF_t-\bar{\dmu}_{t|t-1}\bra{\dtheta_0^\mu}$, with $t\in\Zz$. Because the first-stage QLE in \eqref{eq:MLE_mu_full_model} and \eqref{eq:QLE_mu_full_model} ignores the time-variation in the conditional covariance matrix, the residuals $\big\{\doublehat{\depsi}_t(\dthetamle^{\,\mu})\big\}_{t=1}^T$ still retain conditional heteroskedasticity. This is precisely the feature that makes them suitable for the second-stage estimation of the conditional variance process. Indeed, once the time-varying location component has been filtered out, the residuals can be treated as the relevant inputs for the variance recursion, so that the latter becomes feasible after replacing the unobserved process $\big\{\dmutti\big\}_{t\in\mathbb Z}$ with its first-stage filtered counterpart $\big\{\dhathatmutti(\dthetamle^{\,\mu})\big\}_{t=1}^T$.

In the second stage, we therefore estimate the parameters driving the conditional variance dynamics by maximizing the full quasi-Gaussian likelihood, conditional on the first-stage estimate of the location process. Let
$\dtheta^\sigma
=
\big(
{\ddelta^\sigma}\trans,
\vect{\dPhi^\sigma}\trans,
\vect{\dK^\sigma}\trans,
\vect{\dUpsilon^\sigma}\trans
\big)\trans
\in
\dTheta^\sigma\subset\Rr^{r+3r^2}$ denote the vector of parameters governing the conditional variance recursion. Then, we define
\begin{align}
\label{eq:MLE_sigma_full_model}
\dthetamle^{\,\sigma}
={}&
\argmax_{\dtheta^\sigma\in\dTheta^\sigma\subset\Rr^{r+3r^2}}
\sum_{t=1}^T
\doublehat{\ell}_{\sigma,\depsi}\bra{ \doublehat{\depsi}_t(\dthetamle^{\,\mu}) ; \dtheta^\sigma}
\\
\label{eq:QLE_sigma_full_model}
\doublehat{\ell}_{\sigma,\depsi}
\bra{
\doublehat{\depsi}_t\bra{\dthetamle^{\,\mu}};
\dtheta^\sigma
}
={}&
-\frac{1}{2}
\Bigg(
\log
\vbra{
\dhathatSigmatti
\bra{
\dthetamle^{\,\mu},
\dtheta^\sigma
}
}
+
\doublehat{\depsi}_t
\bra{
\dthetamle^{\,\mu}
}^{\trans}
\dhathatSigmatti
\bra{
\dthetamle^{\,\mu},
\dtheta^\sigma
}^{-1}
\doublehat{\depsi}_t
\bra{
\dthetamle^{\,\mu}
}
\Bigg).
\end{align}
The sequence $\big\{\dhathatSigmatti\big(\dthetamle^{\,\mu},\dtheta^\sigma\big)\big\}_{t=1}^T$ is generated recursively from \eqref{eq:SD_full_conditional_variance1}--\eqref{eq:SD_full_conditional_variance2} using the estimated residuals $\big\{\doublehat{\depsi}_t\big(\dthetamle^{\,\mu}\big)\big\}_{t=1}^T$ defined in \eqref{eq:SD_resid_full_model}. Equivalently, the second-stage filter is obtained by injecting into the variance recursion both the pre-estimated factors and the filtered conditional location process obtained in the first-stage. In this way, after retrieving the factors $\big\{\dhatF_t\big\}_{t=1}^T$ from the first-step PC estimator, the original joint estimation problem is decomposed into two additional feasible and interpretable steps. First, filtering and estimating the conditional location dynamics, and second, filtering and estimating the conditional variance dynamics of the resulting mean-corrected factors.

As a result, we obtain the sequential-QMLE: 
$
\dthetamle^J
=
\big(
{\dthetamle^{\,\mu \top}},
{\dthetamle^{\,\sigma \top}}
\big)\trans,
$
and, accordingly, we denote  its true counterpart as $\dtheta_0^J=\big( {\dtheta_0^{\mu \top}},{\dtheta_0^{\sigma \top}}\big)\trans\in\dTheta^\mu\times \dTheta^\sigma\subset \mathbb R^d$ with $d=r+5r^2$. This sequential approach is analogous in spirit to a pseudo-likelihood procedure of \cite{Ling2003} for their vector ARMA-GARCH model. The first stage uses a misspecified working likelihood to recover the dynamic conditional mean, while the second stage uses the full quasi-Gaussian likelihood to model the remaining conditional heteroskedasticity. Therefore, although the two components are jointly present in the data-generating mechanism, estimation remains feasible by proceeding sequentially, with the first-stage filtered conditional location process acting as an input into the second-stage conditional variance filter.

The first-stage criterion in \eqref{eq:QLE_mu_full_model} is a quasi-likelihood for the conditional location process. We therefore impose Assumptions \ref{ass:smooth_domination_likelihood}--\ref{ass:truePar} on this criterion, with $\ell_{\dF}$, $\dtheta$, and $\dTheta$ replaced by $\ell_{\dmu,\dF}$, $\dtheta^\mu$, and $\dTheta^\mu$, respectively. Theorem \ref{thm:consistency} can therefore be applied to the first-stage estimator $\dthetamle^{\,\mu}$. Under the additional conditions required in Theorem \ref{thm:asy_norm}, including $\sqrt T/N\rightarrow0$, the corresponding asymptotic normality result also applies.

Relative to the framework developed in Section \ref{sec_estimator}, the sequential procedure introduces an additional source of estimation error, because the second-stage variance filter is driven by the estimated residuals in \eqref{eq:SD_resid_full_model}. The following assumption, which extends Assumptions \ref{ass:contraction_cond_I} and \ref{ass:contraction_cond} to the joint conditional location-variance process, provides sufficient conditions for controlling the propagation of this error through the conditional location and variance recursions.

\begin{assumption}
\label{ass:joint_location_variance}
\begin{inparaenum}
\item[(i)]
The parameter spaces $\dTheta^\mu$ and $\dTheta^\sigma$ are compact subsets of $\Rr^{2r^2}$ and $\Rr^{r+3r^2}$, respectively.
\item[(ii)]
For positive reals $\calX_1^\mu<1$ and $\calY_1^\mu<\infty$, $\sup_{\dtheta^\mu\in\dTheta^\mu} \vvbra{\dPhi^\mu-\dK^\mu} \leq\calX_1^\mu$
and $ \sup_{\dtheta^\mu\in\dTheta^\mu} \vvbra{\dK^\mu} \leq\calY_1^\mu $.
\item[(iii)]
Let
$
\mathcal G_r
=
\cbra{
\boldsymbol G=\diag\bra{g_1,\ldots,g_r}:
g_j\in\cbra{0,1},\ j=1,\ldots,r
}.
$
Then, for every $\dtheta^\sigma\in\dTheta^\sigma$ and
$\boldsymbol G\in\mathcal G_r$, $
\dK^\sigma+\dUpsilon^\sigma\boldsymbol G
\geq\boldsymbol 0_{r\times r}
$,
and
$
\dPhi^\sigma-\dK^\sigma-\dUpsilon^\sigma\boldsymbol G
\geq\boldsymbol 0_{r\times r}
$,
where the matrix inequalities are understood componentwise.
\item[(iv)]
For some $k\geq1$ and some finite positive real
$\calX_k^\sigma<1$,
\begin{align*}
\sup_{\dtheta^\sigma\in\dTheta^\sigma}
\max_{\boldsymbol G_1,\ldots,\boldsymbol G_k\in\mathcal G_r}
\vvbra{
\prod_{j=1}^{k}
\bra{
\dPhi^\sigma
-
\dK^\sigma
-
\dUpsilon^\sigma\boldsymbol G_j
}
}
\leq
\calX_k^\sigma.
\end{align*}
\item[(v)]
Let
$\dZ_t\bra{\dtheta^\mu,\dtheta^\sigma}
=
\bra{
\dK^\sigma
+
\dUpsilon^\sigma
\boldsymbol I\bra{\dFt}
}
\dW_r\trans\bra{
\boldsymbol I_r\otimes\depsi_t(\dtheta^\mu)
+
\depsi_t(\dtheta^\mu)\otimes\boldsymbol I_r
}.
$
For the same $p>2$ as in Assumption \ref{ass:contraction_cond}, there exists a finite positive real $\calZ_k^\sigma$, independent of $t$, such that
$$
\vvbra{
\sum_{j=1}^{k}
\sup_{\substack{
\dtheta^\mu\in\dTheta^\mu\\
\dtheta^\sigma\in\dTheta^\sigma
}}
\vvbra{
\dZ_{t-j+1}
\bra{
\dtheta^\mu,\dtheta^\sigma
}
}
}_p
\leq
\calZ_k^\sigma.
$$
\item[(vi)]
There exist finite positive reals $C_I$ and $c_H$ such that, for every sufficiently small $\eta>0$,
\[
\sup_{\substack{
\dH\in GL\bra{r}:\\
\vvbra{\dH-\dH_0}\leq c_H
}}
\sup_{t\in\Nn}
\max_{1\leq i\leq r}
\Pp\bra{
\vbra{
\mathbf e_i\trans\dH\dbarF_t
}
\leq\eta
}
\leq
C_I\eta.
\]
\end{inparaenum}
\end{assumption}

Part $(i)$ imposes compactness of the parameter spaces, while part $(ii)$ ensures contraction of the conditional location filter. Parts $(iii)$ and $(iv)$ guarantee positivity and contraction of the conditional variance filter for every possible sequence of indicator realizations, and part $(v)$ controls the propagation of the estimated-location error through the quadratic residuals. Part $(vi)$ is a uniform $1$-continuity condition for the leverage indicator; see \citet[][Definition~2.10, Example~2.1(ii), and Proposition~2.11]{Borovkova2001}. It controls the indicator under factor approximation and preserves geometrically decreasing approximation coefficients under truncation, as required for part $(d)$ of Proposition \ref{prop:rate_psit_decomp}. 

With the next proposition, we show that the filter approximation results established in Proposition \ref{prop:rate_psit_decomp} extend to the joint conditional location-variance model after accounting for the additional perturbation generated by the estimated conditional location process. 

\begin{prop}
\label{prop:perturb_full_model}
Under Assumptions \ref{ass:noise}--\ref{ass:contraction_cond} and \ref{ass:joint_location_variance}, with $p>2$, for any fixed initial conditions of the conditional location and variance filters, as $N,T\rightarrow\infty$, statements $(a)$--$(e)$ of Proposition \ref{prop:rate_psit_decomp} continue to hold for the joint process
$
\big\{
\dpsitti
\big\}_{t\in\Zz}
\equiv
\big\{
(
\dmutti\trans,
{\dsigmatti}\trans
)^{\trans}
\big\}_{t\in\Zz},
$
uniformly over
$\bra{\dtheta^\mu,\dtheta^\sigma}\in
\dTheta^\mu\times\dTheta^\sigma$.
\end{prop}
Proposition \ref{prop:perturb_full_model} extends the filter approximation rates and the near-epoch-dependence result of Proposition \ref{prop:rate_psit_decomp} to the joint conditional location-variance process. Proposition \ref{lemma:ULLN} and Theorem \ref{thm:consistency} can therefore be applied sequentially to \eqref{eq:QLE_mu_full_model} and \eqref{eq:QLE_sigma_full_model}, yielding the set-consistency of the sequential-QMLE. More formally, as $N,T\rightarrow\infty$, it holds that
$
\big\Vert
\boldsymbol r_{\widehat\dH}^{J,-1}
\big(
\dthetamle^J
\big)
-
\dtheta_0^J
\big\Vert
=
o_{\Pp}(1),
$
where the joint realignment mapping is defined blockwise as 
$\boldsymbol r_{\dH}^J\big(\dtheta^J\big)
=
\big(
\boldsymbol r^\mu_{\dH}\bra{\dtheta^\mu}\trans,
\boldsymbol r^\sigma_{\dH}\bra{\dtheta^\sigma}\trans
\big)\trans$, with inverse defined analogously, see Appendix \ref{appendix_parameter_identifications} for further details.

We therefore focus directly on asymptotic normality. At the root-$T$ scale, the first-stage estimation error enters the second-stage score through the estimated residuals and the variance filter. Since the second-stage criterion is quasi-Gaussian and the leverage indicator is independent of $\dtheta^\mu$ and $\dtheta^\sigma$, the required mixed-score and Hessian expansions can be established directly. The following proposition provides these additional results. 

\begin{prop}
\label{prop:sequential_QMLE_expansion}
Under Assumptions \ref{ass:noise}--\ref{ass:joint_location_variance}, with $p>2$, for every fixed initialization of the conditional location
and variance filters, as $N,T\rightarrow\infty$, if $\sqrt T/N\rightarrow0$, then
\begin{align}
\sqrt T
\bra{
\dthetamle^J
-
\boldsymbol r_{\widehat\dH}^J
\big({\dtheta_0^J}\big)
}
\Rightarrow
\calN\bra{
\boldsymbol 0,
\limitcov_{J,\dH_0}^{-1}
\big({\dtheta_0^J}\big)
\dV_{J,\dH_0}
\big({\dtheta_0^J}\big)
\limitcov_{J,\dH_0}^{-1}
\big({\dtheta_0^J}\big)\trans
},\nonumber
\end{align}
with
$
\limitcov_{J,\dH_0}\big({\dtheta_0^J}\big)
=
{\boldsymbol R_{\dH_0}^{J\,\, \top}}
\limitcov_J\big({\dtheta_0^J}\big)
\boldsymbol R_{\dH_0}^J,
$
$
\dV_{J,\dH_0}\big({\dtheta_0^J}\big)
=
{\boldsymbol R_{\dH_0}^{J\,\, \top}}
\dV_J\big({\dtheta_0^J}\big)
\boldsymbol R_{\dH_0}^J,
$
\begin{equation*}
\boldsymbol R_{\dH}^J
=
\begin{pmatrix}
\boldsymbol R_{\dH}^\mu & \boldsymbol 0
\\
\boldsymbol 0 & \boldsymbol R_{\dH}^\sigma
\end{pmatrix},
\qquad
\boldsymbol R_{\dH}^\mu
=
\frac{\partial}{\partial{\dtheta^\mu}\trans}
\boldsymbol r_{\dH}^{\mu,-1}\bra{\dtheta_0^\mu},
\qquad
\boldsymbol R_{\dH}^\sigma
=
\frac{\partial}{\partial{\dtheta^\sigma}\trans}
\boldsymbol r_{\dH}^{\sigma,-1}\bra{\dtheta_0^\sigma}.
\end{equation*}
and where
\begin{equation*}
\limitcov_J\big({\dtheta_0^J}\big)
=
\begin{pmatrix}
\limitcov_{\mu\mu}\bra{\dtheta_0^\mu}
&
\boldsymbol 0
\\
\limitcov_{\sigma\mu}\big({\dtheta_0^J}\big)
&
\limitcov_{\sigma\sigma}\big({\dtheta_0^J}\big)
\end{pmatrix},
\end{equation*}
with
$
\limitcov_{\mu\mu}\bra{\dtheta_0^\mu}
=
\lim_{T\rightarrow\infty}
-
\frac{1}{T}
\Ee\big[{
\nabla^{\dtheta^\mu\dtheta^\mu}
\ell_{\dmu,\dF}
\bra{
\dbarF;
\dtheta_0^\mu
}
}\big],$
$
\limitcov_{\sigma\sigma}\big({\dtheta_0^J}\big)
=
\lim_{T\rightarrow\infty}
-
\frac{1}{T}
\Ee\big[{
\nabla^{\dtheta^\sigma\dtheta^\sigma}
\ell_{\sigma,\depsi}
\big({
\bar{\depsi}\bra{\dtheta_0^\mu};
\dtheta_0^\sigma
}\big)
}\big],
$
$
\limitcov_{\sigma\mu}\big({\dtheta_0^{J}}\big)
=
\lim_{T\rightarrow\infty}
-
\frac{1}{T}
\Ee\big[{
\nabla^{\dtheta^\sigma\dtheta^\mu}
\ell_{\sigma,\depsi}
\big({
\bar{\depsi}\bra{\dtheta_0^\mu};
\dtheta_0^\sigma
}\big)
}\big],
$
\begin{equation*}
\dV_J\bra{\dtheta_0^J}
=
\lim_{T\rightarrow\infty}
\frac{1}{T}
\Ee\sbra{
\begin{pmatrix}
\nabla^{\dtheta^\mu}
\ell_{\dmu,\dF}
\bra{
\dbarF;
\dtheta_0^\mu
}
\\
\nabla^{\dtheta^\sigma}
\ell_{\sigma,\depsi}
\bra{
\bar{\depsi}\bra{\dtheta_0^\mu};
\dtheta_0^\sigma
}
\end{pmatrix}
\begin{pmatrix}
\nabla^{\dtheta^\mu}
\ell_{\dmu,\dF}
\bra{
\dbarF;
\dtheta_0^\mu
}
\\
\nabla^{\dtheta^\sigma}
\ell_{\sigma,\depsi}
\bra{
\bar{\depsi}\bra{\dtheta_0^\mu};
\dtheta_0^\sigma
}
\end{pmatrix}^{\trans}
}.
\end{equation*}
\end{prop}


This result shows that the implied consistency rate of the joint QMLE is still $O_\Pp(\max(N^{-1},T^{-1/2}))$. Furthermore, it is clear that Theorems \ref{thm:filter_approx_error_asy_norm} and \ref{thm:CLT_Gaussian_models}
extend directly to the joint model by replacing the QMLE covariance block in the bottom $d\times d$ diagonal block of $\dOmega\bra{\dtheta_0}$ with the asymptotic covariance \linebreak
$\limitcov_J^{-1}\big(\dtheta_0^J\big) \dV_J\big({\dtheta_0^J}\big) \limitcov_J^{-1}\big({\dtheta_0^J}\big)\trans$
as defined in Proposition \ref{prop:sequential_QMLE_expansion}. 
For the conditional location, only its upper-left block enters, because $\dmutti$ depends only on $\dtheta^\mu$. For the conditional variance, the full matrix, including its cross-blocks, is required because $\dsigmatti$ depends on both $\dtheta^\mu$ and $\dtheta^\sigma$.  In Theorem \ref{thm:CLT_Gaussian_models}, the  matrix $\calV_t\bra{\dtheta_0^\mu}$ remains unchanged, whereas $\calW_t\bra{\dtheta_0}$ is replaced by
\begin{align*}
&\calW_t^J\bra{\dtheta_0^J}
=
\sum_{k=0}^{t-1}
\sum_{h=0}^{t-1}
\bra{
\dW_r
\bra{
\dPhi_0^\sigma-\dK_0^\sigma
}
\dW_r\trans
}^{k}
\dW_r\dK_0^\sigma\dW_r\trans
\\
&\times
\bra{
\boldsymbol I_r\otimes
{\depsi}_{t-k-1}\bra{\dtheta_0^\mu}
+
{\depsi}_{t-k-1}\bra{\dtheta_0^\mu}
\otimes\boldsymbol I_r
}
\dPi^{\depsi}_{t-k-1,t-h-1}
\bra{
\boldsymbol I_r\otimes
{\depsi}_{t-h-1}\bra{\dtheta_0^\mu}
+
{\depsi}_{t-h-1}\bra{\dtheta_0^\mu}
\otimes\boldsymbol I_r
}^{\trans}
\\
&\times \dW_r{\dK_0^\sigma}\trans\dW_r\trans
\bra{
\dW_r
\bra{
\dPhi_0^\sigma-\dK_0^\sigma
}
\dW_r\trans
}^{h\trans},
\end{align*}
where
\begin{align*}
\dPi^{\depsi}_{s,u}
=&
\lim_{N\rightarrow\infty}
\frac{1}{N}
\sum_{i=1}^{N}
\sum_{j=1}^{N}
\Ee\Bigg[
\bigg(
\dSigma_\Lambda\inv
\boldsymbol\lambda_{0i}e_{is}
-
\sum_{a=0}^{s-1}
\bra{
\dPhi_0^\mu-\dK_0^\mu
}^{a}
\dK_0^\mu
\dSigma_\Lambda\inv
\boldsymbol\lambda_{0i}e_{i,s-a-1}
\bigg)
\\
&\qquad\qquad\qquad\qquad\qquad\qquad\times
\bigg(
\dSigma_\Lambda\inv
\boldsymbol\lambda_{0j}e_{ju}
-
\sum_{b=0}^{u-1}
\bra{
\dPhi_0^\mu-\dK_0^\mu
}^{b}
\dK_0^\mu
\dSigma_\Lambda\inv
\boldsymbol\lambda_{0j}e_{j,u-b-1}
\bigg)^{\trans}
\Bigg],
\end{align*}
is the limiting covariance of the cross-sectional PC error after subtracting the part propagated through the conditional-location filter. 

\section{Monte Carlo study}\label{sec:MCsim}

\paragraph{Data generating process.} For each replication and for \(t=1,\dots,T\), the observed \(N\)-dimensional time series $\cbra{\dxt }_{t\in\Zz}$ is generated by a static factor model as in equation \eqref{eq:Model_xt}. The matrix of factor loadings is generated according to a Gaussian random vector, such that $\boldsymbol \lambda_{i} \sim \calN\bra{\boldsymbol 0_r, \boldsymbol S_\Lambda}$, for each $i=1,\dots,N$, independently across \(i\) and where $\boldsymbol S_\Lambda$ is a diagonal $r\times r$ matrix with distinct entries in $[0.6, 1.6]$. We consider the cases with $r\in\{1,3\}$ factors.
The idiosyncratic term is generated purely as cross-sectionally correlated Gaussian noise with no time dynamics, and then de-meaned. In particular, we set $\dxit \sim \calN\bra{\dzeroN, \dSigma_{\boldsymbol e}}$, where $\dSigma_{\boldsymbol e} = \boldsymbol \Gamma_1^{1/2}\boldsymbol \Gamma_2\boldsymbol \Gamma_1^{1/2},$ with \(\boldsymbol \Gamma_1 = \diag(s_1,\dots,s_N)\) with \(s_i \sim \text{Unif}(0.5,1.5)\), and \(\boldsymbol \Gamma_2\) is Toeplitz with entries \(\sbra{\boldsymbol \Gamma_2}_{ij}=0.5^{\vbra{i-j}}\,\mathbbm{1}\cbra{\vbra{i-j}\le 10}\). 

In the conditional location case, the latent factors satisfy equation \eqref{eq:SD_Model_location}, and the conditional location process follows the score-driven recursion in equation \eqref{eq:SD_Model_mut}. We consider alternative distributions for the innovation term. Specifically, we set \(\dv_t \sim \mathcal{N}(\boldsymbol 0_r,\dSigma)\) in the Gaussian case, and \(\dvt \sim t_{\nu}(\boldsymbol 0_r,\dSigma)\) in the heavy-tailed case, with \(\nu \in \{4,5\}\), so that the score-driven recursion satisfies the Gaussian updates case, in the former, or Student's \emph{t} updates in the latter. We fix the score-recursion parameters ex ante by setting \(\dPhi=0.8\cdot \boldsymbol I_r\), \(\dK=0.15 \cdot \boldsymbol I_r\), and drawing \(\dSigma=\diag\bra{10 \cdot \vbra{z_1},\dots,10 \cdot \vbra{z_r}}\) with \(z_l\sim\calN(0,1)\) for $l = 1,\dots,r$.
In the conditional variance case, the latent factors satisfy equation \eqref{eq:SD_Model_covariance}, and the conditional variance process follows the score-driven recursion in equation \eqref{eq:SD_Model_sigmat}. Analogously to the previous case, we consider two distributions for the innovation term, namely \(\dv_t \sim \mathcal{N}(\boldsymbol 0_r,\boldsymbol I_r)\) in the Gaussian case, and \(\dvt \sim t_{\nu}(\boldsymbol 0_r,\boldsymbol I_r)\) in the Student's \emph{t} case, with \(\nu \in \{4,5\}\), and the score-driven recursion modifies accordingly with the Gaussian updates, in the former, and Student's \emph{t} updates in the latter. For the score-driven parameters, we set \(\dPhi=0.8\cdot \boldsymbol I_r \) , \(\dK=0.15\cdot \boldsymbol I_r \) and fix \(\ddelta = \diag\bra{\boldsymbol I_r - \dPhi } \).



\begin{table}[!t]
	\centering
	\caption{Monte Carlo simulation results for the QMLE of the conditional
		location and variance models}
	\label{tab:sim_qmle_cond_loc_var}
	\scriptsize
	\begingroup
	\setlength{\tabcolsep}{3pt}
	\begin{tabular}{@{}lccrrrrrrrr@{}}
		\toprule
		& & &
		\multicolumn{4}{c}{Conditional location model} &
		\multicolumn{4}{c}{Conditional variance model} \\
		\cmidrule(lr){4-7}\cmidrule(lr){8-11}
		& & &
		\multicolumn{2}{c}{One factor} &
		\multicolumn{2}{c}{Three factors} &
		\multicolumn{2}{c}{One factor} &
		\multicolumn{2}{c}{Three factors} \\
		\cmidrule(lr){4-5}\cmidrule(lr){6-7}
		\cmidrule(lr){8-9}\cmidrule(lr){10-11}
		Distribution & $N$ & $T$
		& Bias$^2$ & Var.
		& Bias$^2$ & Var.
		& Bias$^2$ & Var.
		& Bias$^2$ & Var. \\
		\midrule
		Gaussian
		& 100 & 200
		& 0.00017 & 0.00226 & 0.00295 & 0.03077
		& 0.00664 & 0.03408 & 0.01221 & 0.04470 \\
		& 100 & 500
		& 0.00005 & 0.00203 & 0.00203 & 0.03780
		& 0.00401 & 0.02515 & 0.00327 & 0.01727 \\
		& 200 & 200
		& 0.00018 & 0.00230 & 0.00680 & 0.16029
		& 0.00760 & 0.04016 & 0.01117 & 0.03901 \\
		& 200 & 500
		& 0.00004 & 0.00183 & 0.00165 & 0.01552
		& 0.05253 & 0.07909 & 0.00149 & 0.01267 \\
		& 500 & 200
		& 0.00009 & 0.00235 & 0.01079 & 0.14950
		& 0.00449 & 0.02850 & 0.00787 & 0.03727 \\
		& 500 & 500
		& 0.00011 & 0.00221 & 0.00042 & 0.00405
		& 0.00197 & 0.01390 & 0.00116 & 0.01089 \\
		& 500 & 1000
		& 0.00004 & 0.00151 & 0.00026 & 0.00312
		& 0.00077 & 0.00549 & 0.00028 & 0.00363 \\
		\addlinespace
		
		Student's $t_4$
		& 100 & 200
		& 0.00048 & 0.00229 & 0.00088 & 0.04345
		& 0.00757 & 0.03751 & 0.01200 & 0.03995 \\
		& 100 & 500
		& 0.00008 & 0.00167 & 0.00470 & 0.12221
		& 0.00045 & 0.01206 & 0.00278 & 0.01746 \\
		& 200 & 200
		& 0.00063 & 0.00265 & 0.00322 & 0.28083
		& 0.00607 & 0.03068 & 0.01063 & 0.03714 \\
		& 200 & 500
		& 0.00011 & 0.00192 & 0.03426 & 0.28979
		& 0.00228 & 0.01255 & 0.00223 & 0.01373 \\
		& 500 & 200
		& 0.00047 & 0.00347 & 0.00029 & 0.01524
		& 0.01416 & 0.05253 & 0.01176 & 0.03643 \\
		& 500 & 500
		& 0.00033 & 0.00191 & 0.00986 & 0.19619
		& 0.00410 & 0.01884 & 0.00245 & 0.01510 \\
		& 500 & 1000
		& 0.00016 & 0.00216 & 0.00039 & 0.00355
		& 0.00052 & 0.00540 & 0.00038 & 0.00363 \\
		\addlinespace
		
		Student's $t_5$
		& 100 & 200
		& 0.00015 & 0.00258 & 0.00022 & 0.02218
		& 0.00441 & 0.03707 & 0.01175 & 0.05731 \\
		& 100 & 500
		& 0.00012 & 0.00193 & 0.00026 & 0.00582
		& 0.00405 & 0.02541 & 0.00430 & 0.02313 \\
		& 200 & 200
		& 0.00042 & 0.00223 & 0.00021 & 0.01130
		& 0.00483 & 0.02926 & 0.01270 & 0.07075 \\
		& 200 & 500
		& 0.00006 & 0.00222 & 0.00042 & 0.02285
		& 0.00061 & 0.01579 & 0.00212 & 0.01548 \\
		& 500 & 200
		& 0.00017 & 0.00220 & 0.00021 & 0.01032
		& 0.00309 & 0.02473 & 0.01088 & 0.04236 \\
		& 500 & 500
		& 0.00014 & 0.00209 & 0.01408 & 0.28669
		& 0.00091 & 0.01057 & 0.00180 & 0.01378 \\
		& 500 & 1000
		& 0.00001 & 0.00151 & 0.00130 & 0.03564
		& 0.00013 & 0.00452 & 0.00037 & 0.00380 \\
		\bottomrule
	\end{tabular}
	\endgroup
	
	\vspace{0.4em}
	\parbox{0.98\linewidth}{\footnotesize
		Monte Carlo averages of the squared bias and variance of the QMLE.
		For the conditional location models, entries are arithmetic means of
		the coordinate-specific values for the common dynamic parameters
		$\operatorname{vec}(\dPhi)$ and $\operatorname{vec}(\dK)$.
		For the conditional variance models, the averages additionally include
		the intercept parameters $\ddelta$. Results are based on 100
		replications.}
\end{table}

\vspace{-.5cm}
\paragraph{Results.}
In the following, we report results averaged over the Monte Carlo replications.~Factors and loadings are always estimated via PCA. Since the numerical performance of PCA has been thoroughly investigated in many works, we do not report results on estimation of the factors but we directly focus on the QMLE of the score driven parameters and the implied filtered processes and prediction intervals. Results obtained using the spatial Kendall's tau PCs approach by \citet{he2020elliptic} are in Appendix \ref{app:kendall}.

First, we verify consistency of QMLE. According to Theorem \ref{thm:consistency} the QMLE estimator  $\dthetamle$ of the score-driven parameters is a consistent estimator of $\boldsymbol r_{\dH_0}
\bra{
\dtheta_0
}$.
Table \ref{tab:sim_qmle_cond_loc_var} report the squared bias and variance for $\dthetamle$ averaged over all replications, where, in agreement with to Proposition \ref{prop:rate_yt_decomp},  $\dH_0$ is estimated as the coefficient $\dhatH$ of the linear projection of $\dhatF$ onto $\dF$ and the mapping $\boldsymbol r_{\dH_0}(\cdot)$ is given in Appendix \ref{appendix_parameter_identifications}. 

Second, we verify Theorems \ref{thm:filter_approx_error_asy_norm} and \ref{thm:CLT_Gaussian_models} by computing the empirical coverage of the corresponding 95\% asymptotic prediction intervals. For the Gaussian models, the intervals are constructed using Theorem \ref{thm:CLT_Gaussian_models}, whereas for the Student's $t$ models they are constructed using Theorem \ref{thm:filter_approx_error_asy_norm}. The asymptotic covariance matrices are estimated as described at the end of Section \ref{sec:API}. Table \ref{tab:sim_cond_loc_var_pca_picp} reports the coverage probabilities for both the common conditional locations and
the common conditional variances.

\begin{table}[!t]
	\centering
	\caption{Monte Carlo coverage probabilities for the common conditional
		locations and variances}
	\label{tab:sim_cond_loc_var_pca_picp}
	\scriptsize
	\begingroup
	\setlength{\tabcolsep}{5pt}
	\begin{tabular}{@{}lccrrrr@{}}
		\toprule
		& & &
		\multicolumn{2}{c}{$\mathrm{PICP}_{C}$} &
		\multicolumn{2}{c}{$\mathrm{PICP}_{P}$} \\
		\cmidrule(lr){4-5}\cmidrule(lr){6-7}
		Distribution & $N$ & $T$
		& One factor & Three factors
		& One factor & Three factors \\
		\midrule
		Gaussian
		& 200 & 200
		& 0.95 & 0.96
		& 0.95 & 0.94 \\
		& 200 & 500
		& 0.95 & 0.96
		& 0.95 & 0.94 \\
		& 500 & 200
		& 0.95 & 0.96
		& 0.94 & 0.94 \\
		& 500 & 500
		& 0.95 & 0.96
		& 0.94 & 0.94 \\
		& 500 & 1000
		& 0.95 & 0.95
		& 0.95 & 0.94 \\
		& 1000 & 500
		& 0.94 & 0.96
		& 0.94 & 0.94 \\
		& 1000 & 1000
		& 0.95 & 0.95
		& 0.95 & 0.95 \\
		\addlinespace
		
		Student's $t_4$
		& 200 & 200
		& 0.92 & 0.92
		& 0.90 & 0.90 \\
		& 200 & 500
		& 0.94 & 0.93
		& 0.92 & 0.92 \\
		& 500 & 200
		& 0.96 & 0.93
		& 0.96 & 0.96 \\
		& 500 & 500
		& 0.94 & 0.96
		& 0.93 & 0.93 \\
		& 500 & 1000
		& 0.94 & 0.94
		& 0.92 & 0.92 \\
		& 1000 & 500
		& 0.96 & 0.95
		& 0.94 & 0.94 \\
		& 1000 & 1000
		& 0.95 & 0.96
		& 0.94 & 0.94 \\
		\addlinespace
		
		Student's $t_5$
		& 200 & 200
		& 0.94 & 0.93
		& 0.92 & 0.92 \\
		& 200 & 500
		& 0.94 & 0.95
		& 0.92 & 0.92 \\
		& 500 & 200
		& 0.96 & 0.93
		& 0.93 & 0.93 \\
		& 500 & 500
		& 0.96 & 0.96
		& 0.95 & 0.94 \\
		& 500 & 1000
		& 0.94 & 0.96
		& 0.93 & 0.93 \\
		& 1000 & 500
		& 0.95 & 0.96
		& 0.94 & 0.94 \\
		& 1000 & 1000
		& 0.95 & 0.95
		& 0.95 & 0.95 \\
		\bottomrule
	\end{tabular}
	\endgroup
	
	\medskip
	\begin{minipage}{0.96\textwidth}
		\footnotesize
		The columns under conditional locations report
		$\mathrm{PICP}_{C}$, the empirical coverage probability of the
		prediction interval for the common conditional location
		$C_{i,t|t-1}$. The columns under conditional variances report
		$\mathrm{PICP}_{P}$, the empirical coverage probability of the
		prediction interval for the common conditional variance
		$P_{i,t|t-1}$. The nominal coverage probability is $0.95$.
		Results are based on 300 replications.
	\end{minipage}
\end{table}

\begin{table}[!h]
	\centering
	\caption{Monte Carlo coverage probabilities for the joint conditional
		location and variance models}
	\label{tab:sim_joint_pca_picp}
	\scriptsize
	\begingroup
	\setlength{\tabcolsep}{5pt}
	\begin{tabular}{@{}lccrrrr@{}}
		\toprule
		& & &
		\multicolumn{2}{c}{One factor} &
		\multicolumn{2}{c}{Three factors} \\
		\cmidrule(lr){4-5}\cmidrule(lr){6-7}
		Distribution & $N$ & $T$
		& $\mathrm{PICP}_{C}$ & $\mathrm{PICP}_{P}$
		& $\mathrm{PICP}_{C}$ & $\mathrm{PICP}_{P}$ \\
		\midrule
		Gaussian
		& 200 & 200  & 0.94 & 0.91 & 0.94 & 0.91 \\
		& 200 & 500  & 0.95 & 0.92 & 0.95 & 0.92 \\
		& 500 & 200  & 0.95 & 0.93 & 0.96 & 0.92 \\
		& 500 & 500  & 0.95 & 0.94 & 0.95 & 0.94 \\
		& 500 & 1000 & 0.95 & 0.95 & 0.95 & 0.94 \\
		& 1000 & 500  & 0.95 & 0.94 & 0.96 & 0.95 \\
		& 1000 & 1000 & 0.95 & 0.95 & 0.95 & 0.96 \\
		\addlinespace
		
		Student's $t_4$
		& 200 & 200  & 0.93 & 0.89 & 0.92 & 0.87 \\
		& 200 & 500  & 0.95 & 0.91 & 0.94 & 0.91 \\
		& 500 & 200  & 0.95 & 0.91 & 0.93 & 0.90 \\
		& 500 & 500  & 0.95 & 0.90 & 0.96 & 0.90 \\
		& 500 & 1000 & 0.94 & 0.91 & 0.93 & 0.91 \\
		& 1000 & 500  & 0.93 & 0.90 & 0.92 & 0.90 \\
		& 1000 & 1000 & 0.95 & 0.93 & 0.95 & 0.92 \\
		\addlinespace
		
		Student's $t_5$
		& 200 & 200  & 0.94 & 0.90 & 0.94 & 0.89 \\
		& 200 & 500  & 0.95 & 0.90 & 0.96 & 0.90 \\
		& 500 & 200  & 0.96 & 0.92 & 0.96 & 0.91 \\
		& 500 & 500  & 0.95 & 0.92 & 0.94 & 0.92 \\
		& 500 & 1000 & 0.95 & 0.93 & 0.94 & 0.93 \\
		& 1000 & 500  & 0.95 & 0.93 & 0.96 & 0.93 \\
		& 1000 & 1000 & 0.95 & 0.94 & 0.95 & 0.94 \\
		\bottomrule
	\end{tabular}
	\endgroup
	
	\medskip
	\begin{minipage}{0.96\textwidth}
		\footnotesize
		$\mathrm{PICP}_{C}$ is the empirical coverage probability of the
		prediction interval for the common conditional location
		$C_{i,t|t-1}$, whereas $\mathrm{PICP}_{P}$ is the empirical coverage
		probability of the prediction interval for the common conditional
		variance $P_{i,t|t-1}$. The nominal coverage probability is $0.95$.
		Results are based on 300 replications.
	\end{minipage}
\end{table}

Third, we compute the asymptotic prediction intervals implied by Proposition \ref{prop:sequential_QMLE_expansion} when using the joint conditional location--variance estimator. Table \ref{tab:sim_joint_pca_picp} reports the coverage probabilities for the joint common conditional locations and
the common conditional variances. In Appendix \ref{appendix_MC} we report additional results on the goodness of the recovery of the time-varying parameters $\big\{\dhathatpsithetatti\big\}_{t\in\Nn}$.

\newpage

\bibliographystyle{Chicago}

\bibliography{biblio}


\newpage
\small
\appendix

\section{Score for the conditional variance with Student's $t$ updates}
\label{app:creal}
Following the proof of \citet[Theorem 1]{Creal2011} 
\begin{align}
\dut&= \frac{1}{2}
 \dSt\inv
 \dW_r\trans \bra{\mathbf J_t\trans\otimes \mathbf J_t\trans} 
 \bra{
 \frac{1+\nu^{-1}r}{1+\nu^{-1} \bra{\dFt\trans\mathbf J_t\trans \mathbf J_t\dFt -2}} \bra{\mathbf J_t\dFt\otimes \mathbf J_t\dFt}
 -\vect\dIdentityr
 },\nonumber
\end{align}
where $\mathbf J_t$ is such that $\dSigmatti\inv=\mathbf J_t\trans\mathbf J_t$ and $\mathbf J_t$ can be obtained from any convenient matrix decomposition procedure. Since $\dSigmatti$ is diagonal and positive definite we can set $\mathbf J_t=\dSigmatti^{-1/2}$ and $\mathbf J_t\trans=\dSigmatti^{-1/2}$, hence,
\begin{align}
\!\!\dut&= \frac{1}{2}
 \dSt\inv
 \dW_r\trans \bra{\dSigmatti^{-1/2}\otimes\dSigmatti^{-1/2}} 
 \bra{
 \frac{1+\nu^{-1}r}{1+\nu^{-1} \bra{\dFt\trans\dSigmatti^{-1}\dFt -2}} \bra{\dSigmatti^{-1/2}\dFt\otimes\dSigmatti^{-1/2} \dFt}
 -\vect\dIdentityr
 },\nonumber
\end{align}
or, equivalently,
\begin{equation*}
\dut = 
\frac{1}{2}
 \dSt\inv
 \dW_r\trans
\bra{\dSigmatti\inv \otimes \dSigmatti\inv}
\bra{
 \frac{\nu + r }{\nu - 2 + \dFt\trans\dSigmatti\inv\dFt} \bra{\dFt \otimes \dFt} - \vect \dSigmatti
}
,
\end{equation*}
{Furthermore, by adopting the rescaling 
\[
 \dSt\inv
=2\dW_N\trans\bra{\dSigmatti \otimes \dSigmatti} \dW_N,
\]
we have
\begin{align}
\dut&=\dW_r\trans \bra{
 \frac{\nu + r }{\nu - 2 + \dFt\trans\dSigmatti\inv\dFt} \bra{\dFt \otimes \dFt} - \vect \dSigmatti
}.\nonumber
\end{align}
And further calculations show that
\[
 \dSt = \frac 12 \dW_r\trans
\bra{\dSigmatti\inv \otimes \dSigmatti\inv} \dW_r,
\] 
which is proportional to the conditional Fisher Information matrix used by  \citet[Theorem 1]{Creal2011}} who show that:
\begin{equation*}
\dSt =
\frac{1}{4}
\dW_r\trans
\bra{\dSigmatti^{-1/2} \otimes \dSigmatti^{-1/2}}
\bra{
g \dG - \vect\dIdentityr \vect\dIdentityr\trans
}
\bra{\dSigmatti^{-1/2} \otimes \dSigmatti^{-1/2}}
\dW_r,
\end{equation*}
with $g = \frac{\nu+r}{\nu+r+2}$ and:
$
\dG\sbra{\bra{i-1} N + \ell, \bra{j-1} r + m}
=
\delta_{ij}\delta_{\ell m}
+
\delta_{i\ell}\delta_{j m}
+
\delta_{im}\delta_{j \ell}
,
$
where $\delta_{ij}$ is the Kronecker delta.

\section{Conditional variance models with leverage effects}
\label{app:leverage_effects}

The time-varying variance models introduced in Section \ref{sec_model} can be extended to incorporate leverage effects, in the spirit of the GJR-GARCH model of \citet{Glosten1993} and the asymmetric vector GARCH model of \citet{Mcaleer2009}. In particular, for the multivariate multiplicative model in \eqref{eq:SD_Model_covariance}, we augment the updating equation in \eqref{eq:SD_Model_sigmat} to allow for asymmetric effects across equations, so that the model becomes 
\begin{align}
\label{eq:SD_Model_asymmetric_sigmat1}
&\dFt = \boldsymbol q\bra{\dpsitti, \dvt} =\dsqrtSigmatti\dvt, 
\quad
\vect\dSigmatti = \dW_r \dsigmatti,
\\
\label{eq:SD_Model_asymmetric_sigmat3}
&\dsigmatpt = \ddelta + \dPhi \dsigmatti + \bra{\dK +\dUpsilon \boldsymbol{I}\bra{\dFt}} \dut,
\end{align}
where $\dUpsilon \in \Rr^{r\times r}$ is the additional matrix of static parameters and $\dI\bra{\dFt} = \diag\big(\dIndicator_{\cbra{F_{1t} \leq 0}},\ldots, \dIndicator_{\cbra{F_{rt} \leq 0}}\big)$ is an $r\times r$ diagonal matrix with indicator functions on the main diagonal. 
 
The new specification for the time-varying conditional variances in \eqref{eq:SD_Model_asymmetric_sigmat3} is valid for both the Gaussian or Student's $t$ updates, that is, by either assuming that noise process $\cbra{\dvt}_{t\in\Zz}$ satisfies $\dvt \sim \calN(\boldsymbol{0}_r, \boldsymbol{I}_r)$ or $\dvt \sim t_\nu(\boldsymbol{0}_r, \boldsymbol{I}_r)$, respectively. Consequently, once the factors has been retrieved via PCA, the unknown parameter vector, which is given by $\dtheta = \big(\ddelta\trans, \vect \dPhi\trans, \vect \dK\trans, \vect\dUpsilon\trans\big)\trans$ for Gaussian updates or is defined as $\dtheta = \big(\nu, \ddelta\trans, \vect \dPhi\trans, \vect \dK\trans, \vect\dUpsilon\trans\big)\trans$ for Student's $t$ updates, can be easily estimated using the QMLE in \eqref{eq:MLE}, by employing either the $\log$-likelihood in \eqref{eq:log_likelihood_gaussian_covariances} or in \eqref{eq:log_likelihood_student_covariances}, respectively. Therefore, the two-stage estimation strategy outlined in Section \ref{sec_estimator} can be readily employed, and the asymptotic theory developed in Section \ref{sec:asymptotic_theory} applies straightforwardly.


\section{Parameter identifications and the mapping $\boldsymbol r_{\dH}(\cdot)$}
\label{appendix_parameter_identifications}

For a generic $r\times r$ invertible matrix $\dH$ consider the mapping
$\dtheta_{\dH} = \boldsymbol r_{\dH}(\dtheta_0)$,
such that $\dtheta_{\dH} $ are the parameters governing the dynamics of the transformed factors $\dH \dF_t$. We give below the explicit forms of the  time-varying means or variances when computed using the transformed parameters, i.e., $\dpsitti (\dtheta_{\dH})$ which is the solution of  \eqref{eq:SD_Model_psit0} when using as parameters $\dtheta_{\dH}$ instead of $\dtheta_0$.

When using $\dH \dF_t$, then \eqref{eq:Model_yt0} becomes
\begin{align}
&\dH\dFt =\dH \boldsymbol q\bra{\dpsitti\bra{\dtheta_0}, \dvt}=\left\{
\begin{array}{ll}
\dH \dmutti(\dtheta_0) + \dH\dvt, & \dvt\sim\bra{\boldsymbol 0_r,\dSigma_0},\\
\dH \dSigmatti^{1/2}(\dtheta_0)\dv_t,& \dvt\sim\bra{\boldsymbol 0_r,\boldsymbol I_r}.
\end{array}
\right.\label{eq:Model_ytH}
\end{align}


\subsection{Time-varying locations}
From \eqref{eq:Model_ytH},
\begin{equation}\label{eq:transform_mu}
\dmutti(\dtheta_{\dH}) = \dH \dmutti (\dtheta_0).
\end{equation}
Therefore, for Gaussian updates
\begin{align*}
\dmutpt(\dtheta_{\dH}) &= \underbrace{\bra{\dH\dPhi_0 \dH\inv}}_{\dPhi_{\dH}} \underbrace{\dH\dmutti(\dtheta_0)}_{\dmutti(\dtheta_{\dH})} + \underbrace{\bra{\dH \dK_0 \dH\inv}}_{\dK_{\dH}} \dH \bra{\dFt - \dmutti(\dtheta_0)}\nonumber\\
&={\dPhi_{\dH}} {\dmutti(\dtheta_{\dH})} + {\dK_{\dH}}  \bra{\dH\dFt - \dmutti(\dtheta_{\dH})},
\end{align*}
and $\dSigma_{\dH} = \dH\dSigma_0\dH\trans$. 
While, for Student's $t$ updates
\begin{align*}
\dmutpt(\dtheta_{\dH})
=&\,
 \underbrace{\bra{\dH\dPhi_0 \dH\inv}}_{\dPhi_{\dH}} \underbrace{\dH\dmutti(\dtheta_0)}_{\dmutti(\dtheta_{\dH})}\nonumber\\ 
&+ \underbrace{\bra{\dH \dK_0 \dH\inv}}_{\dK_{\dH}} \dH \frac{\bra{\dFt - \dmutti(\dtheta_0)}}{1 + \bra{\dFt - \dmutti(\dtheta_0)}\trans \dH\trans 
(\underbrace{\dH \dSigma_0 \dH\trans}_{\dSigma_{\dH}})\inv  \dH \bra{\dFt - \dmutti(\dtheta_0)}/\nu_0}\nonumber\\
=&\,
{\dPhi_{\dH}} {\dmutti(\dtheta_{\dH})} + {\dK_{\dH}} \frac{ \bra{\dH\dFt - \dmutti(\dtheta_{\dH})}}{1 + \bra{\dH\dFt - \dmutti(\dtheta_{\dH})}\trans  
\dSigma_{\dH}\inv   \bra{\dH\dFt - \dmutti(\dtheta_{\dH})}/\nu_{\dH}},
\end{align*}
and $\nu_{\dH}=\nu_0$.

\subsection{Time-varying covariances}
In this case, by taking the expectation of the outer product of \eqref{eq:Model_ytH}, we have
$
{\dSigmatti}(\dtheta_{\dH}) = \dH {\dSigmatti}(\dtheta_0)\dH\trans
$, so
\[
\vect{\dSigmatti}(\dtheta_{\dH}) =\dW_r {\dsigmatti}(\dtheta_{\dH}) =\bra{\dH\otimes\dH}\dW_r {\dsigmatti}(\dtheta_0),
\]
which implies
\begin{equation}\label{eq:transform_sigma}
{\dsigmatti}(\dtheta_{\dH})=\dW_r\trans  \bra{\dH\otimes \dH} \dW_r \dsigmatti (\dtheta_0).
 \end{equation}
 
 Therefore, for Gaussian updates
   \begin{align}
\dsigmatpt(\dtheta_{\dH})
=&\,\underbrace{\dW_r\trans  \bra{\dH\otimes \dH} \dW_r\ddelta_0}_{\ddelta_{\dH}} \nonumber\\
&+\underbrace{\dW_r\trans  \bra{\dH\otimes \dH} \dW_r\dPhi_0 \cbra{\dW_r\trans  \bra{\dH\otimes \dH} \dW_r}\inv}_{\dPhi_{\dH}} \dW_r\trans  \bra{\dH\otimes \dH} \dW_r \dsigmatti(\dtheta_0)\nonumber\\
&+\underbrace{\dW_r\trans  \bra{\dH\otimes \dH} \dW_r \dK_0 \cbra{\dW_r\trans  \bra{\dH\otimes \dH} \dW_r}\inv}_{\dK_{\dH}} \nonumber\\
&\qquad\times\dW_r\trans \bra{\dH\otimes \dH} \dW_r\dW_r\trans \bra{\bra{\dFt\otimes \dFt} -  \dW_r \dsigmatti(\dtheta_0)}\nonumber\\
=&\, \ddelta_{\dH}+\dPhi_{\dH}\dsigmatti(\dtheta_{\dH})+\dK_{\dH} \dW_r\trans \bra{\bra{\dH\dFt\otimes \dH\dFt} -  \vect \dSigmatti(\dtheta_{\dH})}.\nonumber
 \end{align}
While, for Student's $t$ updates
  \begin{align}
\dsigmatpt(\dtheta_{\dH})
=&\,\underbrace{\dW_r\trans  \bra{\dH\otimes \dH} \dW_r\ddelta_0}_{\ddelta_{\dH}} \nonumber\\
&+\underbrace{\dW_r\trans  \bra{\dH\otimes \dH} \dW_r\dPhi_0 \cbra{\dW_r\trans  \bra{\dH\otimes \dH} \dW_r}\inv}_{\dPhi_{\dH}} \dW_r\trans  \bra{\dH\otimes \dH} \dW_r \dsigmatti(\dtheta_0)\nonumber\\
&+\underbrace{\dW_r\trans  \bra{\dH\otimes \dH} \dW_r \dK_0 \cbra{\dW_r\trans  \bra{\dH\otimes \dH} \dW_r}\inv}_{\dK_{\dH}} \dW_r\trans \bra{\dH\otimes \dH} \dW_r
 \nonumber\\
&\qquad \times \dW_r\trans
\bra{
\frac{\nu + N }{\nu - 2 +\dFt\trans\dH\trans \bra{\dH \dSigmatti(\dtheta_0) \dH\trans}\inv\dH \dFt} \bra{\dFt \otimes \dFt} - \dW_r \dsigmatti(\dtheta_0)\nonumber
}\\
=&\, \ddelta_{\dH}+\dPhi_{\dH}\dsigmatti(\dtheta_{\dH})\nonumber\\
&+\dK_{\dH} \dW_r\trans \bra{\frac{\nu + N }{\nu - 2 +\dFt\trans\dH\trans \bra{\dSigmatti(\dtheta_{\dH})}\inv\dH \dFt}\bra{\dH\dFt\otimes \dH\dFt} -  \vect \dSigmatti(\dtheta_{\dH})}.\nonumber
 \end{align}


\section{Proofs}
\label{appendix_proofs}

\subsubsection*{Proof of Proposition \ref{prop:rate_yt_decomp}}

First, let us notice that all assumptions in \citet{Bai2003} hold also in our setting. Indeed,
\begin{inparaenum}
\item Assumption A in \citet{Bai2003} is equivalent to our Assumption \ref{ass:pram_SE}$(iii)$ plus Remark \ref{rem:birk};
\item Assumption B in \citet{Bai2003}  is equivalent to our Assumption \ref{ass:factor_identifiability}; 
\item Assumption C.1 in \citet{Bai2003} is not necessary since,  from closer inspection of the original proofs, we see that we just need $\Ee[e_{it}^4]<\infty$ and not finite 8th moments and this follows from our Assumption \ref{ass:idiosyncratic}$(iv)$;
\item Assumptions C.2-C.4 in \citet{Bai2003} are implied by our Assumptions \ref{ass:idiosyncratic}$(i)$-$(ii)$;
\item Assumption C.5 in \citet{Bai2003}  is not necessary since, from closer inspection of the original proofs, we see that we just need it to hold with power 2 and not 4 and this follows from our Assumption \ref{ass:idiosyncratic}$(v)$;
\item Assumption D in \citet{Bai2003} is implied by our Assumption \ref{ass:idiosyncraticY}$(i)$;
\item Assumptions E.1-E2 in \citet{Bai2003} are implied by our Assumptions \ref{ass:idiosyncratic}$(i)$-$(ii)$;
\item Assumption F.1, F.2, and F.4 in \citet{Bai2003} are equivalent to our Assumptions \ref{ass:idiosyncraticY}$(ii)$-$(iv)$;
\item Assumption F.3 in \citet{Bai2003} is equivalent to our Assumption \ref{ass:idiosyncratic}(iii);
\item Assumption G in \citet{Bai2003} is equivalent to  our Assumption \ref{ass:idiosyncratic}$(v)$.
\end{inparaenum}

Consider the $T\times N$ data matrix $\boldsymbol X=(\dx_1\cdots \dx_T)\trans$ so that in matrix notation the factor model \eqref{eq:Model_xt0} is 
\begin{equation}
\label{eq:Model_xt_mat}
\boldsymbol X = \boldsymbol F\dLambda_0\trans + \boldsymbol E,  
\end{equation}
where
$\boldsymbol F=(\dF_1\cdots \dF_T)\trans$ is $T\times r$ 
and
$\boldsymbol E=(\dxi_1\cdots \dxi_T)\trans$ is $T\times N$. 

By definition the PC estimators are such that they solve the following minimization:
\begin{equation}\label{eq:minimizza}
\min_{\substack{\boldsymbol L\in\mathbb R^{N\times r}\\ \boldsymbol G\in\mathbb R^{T\times r}}} \frac 1{NT} \left\Vert\boldsymbol X-\boldsymbol G\boldsymbol L\trans\right\Vert^2,
\;\text{ s.t. }\; \frac{\boldsymbol L\trans\boldsymbol L} N= \boldsymbol I_r \;\text{ and }\;\frac{\boldsymbol G\trans \boldsymbol G}T\; \text{is diagonal},
\end{equation}
where the constraints are imposed in order to identify the minimum. The solution is obtained by noticing that for given loadings the factors are just the linear projection of the loadings onto the data. Hence, the problem in \eqref{eq:minimizza} can be solved by first solving
\[
\widehat{\dLambda}=\arg\!\!\!\!\!\min_{\boldsymbol L\in\mathbb R^{N\times r}} \frac 1{NT} \left\Vert\boldsymbol X-\boldsymbol X\boldsymbol L\bra{\boldsymbol L\trans\boldsymbol L}^{-1}\boldsymbol L\trans\right\Vert^2, \;\text{ s.t. }\; \frac{\boldsymbol L\trans\boldsymbol L} N= \boldsymbol I_r,
\]
which is solved by the estimated loadings given in \eqref{eq:loadPCA}, i.e., 
$\widehat{\boldsymbol \Lambda}=\sqrt N\widehat{\boldsymbol P}$, where $\widehat{\boldsymbol P}$ are the eigenvectors of $T^{-1}\boldsymbol X\trans \boldsymbol X$ corresponding to its $r$ largest eigenvalues. The estimated factors are obtained by linear projection as in \eqref{eq:factor_lin_proj}, i.e.,
$\widehat{\boldsymbol F} = \widehat{\boldsymbol X} \widehat{\boldsymbol \Lambda}(\widehat{\boldsymbol \Lambda}\trans \widehat{\boldsymbol \Lambda})^{-1} = N^{-1/2} \widehat{\boldsymbol X} \widehat{\boldsymbol P}$.
Note that, as required, $\widehat{\boldsymbol \Lambda}\trans \widehat{\boldsymbol \Lambda}=N \boldsymbol I_r$ and 
$T^{-1}{\widehat{\boldsymbol F}\trans\widehat{\boldsymbol F}} = {N}^{-1} \widehat{\boldsymbol M}$,
where $\widehat{\boldsymbol M}$ is the $r\times r$ diagonal matrix of the largest $r$ eigenvalues of $T^{-1}\boldsymbol X\trans \boldsymbol X$.

An equivalent solution of \eqref{eq:minimizza} same solution is however obtained also by noticing that for given factors the loadings are just the linear projection of the factors onto the data. Hence, the problem in \eqref{eq:minimizza} can be also solved by first solving
\[
\widehat{\boldsymbol F}=\arg\!\!\!\!\!\min_{\boldsymbol G\in\mathbb R^{T\times r}} \frac 1{NT} \left\Vert\boldsymbol X-\boldsymbol G\bra{\boldsymbol G\trans\boldsymbol G}^{-1}\boldsymbol G\trans \boldsymbol X\right\Vert^2, \;\text{ s.t. }\; \frac{\boldsymbol G\trans\boldsymbol G} T\; \text{is diagonal},
\]
which gives
\begin{equation}\label{eq:tildePC1}
\widehat{\boldsymbol F}= \widetilde{\boldsymbol P}\widetilde{\boldsymbol M}^{1/2}  ,
\end{equation}
where $\widetilde{\boldsymbol P}$ are the eigenvectors of $N^{-1}\boldsymbol X \boldsymbol X\trans$ corresponding to its $r$ largest eigenvalues which are collected in the $r\times r$ diagonal matrix $\widetilde{\boldsymbol M}$. Notice that, since $(NT)^{-1}\boldsymbol X \boldsymbol X\trans$ and $(NT)^{-1}\boldsymbol X\trans \boldsymbol X$ have the same eigenvalues, then
\begin{equation}\label{eq:sameval}
\frac{\widehat{\boldsymbol F}\trans \widehat{\boldsymbol F}}{T} =\frac {\widetilde{\boldsymbol M}^{1/2}  \widetilde{\boldsymbol P}\trans \widetilde{\boldsymbol P}\widetilde{\boldsymbol M}^{1/2}}T  =
\frac{\widetilde{\boldsymbol M}}{T}=\frac{\widehat{\boldsymbol M}}{N},
\end{equation}
which is diagonal as required. The estimated loadings are then obtained by linear projection, i.e.,
\begin{equation}\label{eq:tildePC2}
\widehat {\boldsymbol \Lambda} = \boldsymbol X\trans\widehat {\boldsymbol F}
\bra{\widehat {\boldsymbol F}\trans \widehat {\boldsymbol F}}^{-1} = \frac{\boldsymbol X\trans \widehat{\boldsymbol F}}{T}\bra{\frac{\widehat{\boldsymbol M}}{N}}^{-1},
\end{equation}
where we used the last equality in \eqref{eq:sameval}. It follows, that, as requested
\begin{align}
\frac{\widehat{\boldsymbol \Lambda}\trans \widehat{\boldsymbol \Lambda}}{N}&=  
\bra{\frac{\widehat{\boldsymbol M}}{N}}^{-1}
\frac{ \widehat{\boldsymbol F}\trans  {\boldsymbol X}  {\boldsymbol X}\trans  \widehat{\boldsymbol F}}{NT^2}
\bra{\frac{\widehat{\boldsymbol M}}{N}}^{-1}
= 
\bra{\frac{\widetilde{\boldsymbol M}}{T}}^{-1}
\frac{
\widetilde{\boldsymbol M}^{1/2}
\widetilde{\boldsymbol P}\trans
 {\boldsymbol X}  {\boldsymbol X}\trans
\widetilde{\boldsymbol P}
\widetilde{\boldsymbol M}^{1/2}
}{NT^2}
\bra{\frac{\widetilde{\boldsymbol M}}{T}}^{-1}\nonumber\\
&=\bra{\frac{\widetilde{\boldsymbol M}}{T}}^{-1}
\frac{
\widetilde{\boldsymbol M}^{1/2}
\widetilde{\boldsymbol P}\trans
\widetilde{\boldsymbol P}
\widetilde{\boldsymbol M}
\widetilde{\boldsymbol M}^{1/2}
}{T^2}
\bra{\frac{\widetilde{\boldsymbol M}}{T}}^{-1}=\dIdentityN.\nonumber
\end{align}
Given the above, hereafter, we can use the formulations of the PC estimators in \eqref{eq:tildePC1} and \eqref{eq:tildePC2}, instead of \eqref{eq:factor_lin_proj} and \eqref{eq:loadPCA}, to study their asymptotic properties. 

\noindent
$(a)$ Starting from the factors, as formulated in \eqref{eq:tildePC1}, by definition of eigenvectors we have
\begin{align}\label{eq:evecXX}
\frac{\boldsymbol X\boldsymbol X\trans}{NT} \widehat{\boldsymbol F}=\widehat{\boldsymbol F}\frac{\widetilde{\boldsymbol M}}{T}.
\end{align}
By substituting \eqref{eq:Model_xt_mat} in \eqref{eq:evecXX}, using  the last equality in \eqref{eq:sameval}, and rearranging we have
\begin{align}\label{eq:evecXX2}
\widehat{\boldsymbol F} - \boldsymbol F \sbra{\frac{\dLambda_0\trans\dLambda_0}{N} \frac{\boldsymbol F\trans \widehat{\boldsymbol F} }{T} \bra{\frac{\widehat {\boldsymbol M}}N}^{-1}}&= \cbra{
\frac{\boldsymbol E\dLambda_0\boldsymbol F\trans\widehat{\boldsymbol F}}{NT}
+
\frac{\boldsymbol F\dLambda_0\trans \boldsymbol E\trans \widehat{\boldsymbol F}}
{NT}
+ \frac{\boldsymbol E\boldsymbol E\trans \widehat{\boldsymbol F}}{NT}
}
\bra{\frac{\widehat {\boldsymbol M}}N}^{-1}.
\end{align}
By letting
\begin{equation}\label{eq:ACCA}
\widehat{\dH} = \bra{\frac{\widehat {\boldsymbol M}}N}^{-1} \frac{\widehat{\boldsymbol F}\trans \boldsymbol F }{T} \frac{\dLambda_0\trans\dLambda_0}{N},
\end{equation}
and by taking the $t$-th row of \eqref{eq:evecXX2},  we have
\begin{align}\label{eq:evecXX3}
\widehat{\dF}_t - \widehat{\dH} \dF_t =&\,\, \bra{\frac{\widehat {\boldsymbol M}}N}^{-1}\left\{
\underbrace{\frac 1 {NT} \sum_{s=1}^T \widehat{\dF}_s\dF_s\trans {\dLambda}_0\trans\dxi_t}_{A} +
\underbrace{\frac 1 {NT} \sum_{s=1}^T\widehat{\dF}_s\dF_t\trans {\dLambda}_0\trans\dxi_s}_{B}\right.\nonumber\\ 
&\left.+
\underbrace{\frac 1 {NT} \sum_{s=1}^T\widehat{\dF}_s\Ee[\dxi_s\trans \dxi_t]}_{C}+
\underbrace{\frac 1 {NT} \sum_{s=1}^T\widehat{\dF}_s\bra{\dxi_s\trans \dxi_t-\Ee[\dxi_s\trans \dxi_t]}}_{D}
\right\},
\end{align}
which is the same expression as in \citet[equation (A.1)]{Bai2003}. Since all assumptions in \citet{Bai2003} hold also in this paper, then we can apply Lemma A.2 therein to \eqref{eq:evecXX3} and obtain 
\begin{align}\label{eq:consF}
\Vert A\Vert &= O_{\mathbb P}\bra{\frac 1{\sqrt N}},\qquad \Vert B\Vert = O_{\mathbb P}\bra{\frac 1{\sqrt N \min\bra{ \sqrt N,  \sqrt T}}},\\
\Vert C\Vert &= O_{\mathbb P}\bra{\frac 1{\sqrt T \min\bra{ \sqrt N,  \sqrt T}}}, \qquad \Vert D\Vert = O_{\mathbb P}\bra{\frac 1{\sqrt N \min\bra{ \sqrt N,  \sqrt T}}},\nonumber
\end{align}
Moreover, from Lemma A.3 in \citet{Bai2003} $\Vert (N^{-1}{\widehat {\boldsymbol M}})^{-1}\Vert = O_{\mathbb P}(1)$. By using \eqref{eq:consF} into \eqref{eq:evecXX3} we prove part (a).

\noindent
$(b)$ Now, consider the following spectral decomposition: $\dSigma_\Lambda^{1/2}\boldsymbol\Gamma_F\dSigma_\Lambda^{1/2}= \boldsymbol {\Upsilon}\boldsymbol V \boldsymbol {\Upsilon}\trans$,
where $\boldsymbol V$ is the $N\times N$ diagonal matrix of eigenvalues and the columns of $\boldsymbol {\Upsilon}$ are the corresponding eigenvectors. Denote as $\boldsymbol V_N$ the eigenvalues of $N^{-1}\boldsymbol\Lambda \boldsymbol\Gamma_F \boldsymbol\Lambda^\prime$. Then, because of Assumption \ref{ass:factor_identifiability} and Weyl's inequality \citep[Theorem 1]{MK04}
\begin{equation}\label{eq:BaiLemmaA3N}
\left\Vert\boldsymbol V_N - \boldsymbol V\right\Vert\le  
\left\Vert\frac {\boldsymbol\Lambda_0\trans\boldsymbol\Lambda_0}N   \boldsymbol\Gamma_F  -  \dSigma_\Lambda \boldsymbol\Gamma_F\right\Vert
= O_{\mathbb P}\bra{\frac 1{\sqrt N}},
\end{equation}
since $\boldsymbol V$ are also eigenvalues of $\dSigma_\Lambda \boldsymbol\Gamma_F$. 
Moreover, since (see e.g. \citealp{Birkhoff1931}),
\begin{equation}\label{eq:birk}
\left \Vert \frac{\boldsymbol F\trans\boldsymbol F}T-\boldsymbol\Gamma^F\right\Vert=O_{\Pp}\bra{\frac1{\sqrt T}}, 
\end{equation}
then, for all $N\in\Nn$, we have
\begin{equation}\label{eq:BaiLemmaA3T}
\left\Vert\frac{\widehat {\boldsymbol M}}N - \boldsymbol V_N\right\Vert \le 
\left \Vert \frac {\boldsymbol\Lambda_0\trans\boldsymbol\Lambda_0}N \frac{\boldsymbol F\trans\boldsymbol F}T-\frac {\boldsymbol\Lambda_0\trans\boldsymbol\Lambda_0}N\boldsymbol\Gamma^F\right \Vert= O_{\Pp}\bra{\frac1{\sqrt T}},
\end{equation}
where we used again Weyl's inequality.
Then, from \eqref{eq:BaiLemmaA3N} and \eqref{eq:BaiLemmaA3T},
\begin{equation}\label{eq:BaiLemmaA3}
\left\Vert\frac{\widehat {\boldsymbol M}}N - \boldsymbol V\right\Vert = O_{\Pp}\bra{\max \bra{\frac1{\sqrt N},\frac1{\sqrt T}}}.
\end{equation}
Denote as $\boldsymbol {\Upsilon}_{NT}$ the matrix of normalized eigenvectors of $(N^{-1}\boldsymbol\Lambda_0\trans\boldsymbol\Lambda_0)^{1/2}(T^{-1}\boldsymbol F\trans\boldsymbol F)(N^{-1}\boldsymbol\Lambda_0\trans\boldsymbol\Lambda_0)^{1/2}$, then by Davis Kahan Theorem \citep[Corollary 1]{yu15}, Assumption \ref{ass:factor_identifiability}, and \eqref{eq:birk},
\begin{equation}\label{eq:BaiLemmaA3evec}
\left\Vert
\boldsymbol {\Upsilon}_{NT}-\boldsymbol {\Upsilon}{\boldsymbol J}_{NT}
\right\Vert\le C 
\left\Vert
\bra{\frac{\boldsymbol\Lambda_0\trans\boldsymbol\Lambda_0}N}^{1/2}\bra{\frac{\boldsymbol F\trans\boldsymbol F}T}\bra{\frac{\boldsymbol\Lambda_0\trans\boldsymbol\Lambda_0}N}^{1/2}-\dSigma_\Lambda^{1/2}\boldsymbol\Gamma_F\dSigma_\Lambda^{1/2}
\right\Vert
\end{equation}
for some finite constant $C>0$ independent of $N$ and $T$ and where ${\boldsymbol J}_{NT}$ is a diagonal matrix with entries $\pm 1$ which depends on $N$ and $T$ accounting for the possible sign discordance between sample and population eigenvectors.  

Now, from Proposition 1 in  \citet{Bai2003} and using \eqref{eq:sameval}, we have
\begin{equation}\label{eq:BaiProp1}
\left\Vert \bra{\frac{\widehat{\boldsymbol M}}{N}}^{-1/2}  \frac {\widehat{\boldsymbol F}\trans\boldsymbol F } {T}  - \boldsymbol V^{1/2}\boldsymbol {\mathcal J}\boldsymbol {\Upsilon}\trans \dSigma_\Lambda^{-1/2} \right\Vert = \left\Vert \bra{\frac{\widetilde{\boldsymbol M}}{T}}^{-1/2}  \frac {\widehat{\boldsymbol F}\trans\boldsymbol F } {T}  - \boldsymbol V^{1/2}\boldsymbol {\mathcal J}\boldsymbol {\Upsilon}\trans \dSigma_\Lambda^{-1/2} \right\Vert =o_{\mathbb P}(1),
\end{equation}
where $\boldsymbol {\mathcal J}=\Pp\text{-}\lim_{N,T\to\infty} {\boldsymbol J}_{NT}$ is a diagonal matrix with entries $\pm 1$ independent of $N$ and $T$, which in the proof of \citet{Bai2003} is set to $\boldsymbol {\mathcal J}=\boldsymbol I_r$ for convention. Inspection of the proof of Proposition 1 in  \citet{Bai2003} shows that the leading rate in \eqref{eq:BaiProp1} is determined by  \eqref{eq:BaiLemmaA3} and  \eqref{eq:BaiLemmaA3evec}. Hence, from \eqref{eq:BaiLemmaA3} and \eqref{eq:BaiProp1}
\begin{equation}\label{eq:BaiProp1bis}
\left\Vert   \frac {\widehat{\boldsymbol F}\trans\boldsymbol F } {T}  - \boldsymbol V\boldsymbol {\mathcal J}\boldsymbol {\Upsilon}\trans \dSigma_\Lambda^{-1/2} \right\Vert =O_{\Pp}\bra{\max \bra{\frac1{\sqrt N},\frac1{\sqrt T}}}.
\end{equation}
Let $\dH_0=\boldsymbol {\mathcal J} \boldsymbol {\Upsilon}\trans  \dSigma_\Lambda^{1/2}$ and notice that it is finite and positive definite because of Assumptions \ref{ass:pram_SE}(i) and \ref{ass:factor_identifiability}, which imply that $\dSigma_\Lambda^{1/2}\boldsymbol\Gamma_F\dSigma_\Lambda^{1/2}$ has finite and positive eigenvalues. Then, by using Assumption \ref{ass:factor_identifiability}, \eqref{eq:BaiProp1bis}, and \eqref{eq:BaiLemmaA3}, into \eqref{eq:ACCA},
\begin{equation}\label{eq:ACCAlim}
\left \Vert  \widehat{\dH} -\dH_0 \right\Vert=\left \Vert  \widehat{\dH} - \boldsymbol {\mathcal J} \boldsymbol {\Upsilon}\trans  \dSigma_\Lambda^{1/2} \right\Vert=O_{\Pp}\bra{\max \bra{\frac1{\sqrt N},\frac1{\sqrt T}}}.
\end{equation}
Therefore, from \eqref{eq:evecXX3} and \eqref{eq:consF}, for any $t=1,\ldots, T$, if $\sqrt N/T\to 0$, as $N,T\to\infty$,
\begin{align}
\sqrt N \bra{\widehat{\dF}_t - \widehat{\dH} \dF_t}  &=\bra{\frac{\widehat {\boldsymbol M}}N}^{-1} \bra{\frac {\widehat{\boldsymbol F}\trans \boldsymbol F } {T} } \frac {{\dLambda}_0\trans\dxi_t}{\sqrt N}  +o_{\mathbb P}(1)= \widehat{\dH} \bra{\frac{\dLambda_0\trans\dLambda_0}{N}}^{-1}\bra{\frac 1{\sqrt N}\sum_{i=1}^N \boldsymbol{\lambda}_{0i} e_{it}}+o_{\mathbb P}(1)\nonumber\\
&=  \dH_0 \dSigma_\Lambda\inv \bra{\frac 1{\sqrt N}\sum_{i=1}^N \boldsymbol{\lambda}_{0i} e_{it}}+o_{\mathbb P}(1)\Rightarrow\mathcal N\bra{\boldsymbol 0_N,\dH_0 \dSigma_\Lambda\inv \boldsymbol \Pi_{t,t}  \dSigma_\Lambda\inv \dH_0\trans },\nonumber
\end{align}
where we used also \eqref{eq:ACCA}, \eqref{eq:ACCAlim}, Assumptions \ref{ass:factor_identifiability} and \ref{ass:idiosyncratic}(iii), and Slutsky's theorem. This proves part (b).

\noindent
$(c)$ Consider the estimated loadings, as formulated in \eqref{eq:tildePC2}. Then, using \eqref{eq:Model_xt_mat}  and \eqref{eq:sameval}
\begin{align}\label{eq:sviluppoL}
\widehat{\dLambda}&=\frac{\boldsymbol X\trans \widehat{\boldsymbol F}}{T}\bra{\frac{\widehat{\boldsymbol M}}{N}}^{-1}
=\cbra{ \frac{\dLambda_0\boldsymbol F\trans \widehat{\boldsymbol F}}{T} +
\frac{\boldsymbol E\trans \widehat{\boldsymbol F}}{T}}\bra{\frac{\widehat{\boldsymbol M}}{N}}^{-1}\nonumber\\
&= \dLambda_0 \widehat{\dH}^{-1} +
\cbra{
\frac{\dLambda_0\bra{\boldsymbol F-\widehat{\boldsymbol F} \widehat{\dH}^{-1\top}}\trans \widehat{\boldsymbol F}}{T}+ 
\frac{\boldsymbol E\trans \bra{\widehat{\boldsymbol F}-\boldsymbol F\widehat{\dH}\trans}}{T}
+\frac{\boldsymbol E\trans {\boldsymbol F} \widehat{\dH}\trans}{T}
}
\bra{\frac{\widehat{\boldsymbol M}}{N}}^{-1} ,
\end{align}
where, from \eqref{eq:ACCA}, we have
\begin{equation}\label{eq:ACCAinv}
\widehat{\dH}\inv = 
\bra{\frac{\dLambda_0\trans\dLambda_0}{N}}\inv
\bra{\frac{\widehat{\boldsymbol F}\trans \boldsymbol F }{T}}\inv
\frac{\widehat {\boldsymbol M}}N. 
\end{equation}
By taking the $i$th row of \eqref{eq:sviluppoL} and rearranging,  we have
\begin{align}\label{eq:sviluppoL1}
\widehat{\boldsymbol {\lambda}}_i- \widehat{\dH}^{-1\top} \boldsymbol {\lambda}_{0i} &= \bra{\frac{\widehat{\boldsymbol M}}{N}}^{-1}
\cbra{
\underbrace{\frac{\widehat{\dH}\boldsymbol F\trans \boldsymbol e_i}{T}}_{E}+
\underbrace{\frac{\bra{\widehat{\boldsymbol F}-{\boldsymbol F}\widehat{\dH}^{\top}}\trans \boldsymbol e_i}{T}}_{F}+
\underbrace{\frac{\widehat{\boldsymbol F}\trans\bra{\boldsymbol F-\widehat{\boldsymbol F}\widehat{\dH}^{-1\top}}\boldsymbol \lambda_{0i}}{T}}_{G}
},
\end{align}
where $\boldsymbol e_i=(e_{i1}\cdots e_{iT})\trans$. The expression in \eqref{eq:sviluppoL1} coincides with the one in \citet[equation just before equation (B.2)]{Bai2003} but for the first term. By Assumption \ref{ass:idiosyncraticY}(iv)
\begin{equation}\label{eq:termineE}
\Vert E\Vert = O_{\mathbb P}\bra{\frac 1{\sqrt T}},
\end{equation}
while Lemmas B.1 and B.3 in \citet{Bai2003} imply that
\begin{equation}\label{eq:termineFG}
\Vert F\Vert = O_{\mathbb P}\bra{\frac 1{\min(N, T)}}\;\text{ and }\; \Vert G\Vert = O_{\mathbb P}\bra{\frac 1{\min(N, T)}}.
\end{equation}
By using \eqref{eq:termineE} and \eqref{eq:termineFG} into \eqref{eq:sviluppoL1} and since $\Vert (N^{-1}{\widehat {\boldsymbol M}})^{-1}\Vert = O_{\mathbb P}(1)$ by \citet[Lemma A.3]{Bai2003}, we prove part (c).

\noindent
$(d)$ From \eqref{eq:sviluppoL1}, \eqref{eq:termineE}, and \eqref{eq:termineFG}, for any $i=1,\ldots, N$, if $\sqrt T/N\to 0$, as $N,T\to\infty$,
\begin{align}
\sqrt T\bra{\widehat{\boldsymbol {\lambda}}_i- \widehat{\dH}^{-1\top} \boldsymbol {\lambda}_{0i}} &= \bra{\frac{\widehat{\boldsymbol M}}{N}}^{-1} \frac{\widehat{\dH}\boldsymbol F\trans \boldsymbol e_i}{\sqrt T}+o_{\mathbb P}(1)= \bra{\frac{\widehat{\boldsymbol F}\trans \widehat{\boldsymbol F}}{T}}\inv \widehat{\dH}\bra{\frac 1{\sqrt T}\sum_{t=1}^T \dF_te_{it}}+o_{\mathbb P}(1)\nonumber\\
&= \bra{\frac{\widehat{\dH}{\boldsymbol F}\trans {\boldsymbol F}\widehat{\dH}\trans}{T}}\inv \widehat{\dH}\bra{\frac 1{\sqrt T}\sum_{t=1}^T \dF_te_{it}}+o_{\mathbb P}(1)\nonumber\\
&= \widehat{\dH}^{-1\top} \bra{\frac{{\boldsymbol F}\trans {\boldsymbol F}}{T}}\inv \bra{\frac 1{\sqrt T}\sum_{t=1}^T \dF_te_{it}}+o_{\mathbb P}(1)\label{eq:questaservePCA}\\
&= {\dH}_0^{-1\top} \boldsymbol \Gamma_F\inv \bra{\frac 1{\sqrt T}\sum_{t=1}^T \dF_te_{it}}+o_{\mathbb P}(1)
\Rightarrow \mathcal N\bra{\boldsymbol 0_N,
{\dH}_0^{-1\top} \boldsymbol \Gamma_{F}\inv
\boldsymbol \Xi_i
\boldsymbol \Gamma_{F}\inv
{\dH}_0^{-1}
},\nonumber
\end{align}
where we used also 
\eqref{eq:sameval} and part (a) which, together with Lemma B.3 in \citet{Bai2003}, implies $\Vert T^{-1} \widehat{\boldsymbol F}\trans \widehat{\boldsymbol F}-T^{-1}\widehat{\dH}{\boldsymbol F}\trans {\boldsymbol F}\widehat{\dH}\trans \Vert =o_{\mathbb P}(1)$, \eqref{eq:ACCAlim} together with the continuous mapping theorem, 
Assumptions \ref{ass:pram_SE}(i) and \ref{ass:idiosyncraticY}(iv), and Slutsky's theorem. This proves part (d).

\noindent
$(e)$ From parts (a) and (c) combined with \eqref{eq:ACCAlim} we prove part (e) and complete the proof. \qed

\subsubsection*{Proof of Proposition \ref{prop:rate_psit_decomp}}

$(a)$ To prove the first claim, let us embed the nonstationary filtered sequence $\cbra{\dhathatpsithetatti}_{t\in\Nn}$ in the following dynamical system
\begin{align}
\label{eq:gt1}
\dhathatpsithetatpt = g_t\bra{\dhathatpsithetatti, \dhatFt, \dtheta},
\end{align}
where by using \eqref{eq:SD_Model_psit}, the stochastic function $g_t:\calP \times \Rr^r \times \dTheta \to \calP$ is given by 
\begin{align}
\label{eq:gt2}
g_t\bra{\dhathatpsithetatti, \dhatFt, \dtheta}
=
\ddelta 
+
\dPhi \dhathatpsithetatti +\dK\dhathatut(\dtheta),
\end{align}
where $\dhathatut(\dtheta)$ is the scaled-score function evaluated at $\dhathatpsithetatti$ which is, in fact, a nonlinear function of $\dhatFt$.

Hence, for any fixed and non-stochastic starting point $\dpsi_{1|0}$, define the functions $g^{(k)}_t$, representing the $k$-iterations of the dynamic system, by the following recursions 
\begin{align*}
g_t^{(k+1)}\bra{\dpsi_{1|0},  \dhatF_1, \dots, \dhatF_{k+1}, \dtheta}
=
g_{t+k}\bra{ g_t^{(k)}(\dpsi_{1|0}, \dhatF_1, \dots, \dhatF_k, \dtheta), \dhatF_{k+1}, \dtheta }
,
\end{align*}
such that $g_t = g_t^{(1)}$. Note that, substituting $\dhatFt$ with $\widehat\dH\dFt$ in equations \eqref{eq:gt1} and \eqref{eq:gt2}, one obtains the analogous dynamical system for the sequence $\cbra{\dhatpsithetatti}_{t\in\Nn}$. Thus, by repeated applications of the mean value theorem, H\"{o}lder's and Minkowski's inequalities, we can show that
\begin{align*}
&\vvbra{
g_t^{(k)}\bra{\dpsi_{1|0},  \dhatF_1, \dots, \dhatF_{k}, \dtheta}
-
g_t^{(k)}\bra{\dpsi_{1|0}^\star,  \widehat\dH\dF_1, \dots, \widehat\dH\dF_{k}, \dtheta}
}
\\
\leq&
\sup_{\dpsi \in {\calP}}
\vvbra{\prod_{j=1}^k \boldsymbol{X}_{k-j+1}(\dtheta)}
\vvbra{\dpsi_{1|0} - \dpsi_{1|0}^\star}
+
\vbra{
\sum_{j=1}^{k}
\sup_{\dpsi \in {\calP}}
\vvbra{\boldsymbol{Y}_{k-j+1}(\dtheta)}
\vvbra{\dhatF_{k-j+1} - \widehat\dH\dF_{k-j+1}}
}
.
\end{align*}
Now, proceeding recursively for $t\in\Nn$, and using Assumption \ref{ass:contraction_cond} together with Proposition \ref{prop:rate_yt_decomp}, we obtain
\begin{align*}
&\sup_{\dtheta\in\dTheta}
\vvbra{
\dhathatpsithetatpt
-
\dhatpsithetatpt
}
\\
\leq&
\sup_{\dtheta\in\dTheta}
\vbra{
X_{t,k}(\dtheta)}
\sup_{\dtheta\in\dTheta}
\vvbra{\doublehat{\boldsymbol{\psi}}_{t-k+1|t-k}(\dtheta) - \widehat{\boldsymbol{\psi}}_{t-k+1|t-k}(\dtheta)}
+
\vbra{
\sum_{j=1}^{k}
\sup_{\dtheta\in\dTheta}
\vbra{{Y}_{t-j+1}(\dtheta)}
\vvbra{\dhatF_{t-j+1} - \widehat\dH\dF_{t-j+1}}
}.
\end{align*}
Thus, by recursively unfolding the bound back to $t=1$, and applying Assumption \ref{ass:contraction_cond}, we obtain, 
\begin{align}
&\sup_{\dtheta\in\dTheta}
\vvbra{
\dhathatpsithetatpt
-
\dhatpsithetatpt
}
\nonumber
\\
\leq&
\calX_k^{\lfloor t/k \rfloor}
\,
\vvbra{{\boldsymbol{\psi}}_{1|0}- {\boldsymbol{\psi}}^\star_{1|0}}
+
\sum_{j=1}^{t}
\calX^{\lfloor (j-1)/k \rfloor}_{k}
\vbra{
\sup_{\dtheta\in\dTheta}
\vbra{
{Y}_{t-j+1}(\dtheta)
}
\vvbra{
\dhatF_{t-j+1} - \widehat\dH\dF_{t-j+1}
}
}
\nonumber
\\
\label{eq:ineq_prop_inv}
\leq &
\calX_k^{\lfloor t/k \rfloor}
\, 
\Delta_1 
 +   
 k \Delta_2/\bra{1-\calX_{k}},
\end{align}
where $0<\Delta_1<\infty$ is defined as $\Delta_1 = \vvbra{{\boldsymbol{\psi}}_{1|0}- {\boldsymbol{\psi}}^\star_{1|0}}$. At each recursive step of \eqref{eq:ineq_prop_inv}, factor-approximation errors inherited from previous iterations are multiplied by an additional contraction coefficient. Consequently, an error introduced $j-1$ recursive steps earlier is weighted by $\calX^{\lfloor (j-1)/k \rfloor}_{k}$ for $j=1,\dots,t$.

An inspection of \eqref{eq:evecXX3} and \eqref{eq:consF} in the proof
of Proposition \ref{prop:rate_yt_decomp}$(a)$ shows that, for every
fixed $s\geq1$,
\[
\vvbra{
\dhatF_s-\widehat\dH\dF_s
}
=
O_{\Pp}\bra{
\frac{1}{\min\bra{\sqrt N,T}}
}.
\]
Since $t$ is fixed, an application of the union bound shows that
\begin{align*}
&
\lim_{M\rightarrow\infty}
\limsup_{N,T\rightarrow\infty}
\Pp\bra{
\min\bra{\sqrt N,T}
\max_{1\leq s\leq t}
\vvbra{
\dhatF_s-\widehat\dH\dF_s
}
>
M
}
\\
\leq{}&
\sum_{s=1}^{t}
\lim_{M\rightarrow\infty}
\limsup_{N,T\rightarrow\infty}
\Pp\bra{
\min\bra{\sqrt N,T}
\vvbra{
\dhatF_s-\widehat\dH\dF_s
}
>
M
}
=
0,
\end{align*}
and, consequently,
\begin{align}
\label{eq:fixed_maximum_factor_error}
\max_{1\leq s\leq t}
\vvbra{
\dhatF_s-\widehat\dH\dF_s
}
=
O_{\Pp}\bra{
\frac{1}{\min\bra{\sqrt N,T}}
}.
\end{align}
Moreover, Assumption \ref{ass:contraction_cond}$(ii)$ and Markov's
inequality imply that, for every fixed $t$,
\[
\max_{1\leq s\leq t}
\sup_{\dtheta\in\dTheta}
\vbra{
Y_s(\dtheta)
}
=
O_{\Pp}(1).
\]
It therefore follows from \eqref{eq:fixed_maximum_factor_error} that
\begin{align}
\label{eq:ineq_prop_ybound}
\Delta_2
&=
\max_{1\leq s\leq t}
\bra{
\sup_{\dtheta\in\dTheta}
\vbra{
Y_s(\dtheta)
}
\vvbra{
\dhatF_s-\widehat\dH\dF_s
}
}
\nonumber
\\
&\leq
\max_{1\leq s\leq t}
\sup_{\dtheta\in\dTheta}
\vbra{
Y_s(\dtheta)
}
\max_{1\leq s\leq t}
\vvbra{
\dhatF_s-\widehat\dH\dF_s
}
=
O_{\Pp}\bra{
\frac{1}{\min\bra{\sqrt N,T}}
}.
\end{align}
Finally, since $\Delta_1<\infty$, $k$ is fixed, and
$0<\calX_k<1$, substituting \eqref{eq:ineq_prop_ybound} into
\eqref{eq:ineq_prop_inv} gives
\begin{align}
\label{eq:asMARKOV}
\sup_{\dtheta\in\dTheta}
\vvbra{
\dhathatpsithetatpt-\dhatpsithetatpt
}
=
O_{\Pp}\bra{
\frac{1}{\min\bra{\sqrt N,T}}
}
+
O_{a.s.}\bra{
\calX_k^{\lfloor t/k\rfloor}
},
\end{align}
as $N,T\rightarrow\infty$.

\noindent
$(b)$ For the second claim, we note that the sequence $\cbra{\dtildepsithetatti}_{t\in\Nn}$ satisfies \eqref{eq:gt1} and \eqref{eq:gt2}, but replacing $\dhatFt$ with $\dH_0 \dbarF_t$. Therefore, by the same arguments discussed in point $(a)$, we immediately obtain
\begin{align*}
&\sup_{\dtheta\in\dTheta}
\vvbra{
g_t^{(k)}\bra{\dpsi^\star_{1|0},  \widehat\dH\dF_1, \dots, \widehat\dH\dF_{k}, \dtheta}
-
g_t^{(k)}\bra{\dpsi^\star_{1|0},  \dH_0\dbarF_1, \dots, \dH_0\dbarF_{k}, \dtheta}
}
\\
\leq{}&
\vbra{
\sum_{j=1}^{k}
\sup_{\dtheta\in\dTheta}
\vbra{{Y}_{k-j+1}(\dtheta)}
\vvbra{\widehat\dH\dF_{k-j+1}-\dH_0\dbarF_{k-j+1}}
}.
\end{align*}
Thus, using the decomposition $\widehat\dH\dF_{t} - \dH_0\dbarF_{t} = \bra{\widehat\dH-\dH_0}\dF_{t} + \dH_0\bra{\dF_{t}-\dbarF_{t}}$, \eqref{eq:ineq_prop_inv} becomes
\begin{align}
&\sup_{\dtheta\in\dTheta}
\vvbra{
\dhatpsithetatpt
-
\dtildepsithetatpt
}
\nonumber
\\
\leq{}&
\sum_{j=1}^{t}
\calX^{\lfloor (j-1)/k \rfloor}_{k}
\vbra{
\sup_{\dtheta\in\dTheta}
\vbra{
{Y}_{t-j+1}(\dtheta)
}\vvbra{
\widehat\dH\dF_{t-j+1} - \dH_0\dbarF_{t-j+1}
}}
\nonumber
\\
\leq{}&
\sum_{j=1}^{t}
\calX_k^{\lfloor (j-1)/k\rfloor}
\vbra{
\sup_{\dtheta\in\dTheta}
\vbra{Y_{t-j+1}(\dtheta)}
\vvbra{
\bra{\widehat\dH-\dH_0}\dF_{t-j+1}
}
}
\nonumber
\\
&+
\sum_{j=1}^{t}
\calX_k^{\lfloor (j-1)/k\rfloor}
\vbra{
\sup_{\dtheta\in\dTheta}
\vbra{Y_{t-j+1}(\dtheta)}
\vvbra{
\dH_0\bra{\dF_{t-j+1}-\dbarF_{t-j+1}}
}
}.
\label{eq:ineq2_prop_inv}
\end{align}
For every fixed $t$, the first term on the right-hand side of \eqref{eq:ineq2_prop_inv} can be bounded by
\begin{align*}
&\sum_{j=1}^{t}
\calX_k^{\lfloor (j-1)/k\rfloor}
\vbra{
\sup_{\dtheta\in\dTheta}
\vbra{Y_{t-j+1}(\dtheta)}
\vvbra{
\bra{\widehat\dH-\dH_0}\dF_{t-j+1}
}
}
\\
\leq{}&
\Delta_3
\sum_{j=1}^{t}
\calX_k^{\lfloor (j-1)/k\rfloor}
\leq
{k\Delta_3}/\bra{1-\calX_k},
\end{align*}
where using \eqref{eq:ACCAlim}, Assumptions \ref{ass:noise}, \ref{ass:pram_SE} and \ref{ass:contraction_cond},
\begin{equation}
\label{eq:Delta3}
\Delta_3
=
\max_{1\leq s\leq t}
\vbra{
\sup_{\dtheta\in\dTheta}
\vbra{Y_s(\dtheta)}
\vvbra{
\bra{\widehat\dH-\dH_0}\dF_s
}
} = O\bra{\frac 1{\min\bra{\sqrt N,  \sqrt T}}}
.
\end{equation}
To control the second term on the right-hand side of \eqref{eq:ineq2_prop_inv}, we use the fact that $\dF_t$ and $\dbarF_t$ admit the model representation in \eqref{eq:Model_yt0}. Assumptions \ref{ass:noise} and \ref{ass:pram_SE} therefore allow us to exploit the geometric contraction governing the difference between the stationary and nonstationary factor processes. 

In particular, we have 
\begin{align*}
&\vbra{
\sup_{\dtheta\in\dTheta}
\vbra{
{Y}_{s}(\dtheta)
}
\vvbra{\dH_0}
 \vvbra{\dF_{s} - \dbarF_{s}}
 }
 \\
 \leq{}&
\vbra{
\sup_{\dtheta\in\dTheta}
\vbra{
{Y}_{s}(\dtheta)
}
\vvbra{\dH_0}
 \vvbra{\boldsymbol q\bra{\dpsi_{s|s-1}(\dtheta_0), \dv_s} - \boldsymbol q\bra{\dbarpsi_{s|s-1}(\dtheta_0), \dv_s}}
 }
=
O_{a.s.}\bra{\calX_k^{\lfloor s/k \rfloor}},
\end{align*}
by Assumptions \ref{ass:noise}, \ref{ass:pram_SE} and \ref{ass:contraction_cond}. Substituting $s=t-j+1$, we obtain
\begin{equation}
\label{eq:O_sum_calXk}
\sum_{j=1}^{t}
\calX_k^{\lfloor (j-1)/k\rfloor}
\vbra{
\sup_{\dtheta\in\dTheta}
\vbra{Y_{t-j+1}(\dtheta)}
\vvbra{
\dH_0\bra{\dF_{t-j+1}-\dbarF_{t-j+1}}
}
}
=
O\bra{
\sum_{j=1}^{t}
\calX_k^{
\lfloor (j-1)/k\rfloor
+
\lfloor (t-j+1)/k\rfloor
}
}.
\end{equation}
Since $\left\lfloor\frac{j-1}{k}\right\rfloor+\left\lfloor\frac{t-j+1}{k}\right\rfloor \in\cbra{\left\lfloor\frac{t}{k}\right\rfloor-1, \left\lfloor\frac{t}{k}\right\rfloor}$ and $0 < \calX_k < 1$, we can write $\calX_k^{\lfloor (j-1)/k\rfloor+\lfloor (t-j+1)/k\rfloor}\leq\calX_k^{\lfloor t/k\rfloor-1}$, such that $\sum_{j=1}^t\calX_k^{\lfloor (j-1)/k\rfloor+\lfloor (t-j+1)/k\rfloor} \leq t\cdot\calX_k^{\lfloor t/k\rfloor-1}$. By inserting this result into \eqref{eq:O_sum_calXk} we obtain
\begin{equation}
\label{eq:O_sum_calXk2}
\sum_{j=1}^{t}
\calX_k^{\lfloor (j-1)/k\rfloor}
\vbra{
\sup_{\dtheta\in\dTheta}
\vbra{Y_{t-j+1}(\dtheta)}
\vvbra{
\dH_0\bra{\dF_{t-j+1}-\dbarF_{t-j+1}}
}
}
=
O\bra{
\varrho_k^{\lfloor t/k\rfloor}
}
,
\end{equation}
for some $0< \calX_k < \varrho_k <1$.

Therefore, substituting \eqref{eq:O_sum_calXk2} and \eqref{eq:Delta3} into \eqref{eq:ineq2_prop_inv}, reveals that 
\begin{equation}
\label{eq:asMARKOV2}
\sup_{\dtheta\in\dTheta}
\vvbra{
\dhatpsithetatpt - \dtildepsithetatpt
}
= 
O_{\mathbb P}\bra{\frac 1{\min(\sqrt N,  \sqrt T)}}
+
O_{a.s.}\bra{\varrho_k^{\lfloor t/k\rfloor}}
.
\end{equation}

\noindent
$(c)$ For proving the third claim, we verify the conditions given in Theorem 3.1 of \cite{Bougerol1993}. First, Assumptions \ref{ass:noise} and \ref{ass:pram_SE} imply that the sequence of stochastic functions defining the filtering recursion, driven by the stationary and ergodic process $
\cbra{\dH_0\dbarF_t}_{t\in\Zz}$, is itself stationary and ergodic. Therefore, to establish the existence of a unique stationary and ergodic solution
$\cbra{\dbarpsi_{t|t-1}(\dtheta)}_{t\in\Zz}$ and the convergence of the initialized filter toward this solution, it remains to verify conditions (C.1) and (C.2) of Theorem 3.1 of \cite{Bougerol1993}.

We start by verifying (C.1), that requires the existence of a $\log$-moment for the displacement of the stochastic recurrence $\cbra{\dtildepsitti(\dtheta)}_{t\in\Zz}$ at one fixed point. To prove this, we note that Assumption \ref{ass:contraction_cond_I} entails that
\begin{align}
\label{eq:log_mom_proof}
\Ee\sbra{\log^+ \sup_{\dtheta\in\dTheta}\vvbra{\dtildepsi_{2|1}(\dtheta) - \dpsi_{1|0}} }
\leq{}&
2\log 2 +
 \log^+ \sup_{\dtheta\in\dTheta}\vvbra{\ddelta}
+
\log^+ \sup_{\dtheta\in\dTheta}\vvbra{\dPhi - \dIdentityr}
+
\log^+ \sup_{\dtheta\in\dTheta}\vvbra{\dpsi_{1|0}}
\nonumber
\\
&+
\log^+ \sup_{\dtheta\in\dTheta}\vvbra{\dK}
+
\Ee\sbra{
\log^+ \sup_{\dtheta\in\dTheta}\vvbra{\du_1(\dtheta)}
}
<\infty.
\end{align}
Furthermore, for condition (C.2) in Theorem 3.1 of \cite{Bougerol1993}, it is enough to note that, by repeated applications of the mean value theorem, we get 
\begin{align*}
\sup_{\dtheta \in \dTheta}&\vvbra{\dtildepsithetatpt - \dbarpsitpt(\dtheta)}
\\
=&\,
\sup_{\dtheta \in \dTheta}
\vvbra{
\dPhi + \dK \frac{\partial \dut}{\partial\dpsi\trans}
\bigg|_{\dpsi = \boldsymbol{\psi}^\bullet_{t|t-1}(\dtheta)}
\bra{\dtildepsithetatti - \bar{\boldsymbol{\psi}}_{t|t-1}(\dtheta)}
}
\\
=&\,
\sup_{\dtheta \in \dTheta}
\vvbra{
\prod_{j=1}^k\bra{
\dPhi + \dK \frac{\partial \du_{t-j+1}}{\partial\dpsi\trans}
\bigg|_{\dpsi = \boldsymbol{\psi}^\bullet_{t-j+1|t-j}(\dtheta)}
}
\bra{\widetilde{\boldsymbol{\psi}}_{t-k+1|t-k}(\dtheta) - \bar{\boldsymbol{\psi}}_{t-k+1|t-k}(\dtheta)}
}
\\
\leq& 
\sup_{\dtheta \in \dTheta}\sup_{\dpsi \in \calP}
\vvbra{
\prod_{j=1}^k\boldsymbol{X}_{t-j+1}(\dtheta)
}
\sup_{\dtheta \in \dTheta}
\vvbra{\widetilde{\boldsymbol{\psi}}_{t-k+1|t-k}(\dtheta) - \bar{\boldsymbol{\psi}}_{t-k+1|t-k}(\dtheta)}
\\
\leq& 
\sup_{\dtheta \in \dTheta}
\vbra{
{X}_{t,k}(\dtheta)
}
\sup_{\dtheta \in \dTheta}
\vvbra{\widetilde{\boldsymbol{\psi}}_{t-k+1|t-k}(\dtheta) - \bar{\boldsymbol{\psi}}_{t-k+1|t-k}(\dtheta)}
,
\end{align*}
with $\boldsymbol{\psi}^\bullet_{t-j+1|t-j}(\dtheta)\in\bra{\widetilde{\boldsymbol{\psi}}_{t-j+1|t-j}(\dtheta), \bar{\boldsymbol{\psi}}_{t-j+1|t-j}(\dtheta)}$ for $j=1,\dots,k$ and $k\geq 1$. Now, since under Assumption \ref{ass:pram_SE} both $\cbra{\dbarFt}_{t\in\Zz}$ and $\cbra{{X}_{t,k}(\dtheta)}_{t\in\Nn}$ are strictly stationary and ergodic, it follows by Assumption \ref{ass:contraction_cond_I} that also C.2 of Theorem 3.1 of \cite{Bougerol1993} is satisfied.

Thus, without loss of generality, assuming that $t$ is a multiple of $k$ and proceeding recursively,
\begin{align*}
&\sup_{\dtheta \in \dTheta}\vvbra{\dtildepsithetatpt - \dbarpsitpt(\dtheta)}
\\
\leq& 
\sup_{\dtheta \in \dTheta}
\vbra{
{X}_{t,k}(\dtheta)
}
\sup_{\dtheta \in \dTheta}
\vvbra{\widetilde{\boldsymbol{\psi}}_{t-k+1|t-k}(\dtheta) - \bar{\boldsymbol{\psi}}_{t-k+1|t-k}(\dtheta)}
\\
\leq&
\sup_{\dtheta \in \dTheta}
\vbra{
{X}_{t,k}(\dtheta)
}
\vbra{
{X}_{t-k,k}(\dtheta)
}
\sup_{\dtheta \in \dTheta}
\vvbra{\widetilde{\boldsymbol{\psi}}_{t-2k+1|t-2k}(\dtheta) - \bar{\boldsymbol{\psi}}_{t-2k+1|t-2k}(\dtheta)}
\\
\leq&
\sup_{\dtheta \in \dTheta}
\vbra{
{X}_{t,k}(\dtheta)
}
\vbra{
{X}_{t-k,k}(\dtheta)
}
\vbra{
{X}_{t-2k,k}(\dtheta)
}
\dots
\vbra{
{X}_{k,k}(\dtheta)
}
\sup_{\dtheta \in \dTheta}
\vvbra{{\boldsymbol{\psi}}^\star_{1|0} - \boldsymbol{\psi}_{1|0}(\dtheta)}
.
\end{align*}
Clearly, the process $\cbra{\sup_{\dtheta\in\dTheta}|X_{t-ik,k}(\dtheta)|}_{i\geq0}$ in the last inequality is stationary and ergodic, and from Assumption \ref{ass:contraction_cond_I} we have that $\Ee\sbra{\log\sup_{\dtheta\in\dTheta}\vbra{X_{t,k}(\dtheta)} }<0$. Then, by Lemma 2.4 of \cite{Straumann2006} it holds that $\prod_{i=0}^{\lfloor t/k\rfloor-1}
\sup_{\dtheta\in\dTheta}
\vbra{
{X}_{t-ik,k}(\dtheta)
}\easconv 0$ as $t\rightarrow\infty$
and therefore, this result in combination with the bound obtained in \eqref{eq:log_mom_proof} implies that
\begin{align*}
&\sup_{\dtheta \in \dTheta}\vvbra{\dtildepsithetatpt - \dbarpsitpt(\dtheta)}
\leq
\bra{
\prod_{i=0}^{\lfloor t/k\rfloor-1}
\sup_{\dtheta \in \dTheta}
\vbra{
{X}_{t-ik,k}(\dtheta)
}
}
\sup_{\dtheta \in \dTheta}
\vvbra{{\boldsymbol{\psi}}_{1|0} - \bar{\boldsymbol{\psi}}_{1|0}(\dtheta)}
\easconv 0,
\quad t\rightarrow\infty. 
\end{align*}
This proves that the filtered vector process $\cbra{\dtildepsithetatti}_{t\in\Nn}$ is invertible, and converges exponentially fast almost surely to the unique stationary ergodic sequence $\cbra{ \dbarpsitti(\dtheta) }_{t \in \Zz}$ for any initialization of the filtering recursion, $\dpsi_{1|0} \in \boldsymbol{\mathcal{P}}$ uniformly over $\dTheta$ and so
\begin{align}\label{eq:EXPDEC}
\sup_{\dtheta \in \dTheta}\vvbra{\dtildepsithetatpt - \dbarpsitpt(\dtheta)}
=
O_{a. s.}\bra{\varrho_k^{\lfloor t/k \rfloor}}
,
\end{align}
with $\prod_{i=0}^{\lfloor t/k\rfloor-1}
\sup_{\dtheta\in\dTheta}
\vbra{
{X}_{t-ik,k}(\dtheta)
}  = O_{a.s.}\bra{\varrho_k^{\lfloor t/k \rfloor}}$ for some $0 < \varrho_k < 1$.

\noindent
$(d)$ The near-epoch dependence property of $\cbra{\dbarpsitti(\dtheta_0)}_{t\in\Nn}$  follows by a direct application of Lemma \ref{lemma:NED}.

\noindent
$(e)$ The proof of part (e) follows directly from \eqref{eq:asMARKOV}, \eqref{eq:asMARKOV2} and \eqref{eq:EXPDEC}. This completes the proof. \qed


\subsubsection*{Proof of Proposition \ref{lemma:ULLN}}

By the triangle inequality,
\begin{align}
\label{eq:log_like_decomposition}
&
\sup_{\dtheta\in\dTheta}
\vbra{
\frac{1}{T}
\sum_{t=1}^T
\doublehat{\ell}_{\dF}
\bra{
\dhatFt;
\boldsymbol r_{\widehat\dH}(\dtheta)
}
+
\log\vbra{\widehat\dH}
-
\Ee\sbra{
\ell_{\dF}\bra{\dbarFt;\dtheta}
}
}
\nonumber
\\
\leq{}&
\sup_{\dtheta\in\dTheta}
\vbra{
\frac{1}{T}
\sum_{t=1}^T
\doublehat{\ell}_{\dF}
\bra{
\dhatFt;
\boldsymbol r_{\widehat\dH}(\dtheta)
}
-
\frac{1}{T}
\sum_{t=1}^T
\widehat{\ell}_{\dF}
\bra{
\dFt\widehat\dH\trans;
\boldsymbol r_{\widehat\dH}(\dtheta)
}
}
\nonumber
\\
&+
\sup_{\dtheta\in\dTheta}
\vbra{
\frac{1}{T}
\sum_{t=1}^T
\widehat{\ell}_{\dF}
\bra{
\dFt\widehat\dH\trans;
\boldsymbol r_{\widehat\dH}(\dtheta)
}
+
\log\vbra{\widehat\dH}
-
\frac{1}{T}
\sum_{t=1}^T
\ell_{\dF}\bra{\dFt;\dtheta}
}
\nonumber
\\
&+
\sup_{\dtheta\in\dTheta}
\vbra{
\frac{1}{T}
\sum_{t=1}^T
\ell_{\dF}\bra{\dFt;\dtheta}
-
\frac{1}{T}
\sum_{t=1}^T
\ell_{\dF}\bra{\dbarFt;\dtheta}
}
+
\sup_{\dtheta\in\dTheta}
\vbra{
\frac{1}{T}
\sum_{t=1}^T
\ell_{\dF}\bra{\dbarFt;\dtheta}
-
\Ee\sbra{
\ell_{\dF}\bra{\dbarFt;\dtheta}
}
}.
\end{align}
We now prove that each term on the right-hand side of \eqref{eq:log_like_decomposition} converges in probability to zero
uniformly over $\dtheta\in\dTheta$.

For the first term, we retain the time-averaging structure and write
\begin{align}
\label{eq:log_like_decomposition_term1}
&
\sup_{\dtheta\in\dTheta}
\vbra{
\frac{1}{T}
\sum_{t=1}^T
\doublehat{\ell}_{\dF}
\bra{
\dhatFt;
\boldsymbol r_{\widehat\dH}(\dtheta)
}
-
\frac{1}{T}
\sum_{t=1}^T
\widehat{\ell}_{\dF}
\bra{
\dFt\widehat\dH\trans;
\boldsymbol r_{\widehat\dH}(\dtheta)
}
}
\nonumber
\\
={}&
\sup_{\dtheta\in\dTheta}
\vbra{
\frac{1}{T}
\sum_{t=1}^{T}
\bra{
\doublehat{\ell}_{\dF}
\bra{
\dhatFt;
\boldsymbol r_{\widehat\dH}(\dtheta)
}
-
\widehat{\ell}_{\dF}
\bra{
\widehat\dH\dFt;
\boldsymbol r_{\widehat\dH}(\dtheta)
}
}
}.
\end{align}
Notice that we do not apply the triangle inequality to the sum in \eqref{eq:log_like_decomposition_term1}, since doing so would discard
the time-averaging structure of the factor-estimation error.

For the second term, by the triangle inequality
\begin{align}
\label{eq:log_like_decomposition_term2}
&
\sup_{\dtheta\in\dTheta}
\vbra{
\frac{1}{T}
\sum_{t=1}^T
\widehat{\ell}_{\dF}
\bra{
\dFt\widehat\dH\trans;
\boldsymbol r_{\widehat\dH}(\dtheta)
}
+
\log\vbra{\widehat\dH}
-
\frac{1}{T}
\sum_{t=1}^T
\ell_{\dF}\bra{\dFt;\dtheta}
}
\nonumber
\\
\leq{}&
\frac{1}{T}
\sum_{t=1}^{T}
\sup_{\dtheta\in\dTheta}
\vbra{
\widehat{\ell}_{\dF}
\bra{
\widehat\dH\dFt;
\boldsymbol r_{\widehat\dH}(\dtheta)
}
+
\log\vbra{\widehat\dH}
-
\ell_{\dF}\bra{\dFt;\dtheta}
}.
\end{align}
By equivariance of the conditional likelihood under the alignment mapping (see equations \eqref{eq:conditional_density_equivariance}--\eqref{eq:log_likelihood_equivariance_total} below), the summands on the right-hand side of \eqref{eq:log_like_decomposition_term2} would be zero if the two filters were initialized consistently under the transformation.
Consequently, these summands contain only the effect of the possibly different filter initializations.


For the third term, we similarly obtain
\begin{align}
\label{eq:log_like_decomposition_term3}
&
\sup_{\dtheta\in\dTheta}
\vbra{
\frac{1}{T}
\sum_{t=1}^T
\ell_{\dF}\bra{\dFt;\dtheta}
-
\frac{1}{T}
\sum_{t=1}^T
\ell_{\dF}\bra{\dbarFt;\dtheta}
}
\nonumber
\\
\leq{}&
\frac{1}{T}
\sum_{t=1}^{T}
\sup_{\dtheta\in\dTheta}
\vbra{
\ell_{\dF}\bra{\dFt;\dtheta}
-
\ell_{\dF}\bra{\dbarFt;\dtheta}
}.
\end{align}
Finally, the fourth term is
\begin{align}
\label{eq:log_like_decomposition_term4}
\sup_{\dtheta\in\dTheta}
\vbra{
\frac{1}{T}
\sum_{t=1}^{T}
\bra{
\ell_{\dF}\bra{\dbarFt;\dtheta}
-
\Ee\sbra{
\ell_{\dF}\bra{\dbarFt;\dtheta}
}
}
}.
\end{align}
We now examine the first term on the right-hand side of \eqref{eq:log_like_decomposition}.

\bigskip
\noindent
\textbf{Controlling the term in \eqref{eq:log_like_decomposition_term1}}\\
A first-order Taylor expansion around
$\bra{
\dhatpsitti\bra{\boldsymbol r_{\widehat\dH}(\dtheta)},
\widehat\dH\dFt
}$
with a second-order remainder gives, uniformly over $\dtheta\in\dTheta$,
\begin{align}
\label{eq:likelihood_signed_expansion}
&
\frac{1}{T}
\sum_{t=1}^{T}
\bra{
\doublehat{\ell}_{\dF}
\bra{
\dhatFt;
\boldsymbol r_{\widehat\dH}(\dtheta)
}
-
\widehat{\ell}_{\dF}
\bra{
\widehat\dH\dFt;
\boldsymbol r_{\widehat\dH}(\dtheta)
}
}
\nonumber
\\
={}&
\frac{1}{T}
\sum_{t=1}^{T}
\nabla^{\dpsi}
\widehat{\ell}_{\dF}
\bra{
\widehat\dH\dFt;
\boldsymbol r_{\widehat\dH}(\dtheta)
}^{\trans}
\bra{
\dhathatpsitti
\bra{
\boldsymbol r_{\widehat\dH}(\dtheta)
}
-
\dhatpsitti
\bra{
\boldsymbol r_{\widehat\dH}(\dtheta)
}
}
\nonumber
\\
&+
\frac{1}{T}
\sum_{t=1}^{T}
\nabla^{\dF}
\widehat{\ell}_{\dF}
\bra{
\widehat\dH\dFt;
\boldsymbol r_{\widehat\dH}(\dtheta)
}^{\trans}
\bra{
\dhatFt-\widehat\dH\dFt
}
\nonumber
\\
&+
\frac{1}{T}
\sum_{t=1}^{T}
R_{t,NT}(\dtheta).
\end{align}
Under Assumptions \ref{ass:smooth_domination_likelihood} and \ref{ass:moment_score}, there exists a non-negative process
$\cbra{C_t}_{t\in\Nn}$, with uniformly bounded moments of the required order, such that
\begin{align}
\label{eq:likelihood_remainder_bound}
\sup_{\dtheta\in\dTheta}
\vbra{
R_{t,NT}(\dtheta)
}
\leq
C_t
\bra{
\vvbra{
\dhatF_t-\widehat\dH\dF_t
}^{2}
+
\sup_{\dtheta\in\dTheta}
\vvbra{
\dhathatpsi_{t|t-1}
\bra{
\boldsymbol r_{\widehat\dH}(\dtheta)
}
-
\dhatpsi_{t|t-1}
\bra{
\boldsymbol r_{\widehat\dH}(\dtheta)
}
}^{2}
}
\end{align}
The remainder also contains a cross-product between the factor and filter errors, which is absorbed into \eqref{eq:likelihood_remainder_bound} using
$2\vvbra{\boldsymbol a}\vvbra{\boldsymbol b} \leq \vvbra{\boldsymbol a}^{2} + \vvbra{\boldsymbol b}^{2}$, for any $\boldsymbol a,\boldsymbol b\in\Rr^r$.

To control the two average squared errors appearing on the right-hand side of \eqref{eq:likelihood_remainder_bound}, we can apply
Lemma A.1 of \citet{Bai2003} which implies that
\begin{align}
\label{eq:quadratic_factor_error}
\frac{1}{T}
\sum_{t=1}^{T}
\vvbra{
\dhatF_t-\widehat\dH\dF_t
}^{2}
=
O_{\Pp}\bra{
\frac{1}{N}
+
\frac{1}{T}
}.
\end{align}
Equivalently, the root-mean-square factor-estimation error satisfies
\[
\bra{
\frac{1}{T}
\sum_{t=1}^{T}
\vvbra{
\dhatF_t-\widehat\dH\dF_t
}^{2}
}^{1/2}
=
O_{\Pp}\bra{
\frac{1}{\sqrt N}
+
\frac{1}{\sqrt T}
}.
\]
The filter inherits the same root-mean-square rate. In fact, in a similar way as done in the proof of Proposition \ref{prop:rate_psit_decomp}$(a)$, iterating the one-step filter difference and using Assumption \ref{ass:contraction_cond} we get
\begin{align*}
&
\sup_{\dtheta\in\dTheta}
\vvbra{
\dhathatpsitti
\bra{
\boldsymbol r_{\widehat\dH}(\dtheta)
}
-
\dhatpsitti
\bra{
\boldsymbol r_{\widehat\dH}(\dtheta)
}
}
\\
\leq{}&
\calX_k^{\lfloor(t-1)/k\rfloor}
\vvbra{
\dpsi_{1|0}-\dpsi_{1|0}^{\star}
}
+
\sum_{s=1}^{t-1}
\calX_k^{\lfloor(t-s-1)/k\rfloor}
\vbra{
\sup_{\dtheta\in\dTheta}
\vbra{
\boldsymbol Y_s\bra{\boldsymbol r_{\widehat\dH}(\dtheta)}
}
\vvbra{
\dhatF_s-\widehat\dH\dF_s
}}.
\end{align*}
Hence, by the discrete convolution inequality and the geometric summability of the powers of $\calX_k$,
\begin{align*}
&
\bra{
\frac{1}{T}
\sum_{t=1}^{T}
\sup_{\dtheta\in\dTheta}
\vvbra{
\dhathatpsitti
\bra{
\boldsymbol r_{\widehat\dH}(\dtheta)
}
-
\dhatpsitti
\bra{
\boldsymbol r_{\widehat\dH}(\dtheta)
}
}^{2}
}^{1/2}
\\
={}&
O_{\Pp}
\bra{
\bra{
\frac{1}{T}
\sum_{t=1}^{T}
\vvbra{
\dhatF_t-\widehat\dH\dF_t
}^{2}
}^{1/2}
+
\bra{
\frac{1}{T}
\sum_{t=1}^{T}
\calX_k^{2\lfloor(t-1)/k\rfloor}
}^{1/2}
}
\nonumber
\\
={}&
O_{\Pp}\bra{
\frac{1}{\sqrt N}
+
\frac{1}{\sqrt T}
}.
\end{align*}
Here,
\[
\frac{1}{T}
\sum_{t=1}^{T}
\calX_k^{2\lfloor(t-1)/k\rfloor}
\leq
\frac{k}{T\bra{1-\calX_k^2}}
=
O\bra{\frac{1}{T}},
\]
so the second term accounts for the geometrically decreasing effect of the different filter initializations. By squaring the preceding
root-mean-square bound we therefore get
\begin{align}
\label{eq:quadratic_filter_error}
&
\frac{1}{T}
\sum_{t=1}^{T}
\sup_{\dtheta\in\dTheta}
\vvbra{
\dhathatpsitti
\bra{
\boldsymbol r_{\widehat\dH}(\dtheta)
}
-
\dhatpsitti
\bra{
\boldsymbol r_{\widehat\dH}(\dtheta)
}
}^{2}
=
O_{\Pp}\bra{
\frac{1}{N}
+
\frac{1}{T}
}.
\end{align}
Indeed, $
\bra{
\frac{1}{\sqrt N}
+
\frac{1}{\sqrt T}
}^{2}
=
\frac{1}{N}
+
\frac{1}{T}
+
\frac{2}{\sqrt{NT}}
=
O\bra{
\frac{1}{N}
+
\frac{1}{T}
}$, where the last equality follows from the fact that $\frac{2}{\sqrt{NT}}
\leq
\frac{1}{N}
+
\frac{1}{T}.
$
Consequently, under Assumptions \ref{ass:smooth_domination_likelihood} and \ref{ass:moment_score}, H\"{o}lder's inequality, and \eqref{eq:quadratic_factor_error}--\eqref{eq:quadratic_filter_error} imply
\begin{align}
\label{eq:average_likelihood_remainder_rate}
\frac{1}{T}
\sum_{t=1}^{T}
\sup_{\dtheta\in\dTheta}
\vbra{
R_{t,NT}(\dtheta)
}
=
O_{\Pp}\bra{
\frac{1}{N}
+
\frac{1}{T}
}.
\end{align}
Thus, \eqref{eq:likelihood_signed_expansion} can equivalently be written
as
\begin{align}
&
\frac{1}{T}
\sum_{t=1}^{T}
\bra{
\doublehat{\ell}_{\dF}
\bra{
\dhatFt;
\boldsymbol r_{\widehat\dH}(\dtheta)
}
-
\widehat{\ell}_{\dF}
\bra{
\widehat\dH\dFt;
\boldsymbol r_{\widehat\dH}(\dtheta)
}
}
\nonumber
\\
={}&
\frac{1}{T}
\sum_{t=1}^{T}
\nabla^{\dpsi}
\widehat{\ell}_{\dF}
\bra{
\widehat\dH\dFt;
\boldsymbol r_{\widehat\dH}(\dtheta)
}^{\trans}
\bra{
\dhathatpsitti
\bra{
\boldsymbol r_{\widehat\dH}(\dtheta)
}
-
\dhatpsitti
\bra{
\boldsymbol r_{\widehat\dH}(\dtheta)
}
}
\nonumber
\\
&+
\frac{1}{T}
\sum_{t=1}^{T}
\nabla^{\dF}
\widehat{\ell}_{\dF}
\bra{
\widehat\dH\dFt;
\boldsymbol r_{\widehat\dH}(\dtheta)
}^{\trans}
\bra{
\dhatFt-\widehat\dH\dFt
}
+
O_{\Pp}\bra{
\frac{1}{N}
+
\frac{1}{T}
},
\end{align}
uniformly over $\dtheta\in\dTheta$.

Direct from a direct differentiation of the update maps, together with Assumptions \ref{ass:noise}--\ref{ass:contraction_cond}, and a first-order Taylor expansion with a second-order remainder we therefore obtain
\begin{align}
&
\dhathatpsi_{t+1|t}
\bra{
\boldsymbol r_{\widehat\dH}(\dtheta)
}
-
\dhatpsi_{t+1|t}
\bra{
\boldsymbol r_{\widehat\dH}(\dtheta)
}
\nonumber
\\
={}&
\boldsymbol X_t
\bra{
\boldsymbol r_{\widehat\dH}(\dtheta)
}
\bra{
\dhathatpsitti
\bra{
\boldsymbol r_{\widehat\dH}(\dtheta)
}
-
\dhatpsitti
\bra{
\boldsymbol r_{\widehat\dH}(\dtheta)
}
}
\nonumber
\\
&+
\boldsymbol Y_t
\bra{
\boldsymbol r_{\widehat\dH}(\dtheta)
}
\bra{
\dhatFt-\widehat\dH\dFt
}
+
O_{\Pp}\bra{
\vvbra{
\dhathatpsitti
\bra{\boldsymbol r_{\widehat\dH}(\dtheta)}
-
\dhatpsitti
\bra{\boldsymbol r_{\widehat\dH}(\dtheta)}
}^{2}
+
\vvbra{\dhatFt-\widehat\dH\dFt}^{2}
},
\label{eq:filter_signed_expansion}
\end{align}
uniformly over $\dtheta\in\dTheta$. Then, iterating \eqref{eq:filter_signed_expansion} yields
\begin{align}
&
\dhathatpsi_{t+1|t}
\bra{
\boldsymbol r_{\widehat\dH}(\dtheta)
}
-
\dhatpsi_{t+1|t}
\bra{
\boldsymbol r_{\widehat\dH}(\dtheta)
}
\nonumber
\\
={}&
\bra{
\prod_{h=1}^{t}
\boldsymbol X_{t-h+1}
\bra{
\boldsymbol r_{\widehat\dH}(\dtheta)
}
}
\bra{
\dpsi_{1|0}-\dpsi^\star_{1|0}
}
+
\sum_{s=1}^{t}
\bra{
\prod_{h=s+1}^{t}
\boldsymbol X_h
\bra{
\boldsymbol r_{\widehat\dH}(\dtheta)
}
}
\boldsymbol Y_s
\bra{
\boldsymbol r_{\widehat\dH}(\dtheta)
}
\bra{
\dhatF_s-\widehat\dH\dF_s
}
\nonumber
\\
&{}+
\text{second-order terms},
\label{eq:filter_iterated_expansion}
\end{align}
where all products are ordered with the largest time index on the left, and any product whose lower limit exceeds its upper limit is equal to $\boldsymbol I_r$. By Assumptions \ref{ass:contraction_cond} and \ref{ass:moment_score}, and the geometric summability of the products of $\boldsymbol X_t$, the contribution of the second-order terms in \eqref{eq:filter_iterated_expansion} to the average likelihood is $O_{\Pp}\bra{N^{-1}+T^{-1}}$. Moreover, the contribution of the first term on the right-hand side of \eqref{eq:filter_iterated_expansion} is $O_{\Pp}\bra{T^{-1}}$. Consequently, by substituting \eqref{eq:filter_iterated_expansion} into the first sum in \eqref{eq:likelihood_signed_expansion}, we obtain
\begin{align}
&
\frac{1}{T}
\sum_{t=1}^{T}
\bra{
\doublehat{\ell}_{\dF}
\bra{
\dhatFt;
\boldsymbol r_{\widehat\dH}(\dtheta)
}
-
\widehat{\ell}_{\dF}
\bra{
\widehat\dH\dFt;
\boldsymbol r_{\widehat\dH}(\dtheta)
}
}
\nonumber
\\
={}&
\frac{1}{T}
\sum_{t=1}^{T}
\nabla^{\dpsi}
\widehat{\ell}_{\dF}
\bra{
\widehat\dH\dFt;
\boldsymbol r_{\widehat\dH}(\dtheta)
}^{\trans}
\sum_{s=1}^{t-1}
\bra{
\prod_{h=s+1}^{t-1}
\boldsymbol X_h
\bra{
\boldsymbol r_{\widehat\dH}(\dtheta)
}
}
\boldsymbol Y_s
\bra{
\boldsymbol r_{\widehat\dH}(\dtheta)
}
\bra{
\dhatF_s-\widehat\dH\dF_s
}
\nonumber
\\
&+
\frac{1}{T}
\sum_{t=1}^{T}
\nabla^{\dF}
\widehat{\ell}_{\dF}
\bra{
\widehat\dH\dFt;
\boldsymbol r_{\widehat\dH}(\dtheta)
}^{\trans}
\bra{
\dhatFt-\widehat\dH\dFt
}
+
O_{\Pp}\bra{
\frac{1}{N}
+
\frac{1}{T}
}.
\label{eq:likelihood_filter_double_sum}
\end{align}
Since $\sum_{t=1}^{T}\sum_{s=1}^{t-1}=\sum_{s=1}^{T-1}\sum_{t=s+1}^{T}$, equation \eqref{eq:likelihood_filter_double_sum} can be rewritten as
\begin{align}
&
\frac{1}{T}
\sum_{t=1}^{T}
\bra{
\doublehat{\ell}_{\dF}
\bra{
\dhatFt;
\boldsymbol r_{\widehat\dH}(\dtheta)
}
-
\widehat{\ell}_{\dF}
\bra{
\widehat\dH\dFt;
\boldsymbol r_{\widehat\dH}(\dtheta)
}
}
\nonumber
\\
={}&
\frac{1}{T}
\sum_{s=1}^{T}
\bra{
\nabla^{\dF}
\widehat{\ell}_{\dF}
\bra{
\widehat\dH\dF_s;
\boldsymbol r_{\widehat\dH}(\dtheta)
}
+
\boldsymbol Y_s
\bra{
\boldsymbol r_{\widehat\dH}(\dtheta)
}^{\trans}
\sum_{t=s+1}^{T}
\bra{
\prod_{h=s+1}^{t-1}
\boldsymbol X_h
\bra{
\boldsymbol r_{\widehat\dH}(\dtheta)
}
}^{\trans}
\nabla^{\dpsi}
\widehat{\ell}_{\dF}
\bra{
\widehat\dH\dFt;
\boldsymbol r_{\widehat\dH}(\dtheta)
}
}^{\trans}
\nonumber
\\
&\qquad\qquad \times
\bra{
\dhatF_s-\widehat\dH\dF_s
}
+
O_{\Pp}\bra{\frac{1}{N}+\frac{1}{T}}.
\label{eq:likelihood_weighted_factor_error}
\end{align}
Now, let $\calH\subset GL(r)$ be a fixed compact neighbourhood of $\dH_0$ containing only invertible matrices, with $
GL\bra{r}
=
\cbra{
\dH\in\Rr^{r\times r}:
\vbra{\dH}\neq0
},
$. For each $\dH\in\calH$, let
\begin{align}
\label{eq:effective_likelihood_weight}
\boldsymbol{\cal G}_{s,T}\bra{\dtheta,\dH}
={}&
\nabla^{\dF}
\widehat{\ell}_{\dF}
\bra{
\dH\dF_s;
\boldsymbol r_{\dH}(\dtheta)
}
\nonumber
\\
&+
\boldsymbol Y_s
\bra{
\boldsymbol r_{\dH}(\dtheta)
}^{\trans}
\sum_{t=s+1}^{T}
\bra{
\prod_{h=s+1}^{t-1}
\boldsymbol X_h
\bra{
\boldsymbol r_{\dH}(\dtheta)
}
}^{\trans}
\nabla^{\dpsi}
\widehat{\ell}_{\dF}
\bra{
\dH\dF_t;
\boldsymbol r_{\dH}(\dtheta)
},
\end{align}
and note that, by Assumptions \ref{ass:contraction_cond} and \ref{ass:moment_score}, together with the geometric summability of the
products of $\boldsymbol X_t$, it holds that
\begin{align}
\label{eq:effective_likelihood_weight_moment}
\sup_{T\geq1}
\sup_{1\leq s\leq T}
\Ee\sbra{
\sup_{\dH\in\calH}
\sup_{\dtheta\in\dTheta}
\vvbra{
\boldsymbol{\cal G}_{s,T}
\bra{
\dtheta,\dH
}
}^{2}
}
<\infty.
\end{align}
The same smoothness conditions and the compactness of $\dTheta$ and $\calH$ imply that $\boldsymbol{\cal G}_{s,T}\bra{\dtheta,\dH}$ is locally Lipschitz in $\dtheta$ and $\dH$, uniformly over $\dTheta\times\calH$, with a random Lipschitz coefficient having a bounded second moment. Moreover, \eqref{eq:effective_likelihood_weight_moment}, the fact that $\Pp\bra{\widehat\dH\in\calH}\longrightarrow1$, and Markov's inequality imply
\begin{align}
\label{eq:effective_likelihood_weight_average}
\frac{1}{T}
\sum_{s=1}^{T}
\sup_{\dtheta\in\dTheta}
\vvbra{
\boldsymbol{\cal G}_{s,T}
\bra{
\dtheta,\widehat\dH
}
}^{2}
=
O_{\Pp}(1).
\end{align}
By substituting \eqref{eq:effective_likelihood_weight} into \eqref{eq:likelihood_weighted_factor_error} and using the decomposition
in \eqref{eq:evecXX3} we get 
\begin{align}
\label{eq:weighted_likelihood_PC_expansion}
&
\frac{1}{T}
\sum_{t=1}^{T}
\bra{
\doublehat{\ell}_{\dF}
\bra{
\dhatF_t;
\boldsymbol r_{\widehat\dH}(\dtheta)
}
-
\widehat{\ell}_{\dF}
\bra{
\widehat\dH\dF_t;
\boldsymbol r_{\widehat\dH}(\dtheta)
}
}
\nonumber
\\
={}&
\frac{1}{T}
\sum_{s=1}^{T}
\boldsymbol{\cal G}_{s,T}
\bra{
\dtheta,\widehat\dH
}^{\trans}
\bra{\frac{\widehat{\boldsymbol M}}{N}}^{-1}
\bra{
A_s+B_s+C_s+D_s
}
+
O_{\Pp}\bra{
\frac{1}{N}
+
\frac{1}{T}
},
\end{align}
uniformly over $\dtheta\in\dTheta$, where
\begin{align}
A_s
&=
\bra{
\frac{1}{T}
\sum_{u=1}^{T}
\dhatF_u\dF_u\trans
}
\bra{
\frac{\dLambda_0\trans\dxi_s}{N}
},
\label{eq:weighted_likelihood_A}
\\
B_s
&=
\bra{
\frac{1}{NT}
\sum_{u=1}^{T}
\dhatF_u\dxi_u\trans\dLambda_0
}
\dF_s,
\label{eq:weighted_likelihood_B}
\\
C_s
&=
\frac{1}{NT}
\sum_{u=1}^{T}
\dhatF_u
\Ee\sbra{
\dxi_u\trans\dxi_s
},
\label{eq:weighted_likelihood_C}
\\
D_s
&=
\frac{1}{NT}
\sum_{u=1}^{T}
\dhatF_u
\bra{
\dxi_u\trans\dxi_s
-
\Ee\sbra{
\dxi_u\trans\dxi_s
}
}.
\label{eq:weighted_likelihood_D}
\end{align}
By Lemma A.3 of \citet{Bai2003},
\begin{align}
\label{eq:inverse_PC_eigenvalue_bound}
\vvbra{
\bra{
\frac{\widehat{\boldsymbol M}}{N}
}^{-1}
}
=
O_{\Pp}(1).
\end{align}
We now control separately the four terms on the right-hand side of \eqref{eq:weighted_likelihood_PC_expansion}.

\medskip
\noindent
\emph{Contribution of $A_s$.}
For every nonrandom $\dH\in\calH$, the process $\boldsymbol{\cal G}_{s,T}\bra{\dtheta,\dH}$ is measurable with respect
to the factor process and is therefore independent of the idiosyncratic process. Moreover, Assumptions \ref{ass:factor_identifiability} and \ref{ass:idiosyncratic} imply that, for every $s,u\in\Zz$,
\begin{equation}
\label{eq:projected_idiosyncratic_covariance_lemma}
\vvbra{
\Ee\sbra{
\frac{\dLambda_0\trans\dxi_s}{N}
\frac{\dxi_u\trans\dLambda_0}{N}
}
}
\leq
\frac{C}{N}
\rho^{\vbra{s-u}},
\end{equation}
for some finite positive real $C$ independent of $N$, $s$, and $u$. Since $\dTheta$ is compact and
$\boldsymbol{\cal G}_{s,T}\bra{\dtheta,\dH_0}$ is uniformly Lipschitz in $\dtheta$ with a square-integrable envelope, the standard
$L_2$ maximal inequality for finite-dimensional Lipschitz classes, applied conditionally on the factor process, together with
\eqref{eq:effective_likelihood_weight_moment}, \eqref{eq:projected_idiosyncratic_covariance_lemma}, and Assumption
\ref{ass:idiosyncraticY}$(i)$ implies
\begin{align*}
\Ee\sbra{
\sup_{\dtheta\in\dTheta}
\vvbra{
\frac{1}{T}
\sum_{s=1}^{T}
\frac{\dLambda_0\trans\dxi_s}{N}
\boldsymbol{\cal G}_{s,T}
\bra{
\dtheta,\dH_0
}^{\trans}
}^{2}
}
\leq
\frac{C}{NT^{2}}
\sum_{s=1}^{T}
\sum_{u=1}^{T}
\rho^{\vbra{s-u}}
\leq
\frac{C}{NT},
\end{align*}
and therefore, by Markov's inequality,
\begin{align}
\label{eq:weighted_A_average_rate_H0}
\sup_{\dtheta\in\dTheta}
\vvbra{
\frac{1}{T}
\sum_{s=1}^{T}
\frac{\dLambda_0\trans\dxi_s}{N}
\boldsymbol{\cal G}_{s,T}
\bra{
\dtheta,\dH_0
}^{\trans}
}
=
O_{\Pp}\bra{
\frac{1}{\sqrt{NT}}
}.
\end{align}
To replace $\dH_0$ with $\widehat\dH$, the Cauchy--Schwarz inequality and the uniform local Lipschitz property stated above yield
\begin{align*}
&
\sup_{\dtheta\in\dTheta}
\vvbra{
\frac{1}{T}
\sum_{s=1}^{T}
\frac{\dLambda_0\trans\dxi_s}{N}
\bra{
\boldsymbol{\cal G}_{s,T}
\bra{
\dtheta,\widehat\dH
}
-
\boldsymbol{\cal G}_{s,T}
\bra{
\dtheta,\dH_0
}
}^{\trans}
}
\nonumber
\\
\leq{}&
\bra{
\frac{1}{T}
\sum_{s=1}^{T}
\vvbra{
\frac{\dLambda_0\trans\dxi_s}{N}
}^{2}
}^{1/2}
\bra{
\frac{1}{T}
\sum_{s=1}^{T}
\sup_{\dtheta\in\dTheta}
\vvbra{
\boldsymbol{\cal G}_{s,T}
\bra{
\dtheta,\widehat\dH
}
-
\boldsymbol{\cal G}_{s,T}
\bra{
\dtheta,\dH_0
}
}^{2}
}^{1/2}
\nonumber
\\
={}&
O_{\Pp}\bra{
\frac{1}{\sqrt N}
}
O_{\Pp}\bra{
\vvbra{
\widehat\dH-\dH_0
}
}
=
O_{\Pp}\bra{
\frac{1}{N}
+
\frac{1}{\sqrt{NT}}
},
\end{align*}
where the last equality follows from \eqref{eq:ACCAlim}. Combining this bound with \eqref{eq:weighted_A_average_rate_H0} we get
\begin{align}
\label{eq:weighted_A_average_rate_lemma}
\sup_{\dtheta\in\dTheta}
\Bigg\|
\frac{1}{T}
\sum_{s=1}^{T}
\frac{\dLambda_0\trans\dxi_s}{N}
\boldsymbol{\cal G}_{s,T}
\bra{
\dtheta,\widehat\dH
}^{\trans}
\Bigg\|
=
O_{\Pp}\bra{
\frac{1}{\sqrt{NT}}
+
\frac{1}{N}
}.
\end{align}
Additionally, by the matrix Cauchy--Schwarz inequality,
\begin{align*}
\vvbra{
\frac1T\sum_{u=1}^{T}
\dhatF_u\dF_u\trans
}
&\leq
\bra{
\vvbra{
\frac1T\sum_{u=1}^{T}
\dhatF_u\dhatF_u\trans
}
\vvbra{
\frac1T\sum_{u=1}^{T}
\dF_u\dF_u\trans
}
}^{1/2},
\end{align*}
and using the equivalence in \eqref{eq:sameval},
$
\frac1T\sum_{s=1}^{T}
\dhatF_s\dhatF_s\trans
=
\frac{\widehat{\boldsymbol M}}{N},
$
where $
\vvbra{
\frac{\widehat{\boldsymbol M}}{N}
}
=
O_{\Pp}(1).$ Under Assumption \ref{ass:pram_SE}, we also know that $\vvbra{
\frac1T\sum_{s=1}^{T}
\dF_s\dF_s\trans
}
=
O_{\Pp}(1),
$
and hence,
\begin{equation}
\label{eq:common_bound1}
\vvbra{
\frac1T\sum_{u=1}^{T}
\dhatF_u\dF_u\trans
}
=
O_{\Pp}(1).
\end{equation}
Consequently, using the trace inequality, \eqref{eq:common_bound1}, and \eqref{eq:weighted_A_average_rate_lemma}, the contribution of $A_s$ to \eqref{eq:weighted_likelihood_PC_expansion} is
\begin{align}
\label{eq:weighted_A_contribution_rate_lemma}
\sup_{\dtheta\in\dTheta}
\vbra{
\frac{1}{T}
\sum_{s=1}^{T}
\boldsymbol{\cal G}_{s,T}
\bra{
\dtheta,\widehat\dH
}^{\trans}
\bra{\frac{\widehat{\boldsymbol M}}{N}}^{-1}
A_s
}
=
O_{\Pp}\bra{
\frac{1}{\sqrt{NT}}
+
\frac{1}{N}
}.
\end{align}

\medskip
\noindent
\emph{Contribution of $B_s$.}
We have
\begin{align}
\label{eq:weighted_B_coefficient_decomposition_lemma}
\frac{1}{NT}
\sum_{u=1}^{T}
\dhatF_u\dxi_u\trans\dLambda_0
={}&
\widehat\dH
\bra{
\frac{1}{NT}
\sum_{u=1}^{T}
\dF_u\dxi_u\trans\dLambda_0
}
+
\frac{1}{T}
\sum_{u=1}^{T}
\bra{
\dhatF_u-\widehat\dH\dF_u
}
\frac{\dxi_u\trans\dLambda_0}{N}.
\end{align}
Assumption \ref{ass:idiosyncraticY}$(iii)$ implies that
\begin{align}
\label{eq:weighted_B_leading_rate_lemma}
\vvbra{
\frac{1}{NT}
\sum_{u=1}^{T}
\dF_u\dxi_u\trans\dLambda_0
}
=
O_{\Pp}\bra{
\frac{1}{\sqrt{NT}}
}.
\end{align}
For the second term in \eqref{eq:weighted_B_coefficient_decomposition_lemma}, using the matrix Cauchy--Schwarz inequality we get
\begin{align*}
\vvbra{
\frac{1}{T}
\sum_{u=1}^{T}
\bra{
\dhatF_u-\widehat\dH\dF_u
}
\frac{\dxi_u\trans\dLambda_0}{N}
}
\leq
\bra{
\frac{1}{T}
\sum_{u=1}^{T}
\vvbra{
\dhatF_u-\widehat\dH\dF_u
}^{2}
}^{1/2}
\bra{
\frac{1}{T}
\sum_{u=1}^{T}
\vvbra{
\frac{\dLambda_0\trans\dxi_u}{N}
}^{2}
}^{1/2}.
\end{align*}
Lemma A.1 of \citet{Bai2003} and \eqref{eq:projected_idiosyncratic_covariance_lemma} imply, respectively,
\begin{align*}
\bra{
\frac{1}{T}
\sum_{u=1}^{T}
\vvbra{
\dhatF_u-\widehat\dH\dF_u
}^{2}
}^{1/2}
&=
O_{\Pp}\bra{
\frac{1}{\sqrt N}
+
\frac{1}{\sqrt T}
},
\\
\bra{
\frac{1}{T}
\sum_{u=1}^{T}
\vvbra{
\frac{\dLambda_0\trans\dxi_u}{N}
}^{2}
}^{1/2}
&=
O_{\Pp}\bra{
\frac{1}{\sqrt N}
}.
\end{align*}
It follows that
\begin{align}
\label{eq:weighted_B_coefficient_rate_lemma}
\vvbra{
\frac{1}{NT}
\sum_{u=1}^{T}
\dhatF_u\dxi_u\trans\dLambda_0
}
=
O_{\Pp}\bra{
\frac{1}{\sqrt{NT}}
+
\frac{1}{N}
}.
\end{align}
Furthermore, by \eqref{eq:effective_likelihood_weight_average} and the fact that, under Assumption \ref{ass:pram_SE}, $\frac{1}{T} \sum_{s=1}^{T}\vvbra{\dF_s}^{2}= O_{\Pp}(1)$, we obtain
\begin{align*}
&
\frac{1}{T}
\sum_{s=1}^{T}
\sup_{\dtheta\in\dTheta}
\vvbra{
\boldsymbol{\cal G}_{s,T}
\bra{
\dtheta,\widehat\dH
}
}
\vvbra{\dF_s}
\\
\leq{}&
\bra{
\frac{1}{T}
\sum_{s=1}^{T}
\sup_{\dtheta\in\dTheta}
\vvbra{
\boldsymbol{\cal G}_{s,T}
\bra{
\dtheta,\widehat\dH
}
}^{2}
}^{1/2}
\bra{
\frac{1}{T}
\sum_{s=1}^{T}
\vvbra{\dF_s}^{2}
}^{1/2}
=
O_{\Pp}(1).
\end{align*}
Consequently, the contribution of $B_s$ is
\begin{align}
\label{eq:weighted_B_contribution_rate_lemma}
\sup_{\dtheta\in\dTheta}
\vbra{
\frac{1}{T}
\sum_{s=1}^{T}
\boldsymbol{\cal G}_{s,T}
\bra{
\dtheta,\widehat\dH
}^{\trans}
\bra{\frac{\widehat{\boldsymbol M}}{N}}^{-1}
B_s
}
=
O_{\Pp}\bra{
\frac{1}{\sqrt{NT}}
+
\frac{1}{N}
}.
\end{align}

\medskip
\noindent
\emph{Contribution of $C_s$.}
By Assumption \ref{ass:idiosyncratic}$(ii)$,
\begin{align*}
\vbra{
\Ee\sbra{
\dxi_u\trans\dxi_s
}
}
\leq
NC_e^\prime
\rho^{\vbra{u-s}}.
\end{align*}
Therefore,
\begin{align*}
\vvbra{C_s}
\leq
\frac{C_e^\prime}{T}
\sum_{u=1}^{T}
\rho^{\vbra{u-s}}
\vvbra{\dhatF_u}.
\end{align*}
By the discrete convolution inequality and the fact that $T^{-1}\sum_{u=1}^{T}\vvbra{\dhatF_u}^{2}=O_{\Pp}(1)$, we obtain
\begin{align}
\label{eq:weighted_C_average_rate_lemma}
\bra{
\frac{1}{T}
\sum_{s=1}^{T}
\vvbra{C_s}^{2}
}^{1/2}
&\leq
\frac{C_e^\prime}{T}
\sum_{h=-\infty}^{\infty}
\rho^{\vbra h}
\bra{
\frac{1}{T}
\sum_{u=1}^{T}
\vvbra{\dhatF_u}^{2}
}^{1/2}
=
O_{\Pp}\bra{
\frac{1}{T}
}.
\end{align}
Combining \eqref{eq:effective_likelihood_weight_average} and \eqref{eq:weighted_C_average_rate_lemma} with the Cauchy--Schwarz
inequality shows that the contribution of $C_s$ is
\begin{align}
\label{eq:weighted_C_contribution_rate_lemma}
\sup_{\dtheta\in\dTheta}
\vbra{
\frac{1}{T}
\sum_{s=1}^{T}
\boldsymbol{\cal G}_{s,T}
\bra{
\dtheta,\widehat\dH
}^{\trans}
\bra{\frac{\widehat{\boldsymbol M}}{N}}^{-1}
C_s
}
=
O_{\Pp}\bra{
\frac{1}{T}
}.
\end{align}

\medskip
\noindent
\emph{Contribution of $D_s$.}
From \eqref{eq:weighted_likelihood_D},
\begin{align}
\label{eq:weighted_D_decomposition_lemma}
D_s
={}&
\frac{\widehat\dH}{NT}
\sum_{u=1}^{T}
\dF_u
\bra{
\dxi_u\trans\dxi_s
-
\Ee\sbra{
\dxi_u\trans\dxi_s
}
}
+
\frac{1}{T}
\sum_{u=1}^{T}
\bra{
\dhatF_u-\widehat\dH\dF_u
}
\frac{
\dxi_u\trans\dxi_s
-
\Ee\sbra{
\dxi_u\trans\dxi_s
}
}{N}.
\end{align}
Assumption \ref{ass:idiosyncraticY}$(ii)$ and $\vvbra{\widehat\dH}=O_{\Pp}(1)$ yield
\begin{align}
\label{eq:weighted_D_leading_rate_lemma}
\bra{
\frac{1}{T}
\sum_{s=1}^{T}
\vvbra{
\frac{\widehat\dH}{NT}
\sum_{u=1}^{T}
\dF_u
\bra{
\dxi_u\trans\dxi_s
-
\Ee\sbra{
\dxi_u\trans\dxi_s
}
}
}^{2}
}^{1/2}
=
O_{\Pp}\bra{
\frac{1}{\sqrt{NT}}
}.
\end{align}
For the second term in \eqref{eq:weighted_D_decomposition_lemma}, Cauchy--Schwarz gives, for every $s$,
\begin{align*}
&
\vvbra{
\frac{1}{T}
\sum_{u=1}^{T}
\bra{
\dhatF_u-\widehat\dH\dF_u
}
\frac{
\dxi_u\trans\dxi_s
-
\Ee\sbra{
\dxi_u\trans\dxi_s
}
}{N}
}^{2}
\\
\leq{}&
\bra{
\frac{1}{T}
\sum_{u=1}^{T}
\vvbra{
\dhatF_u-\widehat\dH\dF_u
}^{2}
}
\bra{
\frac{1}{T}
\sum_{u=1}^{T}
\frac{
\vbra{
\dxi_u\trans\dxi_s
-
\Ee\sbra{
\dxi_u\trans\dxi_s
}
}^{2}
}{N^{2}}
}.
\end{align*}
By Assumption \ref{ass:idiosyncratic}$(v)$ and Markov's inequality,
\begin{align}
\label{eq:quadratic_idiosyncratic_average_lemma}
\frac{1}{T^{2}}
\sum_{s=1}^{T}
\sum_{u=1}^{T}
\frac{
\vbra{
\dxi_u\trans\dxi_s
-
\Ee\sbra{
\dxi_u\trans\dxi_s
}
}^{2}
}{N^{2}}
=
O_{\Pp}\bra{
\frac{1}{N}
}.
\end{align}
It follows that
\begin{align}
\label{eq:weighted_D_remainder_rate_lemma}
&
\bra{
\frac{1}{T}
\sum_{s=1}^{T}
\vvbra{
\frac{1}{T}
\sum_{u=1}^{T}
\bra{
\dhatF_u-\widehat\dH\dF_u
}
\frac{
\dxi_u\trans\dxi_s
-
\Ee\sbra{
\dxi_u\trans\dxi_s
}
}{N}
}^{2}
}^{1/2}
\nonumber
\\
={}&
O_{\Pp}\bra{
\bra{
\frac{1}{\sqrt N}
+
\frac{1}{\sqrt T}
}
\frac{1}{\sqrt N}
}
=
O_{\Pp}\bra{
\frac{1}{N}
+
\frac{1}{\sqrt{NT}}
}.
\end{align}
Combining \eqref{eq:effective_likelihood_weight_average}, \eqref{eq:weighted_D_leading_rate_lemma}, and \eqref{eq:weighted_D_remainder_rate_lemma} with the Cauchy--Schwarz inequality shows that the contribution of $D_s$ is
\begin{align}
\label{eq:weighted_D_contribution_rate_lemma}
\sup_{\dtheta\in\dTheta}
\vbra{
\frac{1}{T}
\sum_{s=1}^{T}
\boldsymbol{\cal G}_{s,T}
\bra{
\dtheta,\widehat\dH
}^{\trans}
\bra{\frac{\widehat{\boldsymbol M}}{N}}^{-1}
D_s
}
=
O_{\Pp}\bra{
\frac{1}{\sqrt{NT}}
+
\frac{1}{N}
}.
\end{align}
Therefore, by combining \eqref{eq:weighted_A_contribution_rate_lemma}, \eqref{eq:weighted_B_contribution_rate_lemma}, \eqref{eq:weighted_C_contribution_rate_lemma}, and \eqref{eq:weighted_D_contribution_rate_lemma} with \eqref{eq:weighted_likelihood_PC_expansion} we finally obtain 
\begin{align}
\label{eq:weighted_likelihood_rate}
\sup_{\dtheta\in\dTheta}
\vbra{
\frac{1}{T}
\sum_{t=1}^{T}
\bra{
\doublehat{\ell}_{\dF}
\bra{
\dhatF_t;
\boldsymbol r_{\widehat\dH}(\dtheta)
}
-
\widehat{\ell}_{\dF}
\bra{
\widehat\dH\dF_t;
\boldsymbol r_{\widehat\dH}(\dtheta)
}
}
}
=
O_{\Pp}\bra{
\frac{1}{\sqrt{NT}}
+
\frac{1}{N}
+
\frac{1}{T}
}.
\end{align}

\bigskip
\noindent
\textbf{Controlling the term in \eqref{eq:log_like_decomposition_term2}}\\
Fix an invertible $r\times r$ matrix $\dH$.
By the definition of $\boldsymbol r_{\dH}$, the conditional density of the transformed factors satisfies
\begin{align}
\label{eq:conditional_density_equivariance}
f_{\dH\dF}
\bra{
\dH\dF_t
\mid
\calF_{t-1};
\boldsymbol r_{\dH}(\dtheta)
}
=
\frac{1}{\vbra{\dH}}
f_{\dF}
\bra{
\dF_t
\mid
\calF_{t-1};
\dtheta
}.
\end{align}
For the time-varying location model, this follows from \eqref{eq:transform_mu}; for the time-varying covariance model, it follows from \eqref{eq:transform_sigma}. Thus, if the two filters are initialized equivariantly, then, for every $t\geq1$,
\begin{align}
\label{eq:log_likelihood_equivariance_contribution}
\ell_{\dF}
\bra{
\dH\dF_t;
\boldsymbol r_{\dH}(\dtheta)
}
+
\log\vbra{\dH}
=
\ell_{\dF}
\bra{
\dF_t;
\dtheta
}.
\end{align}
Summing \eqref{eq:log_likelihood_equivariance_contribution} over $t=1,\ldots,T$ 
\begin{align}
\label{eq:log_likelihood_equivariance_total}
\ell_{\dF}
\bra{
\dH\dF;
\boldsymbol r_{\dH}(\dtheta)
}
+
T\log\vbra{\dH}
=
\ell_{\dF}
\bra{
\dF;
\dtheta
}.
\end{align}
It remains to control the difference between $\widehat{\ell}_{\dF}$ and $\ell_{\dF}$ generated by their possibly
different filter initializations. Let $\cbra{\dhatpsi_{t|t-1}\bra{\boldsymbol r_{\dH}(\dtheta) }}_{t\in\Nn}$ denote the filter used to construct $\widehat{\ell}_{\dF} \bra{\dH\dF;\boldsymbol r_{\dH}(\dtheta)}$, initialized at $\dpsi^\star_{1|0}$, and let $\cbra{ \dpsi_{t|t-1}\bra{\boldsymbol r_{\dH}(\dtheta) }}_{t\in\Nn}$ denote the filter driven by the same transformed factors and initialized at the value obtained by applying the transformation associated with $\dH$ to the initial condition of the untransformed filter. By construction, the latter filter is the one entering \eqref{eq:log_likelihood_equivariance_contribution}.

For each $\dH\in\calH$, with $\calH\subset GL(r)$ as defined above
equation \eqref{eq:likelihood_weighted_factor_error}, the continuity of
the transformation associated with $\boldsymbol r_{\dH}$ and the
compactness of $\dTheta$ and $\calH$ imply
\begin{align}
\label{eq:rotated_initial_condition_bound}
\sup_{\dH\in\calH}
\sup_{\dtheta\in\dTheta}
\vvbra{
\dpsi_{1|0}^\star
-
\dpsi_{1|0}
\bra{
\boldsymbol r_{\dH}(\dtheta)
}
}
\leq
\Delta_1,
\end{align}
for some finite positive real $\Delta_1$ independent of $N$, $T$, and $t$. Now, following the same approach of the Proof of Proposition \ref{prop:rate_psit_decomp}, by repeated applications of the mean value theorem to the two filters and Assumption \ref{ass:contraction_cond}, there exists a finite positive real $C$, independent of $N$, $T$, $t$, $\dtheta$, and $\dH$, such that
\begin{align}
\label{eq:rotated_filter_initialization_rate}
\sup_{\dH\in\calH}
\sup_{\dtheta\in\dTheta}
\vvbra{
\dhatpsi_{t|t-1}
\bra{
\boldsymbol r_{\dH}(\dtheta)
}
-
\dpsi_{t|t-1}
\bra{
\boldsymbol r_{\dH}(\dtheta)
}
}
\leq
C
\calX_k^{\lfloor(t-1)/k\rfloor}
\Delta_1,
\end{align}
almost surely. An application of the mean value theorem to each likelihood contribution, together with Assumption \ref{ass:moment_score}, now implies that
\begin{align}
\label{eq:rotated_likelihood_initialization_bound}
&
\sup_{\dH\in\calH}
\sup_{\dtheta\in\dTheta}
\vbra{
\frac{1}{T}
\sum_{t=1}^T
\widehat{\ell}_{\dF}
\bra{
\dH\dFt;
\boldsymbol r_{\dH}(\dtheta)
}
-
\frac{1}{T}
\sum_{t=1}^T
\ell_{\dF}
\bra{
\dH\dFt;
\boldsymbol r_{\dH}(\dtheta)
}
}
\leq
\frac{C\Delta_1}{T}
\sum_{t=1}^{T}
D_{2,\dpsitti}
\calX_k^{\lfloor(t-1)/k\rfloor},
\end{align}
almost surely.

Taking expectations in \eqref{eq:rotated_likelihood_initialization_bound} and using the geometric-sum bound yields
\begin{align*}
&
\Ee\sbra{
\sup_{\dH\in\calH}
\sup_{\dtheta\in\dTheta}
\vbra{
\frac{1}{T}
\sum_{t=1}^T
\widehat{\ell}_{\dF}
\bra{
\dH\dFt;
\boldsymbol r_{\dH}(\dtheta)
}
-
\frac{1}{T}
\sum_{t=1}^T
\ell_{\dF}
\bra{
\dH\dFt;
\boldsymbol r_{\dH}(\dtheta)
}
}
}
\nonumber
\\
\leq{}&
\frac{C\Delta_1}{T}
\sup_{t\geq1}
\Ee\sbra{
D_{2,\dpsitti}
}
\sum_{t=1}^{T}
\calX_k^{\lfloor(t-1)/k\rfloor}
\nonumber
\\
\leq{}&
\frac{C\Delta_1}{T}
\sup_{t\geq1}
\Ee\sbra{
D_{2,\dpsitti}
}
\frac{k}{1-\calX_k}
=
O\bra{\frac{1}{T}}.
\end{align*}
Markov's inequality therefore implies
\begin{align}
\label{eq:rotated_likelihood_initialization_rate}
&
\sup_{\dH\in\calH}
\sup_{\dtheta\in\dTheta}
\Bigg|
\frac{1}{T}
\sum_{t=1}^T
\widehat{\ell}_{\dF}
\bra{
\dH\dFt;
\boldsymbol r_{\dH}(\dtheta)
}
-
\frac{1}{T}
\sum_{t=1}^T
\ell_{\dF}
\bra{
\dH\dFt;
\boldsymbol r_{\dH}(\dtheta)
}
\Bigg|
=
O_{\Pp}\bra{\frac{1}{T}}.
\end{align}
Since, $\widehat\dH-\dH_0=o_{\Pp}(1)$ by Proposition \ref{prop:rate_yt_decomp}, and $\dH_0$ belongs to the interior
of $\calH$, we have $\Pp\bra{\widehat\dH\in\calH}\rightarrow1$. Hence, \eqref{eq:rotated_likelihood_initialization_rate} also holds at
$\dH=\widehat\dH$. Combining this result with \eqref{eq:log_likelihood_equivariance_total} proves
\begin{align}
\label{eq:rotation_equivariance_rate}
\sup_{\dtheta\in\dTheta}
\vbra{
\frac{1}{T}
\sum_{t=1}^{T}
\widehat{\ell}_{\dF}
\bra{
\dFt\widehat\dH\trans;
\boldsymbol r_{\widehat\dH}(\dtheta)
}
+
\log\vbra{\widehat\dH}
-
\frac{1}{T}
\sum_{t=1}^{T}
\ell_{\dF}
\bra{
\dFt;
\dtheta
}
}
=
O_{\Pp}\bra{\frac{1}{T}}.
\end{align}

\bigskip
\noindent
\textbf{Controlling the term in \eqref{eq:log_like_decomposition_term3}}\\
We first collect the two exponential-convergence results needed below. By Assumptions \ref{ass:noise}, \ref{ass:pram_SE}, and \ref{ass:contraction_cond}, and using Proposition \ref{prop:rate_psit_decomp}$(c)$, for some $0<\calX_k<\varrho_k<1$,
\begin{align}
\label{eq:stationary_factor_geometric_rate}
&\vvbra{
\dFt-\dbarFt
}
=
O_{a.s.}\bra{
\varrho_k^{\lfloor(t-1)/k\rfloor}
},
\\
\label{eq:stationary_filter_geometric_rate}
&\sup_{\dtheta\in\dTheta}
\vvbra{
\dpsithetatti
-
\dbarpsitti(\dtheta)
}
=
O_{a.s.}\bra{
\varrho_k^{\lfloor(t-1)/k\rfloor}
}.
\end{align}
For completeness, \eqref{eq:stationary_filter_geometric_rate} follows by inserting between the two filters the filter driven by
$\cbra{\dbarFt}_{t\in\Zz}$ and initialized at $\dpsi_{1|0}$. The difference between the first two filters is controlled by the same
iteration used in \eqref{eq:O_sum_calXk2}, whereas the difference between the latter filter and $\cbra{\dbarpsitti(\dtheta)}_{t\in\Zz}$ is controlled by Proposition \ref{prop:rate_psit_decomp}$(c)$, or equivalently by \eqref{eq:EXPDEC}. Enlarging $\varrho_k$, if necessary, allows the two bounds to be written with the same geometric rate.

The mean value theorem and Assumption \ref{ass:moment_score} then imply
\begin{align}
\label{eq:stationary_likelihood_contribution_bound}
\sup_{\dtheta\in\dTheta}
\vbra{
\ell_{\dF}
\bra{
\dFt;
\dtheta
}
-
\ell_{\dF}
\bra{
\dbarFt;
\dtheta
}
}
\leq
D_{2,\dpsitti}
\sup_{\dtheta\in\dTheta}
\vvbra{
\dpsithetatti
-
\dbarpsitti(\dtheta)
}
+
D_{2,\dFt}
\vvbra{
\dFt-\dbarFt
},
\end{align}
almost surely. Here, the two derivative envelopes are evaluated along the corresponding line segments in $\boldsymbol{\calP}$ and $\Rr^r$, as allowed by Assumption \ref{ass:moment_score}. It remains to verify that the derivative envelopes do not alter the geometric rate. Let $q=2p/(p-2)>1$, as in Assumption \ref{ass:moment_score}, and choose a real number $a$ such that $1<a<\frac{1}{\varrho_k}$. By Markov's inequality and Assumption \ref{ass:moment_score},
\begin{align}
\label{eq:derivative_envelope_summability}
&
\sum_{t=1}^{\infty}
\Pp\bra{
D_{2,\dpsitti}
>
a^{\lfloor(t-1)/k\rfloor}
}
+
\sum_{t=1}^{\infty}
\Pp\bra{
D_{2,\dFt}
>
a^{\lfloor(t-1)/k\rfloor}
}
\nonumber
\\
\leq{}&
\sup_{t\geq1}
\Ee\sbra{
D_{2,\dpsitti}^{\,q}
+
D_{2,\dFt}^{\,q}
}
\sum_{t=1}^{\infty}
a^{-q\lfloor(t-1)/k\rfloor}
<\infty.
\end{align}
By the Borel-Cantelli lemma we therefore get
\begin{align}
\label{eq:derivative_envelope_geometric_growth}
D_{2,\dpsitti}
+
D_{2,\dFt}
=
O_{a.s.}\bra{
a^{\lfloor(t-1)/k\rfloor}
}.
\end{align}
Combining \eqref{eq:stationary_factor_geometric_rate}, \eqref{eq:stationary_filter_geometric_rate}, \eqref{eq:stationary_likelihood_contribution_bound}, and \eqref{eq:derivative_envelope_geometric_growth}, we obtain
\begin{align}
\label{eq:stationary_likelihood_contribution_geometric_rate}
\sup_{\dtheta\in\dTheta}
\vbra{
\ell_{\dF}
\bra{
\dFt;
\dtheta
}
-
\ell_{\dF}
\bra{
\dbarFt;
\dtheta
}
}
=
O_{a.s.}\bra{
\bra{a\varrho_k}^{\lfloor(t-1)/k\rfloor}
}.
\end{align}
Since $a\varrho_k<1$, the right-hand side is summable, and hence,
\begin{align}
&
\sup_{\dtheta\in\dTheta}
\vbra{
\frac{1}{T}
\sum_{t=1}^T
\ell_{\dF}
\bra{
\dFt;
\dtheta
}
-
\frac{1}{T}
\sum_{t=1}^T
\ell_{\dF}
\bra{
\dbarFt;
\dtheta
}
}
\nonumber
\\
\leq{}&
\frac{1}{T}
\sum_{t=1}^{T}
\sup_{\dtheta\in\dTheta}
\vbra{
\ell_{\dF}
\bra{
\dFt;
\dtheta
}
-
\ell_{\dF}
\bra{
\dbarFt;
\dtheta
}
}
=
O_{a.s.}\bra{
\frac{1}{T}
\sum_{t=1}^{\infty}
\bra{
a\varrho_k
}^{\lfloor(t-1)/k\rfloor}
}
=
O_{a.s.}\bra{\frac{1}{T}},
\label{eq:stationary_likelihood_Cesaro_rate}
\end{align}
as $\sum_{t=1}^{\infty} \bra{ a\varrho_k }^{\lfloor(t-1)/k\rfloor} = \frac{k}{1-a\varrho_k} <\infty$. Equation \eqref{eq:stationary_likelihood_Cesaro_rate} implies
\begin{align}
\label{eq:stationary_likelihood_rate}
&
\sup_{\dtheta\in\dTheta}
\vbra{
\frac{1}{T}
\sum_{t=1}^T
\ell_{\dF}
\bra{
\dFt;
\dtheta
}
-
\frac{1}{T}
\sum_{t=1}^T
\ell_{\dF}
\bra{
\dbarFt;
\dtheta
}
}
=
O_{\Pp}\bra{\frac{1}{T}}.
\end{align}

\bigskip
\noindent
\textbf{Controlling the term in \eqref{eq:log_like_decomposition_term4}}\\
For every $\dtheta\in\dTheta$, define
\begin{align}
\label{eq:centered_stationary_likelihood_contribution}
Z_t(\dtheta)
=
\ell_{\dF}
\bra{
\dbarFt;
\dtheta
}
-
\Ee\sbra{
\ell_{\dF}
\bra{
\dbarFt;
\dtheta
}
}.
\end{align}
By construction, $\Ee\sbra{Z_t(\dtheta)}=0$ for every $\dtheta\in\dTheta$.

We first verify the dependence condition for the entire likelihood family. For any $\tau\geq1$, consider the jointly truncated factor and
filter processes, obtained by initializing the score-driven recursion at $t-\tau$ and subsequently using $\dvt,\ldots,\dv_{t-\tau}$. Thus, $\dFt^{(\tau)}$ and $\dpsi_{t|t-1}^{(\tau)}(\dtheta)$ are measurable functions of $\dvt,\ldots,\dv_{t-\tau}$. The proof of Lemma \ref{lemma:NED}, together with the uniform contraction in Assumption \ref{ass:contraction_cond}, applies uniformly over $\dtheta\in\dTheta$ and yields
\begin{align}
\label{eq:uniform_NED_stationary_pair}
\sup_{t\in\Nn}
\cbra{
\vvbra{
\dbarFt-\dFt^{(\tau)}
}_2
+
\vvbra{
\sup_{\dtheta\in\dTheta}
\vvbra{
\dbarpsitti(\dtheta)
-
\dpsi_{t|t-1}^{(\tau)}(\dtheta)
}
}_2
}
=
O\bra{
\calX_k^{\lfloor\tau/k\rfloor}
}.
\end{align}
Let $Z_t^{(\tau)}(\dtheta)$ denote the likelihood contribution computed from $\dFt^{(\tau)}$ and $\dpsi_{t|t-1}^{(\tau)}(\dtheta)$, after subtracting the same unconditional mean as in \eqref{eq:centered_stationary_likelihood_contribution}. Therefore, $Z_t^{(\tau)}(\dtheta)$ is measurable with respect to $\sigma\bra{\dvt,\ldots,\dv_{t-\tau}}$. Since the same unconditional mean is subtracted from $Z_t(\dtheta)$ and $Z_t^{(\tau)}(\dtheta)$, their difference is equal to the difference between the corresponding likelihood contributions. Therefore, by the mean value theorem,
\begin{align}
&
\sup_{\dtheta\in\dTheta}
\vbra{
Z_t(\dtheta)
-
Z_t^{(\tau)}(\dtheta)
}
\nonumber
\\
\leq{}&
D_{2,\dpsitti}
\sup_{\dtheta\in\dTheta}
\vvbra{
\dbarpsitti(\dtheta)
-
\dpsi_{t|t-1}^{(\tau)}(\dtheta)
}
+
D_{2,\dFt}
\vvbra{
\dbarFt-\dFt^{(\tau)}
},
\label{eq:truncated_likelihood_mean_value_bound}
\end{align}
almost surely, where the derivative envelopes are evaluated along the corresponding
line segments in $\Rr^r$ and $\boldsymbol{\calP}$, as allowed by
Assumption \ref{ass:moment_score}.

Taking the $L_1$-norm on both sides of
\eqref{eq:truncated_likelihood_mean_value_bound} and applying the
Cauchy--Schwarz inequality we get
\begin{align}
&
\vvbra{
\sup_{\dtheta\in\dTheta}
\vbra{
Z_t(\dtheta)
-
Z_t^{(\tau)}(\dtheta)
}
}_1
\nonumber
\\
\leq{}&
\vvbra{
D_{2,\dpsitti}
}_2
\vvbra{
\sup_{\dtheta\in\dTheta}
\vvbra{
\dbarpsitti(\dtheta)
-
\dpsi_{t|t-1}^{(\tau)}(\dtheta)
}
}_2
+
\vvbra{
D_{2,\dFt}
}_2
\vvbra{
\dbarFt-\dFt^{(\tau)}
}_2
.
\label{eq:truncated_likelihood_L1_bound}
\end{align}
Since $q=2p/(p-2)>2$, Assumption \ref{ass:moment_score}
implies that the two derivative envelopes have uniformly bounded
second moments. Hence, combining
\eqref{eq:truncated_likelihood_L1_bound} with
\eqref{eq:uniform_NED_stationary_pair}, we obtain
\begin{align}
\label{eq:uniform_L1_truncated_likelihood_bound}
\sup_{t\in\Nn}
\vvbra{
\sup_{\dtheta\in\dTheta}
\vbra{
Z_t(\dtheta)
-
Z_t^{(\tau)}(\dtheta)
}
}_1
=
O\bra{
\calX_k^{\lfloor\tau/k\rfloor}
}.
\end{align}
In addition, Assumption
\ref{ass:smooth_domination_likelihood}, applied to both the stationary
and the truncated likelihood contributions, and Minkowski's
inequality imply that
\begin{align}
\label{eq:uniform_Lp_truncated_likelihood_bound}
\sup_{\tau\geq1}
\sup_{t\in\Nn}
\vvbra{
\sup_{\dtheta\in\dTheta}
\vbra{
Z_t(\dtheta)
-
Z_t^{(\tau)}(\dtheta)
}
}_p
<\infty.
\end{align}
We can now interpolate between
\eqref{eq:uniform_L1_truncated_likelihood_bound} and
\eqref{eq:uniform_Lp_truncated_likelihood_bound}. In particular,
applying H\"{o}lder's inequality to the underlying expectation yields
\begin{align*}
&
\Ee\sbra{
\bra{
\sup_{\dtheta\in\dTheta}
\vbra{
Z_t(\dtheta)
-
Z_t^{(\tau)}(\dtheta)
}
}^2
}
\\
={}&
\Ee\sbra{
\bra{
\sup_{\dtheta\in\dTheta}
\vbra{
Z_t(\dtheta)
-
Z_t^{(\tau)}(\dtheta)
}
}^{
\frac{p-2}{p-1}
}
\bra{
\sup_{\dtheta\in\dTheta}
\vbra{
Z_t(\dtheta)
-
Z_t^{(\tau)}(\dtheta)
}
}^{
\frac{p}{p-1}
}
}
\\
\leq{}&
\bra{
\Ee\sbra{
\sup_{\dtheta\in\dTheta}
\vbra{
Z_t(\dtheta)
-
Z_t^{(\tau)}(\dtheta)
}
}
}^{
\frac{p-2}{p-1}
}
\bra{
\Ee\sbra{
\bra{
\sup_{\dtheta\in\dTheta}
\vbra{
Z_t(\dtheta)
-
Z_t^{(\tau)}(\dtheta)
}
}^p
}
}^{
\frac{1}{p-1}
}.
\end{align*}
Rewriting these expectations back into their respective $L^1$ and $L^p$ norm representations 
\begin{align*}
&
\vvbra{
\sup_{\dtheta\in\dTheta}
\vbra{
Z_t(\dtheta)
-
Z_t^{(\tau)}(\dtheta)
}
}_2^2
\\
\leq{}&
\vvbra{
\sup_{\dtheta\in\dTheta}
\vbra{
Z_t(\dtheta)
-
Z_t^{(\tau)}(\dtheta)
}
}_1^{
\frac{p-2}{p-1}
}
\vvbra{
\sup_{\dtheta\in\dTheta}
\vbra{
Z_t(\dtheta)
-
Z_t^{(\tau)}(\dtheta)
}
}_p^{
\frac{p}{p-1}
}.
\end{align*}
Taking the square root of both sides then yields the final statement
\begin{align}
\label{eq:interp_conv}
&
\vvbra{
\sup_{\dtheta\in\dTheta}
\vbra{
Z_t(\dtheta)
-
Z_t^{(\tau)}(\dtheta)
}
}_2
\nonumber
\\
\leq{}&
\vvbra{
\sup_{\dtheta\in\dTheta}
\vbra{
Z_t(\dtheta)
-
Z_t^{(\tau)}(\dtheta)
}
}_1^{
\frac{p-2}{2\bra{p-1}}
}
\vvbra{
\sup_{\dtheta\in\dTheta}
\vbra{
Z_t(\dtheta)
-
Z_t^{(\tau)}(\dtheta)
}
}_p^{
\frac{p}{2\bra{p-1}}
},
\end{align}
where $ \frac12 = \frac{p-2}{2\bra{p-1}} + \frac{p}{2\bra{p-1}}\frac1p$.
%

Consequently,
\begin{align}
\label{eq:uniform_truncated_likelihood_bound}
\sup_{t\in\Nn}
\vvbra{
\sup_{\dtheta\in\dTheta}
\vbra{
Z_t(\dtheta)
-
Z_t^{(\tau)}(\dtheta)
}
}_2
=
O\bra{
\calX_k^{
\frac{p-2}{2\bra{p-1}}
\lfloor\tau/k\rfloor
}
}.
\end{align}
Since $Z_t^{(\tau)}(\dtheta)$ is measurable with respect to $\sigma\bra{\dvt,\ldots,\dv_{t-\tau}}$, we have
\begin{align*}
&
Z_t(\dtheta)
-
\Ee\sbra{
Z_t(\dtheta)
\mid
\dvt,\ldots,\dv_{t-\tau}
}
\\
={}&
Z_t(\dtheta)
-
Z_t^{(\tau)}(\dtheta)
-
\Ee\sbra{
\left.
Z_t(\dtheta)
-
Z_t^{(\tau)}(\dtheta)
\right|
\dvt,\ldots,\dv_{t-\tau}
}.
\end{align*}
Therefore, Minkowski's inequality and the contraction property of conditional expectations 
\begin{align*}
&
\vvbra{
\sup_{\dtheta\in\dTheta}
\vbra{
Z_t(\dtheta)
-
\Ee\sbra{
Z_t(\dtheta)
\mid
\dvt,\ldots,\dv_{t-\tau}
}
}
}_2
\\
\leq{}&
2
\vvbra{
\sup_{\dtheta\in\dTheta}
\vbra{
Z_t(\dtheta)
-
Z_t^{(\tau)}(\dtheta)
}
}_2.
\end{align*}
Combining this inequality with
\eqref{eq:uniform_truncated_likelihood_bound} yields
\begin{align}
\label{eq:uniform_NED_stationary_likelihood}
\sup_{t\in\Nn}
\vvbra{
\sup_{\dtheta\in\dTheta}
\vbra{
Z_t(\dtheta)
-
\Ee\sbra{
Z_t(\dtheta)
\mid
\dvt,\ldots,\dv_{t-\tau}
}
}
}_2
=
O\bra{
\calX_k^{
\frac{p-2}{2\bra{p-1}}
\lfloor\tau/k\rfloor
}
}.
\end{align}
Hence, $\cbra{Z_t(\dtheta)}_{t\in\Zz}$ is uniformly near-epoch dependent of size $-1$ on $\cbra{\dvt}_{t\in\Zz}$, uniformly over
$\dtheta\in\dTheta$. Moreover, by Assumption \ref{ass:smooth_domination_likelihood} 
\begin{align}
\label{eq:stationary_likelihood_uniform_moment}
&
\sup_{t\in\Zz}
\Ee\sbra{
\sup_{\dtheta\in\dTheta}
\vbra{
Z_t(\dtheta)
}^{p}
}
<\infty
\qquad
\text{for some }p>2.
\end{align}
To obtain the root-$T$ bound uniformly over $\dTheta$, we must control the complexity of the class of centered likelihood contributions. To this end, let us fix $2<s<p$, so that by Assumption \ref{ass:smooth_domination_likelihood}, there exists a nonnegative process $\cbra{L_t}_{t\in\Zz}$ such that, almost surely,
\begin{align}
\label{eq:stationary_likelihood_Lipschitz_bound}
\vbra{
\ell_{\dF}
\bra{
\dbarFt;
\dtheta_1
}
-
\ell_{\dF}
\bra{
\dbarFt;
\dtheta_2
}
}
\leq
L_t
\vvbra{
\dtheta_1-\dtheta_2
},
\qquad
\dtheta_1,\dtheta_2\in\dTheta,
\end{align}
and $\sup_{t\in\Zz}\Ee\sbra{L_t^p}<\infty$. It follows from the definition of $Z_t(\dtheta)$ and \eqref{eq:stationary_likelihood_Lipschitz_bound} that, almost surely,
\begin{align}
\label{eq:centered_stationary_likelihood_Lipschitz_bound}
\vbra{
Z_t(\dtheta_1)
-
Z_t(\dtheta_2)
}
\leq
\bra{
L_t+\Ee\sbra{L_t}
}
\vvbra{
\dtheta_1-\dtheta_2
}.
\end{align}
Moreover, since $2<s<p$,
\begin{align}
\label{eq:centered_stationary_likelihood_Lipschitz_moment}
\sup_{t\in\Zz}
\vvbra{
L_t+\Ee\sbra{L_t}
}_s
\leq
2
\sup_{t\in\Zz}
\vvbra{
L_t
}_s
<
\infty.
\end{align}
Recall that, following \citet[][Section~2]{Doukhan1995},
$
N_{\sbra{\,}}\bra{
\varepsilon,
\cbra{
Z_t\bra{\dtheta}:
\dtheta\in\dTheta
},
\vvbra{\,\cdot\,}_s
}
$
denotes the minimum number of brackets $\sbra{\underline Z_t,\overline Z_t}$ satisfying $\underline Z_t\leq\overline Z_t$ almost surely and
$\vvbra{\overline Z_t-\underline Z_t}_s\leq\varepsilon$ required to cover $\cbra{Z_t\bra{\dtheta}:\dtheta\in\dTheta}$. If $\dtheta_1,\ldots,\dtheta_J$ form a Euclidean $\delta$-cover of $\dTheta$, the inequality in \eqref{eq:centered_stationary_likelihood_Lipschitz_bound}
shows that the brackets of the form
$
\sbra{
Z_t(\dtheta_j)
-
\delta\bra{L_t+\Ee\sbra{L_t}},
\,
Z_t(\dtheta_j)
+
\delta\bra{L_t+\Ee\sbra{L_t}}
},$ for  $j=1,\ldots,J,$ cover the centered likelihood class and have $L_s$-size bounded by
$
2\delta
\sup_{u\in\Zz}
\vvbra{
L_u+\Ee\sbra{L_u}
}_s.
$
Since $\dTheta$ is a compact subset of $\Rr^d$, choosing
$
\delta
=
\frac{\varepsilon}{
2\sup_{u\in\Zz}
\vvbra{
L_u+\Ee\sbra{L_u}
}_s
}
$
gives, for some finite positive real $C_{\dTheta}$ and every $0<\varepsilon\leq1$,
\begin{align}
\label{eq:stationary_likelihood_covering_number}
N_{\sbra{\,}}\bra{
\varepsilon,
\cbra{
Z_t\bra{\dtheta}:
\dtheta\in\dTheta
},
\vvbra{\,\cdot\,}_s
}
\leq
C_{\dTheta}
\varepsilon^{-d}.
\end{align}
Consequently,
\begin{align}
\label{eq:stationary_likelihood_entropy_integral}
&
\int_0^1
\sqrt{
\log
N_{\sbra{\,}}\bra{
\varepsilon,
\cbra{
Z_t\bra{\dtheta}:
\dtheta\in\dTheta
},
\vvbra{\,\cdot\,}_s
}
}
\,\mathrm d\varepsilon
\nonumber
\\
\leq{}&
\int_0^1
\sqrt{
\log C_{\dTheta}
+
d\log\bra{\frac{1}{\varepsilon}}
}
\,\mathrm d\varepsilon
<
\infty.
\end{align}
This is the bracketing entropy condition used in Theorem~1 and Application~1 of \citet{Doukhan1995}.  Next, to control the approximation of the stationary likelihood by its finite-window counterpart in the same $L_s$ norm, we note that since
$
\frac{1}{s}
=
\frac{2\bra{p-s}}{s\bra{p-2}}\frac{1}{2}
+
\frac{p\bra{s-2}}{s\bra{p-2}}\frac{1}{p},
$
by H\"{o}lder's inequality 
\begin{align*}
&
\vvbra{
\sup_{\dtheta\in\dTheta}
\vbra{
Z_t\bra{\dtheta}
-
Z_t^{(\tau)}\bra{\dtheta}
}
}_s
\\
\leq{}&
\vvbra{
\sup_{\dtheta\in\dTheta}
\vbra{
Z_t\bra{\dtheta}
-
Z_t^{(\tau)}\bra{\dtheta}
}
}_2^{
\frac{2\bra{p-s}}{s\bra{p-2}}
}
\vvbra{
\sup_{\dtheta\in\dTheta}
\vbra{
Z_t\bra{\dtheta}
-
Z_t^{(\tau)}\bra{\dtheta}
}
}_p^{
\frac{p\bra{s-2}}{s\bra{p-2}}
}.
\end{align*}
Combining \eqref{eq:uniform_truncated_likelihood_bound} and \eqref{eq:uniform_Lp_truncated_likelihood_bound} therefore yields
\begin{align}
\label{eq:uniform_Ls_truncated_likelihood_bound}
\sup_{t\in\Nn}
\vvbra{
\sup_{\dtheta\in\dTheta}
\vbra{
Z_t\bra{\dtheta}
-
Z_t^{(\tau)}\bra{\dtheta}
}
}_s
=
O\bra{
\calX_k^{
\frac{p-s}{s\bra{p-1}}
\lfloor\tau/k\rfloor
}
}.
\end{align}
For every fixed $\tau$, the process
$\cbra{Z_t^{(\tau)}\bra{\dtheta}}_{t\in\Zz}$ is a measurable function
of the finite blocks
$\cbra{\dvt,\ldots,\dv_{t-\tau}}_{t\in\Zz}$. Hence, for every
$h>\tau$, its $\beta$-mixing coefficients satisfy
$
\beta_{Z^{(\tau)}}\bra{h}
\leq
\beta_{\dv}\bra{h-\tau}.
$
It follows from Assumption \ref{ass:noise}$(i)$ that, for every fixed
$\tau$, the finite-window process is geometrically $\beta$-mixing.

Consider now the centered finite-window decomposition
\begin{align*}
Z_t\bra{\dtheta}
={}&
Z_t^{(1)}\bra{\dtheta}
-
\Ee\sbra{
Z_t^{(1)}\bra{\dtheta}
}
\\
&+
\sum_{j=1}^{\infty}
\bra{
Z_t^{(2^j)}\bra{\dtheta}
-
Z_t^{(2^{j-1})}\bra{\dtheta}
-
\Ee\sbra{
Z_t^{(2^j)}\bra{\dtheta}
-
Z_t^{(2^{j-1})}\bra{\dtheta}
}
},
\end{align*}
which holds in $L_s$ by \eqref{eq:uniform_Ls_truncated_likelihood_bound}. For every$j\geq1$, Minkowski's inequality and \eqref{eq:uniform_Ls_truncated_likelihood_bound} give
\begin{align}
&
\sup_{t\in\Zz}
\vvbra{
\sup_{\dtheta\in\dTheta}
\Bigg|
Z_t^{(2^j)}\bra{\dtheta}
-
Z_t^{(2^{j-1})}\bra{\dtheta}
-
\Ee\sbra{
Z_t^{(2^j)}\bra{\dtheta}
-
Z_t^{(2^{j-1})}\bra{\dtheta}
}
\Bigg|
}_s
\nonumber
\\
\leq{}&
2
\sup_{t\in\Zz}
\vvbra{
\sup_{\dtheta\in\dTheta}
\vbra{
Z_t\bra{\dtheta}
-
Z_t^{(2^j)}\bra{\dtheta}
}
}_s
+
2
\sup_{t\in\Zz}
\vvbra{
\sup_{\dtheta\in\dTheta}
\vbra{
Z_t\bra{\dtheta}
-
Z_t^{(2^{j-1})}\bra{\dtheta}
}
}_s
\nonumber
\\
={}&
O\bra{
\calX_k^{
\frac{p-s}{s\bra{p-1}}
\lfloor 2^{j-1}/k\rfloor
}
}.
\label{eq:finite_window_difference_Ls_bound}
\end{align}
The centered difference in \eqref{eq:finite_window_difference_Ls_bound} is measurable with respect to
$\sigma\bra{\dvt,\ldots,\dv_{t-2^j}}$. Hence, for $h>2^j$, its $\beta$-mixing coefficients are bounded by
$\beta_{\dv}\bra{h-2^j}$.

The same Lipschitz argument used in \eqref{eq:centered_stationary_likelihood_Lipschitz_bound}--\eqref{eq:stationary_likelihood_entropy_integral} gives the
corresponding polynomial $L_s$-bracketing bound for these finite-window differences. Therefore, Theorem~1 of \citet{Doukhan1995}, together with
\eqref{eq:finite_window_difference_Ls_bound}, imply that, for some finite positive real $C$ independent of $j$,
\begin{align*}
&
\limsup_{T\rightarrow\infty}
\Ee\sbra{
\sup_{\dtheta\in\dTheta}
\Bigg|
\frac{1}{\sqrt T}
\sum_{t=1}^{T}
\bra{
Z_t^{(2^j)}\bra{\dtheta}
-
Z_t^{(2^{j-1})}\bra{\dtheta}
-
\Ee\sbra{
Z_t^{(2^j)}\bra{\dtheta}
-
Z_t^{(2^{j-1})}\bra{\dtheta}
}
}
\Bigg|
}
\nonumber
\\
&\leq
C\,2^j
\calX_k^{
\frac{p-s}{s\bra{p-1}}
\lfloor 2^{j-1}/k\rfloor
}.
\end{align*}
Since $0<\calX_k<1$, $
\sum_{j=1}^{\infty}
2^j
\calX_k^{
\frac{p-s}{s\bra{p-1}}
\lfloor 2^{j-1}/k\rfloor
}
<
\infty$, and therefore, applying the same result to $Z_t^{(1)}\bra{\dtheta}-\Ee\sbra{Z_t^{(1)}\bra{\dtheta}}$ and using the centered finite-window decomposition therefore yields
\begin{align}
\label{eq:stationary_likelihood_root_T_maximal_bound}
\sup_{\dtheta\in\dTheta}
\vbra{
\frac{1}{\sqrt T}
\sum_{t=1}^{T}
Z_t\bra{\dtheta}
}
=
O_{\Pp}(1)
\end{align}
as $T\rightarrow\infty$. It follows that
\begin{align}
\label{eq:stationary_uniform_likelihood_rate}
&
\sup_{\dtheta\in\dTheta}
\vbra{
\frac{1}{T}
\sum_{t=1}^{T}
\ell_{\dF}
\bra{
\dbarFt;
\dtheta
}
-
\Ee\sbra{
\ell_{\dF}
\bra{
\dbarFt;
\dtheta
}
}
}
=
O_{\Pp}\bra{
\frac{1}{\sqrt T}
}.
\end{align}

\bigskip
\noindent
\textbf{Concluding the Proof of Proposition \ref{lemma:ULLN}}\\
By combining the rates in \eqref{eq:weighted_likelihood_rate}, \eqref{eq:rotation_equivariance_rate},
\eqref{eq:stationary_likelihood_rate}, and \eqref{eq:stationary_uniform_likelihood_rate}, and taking into account
the triangle inequality in \eqref{eq:log_like_decomposition}, we finally obtain
\begin{align*}
\sup_{\dtheta\in\dTheta}
\vbra{
\frac{1}{T}
\sum_{t=1}^{T}
\doublehat{\ell}_{\dF}
\bra{
\dhatFt;
\boldsymbol r_{\widehat\dH}(\dtheta)
}
+
\log\vbra{\widehat\dH}
-
\Ee\sbra{
\ell_{\dF}\bra{\dbarFt;\dtheta}
}
}
=&
O_{\Pp}\bra{
\frac{1}{\sqrt T}
+
\frac{1}{\sqrt{NT}}
+
\frac{1}{N}
+
\frac{1}{T}
}
\\
=&
O_{\Pp}\bra{
\frac{1}{N}
+
\frac{1}{\sqrt T}
}.
\end{align*}
This completes the proof.
\qed

\subsubsection*{Proof of Theorem \ref{thm:consistency}}

Since $\dH_0\in GL\bra{r}$ and $\widehat\dH\pconv\dH_0$ as $N,T\rightarrow\infty$ by Proposition \ref{prop:rate_yt_decomp}, we have
\[
\Pp\bra{
\widehat\dH\in GL\bra{r}
}
\longrightarrow
1,
\qquad
N,T\rightarrow\infty.
\]
The following arguments can therefore be restricted to this event, whose complement is asymptotically negligible. Since $\boldsymbol r_{\widehat\dH}$ is an invertible reparameterization of $\dTheta$ onto itself, the definition of $\dthetamle$ in \eqref{eq:MLE} implies, on the event $\widehat\dH\in GL\bra{r}$, that
\begin{align}
\label{eq:aligned_QMLE_maximizer}
\boldsymbol r_{\widehat\dH}^{-1}
\bra{
\dthetamle
}
\in
\argmax_{\dtheta\in\dTheta}
\cbra{
\frac{1}{T}
\sum_{t=1}^T
\doublehat{\ell}_{\dF}
\bra{
\dhatFt;
\boldsymbol r_{\widehat\dH}(\dtheta)
}
+
\log\vbra{\widehat\dH}
},
\end{align}
almost surely, where $\log\vbra{\widehat\dH}$ does not affect the maximizer because it does not depend on $\dtheta$.

Define the uniform approximation error
\begin{align}
\label{eq:criterion_uniform_error}
\delta_{NT}
=
\sup_{\dtheta\in\dTheta}
\vbra{
\frac{1}{T}
\sum_{t=1}^T
\doublehat{\ell}_{\dF}
\bra{
\dhatFt;
\boldsymbol r_{\widehat\dH}(\dtheta)
}
+
\log\vbra{\widehat\dH}
-
\Ee\sbra{
\ell_{\dF}
\bra{
\dbarFt;
\dtheta
}
}
}.
\end{align}
Thus, for every $\dtheta\in\dTheta$,
\begin{align}
\label{eq:criterion_uniform_error_bound}
&
-\delta_{NT}
\leq
\frac{1}{T}
\sum_{t=1}^T
\doublehat{\ell}_{\dF}
\bra{
\dhatFt;
\boldsymbol r_{\widehat\dH}(\dtheta)
}
+
\log\vbra{\widehat\dH}
-
\Ee\sbra{
\ell_{\dF}
\bra{
\dbarFt;
\dtheta
}
}
\leq
\delta_{NT},
\end{align}
almost surely. By Proposition \ref{lemma:ULLN}, as $N,T\rightarrow\infty$, $\min\bra{N,\sqrt T}\delta_{NT} =O_{\Pp}(1)$, and since $\min\bra{N,\sqrt T} \longrightarrow\infty$ as $N,T\rightarrow\infty$, it follows that
\begin{equation}
\label{eq:criterion_uniform_error_op}
\delta_{NT}
=
o_{\Pp}(1),
\qquad
N,T\rightarrow\infty.
\end{equation} 
For every $\dvartheta\in\dTheta_{\dH}$, \eqref{eq:aligned_QMLE_maximizer} and \eqref{eq:criterion_uniform_error_bound} imply, almost surely on the event $\widehat\dH\in GL\bra{r}$, that
\begin{align}
&
\Ee\sbra{
\ell_{\dF}
\bra{
\dbarFt;
\boldsymbol r_{\widehat\dH}^{-1}
\bra{
\dthetamle
}
}
}
\nonumber
\\
\geq{}&
\frac{1}{T}
\sum_{t=1}^T
\doublehat{\ell}_{\dF}
\bra{
\dhatFt;
\boldsymbol r_{\widehat\dH}
\bra{
\boldsymbol r_{\widehat\dH}^{-1}
\bra{
\dthetamle
}
}
}
+
\log\vbra{\widehat\dH}
-
\delta_{NT}
\nonumber
\\
\geq{}&
\frac{1}{T}
\sum_{t=1}^T
\doublehat{\ell}_{\dF}
\bra{
\dhatFt;
\boldsymbol r_{\widehat\dH}
\bra{
\dvartheta
}
}
+
\log\vbra{\widehat\dH}
-
\delta_{NT}
\nonumber
\\
\geq{}&
\Ee\sbra{
\ell_{\dF}
\bra{
\dbarFt;
\dvartheta
}
}
-
2\delta_{NT},
\label{eq:population_comparison}
\end{align}
where the first and third inequalities follow from \eqref{eq:criterion_uniform_error_bound}, whereas the second follows from the maximizing property in \eqref{eq:aligned_QMLE_maximizer}.

By Assumption \ref{ass:regular_level_set}$(iii)$ and the definition of $K_T$, every $\dvartheta\in\dTheta_{\dH}$ satisfies
$
\Ee\sbra{
\ell_{\dF}
\bra{
\dbarFt;
\dvartheta
}
}
=
K_T.$
In fact, Assumption \ref{ass:regular_level_set}$(iii)$ gives
\[
\dTheta_{\dH}
=
\cbra{
\dtheta\in\dTheta:
\Ee\sbra{
\ell_{\dF}
\bra{
\dbarFt;
\dtheta
}
}
\geq
K_T
},
\]
whereas
$
K_T
=
\sup_{\dtheta\in\dTheta}
\Ee\sbra{
\ell_{\dF}
\bra{
\dbarFt;
\dtheta
}
}.$
Therefore, the inequality defining the upper level set must hold with
equality for every $\dtheta\in\dTheta_{\dH}$. In particular, it holds that
$
\inf_{\dvartheta\in\dTheta_{\dH}}
\Ee\sbra{
\ell_{\dF}
\bra{
\dbarFt;
\dvartheta
}
}
=
K_T.
$
Consequently, \eqref{eq:population_comparison} implies, on the event $\widehat\dH\in GL\bra{r}$, that
\begin{align}
\label{eq:population_almost_maximizer}
\Ee\sbra{
\ell_{\dF}
\bra{
\dbarFt;
\boldsymbol r_{\widehat\dH}^{-1}
\bra{
\dthetamle
}
}
}
\geq
K_T
-
2\delta_{NT},
\end{align}
almost surely.

Hereafter, let $m_{\dTheta}\bra{\dthetamle,\dTheta_{\dH}} = \inf_{\dvartheta\in\dTheta_{\dH}}\vvbra{\dthetamle - \dvartheta}$.
Now, fix $\varepsilon>0$, then, by Assumption \ref{ass:regular_level_set}$(iv)$, there exist $\eta_{\varepsilon}>0$ and $T_{\varepsilon}<\infty$ such that, for every $T\geq T_{\varepsilon}$,
\begin{align}
\label{eq:population_level_set_gap}
K_T
-
\sup_{{
\dtheta\in\dTheta:
m_{\dTheta}
\bra{
\dtheta,
\dTheta_{\dH}
}
\geq
\varepsilon
}}
\Ee\sbra{
\ell_{\dF}
\bra{
\dbarFt;
\dtheta
}
}
\geq
\eta_{\varepsilon}.
\end{align}
Hence, the population criterion equals $K_T$ on $\dTheta_{\dH}$ and is uniformly smaller than $K_T$ outside every fixed neighbourhood of
$\dTheta_{\dH}$. Then, on the event $\cbra{m_{\dTheta}\bra{\boldsymbol r_{\widehat\dH}^{-1}\bra{\dthetamle},\dTheta_{\dH}}\geq\varepsilon}$, the definition of the supremum in \eqref{eq:population_level_set_gap} implies that
\begin{align}
&
\Ee\sbra{
\ell_{\dF}
\bra{
\dbarF_t;
\boldsymbol r_{\widehat\dH}^{-1}
\bra{
\dthetamle
}
}
}
\nonumber
\\
\leq{}&
\sup_{{
\dtheta\in\dTheta:
m_{\dTheta}
\bra{
\dtheta,
\dTheta_{\dH}
}
\geq
\varepsilon
}}
\Ee\sbra{
\ell_{\dF}
\bra{
\dbarF_t;
\dtheta
}
}
\nonumber
\\
\leq{}&
K_T-\eta_{\varepsilon},
\label{eq:population_outside_orbit}
\end{align}
almost surely. Combining \eqref{eq:population_almost_maximizer} and \eqref{eq:population_outside_orbit} gives, on the event $
m_{\dTheta}\bra{ \boldsymbol r_{\widehat\dH}^{-1}\bra{\dthetamle},  \dTheta_{\dH}}\geq\varepsilon$, that $K_T-2\delta_{NT}\leq K_T-\eta_{\varepsilon}$, and hence $2\delta_{NT} \geq\eta_{\varepsilon}$ almost surely. Consequently,
\begin{align*}
\Pp\bra{
m_{\dTheta}
\bra{
\boldsymbol r_{\widehat\dH}^{-1}
\bra{
\dthetamle
},
\dTheta_{\dH}
}
\geq
\varepsilon
}
\leq
\Pp\bra{
2\delta_{NT}
\geq
\eta_{\varepsilon}
}
+
\Pp\bra{
\widehat\dH\notin GL\bra{r}
}
\longrightarrow
0
\end{align*}
as $N,T\rightarrow\infty$, where the convergence follows from \eqref{eq:criterion_uniform_error_op} and the fact that
\[
\Pp\bra{
\widehat\dH\notin GL\bra{r}
}
\longrightarrow
0,
\qquad
N,T\rightarrow\infty.
\]
Therefore,
\begin{equation}
\label{eq:aligned_set_consistency}
m_{\dTheta}
\bra{
\boldsymbol r_{\widehat\dH}^{-1}
\bra{
\dthetamle
},
\dTheta_{\dH}
}
=
o_{\Pp}(1),
\qquad
N,T\rightarrow\infty.
\end{equation}
We now transfer \eqref{eq:aligned_set_consistency} to $\dthetamle$. Since the mapping $\boldsymbol r_{\dH}(\cdot)$  satisfies
\[
\boldsymbol r_{\dH_1}
\circ
\boldsymbol r_{\dH_2}
=
\boldsymbol r_{\dH_1\dH_2},
\qquad \text{and} \qquad
\boldsymbol r_{\dH}^{-1}
=
\boldsymbol r_{\dH^{-1}},
\]
for any $\dH_1, \dH_2\in GL(r)$  they preserve the orbit. In particular, on the event $\widehat\dH\in GL\bra{r}$, we have that
\begin{equation}
\label{eq:rotation_orbit_invariance}
\boldsymbol r_{\widehat\dH}
\bra{
\dTheta_{\dH}
}
=
\dTheta_{\dH},
\end{equation}
almost surely. Indeed, if $\dvartheta=\boldsymbol r_{\dG}(\dtheta_0)\in\dTheta_{\dH}$ for some $\dG\in GL\bra{r}$, then
\[
\boldsymbol r_{\widehat\dH}
\bra{
\dvartheta
}
=
\boldsymbol r_{\widehat\dH}
\bra{
\boldsymbol r_{\dG}(\dtheta_0)
}
=
\boldsymbol r_{\widehat\dH\dG}(\dtheta_0)
\in
\dTheta_{\dH},
\]
and the reverse inclusion follows by applying the same argument to $\boldsymbol r_{\widehat\dH}^{-1}
=
\boldsymbol r_{\widehat\dH^{-1}}$.
Moreover, $\widehat\dH\pconv\dH_0$ as $N,T\rightarrow\infty$ by Proposition \ref{prop:rate_yt_decomp}, and the mapping $\boldsymbol r_{\dH}(\dtheta)$ is jointly continuous in $\dH$ and $\dtheta$, uniformly over the compact parameter space $\dTheta$ for $\dH$ in a fixed neighbourhood of $\dH_0$. Hence, uniformly for $\widehat\dH$ in that neighbourhood, parameters that are arbitrarily close to $\dTheta_{\dH}$ are mapped into parameters that are arbitrarily close to
$\boldsymbol r_{\widehat\dH}(\dTheta_{\dH})=\dTheta_{\dH}$.

Since, on the event $\widehat\dH\in GL\bra{r}$, $\dthetamle =\boldsymbol r_{\widehat\dH}\bra{\boldsymbol r_{\widehat\dH}^{-1} \bra{\dthetamle}}$ almost surely, equations \eqref{eq:aligned_set_consistency} and \eqref{eq:rotation_orbit_invariance}, together with the uniform continuity of the mapping $\boldsymbol r_{\dH}(\cdot)$, imply
\[
m_{\dTheta}
\bra{
\dthetamle,
\dTheta_{\dH}
}
=
o_{\Pp}(1)
\qquad
\text{as }
N,T\rightarrow\infty,
\]
as claimed.

This proves the first conclusion of the theorem. To establish the second conclusion, fix the limiting factor orientation $\dH=\dH_0$, with $\dH_0$ defined in Proposition \ref{prop:rate_yt_decomp}. By the equivariance of the likelihood, the corresponding population criterion is
$
\Ee\sbra{
\ell_{\dF}
\bra{ \dH_0\dbarF_t; \dtheta }}$. The nondegeneracy of the score-driven process, together with $\vbra{\dK}\neq0$ in Assumption \ref{ass:regular_level_set}$(i)$, implies that this criterion is uniquely maximized at 
$
\boldsymbol r_{\dH_0}
\bra{
\dtheta_0
}
$. Indeed, equality between the population criteria can hold only if the corresponding conditional densities coincide almost surely. The nondegeneracy of the score-driven process then implies equality of the associated filtered processes and, since $\vbra{\dK}\neq0$, equality of their static parameters. Hence,
$
\dtheta
=
\boldsymbol r_{\dH_0}
\bra{
\dtheta_0
}
$. Moreover, Proposition \ref{lemma:ULLN}, the equivariance of the likelihood, $\widehat\dH\pconv\dH_0$ as $N,T\rightarrow\infty$ from \eqref{eq:ACCAlim} in the Proof of Proposition \ref{prop:rate_yt_decomp}, and the uniform continuity of the mapping $\boldsymbol r_{\dH}(\cdot)$ over the compact parameter space imply that
\begin{align*}
\sup_{\dtheta\in\dTheta}
\vbra{
\frac{1}{T}
\sum_{t=1}^T
\doublehat{\ell}_{\dF}
\bra{
\dhatFt;
\dtheta
}
+
\log\vbra{\widehat\dH}
-
\Ee\sbra{
\ell_{\dF}
\bra{ \dH_0\dbarF_t; \dtheta }
}
-
\log\vbra{\dH_0}
}
=
o_{\Pp}(1)
\end{align*}
as $N,T\rightarrow\infty$. Since the two logarithmic terms do not depend on $\dtheta$, they do not affect the corresponding maximizers. Therefore, by the standard argmax argument, together with the unique maximization of the population criterion at $\boldsymbol r_{\dH_0}\bra{\dtheta_0}$, we get
\[
\dthetamle
-
\boldsymbol r_{\dH_0}
\bra{
\dtheta_0
}
=
o_{\Pp}(1)
\]
as $N,T\rightarrow\infty$. This proves the second conclusion.

Finally,  since $\widehat\dH\pconv\dH_0$ as $N,T\rightarrow\infty$ by Proposition \ref{prop:rate_yt_decomp}, and the inverse mapping $\boldsymbol r_{\dH}\inv(\cdot)$ is continuous, the above also implies that
\[
\boldsymbol r_{\widehat\dH}^{-1}
\bra{
\dthetamle
}
-
\dtheta_0
=
o_{\Pp}(1)
\]
as $N,T\rightarrow\infty$. This completes the proof.
\qed

\subsubsection*{Proof of Theorem \ref{thm:asy_norm}}

Since $\dtheta_0$ lies in the interior of $\dTheta$ by Assumption \ref{ass:truePar}, Theorem \ref{thm:consistency} implies that the first-order condition holds with probability approaching one as $N,T\rightarrow\infty$.
By Lemma \ref{lemma:U_convergence_score_vector}, uniformly over
$\dtheta\in\dTheta$,
\begin{align*}
\frac{1}{\sqrt T}
\vvbra{
\sum_{t=1}^{T}
\nabla^{\dtheta}
\doublehat{\ell}_{\dF}
\bra{
\dhatFt;
\dtheta
}
-
\sum_{t=1}^{T}
{\boldsymbol R}_{\widehat\dH}\trans
\nabla^{\dtheta}
\ell_{\dF}
\bra{
\dFt;
\boldsymbol r_{\widehat\dH}^{-1}
\bra{
\dtheta
}
}
}
=
O_{\Pp}
\bra{
\frac{1}{\sqrt N}
+
\frac{\sqrt T}{N}
+
\frac{1}{\sqrt T}
}
=
o_{\Pp}(1)
\end{align*}
as $N,T\rightarrow\infty$, where the last equality follows from $\sqrt T/N\rightarrow0$. Hence, using the first-order condition, and by the change of variables induced by $\boldsymbol r_{\widehat\dH}^{-1}$, we have
\begin{align*}
\boldsymbol 0_d
=
\frac{1}{\sqrt T}
\sum_{t=1}^{T}
\nabla^{\dtheta}
\doublehat{\ell}_{\dF}
\bra{
\dhatFt;
\dthetamle
}
=
\frac{1}{\sqrt T}
\sum_{t=1}^{T}
{\boldsymbol R}_{\widehat\dH}\trans
\nabla^{\dtheta}
\ell_{\dF}
\bra{
\dFt;
\boldsymbol r_{\widehat\dH}^{-1}
\bra{
\dthetamle
}
}
+
o_{\Pp}(1)
\end{align*}
as $N,T\rightarrow\infty$, where
$
{\boldsymbol R}_{\dH}
=
\frac{\partial}{\partial\dtheta\trans}
\boldsymbol r_{\dH}^{-1}(\dtheta)
$
denotes the Jacobian of the inverse transformation mapping. For every fixed $\dH$ such that $\boldsymbol r_{\dH}$ defines an invertible reparameterization of the maintained model, the explicit transformations reported in Appendix \ref{appendix_parameter_identifications} are linear in $\dtheta$. Therefore, ${\boldsymbol R}_{\dH}$ does not depend on $\dtheta$.

A Taylor expansion around $\boldsymbol r_{\widehat\dH}(\dtheta_0)$, followed by a rescaling by $\sqrt T/T$, yields
\begin{align*}
o_{\Pp}(1)
=
\frac{1}{\sqrt T}
\sum_{t=1}^{T}
{\boldsymbol R}_{\widehat\dH}\trans
\nabla^{\dtheta}
\ell_{\dF}
\bra{
\dFt;
\dtheta_0
}
+
\frac{1}{T}
\sum_{t=1}^{T}
{\boldsymbol R}_{\widehat\dH}\trans
\nabla^{\dtheta\dtheta}
\ell_{\dF}
\bra{
\dFt;
\boldsymbol r_{\widehat\dH}^{-1}
\bra{
\dthetamle^\star
}
}
{\boldsymbol R}_{\widehat\dH}
\sqrt T
\bra{
\dthetamle
-
\boldsymbol r_{\widehat\dH}
\bra{
\dtheta_0
}
},
\end{align*}
where $\dthetamle^\star$ lies between $\dthetamle$ and $\boldsymbol r_{\widehat\dH}\bra{\dtheta_0}$, so that
$
\vvbra{
\boldsymbol r_{\widehat\dH}^{-1}
\bra{
\dthetamle^\star
}
-
\dtheta_0
}
\leq
\vvbra{
\boldsymbol r_{\widehat\dH}^{-1}
\bra{
\dthetamle
}
-
\dtheta_0
}$. Since 
$
\boldsymbol r_{\widehat\dH}^{-1}
\bra{
\dthetamle
}
-
\dtheta_0
=
o_{\Pp}(1)
$ as $N,T\rightarrow\infty$ by Theorem \ref{thm:consistency} and $\widehat\dH\pconv\dH_0$ as $N,T\rightarrow\infty$ by Proposition \ref{prop:rate_yt_decomp}, it also follows that
$
\boldsymbol r_{\widehat\dH}^{-1}
\bra{
\dthetamle^\star
}
\pconv
\dtheta_0
$, as $N,T\rightarrow\infty$.

Moreover, since ${\boldsymbol R}_{\widehat\dH}\pconv{\boldsymbol R}_{\dH_0}$ as $N,T\rightarrow\infty$, and $
\frac{1}{\sqrt T}
\sum_{t=1}^{T}
\nabla^{\dtheta}
\ell_{\dF}
\bra{
\dFt;
\dtheta_0
}
=
O_{\Pp}(1)$ by  Lemma \ref{lemma:CLT_NED_score_vector}, we obtain
\begin{align}
\label{eq:asy_norm_expansion}
&
\frac{1}{\sqrt T}
\sum_{t=1}^{T}
{\boldsymbol R}_{\dH_0}\trans
\nabla^{\dtheta}
\ell_{\dF}
\bra{
\dFt;
\dtheta_0
}
+
o_{\Pp}(1)
\nonumber
\\
={}&
-
\frac{1}{T}
\sum_{t=1}^{T}
{\boldsymbol R}_{\widehat\dH}\trans
\nabla^{\dtheta\dtheta}
\ell_{\dF}
\bra{
\dFt;
\boldsymbol r_{\widehat\dH}^{-1}
\bra{
\dthetamle^\star
}
}
{\boldsymbol R}_{\widehat\dH}
\sqrt T
\bra{
\dthetamle
-
\boldsymbol r_{\widehat\dH}
\bra{
\dtheta_0
}
}.
\end{align}
By Lemma \ref{lemma:ULLN_Hessian_matrix},
\begin{align}
\label{eq:uniform_Hessian_ULLN_theorem2}
&
\sup_{\dtheta\in\dTheta}
\vvbra{
\frac{1}{T}
\sum_{t=1}^{T}
\nabla^{\dtheta\dtheta}
\ell_{\dF}
\bra{
\dFt;
\dtheta
}
-
\Ee\sbra{
\nabla^{\dtheta\dtheta}
\ell_{\dF}
\bra{
\dbarFt;
\dtheta
}
}
}
\asconv
0
\end{align}
as $T\rightarrow\infty$. Moreover, since $\boldsymbol r_{\widehat\dH}^{-1} \bra{\dthetamle^\star}\pconv\dtheta_0$ as $N,T\rightarrow\infty$, stationarity, continuity, and the domination condition in Assumption \ref{ass:smooth_domination_score_Hessian} imply that
\begin{align}
\label{eq:population_Hessian_random_parameter}
&
-
\Ee\sbra{
\nabla^{\dtheta\dtheta}
\ell_{\dF}
\bra{
\dbarFt;
\boldsymbol r_{\widehat\dH}^{-1}
\bra{
\dthetamle^\star
}
}
}
=
-
\Ee\sbra{
\nabla^{\dtheta\dtheta}
\ell_{\dF}
\bra{
\dbarF_t;
\boldsymbol r_{\widehat\dH}^{-1}
\bra{
\dthetamle^\star
}
}
}
\pconv
\limitcov
\bra{
\dtheta_0
}
\end{align}
as $N,T\rightarrow\infty$.

Combining \eqref{eq:uniform_Hessian_ULLN_theorem2} and \eqref{eq:population_Hessian_random_parameter}, and using ${\boldsymbol R}_{\widehat\dH}\pconv
{\boldsymbol R}_{\dH_0}$ as $N,T\rightarrow\infty$, 
\begin{align}
\label{eq:rotated_Hessian_convergence}
&
-
\frac{1}{T}
\sum_{t=1}^{T}
{\boldsymbol R}_{\widehat\dH}\trans
\nabla^{\dtheta\dtheta}
\ell_{\dF}
\bra{
\dFt;
\boldsymbol r_{\widehat\dH}^{-1}
\bra{
\dthetamle^\star
}
}
{\boldsymbol R}_{\widehat\dH}
\pconv
{\boldsymbol R}_{\dH_0}\trans
\limitcov
\bra{
\dtheta_0
}
{\boldsymbol R}_{\dH_0}
=
\limitcov_{\dH_0}
\bra{
\dtheta_0
}
\end{align}
as $N,T\rightarrow\infty$.

Since $\boldsymbol r_{\dH_0}^{-1}$ is continuously differentiable and invertible, ${\boldsymbol R}_{\dH_0}$ has full rank. Hence, the positive definiteness of $\limitcov\bra{\dtheta_0}$ implies that $\limitcov_{\dH_0}\bra{\dtheta_0}$ is positive definite and invertible. We can therefore solve \eqref{eq:asy_norm_expansion} to obtain
\begin{align}
\label{eq:original_QMLE_expansion}
&
\sqrt T
\bra{
\dthetamle
-
\boldsymbol r_{\widehat\dH}
\bra{
\dtheta_0
}
}
=
\limitcov_{\dH_0}\inv
\bra{
\dtheta_0
}
\bra{
\frac{1}{\sqrt T}
\sum_{t=1}^{T}
{\boldsymbol R}_{\dH_0}\trans
\nabla^{\dtheta}
\ell_{\dF}
\bra{
\dFt;
\dtheta_0
}
+
o_{\Pp}(1)
}.
\end{align}
By Lemma \ref{lemma:CLT_NED_score_vector},
\[
\frac{1}{\sqrt T}
\sum_{t=1}^{T}
\nabla^{\dtheta}
\ell_{\dF}
\bra{
\dFt;
\dtheta_0
}
\Rightarrow
\calN
\bra{
\boldsymbol 0_d,
\dV
\bra{
\dtheta_0
}
}
\]
as $T\rightarrow\infty$, and hence,
\[
\frac{1}{\sqrt T}
\sum_{t=1}^{T}
{\boldsymbol R}_{\dH_0}\trans
\nabla^{\dtheta}
\ell_{\dF}
\bra{
\dFt;
\dtheta_0
}
\Rightarrow
\calN
\bra{
\boldsymbol 0_d,
\dV_{\dH_0}
\bra{
\dtheta_0
}
}
\]
as $T\rightarrow\infty$, where
\[
\dV_{\dH_0}
\bra{
\dtheta_0
}
=
{\boldsymbol R}_{\dH_0}\trans
\dV
\bra{
\dtheta_0
}
{\boldsymbol R}_{\dH_0}.
\]
Therefore, by Slutsky's theorem,
\begin{align*}
\sqrt T
\bra{
\dthetamle
-
\boldsymbol r_{\widehat\dH}
\bra{
\dtheta_0
}
}
\Rightarrow
\calN
\bra{
\boldsymbol 0_d,
\limitcov_{\dH_0}\inv
\bra{
\dtheta_0
}
\dV_{\dH_0}
\bra{
\dtheta_0
}
\limitcov_{\dH_0}\inv
\bra{
\dtheta_0
}
}
\end{align*}
as $N,T\rightarrow\infty$.
%
This completes the proof.
\qed


\subsubsection*{Proof of Theorem \ref{thm:filter_approx_error_cons}}
The first-order condition and a mean-value expansion, together with Lemmas \ref{lemma:U_convergence_score_vector} and \ref{lemma:ULLN_Hessian_matrix}, imply, as $N,T\rightarrow\infty$, that
\begin{align*}
\boldsymbol r_{\widehat\dH}^{-1}
\bra{
\dthetamle
}
-
\dtheta_0
=
O_{\Pp}\bra{
\frac{1}{N}
+
\frac{1}{\sqrt T}
}.
\end{align*}
Since the alignment mapping is continuously differentiable with a uniformly bounded Jacobian for $\widehat\dH$ in a neighbourhood of $\dH_0$, it follows that
\begin{align}
\label{eq:QMLE_two_rate_prediction}
\dthetamle
-
\boldsymbol r_{\widehat\dH}
\bra{
\dtheta_0
}
=
O_{\Pp}\bra{
\frac{1}{N}
+
\frac{1}{\sqrt T}
}.
\end{align}
For the conditional mean, consider the difference
\begin{align}
\label{eq:diff_tot_common_component_location}
\hatCittithetamle - \Cittithetaz
=&\,
\widehat{\boldsymbol\lambda}_i\trans \dhathatmuthetamletti
-
\boldsymbol\lambda_{0i}\trans \dmutti(\dtheta_0)
\nonumber
\\
=&\,
\bra{\widehat{\boldsymbol{\lambda}}_i\trans - \boldsymbol{\lambda}_{0i}\trans\widehat{\dH}\inv}  
\dhathatmuthetamletti
+
\boldsymbol{\lambda}_{0i}\trans
\bra{\widehat{\dH}\inv\dhathatmuthetamletti -  \dmutti( \dtheta_0 )  }.
\end{align}
By adding and removing the aligned filters, the first term on the right-hand side of \eqref{eq:diff_tot_common_component_location} can be rewritten as
\begin{align}
\label{eq:diff_termI_common_component_location}
&\bra{\widehat{\boldsymbol{\lambda}}_i\trans - \boldsymbol{\lambda}_{0i}\trans\widehat{\dH}\inv}  
\dhathatmuthetamletti
\nonumber
\\
={}&
\bra{\widehat{\boldsymbol{\lambda}}_i\trans - \boldsymbol{\lambda}_{0i}\trans\widehat{\dH}\inv}  
\widehat{\dH} \dmutti( \dtheta_0 )
\nonumber
\\
&+
\bra{\widehat{\boldsymbol{\lambda}}_i\trans - \boldsymbol{\lambda}_{0i}\trans\widehat{\dH}\inv}  
\bra{
\dhathatmuthetamletti
-
\dhathatmutti
\bra{
\boldsymbol r_{\widehat\dH}(\dtheta_0)
}
}
\nonumber
\\
&+
\bra{\widehat{\boldsymbol{\lambda}}_i\trans - \boldsymbol{\lambda}_{0i}\trans\widehat{\dH}\inv}  
\bra{
\dhathatmutti
\bra{
\boldsymbol r_{\widehat\dH}(\dtheta_0)
}
-
\dhatmutti
\bra{
\boldsymbol r_{\widehat\dH}(\dtheta_0)
}
}
\nonumber
\\
&+
\bra{\widehat{\boldsymbol{\lambda}}_i\trans - \boldsymbol{\lambda}_{0i}\trans\widehat{\dH}\inv}
\bra{
\dhatmutti
\bra{
\boldsymbol r_{\widehat\dH}(\dtheta_0)
}
-
\widehat\dH\dmutti(\dtheta_0)
}
\nonumber
\\
={}&
O_{\Pp}\bra{
\max\bra{
\frac{1}{N},
\frac{1}{\sqrt T}
}
}
+
O_{\Pp}\bra{
\max\bra{
\frac{1}{N},
\frac{1}{\sqrt T}
}
}
O_{\Pp}\bra{
\max\bra{
\frac{1}{\sqrt N},
\frac{1}{\sqrt T}
}
+
\calX_k^{\lfloor t/k\rfloor}
}.
\end{align}
The last equality follows from Proposition \ref{prop:rate_yt_decomp}$(c)$, \eqref{eq:QMLE_two_rate_prediction}, Lemma \ref{lemma:rate_psit_first_derivs}, Proposition \ref{prop:rate_psit_decomp}$(a)$, and Assumption \ref{ass:contraction_cond}. In particular, the four terms on the right-hand side account, respectively, for the loading-estimation error, the parameter-estimation error, the pointwise factor-filter error, and the effect of the possibly different initial conditions.

Similarly, the second term on the right-hand side of \eqref{eq:diff_tot_common_component_location} is
\begin{align}
\label{eq:diff_termII_common_component_location}
&\boldsymbol{\lambda}_{0i}\trans
\bra{\widehat{\dH}\inv\dhathatmuthetamletti -  \dmutti( \dtheta_0 )  }
\nonumber
\\
={}&
\boldsymbol{\lambda}_{0i}\trans\widehat{\dH}\inv
\bra{
\dhathatmuthetamletti
-
\dhathatmutti
\bra{
\boldsymbol r_{\widehat\dH}(\dtheta_0)
}
}
\nonumber
\\
&+
\boldsymbol{\lambda}_{0i}\trans\widehat{\dH}\inv
\bra{
\dhathatmutti
\bra{
\boldsymbol r_{\widehat\dH}(\dtheta_0)
}
-
\dhatmutti
\bra{
\boldsymbol r_{\widehat\dH}(\dtheta_0)
}
}
\nonumber
\\
&+
\boldsymbol{\lambda}_{0i}\trans
\bra{
\widehat\dH^{-1}
\dhatmutti
\bra{
\boldsymbol r_{\widehat\dH}(\dtheta_0)
}
-
\dmutti(\dtheta_0)
}
\nonumber
\\
={}&
O_{\Pp}\bra{
\max\bra{
\frac{1}{\sqrt N},
\frac{1}{\sqrt T}
}
}
+
O_{a.s.}\bra{
\calX_k^{\lfloor t/k\rfloor}
}.
\end{align}
The last equality follows from \eqref{eq:QMLE_two_rate_prediction}, Lemma \ref{lemma:rate_psit_first_derivs}, Proposition \ref{prop:rate_psit_decomp}$(a)$, and Assumption \ref{ass:contraction_cond}. In particular, the three terms on the right-hand side account, respectively, for the parameter-estimation error, the pointwise factor-filter error, and the initialization effect after alignment. The last term decreases geometrically because the two filters are driven by the same sequence $\cbra{\widehat\dH\dF_s}_{s=1}^{t-1}$, and by the mapping in \eqref{eq:transform_mu} we get
$
\widehat\dH^{-1}
\dhatmutti
\bra{
\boldsymbol r_{\widehat\dH}(\dtheta_0)
}
-
\dmutti(\dtheta_0)
=
O_{a.s.}\bra{\calX_k^{\lfloor t/k\rfloor}}.
$

Therefore, substituting \eqref{eq:diff_termI_common_component_location}
and \eqref{eq:diff_termII_common_component_location} into
\eqref{eq:diff_tot_common_component_location}, as
$N,T\rightarrow\infty$,
\begin{align*}
\vbra{
\hatCittithetamle
-
\Cittithetaz
}
=
O_{\Pp}\bra{
\frac{1}{\min\bra{N,\sqrt T}}
+
\frac{1}{\min\bra{\sqrt N,T}}
}
+
O_{a.s.}\bra{
\calX_k^{\lfloor t/k\rfloor}
}.
\end{align*}
Analogously, for the conditional variance, we consider the difference
\begin{align}
\label{eq:diff_tot_common_component_variance}
\hatPittithetamle - \Pittithetaz
={}&
\widehat{\boldsymbol\lambda}_i\trans \dhathatSigmathetamletti\widehat{\boldsymbol\lambda}_i
-
\boldsymbol\lambda_{0i}\trans \dSigmatti(\dtheta_0) \boldsymbol\lambda_{0i}
\nonumber
\\
={}&
\bra{\widehat{\boldsymbol\lambda}_i \otimes \widehat{\boldsymbol\lambda}_i}\trans
\vect\dhathatSigmathetamletti
-
\bra{{\boldsymbol\lambda}_{0i} \otimes {\boldsymbol\lambda}_{0i}}\trans
\vect \dSigmatti(\dtheta_0)
\nonumber
\\
={}&
\bra{
\bra{\widehat{\boldsymbol\lambda}_i\trans \otimes \widehat{\boldsymbol\lambda}_i\trans}
-
\bra{{\boldsymbol\lambda}_{0i}\trans\widehat{\dH}\inv \otimes {\boldsymbol\lambda}_{0i}\trans\widehat{\dH}\inv}
}
\dW_r\dhathatsigmathetamletti
\nonumber
\\
&+
\bra{{\boldsymbol\lambda}_{0i}\trans \otimes {\boldsymbol\lambda}_{0i}\trans}
\bra{ \bra{\widehat{\dH}\inv \otimes \widehat{\dH}\inv} \dW_r\dhathatsigmathetamletti
-
\dW_r \dsigmatti(\dtheta_0)
}
.
\end{align}
Using the same decomposition as for the conditional location, the first term on the right-hand side of \eqref{eq:diff_tot_common_component_variance} can be rewritten as
\begin{align}
\label{eq:diff_termI_common_component_variance}
&\bra{
\bra{\widehat{\boldsymbol\lambda}_i\trans \otimes \widehat{\boldsymbol\lambda}_i\trans}
-
\bra{{\boldsymbol\lambda}_{0i}\trans\widehat{\dH}\inv \otimes {\boldsymbol\lambda}_{0i}\trans\widehat{\dH}\inv}
}
\dW_r\dhathatsigmathetamletti
\nonumber
\\
={}&
\bra{
\bra{\widehat{\boldsymbol\lambda}_i\trans \otimes \widehat{\boldsymbol\lambda}_i\trans}
-
\bra{{\boldsymbol\lambda}_{0i}\trans\widehat{\dH}\inv \otimes {\boldsymbol\lambda}_{0i}\trans\widehat{\dH}\inv}
}
\bra{
\widehat\dH
\otimes
\widehat\dH
}
\dW_r\dsigmatti(\dtheta_0)
\nonumber
\\
&+
\bra{
\bra{\widehat{\boldsymbol\lambda}_i\trans \otimes \widehat{\boldsymbol\lambda}_i\trans}
-
\bra{{\boldsymbol\lambda}_{0i}\trans\widehat{\dH}\inv \otimes {\boldsymbol\lambda}_{0i}\trans\widehat{\dH}\inv}
}
\bra{
\dW_r
\dhathatsigmathetamletti
-
\dW_r
\dhathatsigmatti
\bra{
\boldsymbol r_{\widehat\dH}(\dtheta_0)
}
}
\nonumber
\\
&+
\bra{
\bra{\widehat{\boldsymbol\lambda}_i\trans \otimes \widehat{\boldsymbol\lambda}_i\trans}
-
\bra{{\boldsymbol\lambda}_{0i}\trans\widehat{\dH}\inv \otimes {\boldsymbol\lambda}_{0i}\trans\widehat{\dH}\inv}
}
\bra{
\dW_r
\dhathatsigmatti
\bra{
\boldsymbol r_{\widehat\dH}(\dtheta_0)
}
-
\dW_r
\dhatsigmatti
\bra{
\boldsymbol r_{\widehat\dH}(\dtheta_0)
}
}
\nonumber
\\
&+
\bra{
\bra{\widehat{\boldsymbol\lambda}_i\trans \otimes \widehat{\boldsymbol\lambda}_i\trans}
-
\bra{{\boldsymbol\lambda}_{0i}\trans\widehat{\dH}\inv \otimes {\boldsymbol\lambda}_{0i}\trans\widehat{\dH}\inv}
}
\bra{
\dW_r
\dhatsigmatti
\bra{
\boldsymbol r_{\widehat\dH}(\dtheta_0)
}
-
\bra{
\widehat\dH
\otimes
\widehat\dH
}
\dW_r\dsigmatti(\dtheta_0)
}
\nonumber
\\
={}&
O_{\Pp}\bra{
\max\bra{
\frac{1}{N},
\frac{1}{\sqrt T}
}
}
+
O_{\Pp}\bra{
\max\bra{
\frac{1}{N},
\frac{1}{\sqrt T}
}
}
O_{\Pp}\bra{
\max\bra{
\frac{1}{\sqrt N},
\frac{1}{\sqrt T}
}
+
\calX_k^{\lfloor t/k\rfloor}
}.
\end{align}
The last equality follows from Proposition \ref{prop:rate_yt_decomp}$(c)$, \eqref{eq:QMLE_two_rate_prediction}, Lemma \ref{lemma:rate_psit_first_derivs}, Proposition \ref{prop:rate_psit_decomp}$(a)$, and Assumption \ref{ass:contraction_cond}. Moreover, for any conformable matrices $\boldsymbol A$ and $\boldsymbol B$,
$
\vvbra{
\boldsymbol A\otimes\boldsymbol B
}_{\mathrm{sp}}
=
\vvbra{
\boldsymbol A
}_{\mathrm{sp}}
\vvbra{
\boldsymbol B
}_{\mathrm{sp}},
$
so that Proposition \ref{prop:rate_yt_decomp}$(c)$ also gives the same rate for the corresponding difference between the Kronecker products of the estimated and rotated loadings.

Furthermore, the second term on the right-hand side of \eqref{eq:diff_tot_common_component_variance} is
\begin{align}
\label{eq:diff_termII_common_component_variance}
&\bra{{\boldsymbol\lambda}_{0i}\trans \otimes {\boldsymbol\lambda}_{0i}\trans}
\bra{ \bra{\widehat{\dH}\inv \otimes \widehat{\dH}\inv} \dW_r\dhathatsigmathetamletti
-
\dW_r \dsigmatti(\dtheta_0)
}
\nonumber
\\
={}&
\bra{{\boldsymbol\lambda}_{0i}\trans\widehat{\dH}\inv 
\otimes {\boldsymbol\lambda}_{0i}\trans\widehat{\dH}\inv}
\dW_r
\bra{
\dhathatsigmathetamletti
-
\dhathatsigmatti
\bra{
\boldsymbol r_{\widehat\dH}(\dtheta_0)
}
}
\nonumber
\\
&+
\bra{{\boldsymbol\lambda}_{0i}\trans\widehat{\dH}\inv 
\otimes {\boldsymbol\lambda}_{0i}\trans\widehat{\dH}\inv}
\dW_r
\bra{
\dhathatsigmatti
\bra{
\boldsymbol r_{\widehat\dH}(\dtheta_0)
}
-
\dhatsigmatti
\bra{
\boldsymbol r_{\widehat\dH}(\dtheta_0)
}
}
\nonumber
\\
&+
\bra{{\boldsymbol\lambda}_{0i}\trans
\otimes {\boldsymbol\lambda}_{0i}\trans}
\bra{
\bra{
\widehat\dH^{-1}
\otimes
\widehat\dH^{-1}
}
\dW_r
\dhatsigmatti
\bra{
\boldsymbol r_{\widehat\dH}(\dtheta_0)
}
-
\dW_r\dsigmatti(\dtheta_0)
}
\nonumber
\\
={}&
O_{\Pp}\bra{
\max\bra{
\frac{1}{\sqrt N},
\frac{1}{\sqrt T}
}
}
+
O_{a.s.}\bra{
\calX_k^{\lfloor t/k\rfloor}
}.
\end{align}
The last equality follows from \eqref{eq:QMLE_two_rate_prediction}, Lemma \ref{lemma:rate_psit_first_derivs}, Proposition \ref{prop:rate_psit_decomp}$(a)$, and Assumption \ref{ass:contraction_cond}. The three terms on the right-hand side account, respectively, for the parameter-estimation error, the pointwise factor-filter error, and the initialization effect after alignment. The last term decreases geometrically because the two filters are driven by the same sequence $\cbra{\widehat\dH\dF_s}_{s=1}^{t-1}$, and the mapping in \eqref{eq:transform_sigma} gives
$
\bra{
\widehat\dH^{-1}
\otimes
\widehat\dH^{-1}
}
\dW_r
\dhatsigmatti
\bra{
\boldsymbol r_{\widehat\dH}(\dtheta_0)
}
-
\dW_r\dsigmatti(\dtheta_0)
=
O_{a.s.}\bra{\calX_k^{\lfloor t/k\rfloor}}.
$

Therefore, substituting \eqref{eq:diff_termI_common_component_variance} and \eqref{eq:diff_termII_common_component_variance} into
\eqref{eq:diff_tot_common_component_variance}, as $N,T\rightarrow\infty$,
\begin{align*}
\vbra{
\hatPittithetamle
-
\Pittithetaz
}
=
O_{\Pp}\bra{
\frac{1}{\min\bra{N,\sqrt T}}
+
\frac{1}{\min\bra{\sqrt N,T}}
}
+
O_{a.s.}\bra{
\calX_k^{\lfloor t/k\rfloor}
}.
\end{align*}
Since both stochastic terms are $O\bra{1/\min\bra{\sqrt N,\sqrt T}}$, the rates stated in Theorem \ref{thm:filter_approx_error_cons} follow.
This completes the proof.\qed

\subsubsection*{Proof of Proposition \ref{prop:filter_approx_error_cons_II}}

We first consider the conditional location process. By adding and subtracting
$\widehat\dH^{-1} \dhatmutti\bra{\boldsymbol r_{\widehat\dH}(\dtheta)}$, we obtain
\begin{align}
&
\sup_{\dtheta\in\dTheta}
\vvbra{
\widehat\dH^{-1}
\dhathatmutti
\bra{
\boldsymbol r_{\widehat\dH}(\dtheta)
}
-
\dmutti(\dtheta)
}
\nonumber
\\
\leq{}&
\vvbra{
\widehat\dH^{-1}
}
\sup_{\dtheta\in\dTheta}
\vvbra{
\dhathatmutti
\bra{
\boldsymbol r_{\widehat\dH}(\dtheta)
}
-
\dhatmutti
\bra{
\boldsymbol r_{\widehat\dH}(\dtheta)
}
}
+
\sup_{\dtheta\in\dTheta}
\vvbra{
\widehat\dH^{-1}
\dhatmutti
\bra{
\boldsymbol r_{\widehat\dH}(\dtheta)
}
-
\dmutti(\dtheta)
}.
\label{eq:aligned_mean_filter_decomposition}
\end{align}
The second term on the right-hand side of \eqref{eq:aligned_mean_filter_decomposition} is equal to zero. Indeed, the filter
$\dhatmutti\bra{\boldsymbol r_{\widehat\dH}(\dtheta)}$ is driven by $\widehat\dH\dF_t$, and the  mapping in
\eqref{eq:transform_mu}, together with Assumption \ref{ass:initial_conditions}, implies that
\[
\dhatmutti
\bra{
\boldsymbol r_{\widehat\dH}(\dtheta)
}
=
\widehat\dH
\dmutti(\dtheta)
\]
for every $\dtheta\in\dTheta$ and every $t\geq1$.

Since $\boldsymbol r_{\widehat\dH}$ is an invertible reparameterization of $\dTheta$, by Proposition
\ref{prop:rate_psit_decomp}$(a)$  
\begin{align*}
&
\sup_{\dtheta\in\dTheta}
\vvbra{
\dhathatmutti
\bra{
\boldsymbol r_{\widehat\dH}(\dtheta)
}
-
\dhatmutti
\bra{
\boldsymbol r_{\widehat\dH}(\dtheta)
}
}
=
O_{\Pp}\bra{
\frac{1}{\min\bra{\sqrt N,T}}
}.
\end{align*}
The geometrically decreasing initialization term appearing in
Proposition \ref{prop:rate_psit_decomp}$(a)$ is equal to zero because
the two filters have the same initial condition under Assumption
\ref{ass:initial_conditions}. Moreover,
$\vvbra{\widehat\dH^{-1}}=O_{\Pp}(1)$ by Proposition
\ref{prop:rate_yt_decomp}. Substitution into
\eqref{eq:aligned_mean_filter_decomposition} yields
\[
\sup_{\dtheta\in\dTheta}
\vvbra{
\widehat\dH^{-1}
\dhathatmutti
\bra{
\boldsymbol r_{\widehat\dH}(\dtheta)
}
-
\dmutti(\dtheta)
}
=
O_{\Pp}\bra{
\frac{1}{\min\bra{\sqrt N,T}}
},
\]
which proves the first statement.

For the conditional-variance process, by adding and subtracting 
$
\bra{
\widehat\dH^{-1}
\otimes
\widehat\dH^{-1}
}
\dW_r
\dhatSigmatti
\bra{
\boldsymbol r_{\widehat\dH}(\dtheta)
}
$ we get
\begin{align}
&
\sup_{\dtheta\in\dTheta}
\vvbra{
\bra{
\widehat\dH^{-1}
\otimes
\widehat\dH^{-1}
}
\dW_r
\dhathatsigmatti
\bra{
\boldsymbol r_{\widehat\dH}(\dtheta)
}
-
\dW_r
\dsigmatti(\dtheta)
}
\nonumber
\\
\leq{}&
\vvbra{
\widehat\dH^{-1}
\otimes
\widehat\dH^{-1}
}
\sup_{\dtheta\in\dTheta}
\vvbra{
\dW_r
\dhathatsigmatti
\bra{
\boldsymbol r_{\widehat\dH}(\dtheta)
}
-
\dW_r
\dhatSigmatti
\bra{
\boldsymbol r_{\widehat\dH}(\dtheta)
}
}
\nonumber
\\
&+
\sup_{\dtheta\in\dTheta}
\vvbra{
\bra{
\widehat\dH^{-1}
\otimes
\widehat\dH^{-1}
}
\dW_r
\dhatSigmatti
\bra{
\boldsymbol r_{\widehat\dH}(\dtheta)
}
-
\dW_r
\dsigmatti(\dtheta)
}.
\label{eq:aligned_variance_filter_decomposition}
\end{align}
The second term on the right-hand side of \eqref{eq:aligned_variance_filter_decomposition} is equal to zero. Indeed, by \eqref{eq:transform_sigma} and Assumption
\ref{ass:initial_conditions},
\[
\dW_r
\dhatSigmatti
\bra{
\boldsymbol r_{\widehat\dH}(\dtheta)
}
=
\bra{
\widehat\dH
\otimes
\widehat\dH
}
\dW_r
\dsigmatti(\dtheta)
\]
for every $\dtheta\in\dTheta$ and every $t\geq1$.

Proposition \ref{prop:rate_psit_decomp}$(a)$ therefore implies that
\begin{align*}
&
\sup_{\dtheta\in\dTheta}
\vvbra{
\dW_r
\dhathatsigmatti
\bra{
\boldsymbol r_{\widehat\dH}(\dtheta)
}
-
\dW_r
\dhatSigmatti
\bra{
\boldsymbol r_{\widehat\dH}(\dtheta)
}
}
=
O_{\Pp}\bra{
\frac{1}{\min\bra{\sqrt N,T}}
}.
\end{align*}
Moreover, Proposition \ref{prop:rate_yt_decomp} implies that $\widehat\dH\pconv\dH_0$, where $\dH_0$ is finite and nonsingular. Hence, by the continuous mapping theorem, $\widehat\dH^{-1}\pconv\dH_0^{-1}$ and 
$
\vvbra{
\widehat\dH^{-1}
\otimes
\widehat\dH^{-1}
}_{\mathrm{sp}}
=
\vvbra{
\widehat\dH^{-1}
}_{\mathrm{sp}}^2
=
O_{\Pp}(1)$.
Substitution into \eqref{eq:aligned_variance_filter_decomposition} yields
\[
\sup_{\dtheta\in\dTheta}
\vvbra{
\bra{
\widehat\dH^{-1}
\otimes
\widehat\dH^{-1}
}
\dW_r
\dhathatsigmatti
\bra{
\boldsymbol r_{\widehat\dH}(\dtheta)
}
-
\dW_r
\dsigmatti(\dtheta)
}
=
O_{\Pp}\bra{
\frac{1}{\min\bra{\sqrt N,T}}
},
\]
which proves the second statement.
\qed

\subsubsection*{Proof of Theorem \ref{thm:filter_approx_error_asy_norm}}

For the conditional mean, by Proposition \ref{prop:rate_yt_decomp}$(c)$--$(d)$, Proposition \ref{prop:filter_approx_error_cons_II}, Theorem \ref{thm:asy_norm}, and Lemma \ref{lemma:rate_psit_first_derivs}, and by using the same steps as in \eqref{eq:diff_termI_common_component_location} and \eqref{eq:diff_termII_common_component_location}, we obtain
\begin{align}
\label{eq:diff_tot_common_component_location_sqrt}
\sqrt T&\bra{\hatCittithetamle - \Cittithetaz}
\nonumber
\\
=&\,\sqrt T
\bra{\widehat{\boldsymbol{\lambda}}_i\trans - \boldsymbol{\lambda}_{0i}\trans\widehat{\dH}\inv}  
\dhathatmuthetamletti
+\sqrt T
\boldsymbol{\lambda}_{0i}\trans
\bra{\widehat{\dH}\inv\dhathatmuthetamletti -  \dmutti( \dtheta_0 )  }\nonumber\\
=&\,\sqrt T
\bra{\widehat{\boldsymbol{\lambda}}_i\trans - \boldsymbol{\lambda}_{0i}\trans\widehat{\dH}\inv}  
\dmutti( \boldsymbol r_{\widehat{\dH}}( \dtheta_0 ) )+\sqrt T
\bra{\widehat{\boldsymbol{\lambda}}_i\trans - \boldsymbol{\lambda}_{0i}\trans\widehat{\dH}\inv} 
\bra{\dhathatmuthetamletti - \dmutti(\boldsymbol r_{\widehat{\dH}}( \dtheta_0 )) }
\nonumber
\\
&+ \sqrt T\boldsymbol{\lambda}_{0i}\trans\widehat{\dH}\inv
\bra{\dhathatmuthetamletti
-\dhathatmutti(\boldsymbol r_{\widehat{\dH}}(\dtheta_0))}+\sqrt T \boldsymbol{\lambda}_{0i}\trans\widehat{\dH}\inv\bra{\dhathatmutti(\boldsymbol r_{\widehat{\dH}}(\dtheta_0))-\dmutti(\boldsymbol r_{\widehat{\dH}}(\dtheta_0))}
\nonumber
\\
=&\,
\sqrt T\bra{\widehat{\boldsymbol{\lambda}}_i^{\text{\tiny OLS}}-{\boldsymbol{\lambda}}_{0i}}\trans
\widehat{\dH}^{-1}\dmutti( \boldsymbol r_{\widehat{\dH}}(\dtheta_0) )+{\sqrt T} O_{\Pp}\bra{\frac 1 N}
\nonumber\\
&+\sqrt T
O_{\Pp}\bra{\max\bra{\frac1N, \frac1 {\sqrt{T}} }} \cbra{O_{\Pp}\bra{\max\bra{\frac 1{\sqrt{N}}, \frac 1T} } +
O_{\Pp}\bra{\frac1{\sqrt T}} }
\nonumber
\\
&+ \sqrt T\boldsymbol{\lambda}_{0i}\trans
\bra{
\dmutti
\bra{
\boldsymbol r_{\widehat\dH}^{-1}
\bra{
\dthetamle
}
}
-
\dmutti(\dtheta_0)
}
\nonumber
\\
&+\sqrt T\boldsymbol{\lambda}_{0i}\trans\widehat\dH^{-1}
\bra{
\dhathatmutti(\dthetamle)
-
\widehat\dH
\dmutti
\bra{
\boldsymbol r_{\widehat\dH}^{-1}
\bra{
\dthetamle
}
}
}
\nonumber
\\
=&\,
\sqrt T\bra{\widehat{\boldsymbol{\lambda}}_i^{\text{\tiny OLS}}-{\boldsymbol{\lambda}}_{0i}}\trans \dmutti(\dtheta_0)+\sqrt T\boldsymbol{\lambda}_{0i}\trans
\nabla^{\dtheta}\dmutti(\dtheta_0)
\bra{
\boldsymbol r_{\widehat\dH}^{-1}
\bra{
\dthetamle
}
-
\dtheta_0
+O_{\Pp}\bra{\frac 1T}
}
\nonumber\\
&+\sqrt T\boldsymbol{\lambda}_{0i}\trans\widehat\dH^{-1}
\bra{
\dhathatmutti(\dthetamle)
-
\widehat\dH
\dmutti
\bra{
\boldsymbol r_{\widehat\dH}^{-1}
\bra{
\dthetamle
}
}
}
\nonumber\\
=&\,
\sqrt T\bra{\widehat{\boldsymbol{\lambda}}_i^{\text{\tiny OLS}}-{\boldsymbol{\lambda}}_{0i}}\trans \dmutti(\dtheta_0)+\sqrt T\boldsymbol{\lambda}_{0i}\trans  \nabla^{\dtheta}\dmutti(\dtheta_0)
\bra{
\boldsymbol r_{\widehat\dH}^{-1}
\bra{
\dthetamle
}
-
\dtheta_0
}\nonumber\\
&+\sqrt T\boldsymbol{\lambda}_{0i}\trans\widehat\dH^{-1}
\bra{
\dhathatmutti(\dthetamle)
-
\widehat\dH
\dmutti
\bra{
\boldsymbol r_{\widehat\dH}^{-1}
\bra{
\dthetamle
}
}
},
\end{align}
where we used also the fourth equality in \eqref{eq:questaservePCA} in the proof of Proposition \ref{prop:rate_yt_decomp} with $\widehat{\boldsymbol{\lambda}}_i^{\text{\tiny OLS}}$ being the OLS estimator of ${\boldsymbol{\lambda}}_i$ were the factors known,  and a Taylor expansion of $\dmutti$ in the neighborhood of $\dtheta_0$.

Moreover, by Proposition \ref{prop:filter_approx_error_cons_II}, we also have
\begin{align}
\label{eq:bias_mean}
&
\sqrt T
\boldsymbol{\lambda}_{0i}\trans
\widehat\dH^{-1}
\bra{
\dhathatmutti(\dthetamle)
-
\widehat\dH
\dmutti
\bra{
\boldsymbol r_{\widehat\dH}^{-1}
\bra{
\dthetamle
}
}
}
=
O_{\Pp}\bra{
\sqrt{\frac{T}{N}}
+
\frac{1}{\sqrt T}
}
=
o_{\Pp}(1).
\end{align}
This follows from Proposition \ref{prop:filter_approx_error_cons_II}, Assumption \ref{ass:factor_identifiability}, and the fact that $\Vert \widehat\dH^{-1}\Vert=O_{\Pp}(1)$ by Proposition \ref{prop:rate_yt_decomp} and the continuous mapping theorem.
 Then, if ${T}/N\rightarrow 0$ as $N,T\rightarrow\infty$, from \eqref{eq:diff_tot_common_component_location_sqrt} and \eqref{eq:bias_mean} we get
\begin{align}
\sqrt{T}\bra{\hatCittithetamle - \Cittithetaz}
=\sbra {\dmutti\trans(\dtheta_0) \;\big\vert\; \boldsymbol \lambda_{0i}\trans \nabla^{\dtheta}\dmutti(\dtheta_0)
}\sqrt T\begin{bmatrix}
\widehat{\boldsymbol{\lambda}}_i^{\text{\tiny OLS}}-{\boldsymbol{\lambda}}_{0i}\\
\boldsymbol r_{\widehat\dH}^{-1}
\bra{
\dthetamle
}
-
\dtheta_0
\end{bmatrix}
 +
 o_\Pp(1).\label{eq:CLTC}
\end{align}
By recalling that, by definition of OLS and since ${\mathbb P}\text{-}\lim_{T\to\infty}T^{-1} \sum_{t=1}^T \dFt\dFt\trans = \boldsymbol\Gamma_F$ because of the ergodic theorem \citep{Birkhoff1931}, 
\begin{equation}\label{eq:OLS}
\sqrt T\bra{\widehat{\boldsymbol{\lambda}}_i^{\text{\tiny OLS}}-{\boldsymbol{\lambda}}_{0i}} = \boldsymbol \Gamma_F\inv \bra{\frac 1{\sqrt T}\sum_{s=1}^T \dF_se_{is}} +o_{\Pp}(1),
\end{equation}
and since, conditional on $\widehat\dH$, the inverse alignment mapping is linear in $\dtheta$ (see Appendix \ref{appendix_parameter_identifications}), \begin{align*}
\sqrt T
\bra{
\boldsymbol r_{\widehat\dH}^{-1}
\bra{
\dthetamle
}
-
\dtheta_0
}
=
{\boldsymbol R}_{\widehat\dH}
\sqrt T
\bra{
\dthetamle
-
\boldsymbol r_{\widehat\dH}
\bra{
\dtheta_0
}
}.
\end{align*}
Since
${\boldsymbol R}_{\widehat\dH}\pconv{\boldsymbol R}_{\dH_0}$ as $N,T\rightarrow\infty$, equation \eqref{eq:original_QMLE_expansion} implies that
\begin{align}
\label{eq:asy_norm_expansion2}
\sqrt T
\bra{
\boldsymbol r_{\widehat\dH}^{-1}
\bra{
\dthetamle
}
-
\dtheta_0
}
=
\limitcov\inv
\bra{
\dtheta_0
}
\frac{1}{\sqrt T}
\sum_{t=1}^{T}
\nabla^{\dtheta}
\ell_{\dF}
\bra{
\dFt;
\dtheta_0
}
+
o_{\Pp}(1),
\end{align}
where we used the fact that
$
{\boldsymbol R}_{\dH_0}
\limitcov_{\dH_0}\inv
\bra{
\dtheta_0
}
{\boldsymbol R}_{\dH_0}\trans
=
\limitcov\inv
\bra{
\dtheta_0
}$.

Therefore by \eqref{eq:OLS} and Assumption \ref{ass:idiosyncraticY}$(iv)$, and by \eqref{eq:asy_norm_expansion2}, Theorem \ref{thm:asy_norm} together with an application of Slutsky's theorem, and Lemma \ref{lemma:CLT_NED_score_vector},
\begin{align}
&\sqrt T\bra{\widehat{\boldsymbol{\lambda}}_i^{\text{\tiny OLS}}-{\boldsymbol{\lambda}}_{0i}}| \calF_{t-1}
\Rightarrow 
\calN\bra{\boldsymbol 0_r,  \boldsymbol \Gamma_F^{-1} \boldsymbol \Xi_i  \boldsymbol \Gamma_F^{-1}}, \label{eq:CLTOLS}\\
 &
\sqrt T
\bra{
\boldsymbol r_{\widehat\dH}^{-1}
\bra{
\dthetamle
}
-
\dtheta_0
}
 | \calF_{t-1}
\Rightarrow
\calN\bra{\boldsymbol 0_d, 
\limitcov\inv(\dtheta_0)\dV(\dtheta_0)\limitcov\inv(\dtheta_0)
}
.\label{eq:CLTscore}
\end{align}
Notice that in fact these results are independent of $t$, because of stationarity as guaranteed by Assumptions \ref{ass:pram_SE} and \ref{ass:idiosyncratic}. Likewise the conditional asymptotic covariance of these two terms is in fact independent of $t$ and is given by
\begin{align}
\lim_{T\to\infty}&\Ee\sbra{\bra{\frac 1{\sqrt T}\sum_{s=1}^T \boldsymbol F_s e_{is}}\bra{\frac 1{\sqrt T} \sum_{s=1}^T\nabla^{\dtheta}{\ell}_{\dF}\bra{\dF_s ; \dtheta_0} } \trans  }\nonumber\\
&=\lim_{T\to\infty}\frac 1T\sum_{s_1,s_2=1}^T\Ee\sbra{  \boldsymbol F_{s_1} \bra{\nabla^{\dtheta}{\ell}_{\dF}\bra{\dF_{s_2} ; \dtheta_0}}\trans e_{is_1}}\nonumber\\
&=\lim_{T\to\infty}\frac 1T\sum_{s_1,s_2=1}^T\Ee\sbra{  \boldsymbol F_{s_1} \bra{\nabla^{\dtheta}{\ell}_{\dF}\bra{\dF_{s_2} ; \dtheta_0}}\trans }\Ee[e_{is_1}]=\boldsymbol 0_{r\times d},\label{eq:ACOV}
\end{align}
since the score is just function of the process $\{\boldsymbol F_t\}_{t\in\Zz}$ which is independent of $\{e_{it}\}_{t\in\Zz}$ because of Assumption \ref{ass:idiosyncraticY}$(i)$ and $\Ee[e_{it}]=0$ because of Assumption \ref{ass:idiosyncratic}$(i)$.

By using \eqref{eq:CLTOLS}, \eqref{eq:CLTscore}, and \eqref{eq:ACOV} into \eqref{eq:CLTC}, by the standard Cram\' er-Wold device we get
\begin{align*}
\sqrt T& \bra{\hatCittithetamle - \Cittithetaz} \Bigg \vert \calF_{t-1}\nonumber\\
&\Rightarrow \calN\bra{0, \boldsymbol a_{i,t|t-1}\trans(\dtheta_0)
\begin{bmatrix}
\boldsymbol \Gamma_F^{-1} \boldsymbol \Xi_i  \boldsymbol \Gamma_F^{-1}&\boldsymbol 0_{r\times d}\\
\boldsymbol 0_{d\times r}&\limitcov\inv(\dtheta_0)\dV(\dtheta_0)\limitcov\inv(\dtheta_0)
\end{bmatrix}
\boldsymbol a_{i,t|t-1}(\dtheta_0)},
\end{align*}
where $\boldsymbol a_{i,t|t-1}\trans(\dtheta_0) =\sbra {\dmutti(\dtheta_0)\trans \;\big\vert\; \boldsymbol \lambda_{0i}\trans \nabla^{\dtheta}\dmutti(\dtheta_0)
}$.
This proves the statement for the conditional mean.

For the conditional variance, we first note that, because the Kronecker product is a bilinear and continuous operator, it respects the standard rules for combining asymptotic expansions. Hence as $N,T\rightarrow\infty$, we can use the fact that
\begin{align}\label{eq:preliminare}
&\bra{\widehat{\boldsymbol\lambda}_i\trans \otimes \widehat{\boldsymbol\lambda}_i\trans}
-
\bra{{\boldsymbol\lambda}_{0i}\trans\widehat{\dH}\inv \otimes {\boldsymbol\lambda}_{0i}\trans\widehat{\dH}\inv}
\nonumber\\
=&\,
\bra{
\widehat{\boldsymbol\lambda}_i\trans
-
{\boldsymbol\lambda}_{0i}\trans\widehat{\dH}\inv
}
\otimes 
{\boldsymbol\lambda}_{0i}\trans \widehat{\dH}^{-1}
+
{\boldsymbol\lambda}_{0i}\trans \widehat{\dH}^{-1}
\otimes
\bra{
\widehat{\boldsymbol\lambda}_i\trans
-
{\boldsymbol\lambda}_{0i}\trans\widehat{\dH}\inv
}
\nonumber\\
&+
\bra{
\widehat{\boldsymbol\lambda}_i\trans
-
{\boldsymbol\lambda}_{0i}\trans\widehat{\dH}\inv
}
\otimes 
\bra{
\widehat{\boldsymbol\lambda}_i\trans
-
{\boldsymbol\lambda}_{0i}\trans\widehat{\dH}\inv
}
\nonumber\\
=&\,
\bra{
\widehat{\boldsymbol\lambda}_i\trans
-
{\boldsymbol\lambda}_{0i}\trans\widehat{\dH}\inv
}
\otimes 
{\boldsymbol\lambda}_{0i}\trans \widehat{\dH}^{-1}
+
{\boldsymbol\lambda}_{0i}\trans \widehat{\dH}^{-1}
\otimes
\bra{
\widehat{\boldsymbol\lambda}_i\trans
-
{\boldsymbol\lambda}_{0i}\trans\widehat{\dH}\inv
}
+
O_{\Pp}\bra{\max\bra{\frac 1T,
\frac 1N}}
\nonumber\\
=&\,
\bra{
\bra{
\widehat{\boldsymbol\lambda}_i\trans
-
{\boldsymbol\lambda}_{0i}\trans\widehat{\dH}\inv
}
\otimes 
{\boldsymbol\lambda}_{0i}\trans\widehat{\dH}\inv
}
\bra{\boldsymbol I_{r^2} + \calK_{r^2 \times r^2}}
+
O_{\Pp}\bra{\max\bra{\frac 1T,
\frac 1N}}
,
\end{align}
where in the last equality we used Proposition \ref{prop:rate_yt_decomp}$(d)$. Therefore,  using the identity 
\begin{align*}
&\left(\widehat{\dH}^{-1}\otimes\widehat{\dH}^{-1}\right)
\left(\boldsymbol{I}_{r^2}+\boldsymbol{\mathcal{K}}_{r^2\times r^2}\right)
\left(\widehat{\dH}\otimes \widehat{\dH}\right)
\\
=&\,
\left(\widehat{\dH}^{-1}\otimes \widehat{\dH}^{-1}\right)
\left(\widehat{\dH}\otimes \widehat{\dH}\right)
+
\left(\widehat{\dH}^{-1}\otimes \widehat{\dH}^{-1}\right)
\boldsymbol{\mathcal{K}}_{r^2\times r^2}
\left(\widehat{\dH}\otimes \widehat{\dH}\right)
\\
=&\,
\boldsymbol{I}_{r^2}
+
\left(\widehat{\dH}^{-1}\otimes \widehat{\dH}^{-1}\right)
\left(\widehat{\dH}\otimes\widehat{\dH}\right)
\boldsymbol{\mathcal{K}}_{r^2\times r^2}
\\
=&\,
\boldsymbol{I}_{r^2}+\boldsymbol{\mathcal{K}}_{r^2\times r^2},
\end{align*}
and since $\dsigmatti( \boldsymbol r_{\widehat{\dH}}( \dtheta_0 ) ) = \dW_r\trans \bra{\widehat{\dH} \otimes \widehat{\dH}}\dW_r \dsigmatti( \dtheta_0 )$ by definition (see \eqref{eq:transform_sigma}),
by Proposition \ref{prop:rate_yt_decomp}$(c)$--$(d)$, Proposition \ref{prop:filter_approx_error_cons_II}, Theorem \ref{thm:asy_norm}, and Lemma \ref{lemma:rate_psit_first_derivs}, by using the same steps as in \eqref{eq:diff_termI_common_component_variance} and \eqref{eq:diff_termII_common_component_variance}, we obtain
\begin{align}
\label{eq:diff_tot_common_component_variance_sqrt}
\sqrt{T}&\bra{\hatPittithetamle - \Pittithetaz}
\nonumber
\\
=&\,
\sqrt{T}
\bra{
\bra{\widehat{\boldsymbol\lambda}_i\trans \otimes \widehat{\boldsymbol\lambda}_i\trans}
-
\bra{{\boldsymbol\lambda}_{0i}\trans\widehat{\dH}\inv \otimes {\boldsymbol\lambda}_{0i}\trans\widehat{\dH}\inv}
}
\dW_r\dhathatsigmathetamletti
\nonumber
\\
&+
\sqrt{T}
\bra{{\boldsymbol\lambda}_{0i}\trans \otimes {\boldsymbol\lambda}_{0i}\trans}
\bra{ \bra{\widehat{\dH}\inv \otimes \widehat{\dH}\inv} \dW_r\dhathatsigmathetamletti
-
\dW_r \dsigmatti(\dtheta_0)
}
\nonumber
\\
=&\,
\sqrt{T}
\bra{
\bra{\widehat{\boldsymbol\lambda}_i\trans \otimes \widehat{\boldsymbol\lambda}_i\trans}
-
\bra{{\boldsymbol\lambda}_{0i}\trans\widehat{\dH}\inv \otimes {\boldsymbol\lambda}_{0i}\trans\widehat{\dH}\inv}
}
\dW_r\dsigmatti(\boldsymbol r_{\widehat{\dH}}(\dtheta_0))
\nonumber
\\
&+
\sqrt{T}
\bra{
\bra{\widehat{\boldsymbol\lambda}_i\trans \otimes \widehat{\boldsymbol\lambda}_i\trans}
-
\bra{{\boldsymbol\lambda}_{0i}\trans\widehat{\dH}\inv \otimes {\boldsymbol\lambda}_{0i}\trans\widehat{\dH}\inv}
}
\bra{
\dW_r\dhathatsigmathetamletti
-
\dW_r\dsigmatti(\boldsymbol r_{\widehat{\dH}}(\dtheta_0))
}
\nonumber
\\
&+
\sqrt{T}
\bra{{
\boldsymbol\lambda}_{0i}\trans\widehat{\dH}\inv 
\otimes {\boldsymbol\lambda}_{0i}\trans\widehat{\dH}\inv}
\bra{  \dW_r\dhathatsigmathetamletti
-
\dW_r \dhathatsigmatti(\boldsymbol r_{\widehat{\dH}}(\dtheta_0))
}
\nonumber
\\
&+
\sqrt{T}
\bra{{\boldsymbol\lambda}_{0i}\trans\widehat{\dH}\inv 
\otimes {\boldsymbol\lambda}_{0i}\trans\widehat{\dH}\inv}
\bra{
\dW_r \dhathatsigmatti(\boldsymbol r_{\widehat{\dH}}(\dtheta_0))
-
\dW_r \dsigmatti(\boldsymbol r_{\widehat{\dH}}(\dtheta_0))
}
\nonumber
\\
=&\,
\sqrt{T}\bra{
\bra{\widehat{\boldsymbol{\lambda}}_i^{\text{\tiny OLS}} \otimes \widehat{\boldsymbol{\lambda}}_i^{\text{\tiny OLS}}}\trans
-
\bra{\boldsymbol{\lambda}_{0i} \otimes \boldsymbol{\lambda}_{0i}}\trans
}
\dW_r  \dsigmatti(\dtheta_0)\nonumber
\\
&+\sqrt TO_{\Pp}\bra{\max\bra{\frac 1N,\frac 1T}}
+\sqrt T
O_{\Pp}\bra{\max\bra{\frac1{N},\frac 1{\sqrt{T}} }} 
\cbra{O_{\Pp}\bra{\max\bra{\frac1{\sqrt{N}}, \frac 1T} } +
O_{\Pp}\bra{\frac 1{\sqrt T}} }
\nonumber
\\
&+ \sqrt T 
\bra{{\boldsymbol\lambda}_{0i}\trans\widehat{\dH}\inv 
\otimes {\boldsymbol\lambda}_{0i}\trans\widehat{\dH}\inv}
\bra{
\widehat\dH\otimes\widehat\dH
}
\dW_r
\bra{
\dsigmatti
\bra{
\boldsymbol r_{\widehat\dH}^{-1}
\bra{
\dthetamle
}
}
-
\dsigmatti(\dtheta_0)
}
\nonumber
\\
&+\sqrt T
\bra{{\boldsymbol\lambda}_{0i}\trans\widehat{\dH}\inv 
\otimes {\boldsymbol\lambda}_{0i}\trans\widehat{\dH}\inv}
\bra{
\dW_r\dhathatsigmatti(\dthetamle)
-
\bra{
\widehat\dH\otimes\widehat\dH
}
\dW_r
\dsigmatti
\bra{
\boldsymbol r_{\widehat\dH}^{-1}
\bra{
\dthetamle
}
}
}\nonumber
\\
=&\,
\sqrt{T}\bra{
\bra{\widehat{\boldsymbol{\lambda}}_i^{\text{\tiny OLS}} \otimes \widehat{\boldsymbol{\lambda}}_i^{\text{\tiny OLS}}}\trans
-
\bra{\boldsymbol{\lambda}_{0i} \otimes \boldsymbol{\lambda}_{0i}}\trans
}
\dW_r  \dsigmatti( \dtheta_0 )
\nonumber\\
&+\sqrt T
\bra{\boldsymbol\lambda_{0i} \otimes \boldsymbol\lambda_{0i}}\trans 
 \dW_r
 \nabla^{\dtheta}\dsigmatti(\dtheta_0)
\bra{
\boldsymbol r_{\widehat\dH}^{-1}
\bra{
\dthetamle
}
-
\dtheta_0
+O_{\Pp}\bra{\frac 1T}
}\nonumber\\
&+\sqrt T \bra{{\boldsymbol\lambda}_{0i}\trans\widehat{\dH}\inv 
\otimes {\boldsymbol\lambda}_{0i}\trans\widehat{\dH}\inv}
\bra{
\dW_r\dhathatsigmatti(\dthetamle)
-
\bra{
\widehat\dH\otimes\widehat\dH
}
\dW_r
\dsigmatti
\bra{
\boldsymbol r_{\widehat\dH}^{-1}
\bra{
\dthetamle
}
}
},
\nonumber
\\
=&\,
\sqrt{T}\bra{
\bra{\widehat{\boldsymbol{\lambda}}_i^{\text{\tiny OLS}} \otimes \widehat{\boldsymbol{\lambda}}_i^{\text{\tiny OLS}}}\trans
-
\bra{\boldsymbol{\lambda}_{0i} \otimes \boldsymbol{\lambda}_{0i}}\trans
}
\dW_r  \dsigmatti( \dtheta_0 )
\nonumber\\
&+\sqrt T
\bra{\boldsymbol\lambda_{0i} \otimes \boldsymbol\lambda_{0i}}\trans 
 \dW_r
 \nabla^{\dtheta}\dsigmatti(\dtheta_0)
\bra{
\boldsymbol r_{\widehat\dH}^{-1}
\bra{
\dthetamle
}
-
\dtheta_0
} \nonumber\\
&+\sqrt T \bra{{\boldsymbol\lambda}_{0i}\trans\widehat{\dH}\inv 
\otimes {\boldsymbol\lambda}_{0i}\trans\widehat{\dH}\inv}
\bra{
\dW_r\dhathatsigmatti(\dthetamle)
-
\bra{
\widehat\dH\otimes\widehat\dH
}
\dW_r
\dsigmatti
\bra{
\boldsymbol r_{\widehat\dH}^{-1}
\bra{
\dthetamle
}
}
},
\end{align}
where we used also \eqref{eq:preliminare}, the fourth equality in \eqref{eq:questaservePCA} in the proof of Proposition \ref{prop:rate_yt_decomp},  and a Taylor expansion of $\dsigmatti$ in the neighborhood of $\dtheta_0$.

Moreover, by Proposition \ref{prop:filter_approx_error_cons_II}, we also have
\begin{align}
\label{eq:bias_var}
&
\sqrt T
\bra{
{\boldsymbol\lambda}_{0i}\trans\widehat\dH^{-1}
\otimes
{\boldsymbol\lambda}_{0i}\trans\widehat\dH^{-1}
}
\bra{
\dW_r\dhathatsigmatti(\dthetamle)
-
\bra{
\widehat\dH\otimes\widehat\dH
}
\dW_r
\dsigmatti
\bra{
\boldsymbol r_{\widehat\dH}^{-1}
\bra{
\dthetamle
}
}
}
\nonumber
\\
={}&
O_{\Pp}\bra{
\sqrt{\frac{T}{N}}
+
\frac{1}{\sqrt T}
}
=
o_{\Pp}(1).
\end{align}
This follows from Proposition \ref{prop:filter_approx_error_cons_II}, Assumption \ref{ass:factor_identifiability}, and the fact that $\Vert\widehat\dH^{-1}\otimes\widehat\dH^{-1}\Vert=O_{\Pp}(1)$ by Proposition \ref{prop:rate_yt_decomp} and the continuous mapping theorem.

Hence, if ${T}/N\rightarrow 0$ as $N,T\rightarrow\infty$, from \eqref{eq:diff_tot_common_component_variance_sqrt} and \eqref{eq:bias_var}, we get
\begin{align}
\sqrt{T}&\bra{\hatPittithetamle - \Pittithetaz}
\nonumber
\\
=&
\sbra{
 \dsigmatti(\dtheta_0)\trans \dW_r\trans \;\big\vert\;
\bra{{\boldsymbol\lambda}_{0i}
\otimes
 {\boldsymbol\lambda}_{0i}
 }\trans\dW_r
 \nabla^{\dtheta}\dsigmatti(\dtheta_0)
}
\nonumber
\\
&\qquad\qquad\qquad\qquad\qquad\qquad\qquad\qquad\times
\sqrt T\begin{bmatrix}
\bra{\widehat{\boldsymbol{\lambda}}_i^{\text{\tiny OLS}}\otimes \widehat{\boldsymbol{\lambda}}_i^{\text{\tiny OLS}}}
-
\bra{\boldsymbol{\lambda}_{0i} \otimes \boldsymbol{\lambda}_{0i}}\\
\boldsymbol r_{\widehat\dH}^{-1}
\bra{
\dthetamle
}
-
\dtheta_0
\end{bmatrix}
 +
 o_\Pp(1).
\nonumber
 \\ 
 =&
\sbra{
 \dsigmatti(\dtheta_0)\trans \dW_r\trans \;\big\vert\;
\bra{{\boldsymbol\lambda}_{0i}
\otimes
 {\boldsymbol\lambda}_{0i}
 }\trans\dW_r
 \nabla^{\dtheta}\dsigmatti(\dtheta_0)
}
\begin{bmatrix}
\boldsymbol I_r \otimes \boldsymbol\lambda_{0i} + \boldsymbol\lambda_{0i} \otimes \boldsymbol I_r & \boldsymbol 0_{r^2 \times d}\\
\boldsymbol 0_{d\times r} &  \boldsymbol I_d
\end{bmatrix}
 \nonumber
\\
&\qquad\qquad\qquad\qquad\qquad\qquad\qquad\qquad\times
\sqrt T\begin{bmatrix}
\widehat{\boldsymbol{\lambda}}_i^{\text{\tiny OLS}}
-
\boldsymbol{\lambda}_{0i}\\
\boldsymbol r_{\widehat\dH}^{-1}
\bra{
\dthetamle
}
-
\dtheta_0
\end{bmatrix}
 +
 o_\Pp(1),
 \label{eq:CLTP}
\end{align}
where the second equality follows by an application of the mean value theorem to the nonlinear map $\phi(\dx) = \dx \otimes \dx = \vect (\dx\dx\trans)$, for any generic $r$-dimensional vector $\dx\in\Rr^{r}$.

Therefore, it is now clear that the same arguments used in the discussion to obtain \eqref{eq:CLTOLS}, \eqref{eq:CLTscore}, and \eqref{eq:ACOV} apply here, and hence, using these results into \eqref{eq:CLTP}, by the standard Cram\' er-Wold device we get
\begin{align*}
\sqrt T& \bra{\hatPittithetamle - \Pittithetaz} \Bigg \vert \calF_{t-1}\nonumber\\
&\Rightarrow \calN\bra{0, \boldsymbol b_{i,t|t-1}\trans(\dtheta_0)
\begin{bmatrix}
\boldsymbol \Gamma_F^{-1} \boldsymbol \Xi_i  \boldsymbol \Gamma_F^{-1}&\boldsymbol 0_{r\times d}\\
\boldsymbol 0_{d\times r}&\limitcov\inv(\dtheta_0)\dV(\dtheta_0)\limitcov\inv(\dtheta_0)
\end{bmatrix}
\boldsymbol b_{i,t|t-1}(\dtheta_0)},
\end{align*}
where $$\boldsymbol b_{i,t|t-1}\trans(\dtheta_0) =\sbra{\dsigmatti(\dtheta_0)\trans \dW_r\trans
\bra{\boldsymbol I_r \otimes \boldsymbol\lambda_{0i} + \boldsymbol\lambda_{0i} \otimes \boldsymbol I_r}
\;\big\vert\; \bra{{\boldsymbol\lambda}_{0i}
\otimes
 {\boldsymbol\lambda}_{0i}
 }\trans\dW_r
 \nabla^{\dtheta}\dsigmatti(\dtheta_0)
}.$$
This completes the proof. \qed

\subsubsection*{Proof of Theorem \ref{thm:CLT_Gaussian_models}}
From the result obtained in \eqref{eq:diff_tot_common_component_location_sqrt} we know that 
\begin{align}
\label{eq:common_mean_gauss}
&\min\bra{\sqrt N,\sqrt T}\bra{\hatCittithetamle - \Cittithetaz}
\nonumber
\\
={}&
\min\bra{\sqrt N,\sqrt T}\bra{\widehat{\boldsymbol{\lambda}}_i^{\text{\tiny OLS}}
-
{\boldsymbol{\lambda}}_{0i}}\trans \dmutti(\dtheta_0)
+
\min\bra{\sqrt N,\sqrt T}
\boldsymbol{\lambda}_{0i}\trans
\nabla^{\dtheta}\dmutti(\dtheta_0)
\bra{
\boldsymbol r_{\widehat\dH}^{-1}
\bra{
\dthetamle
}
-
\dtheta_0
}
\nonumber
\\
&+
\min\bra{\sqrt N,\sqrt T}
\boldsymbol{\lambda}_{0i}\trans
\widehat{\dH}^{-1}
\bra{
\dhathatmutti(\dthetamle)
-
\widehat\dH
\dmutti
\bra{
\boldsymbol r_{\widehat\dH}^{-1}
\bra{
\dthetamle
}
}
}
+o_{\Pp}(1).
\end{align}
Now, from \eqref{eq:OLS}, and using the decomposition in equation \eqref{eq:sviluppoL1} in the proof of Proposition \ref{prop:rate_yt_decomp}$(c)$, we get
\begin{align}
\label{eq:RV1_mean_clt}
\min\bra{\sqrt N,\sqrt T}&
\bra{\widehat{\boldsymbol{\lambda}}_i^{\text{\tiny OLS}}
-
{\boldsymbol{\lambda}}_{0i}}
\trans \dmutti(\dtheta_0)
\nonumber
\\
=&
\frac{\min\bra{\sqrt N,\sqrt T}}{\sqrt T}
 \bra{\frac 1{\sqrt T}\sum_{s=1}^T \dF_se_{is}}\trans\boldsymbol \Gamma_F\inv
 \dmutti(\dtheta_0)
 \nonumber
 \\
&+O_{\mathbb P}\bra{\frac {\min\bra{\sqrt N,\sqrt T}}N }
.
\end{align}
Then by Proposition \ref{prop:rate_yt_decomp}$(d)$
\begin{align*}
 \bra{\frac 1{\sqrt T}\sum_{s=1}^T \dF_se_{is}}\trans\boldsymbol \Gamma_F\inv
 \dmutti(\dtheta_0)
 \Big\vert \calF_{t-1}
 \Rightarrow
 \mathcal{N}\bra{0,
\dmutti(\dtheta_0)\trans 
 \boldsymbol \Gamma_F\inv
 \boldsymbol \Xi_i
 \boldsymbol \Gamma_F\inv
 \dmutti(\dtheta_0)
 }
 ,
\end{align*}
where $\boldsymbol \Xi_i$ is defined in Assumption \ref{ass:idiosyncraticY}$(iv)$.

Moreover, applying the first-order condition and the Hessian expansion used in the proof of Theorem \ref{thm:asy_norm}, together with
Lemmas \ref{lemma:U_convergence_score_vector} and \ref{lemma:ULLN_Hessian_matrix}, we obtain, as $N,T\to\infty$,
\begin{align}
\label{eq:aligned_QMLE_two_rate_expansion}
&
\boldsymbol r_{\widehat\dH}^{-1}
\bra{
\dthetamle
}
-
\dtheta_0
=
\limitcov^{-1}(\dtheta_0)
\frac{1}{T}
\nabla^{\dtheta}
\ell_{\dF}
\bra{
\dF;
\dtheta_0
}
+
O_{\Pp}\bra{
\frac{1}{N}
+
\frac{1}{\sqrt{NT}}
+
\frac{1}{T}
}.
\end{align}
Consequently,
\begin{align}
\label{eq:RV2_mean_clt}
\min\bra{\sqrt N,\sqrt T}&
\boldsymbol{\lambda}_{0i}\trans  \nabla^{\dtheta}\dmutti(\dtheta_0)
\bra{
\boldsymbol r_{\widehat\dH}^{-1}
\bra{
\dthetamle
}
-
\dtheta_0
}
\nonumber
\\
=&
\frac{\min\bra{\sqrt N,\sqrt T}}{\sqrt{T}}
\boldsymbol{\lambda}_{0i}\trans  \nabla^{\dtheta}\dmutti(\dtheta_0)
  { \limitcov}(\dtheta_0)\inv\bra{
\frac 1{\sqrt T}\sum_{s=1}^T
 \nabla^{\dtheta}{\ell}_{\dF}\bra{\dF_s ; \dtheta_0}}
 \nonumber
 \\
&+O_{\Pp}\bra{
\frac{\min\bra{\sqrt N,\sqrt T}}{N}
+
\frac{\min\bra{\sqrt N,\sqrt T}}{T}
},
\end{align}
Then by Lemma \ref{lemma:CLT_NED_score_vector} (see also the proof of Theorem \ref{thm:asy_norm})
\begin{align*}
&\boldsymbol{\lambda}_{0i}\trans  \nabla^{\dtheta}\dmutti(\dtheta_0)
  { \limitcov}(\dtheta_0)\inv\bra{
\frac 1{\sqrt T}\sum_{s=1}^T
 \nabla^{\dtheta}{\ell}_{\dF}\bra{\dF_s ; \dtheta_0}}
 \Big\vert\calF_{t-1}
 \\
 &\tab\tab 
 \Rightarrow
 \calN\bra{0, \boldsymbol{\lambda}_{0i}\trans  \nabla^{\dtheta}\dmutti(\dtheta_0)\limitcov(\dtheta_0)\inv
 \dV(\dtheta_0)
 \limitcov(\dtheta_0)\inv \nabla^{\dtheta}\dmutti(\dtheta_0)\trans \boldsymbol{\lambda}_{0i}
 }
 .
\end{align*}
Finally, for the third term on the right-hand side of equation \eqref{eq:common_mean_gauss}, we have 
\begin{align*}
&\widehat{\dH}^{-1}
\bra{
\dhathatmutti(\dthetamle)
-
\widehat\dH
\dmutti
\bra{
\boldsymbol r_{\widehat\dH}^{-1}
\bra{
\dthetamle
}
}
}
\\
=&
\widehat{\dH}\inv\Bigg(
\widehat\dPhi \dhathatmu_{t-1|t-2}(\dthetamle) + \widehat\dK \bra{\dhatF_{t-1} - \dhathatmu_{t-1|t-2}(\dthetamle)} 
-
\widehat\dPhi
\widehat\dH
\dmu_{t-1|t-2}
\bra{
\boldsymbol r_{\widehat\dH}^{-1}
\bra{
\dthetamle
}
}
\\
&\qquad\qquad\qquad\qquad\qquad\qquad\qquad\qquad\qquad-
\widehat\dK
\bra{
\widehat{\dH}\dF_{t-1}
-
\widehat\dH
\dmu_{t-1|t-2}
\bra{
\boldsymbol r_{\widehat\dH}^{-1}
\bra{
\dthetamle
}
}
}
\Bigg)
\\
=&
\widehat{\dH}^{-1}
\bra{\widehat\dPhi-\widehat\dK}
\widehat{\dH}
\widehat{\dH}^{-1}
\bra{
\dhathatmu_{t-1|t-2}(\dthetamle)
-
\widehat\dH
\dmu_{t-1|t-2}
\bra{
\boldsymbol r_{\widehat\dH}^{-1}
\bra{
\dthetamle
}
}
}
+
 \widehat{\dH}\inv\widehat\dK  \bra{\dhatF_{t-1} - \widehat{\dH} \dF_{t-1}} 
 \\
 =&
 \widehat{\dH}^{-1}
 \bra{\widehat\dPhi-\widehat\dK}
 \widehat{\dH}
 \widehat{\dH}^{-1}
 \bra{
 \dhathatmu_{t-1|t-2}(\dthetamle)
 -
 \widehat\dH
 \dmu_{t-1|t-2}
 \bra{
 \boldsymbol r_{\widehat\dH}^{-1}
 \bra{
 \dthetamle
 }
 }
 }
\\
&+
\widehat{\dH}\inv
\widehat{\dK}
\widehat{\dH}
\bra{\frac{\dLambda_0\trans\dLambda_0}{N}}^{-1} \frac {{\dLambda}_0\trans\dxi_{t-1}}{ N}  +O_{\mathbb P}\bra{\frac 1T}
\\
 =&
\bra{ \widehat{\dH}\inv\bra{\widehat\dPhi - \widehat\dK }\widehat{\dH}-\bra{\dPhi_0 -\dK_0}+\bra{\dPhi_0 -\dK_0}}
 \widehat{\dH}^{-1}
 \bra{
 \dhathatmu_{t-1|t-2}(\dthetamle)
 -
 \widehat\dH
 \dmu_{t-1|t-2}
 \bra{
 \boldsymbol r_{\widehat\dH}^{-1}
 \bra{
 \dthetamle
 }
 }
 }
 \\
&+
\bra{\widehat{\dH}\inv
\widehat{\dK}
\widehat{\dH}-\dK_0+
\dK_0}
\bra{\frac{\dLambda_0\trans\dLambda_0}{N}}^{-1} \frac {{\dLambda}_0\trans\dxi_{t-1}}{ N}  +O_{\mathbb P}\bra{\frac 1T}
\\
 =&
\bra{\dPhi_0 -\dK_0}
\widehat{\dH}\inv 
 \bra{
 \dhathatmu_{t-1|t-2}(\dthetamle)
 -
 \widehat\dH
 \dmu_{t-1|t-2}
 \bra{
 \boldsymbol r_{\widehat\dH}^{-1}
 \bra{
 \dthetamle
 }
 }
 }
 \\
 &+
\dK_0
\bra{\frac{\dLambda_0\trans\dLambda_0}{N}}^{-1} \frac {{\dLambda}_0\trans\dxi_{t-1}}{ N}  +O_{\mathbb P}\bra{\frac 1T}+O_{\mathbb P}\bra{\frac 1{\sqrt{NT}}}
\\
=&
\sum_{k=0}^{t-1}
\bra{\dPhi_0 - \dK_0}^k
{\dK}_0
\bra{\frac{\dLambda_0\trans\dLambda_0}{N}}^{-1} \frac {{\dLambda}_0\trans\dxi_{t-k-1}}{ N}  +O_{\mathbb P}\bra{\max\bra{\frac 1T,\frac 1{\sqrt{NT}}}},
\end{align*}
by Theorem \ref{thm:consistency} and Proposition \ref{prop:filter_approx_error_cons_II}. Therefore, as $N,T\to\infty$, 
\begin{align}
\label{eq:RV3_mean_clt}
\min\bra{\sqrt N,\sqrt T}
&\boldsymbol{\lambda}_{0i}\trans
\widehat{\dH}^{-1}
\bra{
\dhathatmutti(\dthetamle)
-
\widehat\dH
\dmutti
\bra{
\boldsymbol r_{\widehat\dH}^{-1}
\bra{
\dthetamle
}
}
}
\nonumber
\\
=&
\frac{\min\bra{\sqrt N,\sqrt T}}{\sqrt {N}}
\boldsymbol{\lambda}_{0i}\trans 
\sum_{k=0}^{t-1}
\bra{\dPhi_0 - \dK_0}^k
{\dK}_0
\bra{\frac{\dLambda_0\trans\dLambda_0}{N}}^{-1} \frac {{\dLambda}_0\trans\dxi_{t-k-1}}{ \sqrt N} \nonumber\\
&+O_{\Pp}\bra{
\frac{\min\bra{\sqrt N,\sqrt T}}{T}
+
\frac{\min\bra{\sqrt N,\sqrt T}}{\sqrt{NT}}
}
.
\end{align}
Then, by the finite-dimensional result in Proposition \ref{prop:rate_yt_decomp}$(b)$ and the Cram\'er--Wold device,
\begin{align*}
&\boldsymbol{\lambda}_{0i}\trans 
\sum_{k=0}^{t-1}
\bra{\dPhi_0 - \dK_0}^k
{\dK}_0
\bra{\frac{\dLambda_0\trans\dLambda_0}{N}}^{-1} \frac {{\dLambda}_0\trans\dxi_{t-k-1}}{ \sqrt N} 
\Big\vert\calF_{t-1}
\\
&\tab\tab 
\Rightarrow
\calN\bra{0,
\boldsymbol{\lambda}_{0i}\trans 
\bra{
\sum_{k=0}^{t-1}
\sum_{h=0}^{t-1}
\bra{\dPhi_0 - \dK_0}^k
{\dK}_0
\dSigma_\Lambda\inv
\dPi_{t-k-1,t-h-1}
\dSigma_\Lambda\inv
{\dK}_0\trans
\bra{\dPhi_0 - \dK_0}^{h\top}
}
\boldsymbol{\lambda}_{0i}
}
.
\end{align*}
The leading term in \eqref{eq:RV1_mean_clt} is the time-series loading term, the leading term in \eqref{eq:RV2_mean_clt} is the QMLE-score term, and the leading term in \eqref{eq:RV3_mean_clt} is the cross-sectional factor-estimation term. Assumptions \ref{ass:idiosyncratic}--\ref{ass:idiosyncraticY}, Proposition
\ref{prop:rate_yt_decomp}$(b)$ and $(d)$, Lemma \ref{lemma:CLT_NED_score_vector}, and the Cram\'er--Wold device imply their joint Gaussian convergence. The asymptotic covariance between  the first two terms is zero by \eqref{eq:ACOV}. The covariance between the QMLE-score term and the factor-estimation term is zero by Assumption \ref{ass:idiosyncraticY}$(i)$ and $\Ee\sbra{e_{it}}=0$, whereas the covariance between the loading term and the factor-estimation term converges to zero by Assumption \ref{ass:idiosyncratic}$(ii)$. Thus, from \eqref{eq:common_mean_gauss},
\begin{align*}
&\min\bra{\sqrt{N}, \sqrt{T}} \bra{\hatCittithetamle - \Cittithetaz}
\\
&\tab\tab
=
\frac{\min\bra{\sqrt N,\sqrt T}}{\sqrt T}
 \bra{\frac 1{\sqrt T}\sum_{s=1}^T \dF_se_{is}}\trans\boldsymbol \Gamma_F\inv
 \dmutti(\dtheta_0)
 \\
 &\tab\tab\quad+
 \frac{\min\bra{\sqrt N,\sqrt T}}{\sqrt{T}}
\boldsymbol{\lambda}_{0i}\trans  \nabla^{\dtheta}\dmutti(\dtheta_0)
  { \limitcov}(\dtheta_0)\inv\bra{
\frac 1{\sqrt T}\sum_{s=1}^T
 \nabla^{\dtheta}{\ell}_{\dF}\bra{\dF_s ; \dtheta_0}}
 \\
 &\tab\tab\quad+
 \frac{\min\bra{\sqrt N,\sqrt T}}{\sqrt {N}}
\boldsymbol{\lambda}_{0i}\trans 
\sum_{k=0}^{t-1}
\bra{\dPhi_0 - \dK_0}^k
{\dK}_0
\bra{\frac{\dLambda_0\trans\dLambda_0}{N}}^{-1} \frac {{\dLambda}_0\trans\dxi_{t-k-1}}{ \sqrt N}
 \nonumber
 \\
&\tab\tab\quad
+O_{\mathbb P}\bra{\frac {1}{\min\bra{\sqrt N,\sqrt T}} }
.
\end{align*}
The preceding decomposition is the analogue of equation (C.1) in the proof of Theorem 3 of \citet{Bai2003}, augmented by the QMLE-score
term. We therefore adapt the two-rate argument used therein. To conclude without imposing a restriction on the relative rate of $N$
and $T$, consider an arbitrary subsequence. It admits a further subsequence along which $N/T$ converges to an element of
$[0,\infty]$. If $N/T\to0$, the $N^{-1/2}$ factor-estimation term is the only non-negligible component under the normalization in the
statement, and $\sqrt N/T\to0$. If $N/T\to\infty$, the $T^{-1/2}$ loading and QMLE-score terms are the only non-negligible
components, and $\sqrt T/N\to0$. If $N/T$ converges to a finite positive real, all three components may contribute and both rate
conditions hold. Hence, along every such subsequence, division by the conditional standard deviation in the statement gives a standard
Gaussian limit. The subsequence principle proves the first result of the theorem as $N,T\to\infty$.

For the conditional variance, we proceed in a similar way. We first use \eqref{eq:diff_tot_common_component_variance_sqrt}, and write
\begin{align}
\label{eq:common_var_gauss}
&\min\bra{\sqrt{N}, \sqrt{T}}\bra{\hatPittithetamle - \Pittithetaz}
\nonumber
\\
=&
\min\bra{\sqrt{N}, \sqrt{T}}\bra{
\bra{\widehat{\boldsymbol{\lambda}}_i^{\text{\tiny OLS}} \otimes \widehat{\boldsymbol{\lambda}}_i^{\text{\tiny OLS}}}\trans
-
\bra{\boldsymbol{\lambda}_{0i} \otimes \boldsymbol{\lambda}_{0i}}\trans
}
\dW_r  \dsigmatti( \dtheta_0 )
\nonumber\\
&+
\min\bra{\sqrt{N}, \sqrt{T}}
\bra{\boldsymbol\lambda_{0i} \otimes \boldsymbol\lambda_{0i}}\trans 
 \dW_r
 \nabla^{\dtheta}\dsigmatti(\dtheta_0)
\bra{
\boldsymbol r_{\widehat\dH}^{-1}
\bra{
\dthetamle
}
-
\dtheta_0
}
\nonumber
\\
&+
\min\bra{\sqrt{N}, \sqrt{T}}
\bra{
{\boldsymbol\lambda}_{0i}\trans\widehat{\dH}^{-1}
\otimes
{\boldsymbol\lambda}_{0i}\trans\widehat{\dH}^{-1}
}
\bra{
\dW_r\dhathatsigmatti(\dthetamle)
-
\bra{
\widehat\dH\otimes\widehat\dH
}
\dW_r
\dsigmatti
\bra{
\boldsymbol r_{\widehat\dH}^{-1}
\bra{
\dthetamle
}
}
}
+o_{\Pp}(1),
\end{align}
and then, for the first term, on the right-hand side of the latter expression we can directly use the same arguments as those applied for the derivation of \eqref{eq:CLTP}, to obtain 
\begin{align}
\label{eq:RV1_var_clt}
&\min\bra{\sqrt{N}, \sqrt{T}}
\bra{
\bra{\widehat{\boldsymbol{\lambda}}_i^{\text{\tiny OLS}} \otimes \widehat{\boldsymbol{\lambda}}_i^{\text{\tiny OLS}}}\trans
-
\bra{\boldsymbol{\lambda}_{0i} \otimes \boldsymbol{\lambda}_{0i}}\trans
}
\dW_r  \dsigmatti( \dtheta_0 )
\nonumber\\
=&
\min\bra{\sqrt{N}, \sqrt{T}}
\bra{
 \widehat{\boldsymbol{\lambda}}_i^{\text{\tiny OLS}}
-
\boldsymbol{\lambda}_{0i}
 }\trans
\bra{ \boldsymbol I_r \otimes \boldsymbol\lambda_{0i} + \boldsymbol\lambda_{0i} \otimes \boldsymbol I_r}\trans
\dW_r  \dsigmatti( \dtheta_0 )
\nonumber\\
=&
\frac{\min\bra{\sqrt N,\sqrt T}}{\sqrt T}
 \bra{\frac 1{\sqrt T}\sum_{s=1}^T \dF_se_{is}}\trans\boldsymbol \Gamma_F\inv
\bra{ \boldsymbol I_r \otimes \boldsymbol\lambda_{0i} + \boldsymbol\lambda_{0i} \otimes \boldsymbol I_r}\trans
\dW_r  \dsigmatti( \dtheta_0 )
 \nonumber
 \\
&+O_{\mathbb P}\bra{\frac {\min\bra{\sqrt N,\sqrt T}}N }
.
\end{align}
Then by Proposition \ref{prop:rate_yt_decomp}$(d)$
\begin{align*}
 &\bra{\frac 1{\sqrt T}\sum_{s=1}^T \dF_se_{is}}\trans\boldsymbol \Gamma_F\inv
 \bra{ \boldsymbol I_r \otimes \boldsymbol\lambda_{0i} + \boldsymbol\lambda_{0i} \otimes \boldsymbol I_r}\trans
\dW_r  \dsigmatti( \dtheta_0 )
 \Big\vert \calF_{t-1}
 \\
 \Rightarrow&
 \mathcal{N}\bigg(0,
 \dsigmatti( \dtheta_0 )\trans\dW_r\trans
\bra{ \boldsymbol I_r \otimes \boldsymbol\lambda_{0i} + \boldsymbol\lambda_{0i} \otimes \boldsymbol I_r}
 \boldsymbol \Gamma_F\inv
 \boldsymbol \Xi_i
 \boldsymbol \Gamma_F\inv
 \\
 &\tab\tab\tab\tab\tab\tab  \times
 \bra{ \boldsymbol I_r \otimes \boldsymbol\lambda_{0i} + \boldsymbol\lambda_{0i} \otimes \boldsymbol I_r}\trans
\dW_r  \dsigmatti( \dtheta_0 )
 \bigg)
 ,
\end{align*}
where $\boldsymbol \Xi_i$ is defined in Assumption
\ref{ass:idiosyncraticY}$(iv)$.

Moreover, by \eqref{eq:aligned_QMLE_two_rate_expansion}, we obtain
\begin{align}
\label{eq:RV2_var_clt}
\min\bra{\sqrt N,\sqrt T}&
\bra{\boldsymbol\lambda_{0i} \otimes \boldsymbol\lambda_{0i}}\trans 
 \dW_r
 \nabla^{\dtheta}\dsigmatti(\dtheta_0)
\bra{
\boldsymbol r_{\widehat\dH}^{-1}
\bra{
\dthetamle
}
-
\dtheta_0
}
\nonumber
\\
=&
\frac{\min\bra{\sqrt N,\sqrt T}}{\sqrt{T}}
\bra{\boldsymbol\lambda_{0i} \otimes \boldsymbol\lambda_{0i}}\trans 
 \dW_r
 \nabla^{\dtheta}\dsigmatti(\dtheta_0)
  { \limitcov}(\dtheta_0)\inv\bra{
\frac 1{\sqrt T}\sum_{s=1}^T
 \nabla^{\dtheta}{\ell}_{\dF}\bra{\dF_s ; \dtheta_0}}
 \nonumber
 \\
&+O_{\Pp}\bra{
\frac{\min\bra{\sqrt N,\sqrt T}}{N}
+
\frac{\min\bra{\sqrt N,\sqrt T}}{T}
},
\end{align}
Then by Lemma \ref{lemma:CLT_NED_score_vector} (see also the proof of Theorem \ref{thm:asy_norm})
\begin{align*}
&\bra{\boldsymbol\lambda_{0i} \otimes \boldsymbol\lambda_{0i}}\trans 
 \dW_r
 \nabla^{\dtheta}\dsigmatti(\dtheta_0)
  { \limitcov}(\dtheta_0)\inv\bra{
\frac 1{\sqrt T}\sum_{s=1}^T
 \nabla^{\dtheta}{\ell}_{\dF}\bra{\dF_s ; \dtheta_0}}
 \Big\vert\calF_{t-1}
 \\ 
 \Rightarrow&
 \calN\bra{0, \bra{\boldsymbol\lambda_{0i} \otimes \boldsymbol\lambda_{0i}}\trans 
 \dW_r
 \nabla^{\dtheta}\dsigmatti(\dtheta_0)
 \limitcov(\dtheta_0)\inv
 \dV(\dtheta_0)
 \limitcov(\dtheta_0)\inv 
  \nabla^{\dtheta}\dsigmatti(\dtheta_0)\trans
  \dW_r\trans
 \bra{\boldsymbol\lambda_{0i} \otimes \boldsymbol\lambda_{0i}}
 }
 .
\end{align*}
In conclusion, for the third term on the right-hand side of equation \eqref{eq:common_var_gauss}, by performing some algebra, we can use the fact that 
\begin{align*}
&
\bra{
\bra{\widehat{\dH}^{-1}\otimes\widehat{\dH}^{-1}}
\dW_r
\dhathatsigmatti(\dthetamle)
-
\dW_r
\dsigmatti
\bra{
\boldsymbol r_{\widehat\dH}^{-1}
\bra{
\dthetamle
}
}
}
\\
=&
\bra{\widehat{\dH}\inv \otimes \widehat{\dH}\inv}
\dW_r
\bra{\widehat\dPhi-\widehat\dK}
\dW_r\trans
\bra{\widehat{\dH} \otimes \widehat{\dH}}
\\&\qquad\qquad\qquad\qquad\qquad\times
\bra{
\bra{\widehat{\dH}^{-1}\otimes\widehat{\dH}^{-1}}
\dW_r
\dhathatsigma_{t-1|t-2}(\dthetamle)
-
\dW_r
\dsigma_{t-1|t-2}
\bra{
\boldsymbol r_{\widehat\dH}^{-1}
\bra{
\dthetamle
}
}
}
\\
&+
\bra{\widehat{\dH}\inv \otimes \widehat{\dH}\inv}
\dW_r
\widehat\dK\dW_r\trans\bra{
\dhatF_{t-1}\otimes \dhatF_{t-1} - \widehat\dH\dF_{t-1} \otimes \widehat\dH\dF_{t-1}}
\\
=&
\bra{\widehat{\dH}\inv \otimes \widehat{\dH}\inv}
\dW_r
\bra{\widehat\dPhi-\widehat\dK}
\dW_r\trans
\bra{\widehat{\dH} \otimes \widehat{\dH}}
\\&\qquad\qquad\qquad\qquad\qquad\times
\bra{
\bra{\widehat{\dH}^{-1}\otimes\widehat{\dH}^{-1}}
\dW_r
\dhathatsigma_{t-1|t-2}(\dthetamle)
-
\dW_r
\dsigma_{t-1|t-2}
\bra{
\boldsymbol r_{\widehat\dH}^{-1}
\bra{
\dthetamle
}
}
}
\\
&+
\bra{\widehat{\dH}\inv \otimes \widehat{\dH}\inv}
\dW_r
\widehat\dK\dW_r\trans
\bra{\widehat{\dH} \otimes \widehat{\dH}}
\bra{
\bra{\widehat{\dH}\inv \otimes \widehat{\dH}\inv}
\bra{\dhatF_{t-1}\otimes \dhatF_{t-1}}
-
\bra{\dF_{t-1}\otimes \dF_{t-1}}
}
\\
=&
\bra{
\bra{\widehat{\dH}\inv \otimes \widehat{\dH}\inv}
\dW_r
\bra{\widehat\dPhi-\widehat\dK}
\dW_r\trans
\bra{\widehat{\dH} \otimes \widehat{\dH}}
-
\dW_r
\bra{\dPhi_0-\dK_0}
\dW_r\trans +
\dW_r
\bra{\dPhi_0-\dK_0}
\dW_r\trans
}
\\&\qquad\qquad\qquad\qquad\qquad\times
\bra{
\bra{\widehat{\dH}^{-1}\otimes\widehat{\dH}^{-1}}
\dW_r
\dhathatsigma_{t-1|t-2}(\dthetamle)
-
\dW_r
\dsigma_{t-1|t-2}
\bra{
\boldsymbol r_{\widehat\dH}^{-1}
\bra{
\dthetamle
}
}
}
\\
&+
\bra{
\bra{\widehat{\dH}\inv \otimes \widehat{\dH}\inv}
\dW_r
\widehat\dK
\dW_r\trans
\bra{\widehat{\dH} \otimes \widehat{\dH}}
-
\dW_r
\dK_0
\dW_r\trans
+
\dW_r
\dK_0
\dW_r\trans
}
\\&\qquad\qquad\qquad\qquad\qquad\times
\bra{\widehat{\dH}\inv \otimes \widehat{\dH}\inv}
\bra{
\dhatF_{t-1}\otimes \dhatF_{t-1}
-
\widehat\dH\dF_{t-1}\otimes \widehat\dH\dF_{t-1}
}
\\
=&
\dW_r
\bra{\dPhi_0-\dK_0}
\dW_r\trans
\bra{
\bra{\widehat{\dH}^{-1}\otimes\widehat{\dH}^{-1}}
\dW_r
\dhathatsigma_{t-1|t-2}(\dthetamle)
-
\dW_r
\dsigma_{t-1|t-2}
\bra{
\boldsymbol r_{\widehat\dH}^{-1}
\bra{
\dthetamle
}
}
}
\\
&+
\dW_r
\dK_0
\dW_r\trans
\bra{\widehat{\dH}\inv \otimes \widehat{\dH}\inv}
\bra{
\dhatF_{t-1}\otimes \dhatF_{t-1}
-
\widehat\dH\dF_{t-1}\otimes \widehat\dH\dF_{t-1}
}
+O_\Pp\bra{
\frac{1}{N}
+
\frac{1}{T}
+
\frac{1}{\sqrt{NT}}
}
\\
=&
\dW_r
\bra{\dPhi_0-\dK_0}
\dW_r\trans
\bra{
\bra{\widehat{\dH}^{-1}\otimes\widehat{\dH}^{-1}}
\dW_r
\dhathatsigma_{t-1|t-2}(\dthetamle)
-
\dW_r
\dsigma_{t-1|t-2}
\bra{
\boldsymbol r_{\widehat\dH}^{-1}
\bra{
\dthetamle
}
}
}
\\
&+
\dW_r
\dK_0
\dW_r\trans
\bra{\widehat{\dH}\inv \otimes \widehat{\dH}\inv}
\bra{
\boldsymbol{I}_r \otimes \widehat\dH\dF_{t-1}
+
\widehat\dH\dF_{t-1} \otimes \boldsymbol{I}_r
}
\bra{
\dhatF_{t-1} - \widehat\dH\dF_{t-1}
}
\\&+
O_\Pp\bra{
\frac{1}{N}
+
\frac{1}{T}
+
\frac{1}{\sqrt{NT}}
}
\\
=&
\dW_r
\bra{\dPhi_0-\dK_0}
\dW_r\trans
\bra{
\bra{\widehat{\dH}^{-1}\otimes\widehat{\dH}^{-1}}
\dW_r
\dhathatsigma_{t-1|t-2}(\dthetamle)
-
\dW_r
\dsigma_{t-1|t-2}
\bra{
\boldsymbol r_{\widehat\dH}^{-1}
\bra{
\dthetamle
}
}
}
\\
&+
\dW_r
\dK_0
\dW_r\trans
\bra{
\boldsymbol{I}_r \otimes \dF_{t-1}
+
\dF_{t-1} \otimes \boldsymbol{I}_r
}
\widehat\dH\inv
\bra{
\dhatF_{t-1} - \widehat\dH\dF_{t-1}
}
+O_\Pp\bra{
\frac{1}{N}
+
\frac{1}{T}
+
\frac{1}{\sqrt{NT}}
}
\\
 =&
\dW_r
\bra{\dPhi_0-\dK_0}
\dW_r\trans
\bra{
\bra{\widehat{\dH}^{-1}\otimes\widehat{\dH}^{-1}}
\dW_r
\dhathatsigma_{t-1|t-2}(\dthetamle)
-
\dW_r
\dsigma_{t-1|t-2}
\bra{
\boldsymbol r_{\widehat\dH}^{-1}
\bra{
\dthetamle
}
}
}
\\
&+
\dW_r
\dK_0
\dW_r\trans
\bra{
\boldsymbol{I}_r \otimes \dF_{t-1}
+
\dF_{t-1} \otimes \boldsymbol{I}_r
}
\bra{\frac{\dLambda_0\trans\dLambda_0}{N}}^{-1} \frac {{\dLambda}_0\trans\dxi_{t-1}}{ N}  
+O_{\Pp}\bra{
\frac{1}{N}
+
\frac{1}{T}
+
\frac{1}{\sqrt{NT}}
}
\\
=&
\sum_{k=0}^{t-1}
\bra{\dW_r\bra{\dPhi_0 - \dK_0}\dW_r\trans}^k
\dW_r
\dK_0
\dW_r\trans
\bra{
\boldsymbol{I}_r \otimes \dF_{t-k-1}
+
\dF_{t-k-1} \otimes \boldsymbol{I}_r
}
\bra{\frac{\dLambda_0\trans\dLambda_0}{N}}^{-1} \frac {{\dLambda}_0\trans\dxi_{t-k-1}}{ N}  
\\
&+
O_{\Pp}\bra{
\frac{1}{N}
+
\frac{1}{T}
+
\frac{1}{\sqrt{NT}}
},
\end{align*}
by Theorem \ref{thm:consistency} and Proposition \ref{prop:filter_approx_error_cons_II}, where we have used the fact that
\begin{align*}
\bra{\widehat{\dH}\inv \otimes \widehat{\dH}\inv}
\bra{\boldsymbol{I}_r \otimes \widehat\dH\dF_{t-1}} \bra{
\dhatF_{t-1} - \widehat\dH\dF_{t-1}
}
=&
\widehat{\dH}\inv 
\bra{
\dhatF_{t-1} - \widehat\dH\dF_{t-1}
}
\otimes  \dF_{t-1}
\\
=&
\bra{\boldsymbol{I}_r \otimes \dF_{t-1}} 
\widehat{\dH}\inv 
\bra{
\dhatF_{t-1} - \widehat\dH\dF_{t-1}
}
.
\end{align*}
Therefore, as $N,T\to\infty$, 
\begin{align}
\label{eq:RV3_var_clt}
\min\bra{\sqrt N,\sqrt T}
&\bra{
\boldsymbol{\lambda}_{0i}\trans
\otimes
\boldsymbol{\lambda}_{0i}\trans
}
\bra{
\bra{
\widehat{\dH}^{-1}\otimes\widehat{\dH}^{-1}
}
\dW_r
\dhathatsigmatti(\dthetamle)
-
\dW_r
\dsigmatti
\bra{
\boldsymbol r_{\widehat\dH}^{-1}
\bra{
\dthetamle
}
}
}
\nonumber\\
=&\, 
\frac{\min\bra{\sqrt N,\sqrt T}}{\sqrt {N}}
\bra{\boldsymbol{\lambda}_{0i}\trans \otimes \boldsymbol{\lambda}_{0i}\trans }
\sum_{k=0}^{t-1}
\bra{\dW_r\bra{\dPhi_0 - \dK_0}\dW_r\trans}^k
\dW_r
\dK_0
\dW_r\trans
\nonumber\\
&\,\times
\bra{
\boldsymbol{I}_r \otimes \dF_{t-k-1}
+
\dF_{t-k-1} \otimes \boldsymbol{I}_r
}
\bra{\frac{\dLambda_0\trans\dLambda_0}{N}}^{-1} \frac {{\dLambda}_0\trans\dxi_{t-k-1}}{ \sqrt N} 
\nonumber\\
&\quad
+O_{\Pp}\bra{
\frac{\min\bra{\sqrt N,\sqrt T}}{N}
+
\frac{\min\bra{\sqrt N,\sqrt T}}{T}
+
\frac{\min\bra{\sqrt N,\sqrt T}}{\sqrt{NT}}
}
.
\end{align}
Then, by the finite-dimensional result in Proposition \ref{prop:rate_yt_decomp}$(b)$ and the Cram\'er--Wold device,
\begin{align*}
&\bra{\boldsymbol{\lambda}_{0i}\trans \otimes \boldsymbol{\lambda}_{0i}\trans }
\sum_{k=0}^{t-1}
\bra{\dW_r\bra{\dPhi_0 - \dK_0}\dW_r\trans}^k
\dW_r
\dK_0
\dW_r\trans
\nonumber\\
&\tab\tab\tab\times
\bra{
\boldsymbol{I}_r \otimes \dF_{t-k-1}
+
\dF_{t-k-1} \otimes \boldsymbol{I}_r
}
\bra{\frac{\dLambda_0\trans\dLambda_0}{N}}^{-1} \frac {{\dLambda}_0\trans\dxi_{t-k-1}}{ \sqrt N} 
\Big\vert\calF_{t-1}
\\ 
\Rightarrow&
\calN\Bigg(0,
\bra{\boldsymbol{\lambda}_{0i}\trans \otimes \boldsymbol{\lambda}_{0i}\trans }
\bigg(
\sum_{k=0}^{t-1}
\sum_{h=0}^{t-1}
\bra{\dW_r\bra{\dPhi_0 - \dK_0}\dW_r\trans}^k
\dW_r
\dK_0
\dW_r\trans
\nonumber\\
&\tab\tab\times
\bra{
\boldsymbol{I}_r \otimes \dF_{t-k-1}
+
\dF_{t-k-1} \otimes \boldsymbol{I}_r
}
\dSigma_\Lambda\inv
\dPi_{t-k-1,t-h-1}
\dSigma_\Lambda\inv
\bra{
\boldsymbol{I}_r \otimes \dF_{t-h-1}
+
\dF_{t-h-1} \otimes \boldsymbol{I}_r
}\trans
\nonumber\\
&\tab\tab\tab\tab\tab \times
\dW_r
\dK_0\trans
\dW_r\trans
\bra{\dW_r\bra{\dPhi_0 - \dK_0}\dW_r\trans}^{h\top}
\bigg)
\bra{\boldsymbol{\lambda}_{0i} \otimes \boldsymbol{\lambda}_{0i} }
\Bigg)
.
\end{align*}
The leading term in \eqref{eq:RV1_var_clt} is the time-series loading term, the leading term in \eqref{eq:RV2_var_clt} is the QMLE-score term, and the leading term in \eqref{eq:RV3_var_clt} is the cross-sectional factor-estimation term. The same Cram\'er--Wold argument used for the conditional mean gives their joint Gaussian convergence. Their pairwise asymptotic covariances are zero by \eqref{eq:ACOV} and Assumptions \ref{ass:idiosyncratic}--\ref{ass:idiosyncraticY}. Thus, from \eqref{eq:common_var_gauss},
\begin{align*}
&\min\bra{\sqrt{N}, \sqrt{T}} \bra{\hatPittithetamle - \Pittithetaz}
\\
=&
\frac{\min\bra{\sqrt N,\sqrt T}}{\sqrt T}
 \bra{\frac 1{\sqrt T}\sum_{s=1}^T \dF_se_{is}}\trans\boldsymbol \Gamma_F\inv
\bra{ \boldsymbol I_r \otimes \boldsymbol\lambda_{0i} + \boldsymbol\lambda_{0i} \otimes \boldsymbol I_r}\trans
\dW_r  \dsigmatti( \dtheta_0 )
 \\
 &+
 \frac{\min\bra{\sqrt N,\sqrt T}}{\sqrt T}
 \bra{\boldsymbol\lambda_{0i} \otimes \boldsymbol\lambda_{0i}}\trans 
 \dW_r
 \nabla^{\dtheta}\dsigmatti(\dtheta_0)
  { \limitcov}(\dtheta_0)\inv\bra{
\frac 1{\sqrt T}\sum_{s=1}^T
 \nabla^{\dtheta}{\ell}_{\dF}\bra{\dF_s ; \dtheta_0}}
 \\
 &+
 \frac{\min\bra{\sqrt N,\sqrt T}}{\sqrt {N}}
\bra{\boldsymbol{\lambda}_{0i}\trans \otimes \boldsymbol{\lambda}_{0i}\trans }
\sum_{k=0}^{t-1}
\bra{\dW_r\bra{\dPhi_0 - \dK_0}\dW_r\trans}^k
\dW_r
\dK_0
\dW_r\trans
\nonumber\\
&\tab\tab\times
\bra{
\boldsymbol{I}_r \otimes \dF_{t-k-1}
+
\dF_{t-k-1} \otimes \boldsymbol{I}_r
}
\bra{\frac{\dLambda_0\trans\dLambda_0}{N}}^{-1} \frac {{\dLambda}_0\trans\dxi_{t-k-1}}{ \sqrt N} 
 \nonumber
 \\
&
+O_{\mathbb P}\bra{\frac {1}{\min\bra{\sqrt N,\sqrt T}} }
.
\end{align*}
This decomposition is likewise the variance counterpart of equation (C.1) in the proof of Theorem 3 of \citet{Bai2003}, augmented by the
QMLE-score term. Applying the same subsequence argument used above and dividing by the conditional standard deviation in the statement
gives a standard Gaussian limit along every subsequence. The subsequence principle therefore proves the second result of the
theorem as $N,T\to\infty$.
\qed

\subsection*{Proof of Proposition \ref{prop:perturb_full_model}}
The only additional term relative to the proof of Proposition \ref{prop:rate_psit_decomp} is generated by the dependence of the conditional variance recursion on the filtered conditional location process. We show that this term does not change any of the rates in Proposition \ref{prop:rate_psit_decomp}.

$(a)$ For every $\dtheta^\mu\in\dTheta^\mu$, define
\begin{align*}
\doublehat{\depsi}_t\bra{\dtheta^\mu}
&=
\dhatFt-
\dhathatmutti\bra{\dtheta^\mu},
\\ 
\widehat{\depsi}_t\bra{\dtheta^\mu}
&=
\widehat\dH\dFt-
\dhatmutti\bra{\dtheta^\mu},
\end{align*}
and note that 
\begin{align}
\label{eq:joint_residual_filter_difference}
\doublehat{\depsi}_t\bra{\dtheta^\mu}
-
\widehat{\depsi}_t\bra{\dtheta^\mu}
=
\dhatFt-\widehat\dH\dFt
-
\bra{
\dhathatmutti\bra{\dtheta^\mu}
-
\dhatmutti\bra{\dtheta^\mu}
},
\end{align}
where the corresponding conditional location filters satisfy
\begin{align*}
\dhathatmutpt\bra{\dtheta^\mu}
&=
\bra{
\dPhi^\mu-\dK^\mu
}
\dhathatmutti\bra{\dtheta^\mu}
+
\dK^\mu\dhatFt,
\\
\dhatmutpt\bra{\dtheta^\mu}
&=
\bra{
\dPhi^\mu-\dK^\mu
}
\dhatmutti\bra{\dtheta^\mu}
+
\dK^\mu\widehat\dH\dFt.
\end{align*}
Therefore, it is straightforward to obtain
\begin{align}
\label{eq:joint_location_filter_difference}
&
\dhathatmutpt\bra{\dtheta^\mu}
-
\dhatmutpt\bra{\dtheta^\mu}
\nonumber
\\
={}&
\bra{
\dPhi^\mu-\dK^\mu
}
\bra{
\dhathatmutti\bra{\dtheta^\mu}
-
\dhatmutti\bra{\dtheta^\mu}
}
+
\dK^\mu
\bra{
\dhatFt-\widehat\dH\dFt
}.
\end{align}
Iterating \eqref{eq:joint_location_filter_difference} and using Assumption \ref{ass:joint_location_variance}$(ii)$ and Proposition
\ref{prop:rate_yt_decomp}$(a)$ we obtain, for any $t\geq1$, as
$N,T\rightarrow\infty$,
\begin{align}
\label{eq:joint_location_filter_rate}
\sup_{\dtheta^\mu\in\dTheta^\mu}
\vvbra{
\dhathatmutpt\bra{\dtheta^\mu}
-
\dhatmutpt\bra{\dtheta^\mu}
}
=
O_{\Pp}\bra{
\frac{1}{\min\bra{\sqrt N,T}}
}
+
O_{a.s.}\bra{
\bra{\calX_1^\mu}^{t}
}.
\end{align}
Now, for every $\dtheta^\sigma\in\dTheta^\sigma$, we have that
\begin{align}
\label{eq:var_recurs1}
\dhathatsigmatpt\bra{\dtheta^\mu, \dtheta^\sigma}
=&
\ddelta^\sigma
+
\dPhi^\sigma\dhathatsigmatti\bra{\dtheta^\mu, \dtheta^\sigma}
\nonumber
\\
&+
\bra{
\dK^\sigma
+
\dUpsilon^\sigma\boldsymbol I\bra{\dhatFt}
}
\dW_r\trans
\bra{
\doublehat{\depsi}_t\bra{\dtheta^\mu}\otimes\doublehat{\depsi}_t\bra{\dtheta^\mu}
-
\vect\dhathatSigmatti\bra{\dtheta^\mu, \dtheta^\sigma}
}
\\
\label{eq:var_recurs2}
\dhatsigmatpt\bra{\dtheta^\mu, \dtheta^\sigma}
=&
\ddelta^\sigma
+
\dPhi^\sigma\dhatsigmatti\bra{\dtheta^\mu, \dtheta^\sigma}
\nonumber
\\
&+
\bra{
\dK^\sigma
+
\dUpsilon^\sigma\boldsymbol I\bra{\widehat\dH\dFt}
}
\dW_r\trans
\bra{
\widehat{\depsi}_t\bra{\dtheta^\mu}\otimes\widehat{\depsi}_t\bra{\dtheta^\mu}
-
\vect\dhatSigmatti\bra{\dtheta^\mu, \dtheta^\sigma}
},
\end{align}
and we let
\begin{align*}
\doublehat{\boldsymbol A}_t^\sigma
=
\dPhi^\sigma
-
\dK^\sigma
-
\dUpsilon^\sigma
\boldsymbol I\bra{\dhatFt},
\qquad 
\doublehat{\boldsymbol B}_t^\sigma
=
\dK^\sigma
+
\dUpsilon^\sigma
\boldsymbol I\bra{\dhatFt}.
\end{align*}
Then, subtracting the variance recursion in \eqref{eq:var_recurs2} from the feasible recursion in \eqref{eq:var_recurs1} driven by $\big\{\dhatFt, \doublehat{\depsi}_t(\dtheta^\mu)\big\}$ yields
\begin{align}
\label{eq:joint_variance_filter_difference}
&
\dhathatsigmatpt
\bra{
\dtheta^\mu,\dtheta^\sigma
}
-
\dhatsigmatpt
\bra{
\dtheta^\mu,\dtheta^\sigma
}
\nonumber
\\
={}&
\doublehat{\boldsymbol A}_t^\sigma
\bra{
\dhathatsigmatti
\bra{
\dtheta^\mu,\dtheta^\sigma
}
-
\dhatsigmatti
\bra{
\dtheta^\mu,\dtheta^\sigma
}
}
\nonumber
\\
&+
\doublehat{\boldsymbol B}_t^\sigma
\dW_r\trans
\bra{
\doublehat{\depsi}_t
\bra{\dtheta^\mu}
\otimes
\doublehat{\depsi}_t
\bra{\dtheta^\mu}
-
\widehat{\depsi}_t
\bra{\dtheta^\mu}
\otimes
\widehat{\depsi}_t
\bra{\dtheta^\mu}
}
\nonumber
\\
&+
\dUpsilon^\sigma
\bra{
\boldsymbol I\bra{\dhatFt}
-
\boldsymbol I\bra{\widehat\dH\dFt}
}
\bra{
\dW_r\trans
\bra{
\widehat{\depsi}_t
\bra{\dtheta^\mu}
\otimes
\widehat{\depsi}_t
\bra{\dtheta^\mu}
}
-
\dhatsigmatti
\bra{
\dtheta^\mu,\dtheta^\sigma
}
}.
\end{align}
The difference between the two quadratic residual terms in
\eqref{eq:joint_variance_filter_difference} satisfies
\begin{align}
\label{eq:joint_quadratic_residual_difference}
&
\doublehat{\depsi}_t
\bra{\dtheta^\mu}
\otimes
\doublehat{\depsi}_t
\bra{\dtheta^\mu}
-
\widehat{\depsi}_t
\bra{\dtheta^\mu}
\otimes
\widehat{\depsi}_t
\bra{\dtheta^\mu}
\nonumber
\\
={}&
\frac{1}{2}
\bra{
\boldsymbol I_r
\otimes
\bra{
\doublehat{\depsi}_t
\bra{\dtheta^\mu}
+
\widehat{\depsi}_t
\bra{\dtheta^\mu}
}
+
\bra{
\doublehat{\depsi}_t
\bra{\dtheta^\mu}
+
\widehat{\depsi}_t
\bra{\dtheta^\mu}
}
\otimes
\boldsymbol I_r
}
\bra{
\doublehat{\depsi}_t
\bra{\dtheta^\mu}
-
\widehat{\depsi}_t
\bra{\dtheta^\mu}
}.
\end{align}
Substituting \eqref{eq:joint_residual_filter_difference} and \eqref{eq:joint_quadratic_residual_difference} into \eqref{eq:joint_variance_filter_difference} and iterating over $k$ periods yields
\begin{align}
\label{eq:joint_variance_filter_k_step}
&
\dhathatsigmatpt
\bra{
\dtheta^\mu,\dtheta^\sigma
}
-
\dhatsigmatpt
\bra{
\dtheta^\mu,\dtheta^\sigma
}
\nonumber
\\
={}&
\bra{
\prod_{h=0}^{k-1}
\doublehat{\boldsymbol A}_{t-h}^\sigma
}
\bra{
\dhathatsigma_{t-k+1|t-k}
\bra{
\dtheta^\mu,\dtheta^\sigma
}
-
\dhatsigma_{t-k+1|t-k}
\bra{
\dtheta^\mu,\dtheta^\sigma
}
}
\nonumber
\\
&+
\frac{1}{2}
\sum_{j=1}^{k}
\bra{
\prod_{h=0}^{j-2}
\doublehat{\boldsymbol A}_{t-h}^\sigma
}
\doublehat{\boldsymbol B}_{t-j+1}^\sigma
\dW_r\trans
\nonumber
\\
&\qquad\times
\bra{
\boldsymbol I_r
\otimes
\bra{
\doublehat{\depsi}_{t-j+1}\bra{\dtheta^\mu}
+
\widehat{\depsi}_{t-j+1}\bra{\dtheta^\mu}
}
+
\bra{
\doublehat{\depsi}_{t-j+1}\bra{\dtheta^\mu}
+
\widehat{\depsi}_{t-j+1}\bra{\dtheta^\mu}
}
\otimes
\boldsymbol I_r
}
\nonumber
\\
&\qquad\times
\bra{
\dhatF_{t-j+1}
-
\widehat\dH\dF_{t-j+1}
-
\bra{
\dhathatmu_{t-j+1|t-j}\bra{\dtheta^\mu}
-
\dhatmu_{t-j+1|t-j}\bra{\dtheta^\mu}
}
}
\nonumber
\\
&+
\sum_{j=1}^{k}
\bra{
\prod_{h=0}^{j-2}
\doublehat{\boldsymbol A}_{t-h}^\sigma
}
\dUpsilon^\sigma
\bra{
\boldsymbol I\bra{\dhatF_{t-j+1}}
-
\boldsymbol I\bra{\widehat\dH\dF_{t-j+1}}
}
\nonumber
\\
&\qquad\times
\bra{
\dW_r\trans
\bra{
\widehat{\depsi}_{t-j+1}\bra{\dtheta^\mu}
\otimes
\widehat{\depsi}_{t-j+1}\bra{\dtheta^\mu}
}
-
\dhatsigma_{t-j+1|t-j}
\bra{
\dtheta^\mu,\dtheta^\sigma
}
},
\end{align}
where the product equals $\boldsymbol I_r$ when $j=1$.

Note that Assumption \ref{ass:joint_location_variance}$(iv)$ implies that
\begin{align}
\label{eq:joint_variance_product_bound}
\sup_{\dtheta^\sigma\in\dTheta^\sigma}
\vvbra{
\prod_{h=0}^{k-1}
\doublehat{\boldsymbol A}_{t-h}^\sigma
}
\leq
\calX_k^\sigma
<
1
\end{align}
almost surely, whereas Assumption \ref{ass:joint_location_variance}$(v)$, together with \eqref{eq:joint_location_filter_rate} and \eqref{eq:joint_residual_filter_difference}, implies that the coefficient multiplying the residual difference in \eqref{eq:joint_variance_filter_k_step} is $O_{\Pp}(1)$ uniformly over $\dTheta^\mu\times\dTheta^\sigma$.

To control the indicator term in \eqref{eq:joint_variance_filter_k_step}, fix $t\geq1$ and let $\eta>0$ be sufficiently small. Componentwise sign changes can occur
only if at least one component of $\dH_0\dF_t$ is within $\eta$ of zero, or if either of the two approximating vectors is more than $\eta$ away from $\dH_0\dF_t$. Therefore, 
\begin{align*}
&
\cbra{
\boldsymbol I\bra{\dhatF_t}
\neq
\boldsymbol I\bra{\widehat\dH\dF_t}
}
\\
\subseteq{}&
\bigcup_{i=1}^{r}
\cbra{
\vbra{
\mathbf e_i\trans\dH_0\dF_t
}
\leq\eta
}
\cup
\cbra{
\vvbra{
\dhatF_t-\dH_0\dF_t
}
>\eta
}
\cup
\cbra{
\vvbra{
\widehat\dH\dF_t-\dH_0\dF_t
}
>\eta
}.
\end{align*}
Indeed, on the complement of the event on the right-hand side, $\dhatF_t$ and $\widehat\dH\dF_t$ are both within $\eta$ of $\dH_0\dF_t$, while every component of $\dH_0\dF_t$ has absolute value larger than $\eta$. Hence, both vectors have the same componentwise signs as $\dH_0\dF_t$.

Taking probabilities and applying the union bound
\begin{align}
\label{eq:joint_indicator_factor_bound}
&
\Pp\bra{
\boldsymbol I\bra{\dhatF_t}
\neq
\boldsymbol I\bra{\widehat\dH\dF_t}
}
\nonumber
\\
\leq{}&
\sum_{i=1}^{r}
\Pp\bra{
\vbra{
\mathbf e_i\trans\dH_0\dF_t
}
\leq\eta
}
+
\Pp\bra{
\vvbra{
\dhatF_t-\dH_0\dF_t
}
>\eta
}
+
\Pp\bra{
\vvbra{
\bra{
\widehat\dH-\dH_0
}
\dF_t
}
>\eta
}
\nonumber
\\
\leq{}&
rC_I\eta
+
\Pp\bra{
\vvbra{
\dhatF_t-\dH_0\dF_t
}
>\eta
}
+
\Pp\bra{
\vvbra{
\bra{
\widehat\dH-\dH_0
}
\dF_t
}
>\eta
},
\end{align}
where the last inequality follows from Assumption \ref{ass:joint_location_variance}$(vi)$ evaluated at $\dH=\dH_0$.

Recall that, for every fixed $\eta>0$, Proposition \ref{prop:rate_yt_decomp}$(e)$ implies that
$
\Pp\bra{
\vvbra{
\dhatF_t-\dH_0\dF_t
}
>\eta
}
\longrightarrow
0
$
as $N,T\rightarrow\infty$. Moreover,
\[
\vvbra{
\bra{
\widehat\dH-\dH_0
}
\dF_t
}
\leq
\vvbra{
\widehat\dH-\dH_0
}
\vvbra{
\dF_t
}
=
o_{\Pp}(1),
\]
because $\widehat\dH\pconv\dH_0$ and $\dF_t=O_{\Pp}(1)$ by Assumption \ref{ass:pram_SE}. Consequently, \eqref{eq:joint_indicator_factor_bound} yields
\[
\limsup_{N,T\rightarrow\infty}
\Pp\bra{
\boldsymbol I\bra{\dhatF_t}
\neq
\boldsymbol I\bra{\widehat\dH\dF_t}
}
\leq
rC_I\eta.
\]
Letting $\eta\downarrow0$, we obtain
$
\Pp\bra{
\boldsymbol I\bra{\dhatF_t}
\neq
\boldsymbol I\bra{\widehat\dH\dF_t}
}
\longrightarrow
0
$
as $N,T\rightarrow\infty$, and hence
$
\boldsymbol I\bra{\dhatF_t}
-
\boldsymbol I\bra{\widehat\dH\dF_t}
=
o_{\Pp}(1).
$

Since $k$ is fixed and $\dTheta^\sigma$ is compact, the coefficient of the indicator difference satisfies
\begin{align}
\label{eq:joint_indicator_coefficient_bound}
&
\max_{1\leq j\leq k}
\sup_{\substack{
\dtheta^\mu\in\dTheta^\mu\\
\dtheta^\sigma\in\dTheta^\sigma
}}
\vvbra{
\bra{
\prod_{h=0}^{j-2}
\doublehat{\boldsymbol A}_{t-h}^\sigma
}
\dUpsilon^\sigma
}
\nonumber
\\
&\qquad\times
\vvbra{
\dW_r\trans
\bra{
\widehat{\depsi}_{t-j+1}\bra{\dtheta^\mu}
\otimes
\widehat{\depsi}_{t-j+1}\bra{\dtheta^\mu}
}
-
\dhatsigma_{t-j+1|t-j}
\bra{
\dtheta^\mu,\dtheta^\sigma
}
}
\nonumber
\\
\leq{}&
C
\max_{1\leq j\leq k}
\sup_{\substack{
\dtheta^\mu\in\dTheta^\mu\\
\dtheta^\sigma\in\dTheta^\sigma
}}
\bra{
1
+
\vvbra{
\widehat{\depsi}_{t-j+1}\bra{\dtheta^\mu}
}^{2}
+
\vvbra{
\dhatsigma_{t-j+1|t-j}
\bra{\dtheta^\mu,\dtheta^\sigma}
}
}
=
O_{\Pp}(1),
\end{align}
where the last equality follows from Assumption \ref{ass:joint_location_variance}$(ii)$--$(v)$. Therefore the last sum in \eqref{eq:joint_variance_filter_k_step} is 
\begin{align}
\label{eq:result_ineq_joint1}
&\sum_{j=1}^{k}
\bra{
\prod_{h=0}^{j-2}
\doublehat{\boldsymbol A}_{t-h}^\sigma
}
\dUpsilon^\sigma
\bra{
\boldsymbol I\bra{\dhatF_{t-j+1}}
-
\boldsymbol I\bra{\widehat\dH\dF_{t-j+1}}
}
\nonumber
\\
&\qquad\times
\bra{
\dW_r\trans
\bra{
\widehat{\depsi}_{t-j+1}\bra{\dtheta^\mu}
\otimes
\widehat{\depsi}_{t-j+1}\bra{\dtheta^\mu}
}
-
\dhatsigma_{t-j+1|t-j}
\bra{
\dtheta^\mu,\dtheta^\sigma
}
}
=
O_{\Pp}\bra{\frac{1}{\min\bra{\sqrt N,T}}}. 
\end{align}
Taking the supremum over $\dTheta^\mu\times\dTheta^\sigma$ in \eqref{eq:joint_variance_filter_k_step}, and using \eqref{eq:joint_location_filter_rate},
\eqref{eq:joint_variance_product_bound}, \eqref{eq:joint_indicator_coefficient_bound}, and \eqref{eq:result_ineq_joint1} we obtain
\begin{align*}
&
\sup_{\substack{
\dtheta^\mu\in\dTheta^\mu\\
\dtheta^\sigma\in\dTheta^\sigma
}}
\vvbra{
\dhathatsigmatpt
\bra{\dtheta^\mu,\dtheta^\sigma}
-
\dhatsigmatpt
\bra{\dtheta^\mu,\dtheta^\sigma}
}
\\
\leq{}&
\calX_k^\sigma
\sup_{\substack{
\dtheta^\mu\in\dTheta^\mu\\
\dtheta^\sigma\in\dTheta^\sigma
}}
\vvbra{
\dhathatsigma_{t-k+1|t-k}
\bra{\dtheta^\mu,\dtheta^\sigma}
-
\dhatsigma_{t-k+1|t-k}
\bra{\dtheta^\mu,\dtheta^\sigma}
}
\\
&+
O_{\Pp}\bra{
\frac{1}{\min\bra{\sqrt N,T}}
}
+
O_{a.s.}\bra{
\bra{\calX_1^\mu}^{t}
}.
\end{align*}
Choose $\varrho_k$ such that
$
\max\cbra{
\calX_k^\sigma,
\bra{\calX_1^\mu}^{k}
}
<\varrho_k<1,
$
then, iterating the preceding inequality yields
\begin{align}
\label{eq:joint_variance_filter_rate}
\sup_{\substack{
\dtheta^\mu\in\dTheta^\mu\\
\dtheta^\sigma\in\dTheta^\sigma
}}
\vvbra{
\dhathatsigmatpt
\bra{\dtheta^\mu,\dtheta^\sigma}
-
\dhatsigmatpt
\bra{\dtheta^\mu,\dtheta^\sigma}
}
=
O_{\Pp}\bra{
\frac{1}{\min\bra{\sqrt N,T}}
}
+
O_{a.s.}\bra{
\varrho_k^{\lfloor t/k\rfloor}
},
\end{align}
thus, equations \eqref{eq:joint_location_filter_rate} and \eqref{eq:joint_variance_filter_rate} prove part $(a)$.

For part $(b)$, the smooth terms are controlled exactly as in the proof of Proposition \ref{prop:rate_psit_decomp}$(b)$. For the indicator term,
we first write
\begin{align*}
&
\boldsymbol I\bra{\widehat\dH\dF_t}
-
\boldsymbol I\bra{\dH_0\dbarF_t}
\\
={}&
\boldsymbol I\bra{\widehat\dH\dF_t}
-
\boldsymbol I\bra{\dH_0\dF_t}
+
\boldsymbol I\bra{\dH_0\dF_t}
-
\boldsymbol I\bra{\dH_0\dbarF_t},
\end{align*}
and hence, since by Proposition \ref{prop:rate_yt_decomp}$(a)$ and $(e)$, it holds that
\begin{align}
\label{eq:bound_smooths}
\vvbra{
\bra{\widehat\dH-\dH_0}\dF_t
}
\leq
\vvbra{
\widehat\dH\dF_t-\dhatF_t
}
+
\vvbra{
\dhatF_t-\dH_0\dF_t
}
=
O_{\Pp}\bra{
\frac{1}{\min\bra{\sqrt N,\sqrt T}}
},
\end{align}
we get, by using Assumption \ref{ass:joint_location_variance}$(vi)$ and the same argument used to obtain \eqref{eq:result_ineq_joint1},
\begin{align}
\label{eq:bound_indicators}
\vvbra{
\boldsymbol I\bra{\widehat\dH\dF_t}
-
\boldsymbol I\bra{\dH_0\dF_t}
}
=
o_{\Pp}\bra{
\frac{1}{\min\bra{\sqrt N,\sqrt T}}
}.
\end{align}
Moreover, Assumptions \ref{ass:noise}$(iii)$ and \ref{ass:contraction_cond} imply that
\begin{align*}
\vvbra{
\dH_0\bra{\dF_t-\dbarF_t}
}_2
\leq
C\calX_k^{\lfloor t/k\rfloor}
\end{align*}
for some finite positive real $C$ independent of $t$. For every sufficiently small $\eta>0$, we have
\begin{align*}
\cbra{
\boldsymbol I\bra{\dH_0\dF_t}
\neq
\boldsymbol I\bra{\dH_0\dbarF_t}
}
\subseteq
\bigcup_{i=1}^{r}
\cbra{
\vbra{
\mathbf e_i\trans\dH_0\dbarF_t
}
\leq\eta
}
\cup
\cbra{
\vvbra{
\dH_0\bra{\dF_t-\dbarF_t}
}
>
\eta
},
\end{align*}
because, outside the event on the right-hand side, every component of $\dH_0\dbarF_t$ is farther than $\eta$ from zero and the perturbation $\dH_0\bra{\dF_t-\dbarF_t}$ has norm at most $\eta$. Hence, $\dH_0\dF_t$ and $\dH_0\dbarF_t$ have the same componentwise signs. Therefore, by the union bound, Assumption \ref{ass:joint_location_variance}$(vi)$, and Markov's inequality,
\begin{align*}
&
\Pp\bra{
\boldsymbol I\bra{\dH_0\dF_t}
\neq
\boldsymbol I\bra{\dH_0\dbarF_t}
}
\\
\leq{}&
\sum_{i=1}^{r}
\Pp\bra{
\vbra{
\mathbf e_i\trans\dH_0\dbarF_t
}
\leq\eta
}
+
\Pp\bra{
\vvbra{
\dH_0\bra{\dF_t-\dbarF_t}
}
>
\eta
}
\\
\leq{}&
rC_I\eta
+
\frac{
C\calX_k^{2\lfloor t/k\rfloor}
}{
\eta^2
}.
\end{align*}
By setting $\eta=\calX_k^{2\lfloor t/k\rfloor/3}$ we get
\begin{align*}
\Pp\bra{
\boldsymbol I\bra{\dH_0\dF_t}
\neq
\boldsymbol I\bra{\dH_0\dbarF_t}
}
\leq
C^\prime
\calX_k^{\frac{2}{3}\lfloor t/k\rfloor},
\end{align*}
for some finite positive real $C^\prime$ independent of $t$. The right-hand side is summable in $t$ and hence, by the Borel--Cantelli lemma, $\boldsymbol I\bra{\dH_0\dF_t}-\boldsymbol I\bra{\dH_0\dbarF_t}=\boldsymbol 0_{r\times r}$ almost surely for all sufficiently large $t$. Consequently, the effect of this indicator difference on the variance recursion decreases at the geometric contraction rate. By combining this result with the bound for $\boldsymbol I\bra{\widehat\dH\dF_t}-\boldsymbol I\bra{\dH_0\dF_t}$ in \eqref{eq:bound_indicators}, \eqref{eq:bound_smooths} and the other smooth terms, we obtain
\begin{align*}
&
\sup_{\substack{
\dtheta^\mu\in\dTheta^\mu\\
\dtheta^\sigma\in\dTheta^\sigma
}}
\vvbra{
\bra{
\dhatmutti\bra{\dtheta^\mu}\trans,
\dhatsigmatti
\bra{\dtheta^\mu,\dtheta^\sigma}\trans
}\trans
-
\bra{
\widetilde{\boldsymbol\mu}_{t|t-1}
\bra{\dtheta^\mu}\trans,
\widetilde{\boldsymbol\sigma}^2_{t|t-1}
\bra{\dtheta^\mu,\dtheta^\sigma}\trans
}\trans
}
\\
={}&
O_{\Pp}\bra{
\frac{1}{\min\bra{\sqrt N,\sqrt T}}
}
+
O_{a.s.}\bra{
\varrho_k^{\lfloor t/k\rfloor}
},
\end{align*}
which proves part $(b)$.

For part $(c)$, both filters are driven by $\dH_0\dbarF_t$, and hence, the factor and indicator differences in \eqref{eq:joint_variance_filter_k_step} are identically zero. The contraction argument used above also implies
\begin{align*}
&
\sup_{\substack{
\dtheta^\mu\in\dTheta^\mu\\
\dtheta^\sigma\in\dTheta^\sigma
}}
\vvbra{
\bra{
\widetilde{\boldsymbol\mu}_{t|t-1}
\bra{\dtheta^\mu}\trans,
\widetilde{\boldsymbol\sigma}^2_{t|t-1}
\bra{\dtheta^\mu,\dtheta^\sigma}\trans
}\trans
-
\bra{
\bar{\boldsymbol\mu}_{t|t-1}
\bra{\dtheta^\mu}\trans,
\bar{\boldsymbol\sigma}^2_{t|t-1}
\bra{\dtheta^\mu,\dtheta^\sigma}\trans
}\trans
}
\\
={}&
O_{a.s.}\bra{
\varrho_k^{\lfloor t/k\rfloor}
},
\end{align*}
which proves part $(c)$.

For part $(d)$, let $\dF_t^{(\tau)}$ be the truncated factor process used in Lemma \ref{lemma:NED}. Then
\begin{align}
\label{eq:joint_factor_truncation_rate}
\sup_{t\in\Nn}
\vvbra{
\dH_0\bra{
\dbarF_t-\dF_t^{(\tau)}
}
}_2
=
O\bra{
\calX_k^{\lfloor\tau/k\rfloor}
}.
\end{align}
For every $i=1,\ldots,r$ and every sufficiently small $\eta>0$, Assumption \ref{ass:joint_location_variance}$(vi)$, \eqref{eq:joint_factor_truncation_rate}, and Markov's inequality we get
\begin{align*}
&
\Pp\bra{
\boldsymbol I\bra{
\mathbf e_i\trans\dH_0\dbarF_t
}
\neq
\boldsymbol I\bra{
\mathbf e_i\trans\dH_0\dF_t^{(\tau)}
}
}
\\
\leq{}&
C_I\eta
+
\frac{
C\calX_k^{2\lfloor\tau/k\rfloor}
}{\eta^2}.
\end{align*}
Setting $\eta=\calX_k^{2\lfloor\tau/k\rfloor/3}$ and summing over the fixed number of components yields
\begin{align}
\label{eq:joint_indicator_truncation_rate}
&
\sup_{t\in\Nn}
\vvbra{
\boldsymbol I\bra{\dH_0\dbarF_t}
-
\boldsymbol I\bra{\dH_0\dF_t^{(\tau)}}
}_2
=
O\bra{
\calX_k^{\frac{1}{3}\lfloor\tau/k\rfloor}
}.
\end{align}
The conditional location truncation error has the same geometric form by Assumption \ref{ass:joint_location_variance}$(ii)$ and  \eqref{eq:joint_factor_truncation_rate}. Therefore, enlarging $\varrho_k$ if necessary so that $\calX_k^{1/3}<\varrho_k<1$, the $k$-step variance recursion and \eqref{eq:joint_variance_product_bound} imply that
\begin{align}
\label{eq:joint_filter_truncation_rate}
&
\sup_{t\in\Nn}
\vvbra{
\bra{
\bar{\boldsymbol\mu}_{t|t-1}
\bra{\dtheta^\mu}\trans,
\bar{\boldsymbol\sigma}^2_{t|t-1}
\bra{\dtheta^\mu,\dtheta^\sigma}\trans
}\trans
-
\bra{
\boldsymbol\mu^{(\tau)}_{t|t-1}
\bra{\dtheta^\mu}\trans,
\boldsymbol\sigma^{2\,(\tau)}_{t|t-1}
\bra{\dtheta^\mu,\dtheta^\sigma}\trans
}\trans
}_2
=
O\bra{
\varrho_k^{\lfloor\tau/k\rfloor}
}.
\end{align}
Since the truncated filter is measurable with respect to $\dvt,\ldots,\dv_{t-\tau}$, from the projection property of conditional expectations and \eqref{eq:joint_filter_truncation_rate} we get
\begin{align*}
&
\sup_{t\in\Nn}
\vvbra{
\bra{
\bar{\boldsymbol\mu}_{t|t-1}
\bra{\dtheta^\mu}\trans,
\bar{\boldsymbol\sigma}^2_{t|t-1}
\bra{\dtheta^\mu,\dtheta^\sigma}\trans
}\trans
-
\Ee\sbra{
\left.
\bra{
\bar{\boldsymbol\mu}_{t|t-1}
\bra{\dtheta^\mu}\trans,
\bar{\boldsymbol\sigma}^2_{t|t-1}
\bra{\dtheta^\mu,\dtheta^\sigma}\trans
}\trans
\right|
\dvt,\ldots,\dv_{t-\tau}
}
}_2
\\
\leq{}&
\sup_{t\in\Nn}
\vvbra{
\bra{
\bar{\boldsymbol\mu}_{t|t-1}
\bra{\dtheta^\mu}\trans,
\bar{\boldsymbol\sigma}^2_{t|t-1}
\bra{\dtheta^\mu,\dtheta^\sigma}\trans
}\trans
-
\bra{
\boldsymbol\mu^{(\tau)}_{t|t-1}
\bra{\dtheta^\mu}\trans,
\boldsymbol\sigma^{2\,(\tau)}_{t|t-1}
\bra{\dtheta^\mu,\dtheta^\sigma}\trans
}\trans
}_2
=
O\bra{
\varrho_k^{\lfloor\tau/k\rfloor}
}.
\end{align*}
Thus, the stationary joint filter is near-epoch dependent of size $-1$ on $\cbra{\dvt}_{t\in\Zz}$, proving part $(d)$. Finally, part $(e)$ follows from parts $(a)$--$(c)$ and the triangle inequality.
\qed

\subsubsection*{Proof of Proposition \ref{prop:sequential_QMLE_expansion}}

Let
\begin{align*}
\widehat{\boldsymbol{\cal S}}_{J,T}
\bra{
\dtheta^\mu,\dtheta^\sigma
}
=
\begin{pmatrix}
\sum_{t=1}^{T}
\nabla^{\dtheta^\mu}
\doublehat{\ell}_{\dmu,\dF}
\bra{
\dhatF_t;
\dtheta^\mu
}
\\
\sum_{t=1}^{T}
\nabla^{\dtheta^\sigma}
\doublehat{\ell}_{\sigma,\depsi}
\bra{
\doublehat{\depsi}_t\bra{\dtheta^\mu};
\dtheta^\sigma
}
\end{pmatrix}
\end{align*}
denote the stacked feasible score. Since the two estimators are consistent for interior points of their respective parameter spaces, their first-order conditions imply that, with probability approaching one,
\begin{align}
\label{eq:sequential_stacked_FOC}
\widehat{\boldsymbol{\cal S}}_{J,T}
\bra{
\dthetamle^{\,\mu},
\dthetamle^{\,\sigma}
}
=
\boldsymbol 0
\end{align}
as $N,T\rightarrow\infty$.

For a generic
$\dtheta^J
=
\bra{
{\dtheta^\mu}\trans,
{\dtheta^\sigma}\trans
}\trans$,
define the stacked stationary score contribution as
\begin{align*}
\boldsymbol s_t^J
\bra{
\dtheta^J
}
=
\begin{pmatrix}
\nabla^{\dtheta^\mu}
\ell_{\dmu,\dF}
\bra{
\dbarF_t;
\dtheta^\mu
}
\\[2mm]
\nabla^{\dtheta^\sigma}
\ell_{\sigma,\depsi}
\bra{
\bar{\depsi}_t\bra{\dtheta^\mu};
\dtheta^\sigma
}
\end{pmatrix}.
\end{align*}
Proposition \ref{prop:perturb_full_model}$(a)$--$(c)$ provides the filter approximation rates required in the proof of Lemma \ref{lemma:U_convergence_score_vector}, while Lemma \ref{lemma:rate_psit_first_derivs}$(a)$--$(c)$ provides the corresponding rates for the first-derivative filters. Moreover, the
leverage indicator does not depend on $\dtheta^\mu$ or $\dtheta^\sigma$, and its approximation is controlled by Assumption \ref{ass:joint_location_variance}$(vi)$ as in the proof of Proposition \ref{prop:perturb_full_model}. Therefore, applying componentwise the score expansion and the weighted PC argument used in the proof of
Lemma \ref{lemma:U_convergence_score_vector} we directly obtain 
\begin{align}
\label{eq:sequential_feasible_score_approximation}
&
\sup_{\dtheta^J\in\dTheta^\mu\times\dTheta^\sigma}
\vvbra{
\frac{1}{\sqrt T}
\widehat{\boldsymbol{\cal S}}_{J,T}
\bra{
\boldsymbol r_{\widehat\dH}^J
\bra{
\dtheta^J
}
}
-
\frac{1}{\sqrt T}
{\boldsymbol R_{\widehat\dH}^J}\trans
\sum_{t=1}^{T}
\boldsymbol s_t^J
\bra{
\dtheta^J
}
}
\nonumber
\\
={}&
O_{\Pp}\bra{
\frac{1}{\sqrt N}
+
\frac{\sqrt T}{N}
+
\frac{1}{\sqrt T}
}
=
o_{\Pp}(1)
\end{align}
as $N,T\rightarrow\infty$, where the last equality follows from $\sqrt T/N\rightarrow0$. In particular, evaluating \eqref{eq:sequential_feasible_score_approximation} at $\dtheta^J=\dtheta_0^J$ yields
\begin{align}
\label{eq:sequential_score_at_true_parameter}
\frac{1}{\sqrt T}
\widehat{\boldsymbol{\cal S}}_{J,T}
\bra{
\boldsymbol r_{\widehat\dH}^J
\bra{
\dtheta_0^J
}
}
=
{\boldsymbol R_{\widehat\dH}^J}\trans
\frac{1}{\sqrt T}
\sum_{t=1}^{T}
\boldsymbol s_t^J
\bra{
\dtheta_0^J
}
+
o_{\Pp}(1)
\end{align}
as $N,T\rightarrow\infty$.

An application of the mean-value expansion to \eqref{eq:sequential_stacked_FOC} around $\boldsymbol r_{\widehat\dH}^J\bra{\dtheta_0^J}$ yields
\begin{align}
\label{eq:sequential_QMLE_score_expansion}
\boldsymbol 0
={}&
\frac{1}{\sqrt T}
\widehat{\boldsymbol{\cal S}}_{J,T}
\bra{
\boldsymbol r_{\widehat\dH}^J
\bra{
\dtheta_0^J
}
}
-
\widehat{\limitcov}_{J,T}
\sqrt T
\bra{
\dthetamle^J
-
\boldsymbol r_{\widehat\dH}^J
\bra{
\dtheta_0^J
}
},
\end{align}
where
\begin{align}
\label{eq:sequential_sample_information}
\widehat{\limitcov}_{J,T}
=
-
\frac{1}{T}
\frac{
\partial
\widehat{\boldsymbol{\cal S}}_{J,T}
\bra{
\dthetamle^{J\star}
}
}{
\partial{\dtheta^J}\trans
},
\end{align}
and the Jacobian evaluated at $\dthetamle^{J\star}$ denotes the Jacobian averaged along the line segment joining $\dthetamle^J$ and $\boldsymbol r_{\widehat\dH}^J\bra{\dtheta_0^J}$.

The Jacobian in \eqref{eq:sequential_sample_information} contains the joint location and variance filters together with their first and second derivatives. From proposition \ref{prop:perturb_full_model}, Lemmas \ref{lemma:rate_psit_first_derivs} and \ref{lemma:rate_psit_second_derivs}, and the uniform Hessian argument of Lemma \ref{lemma:ULLN_Hessian_matrix} we therefore obtain
\begin{align}
\label{eq:sequential_stacked_Hessian_convergence}
\widehat{\limitcov}_{J,T}
\pconv
\boldsymbol R_{\dH_0}^{J\,\trans}
\limitcov_J
\bra{
\dtheta_0^J
}
\boldsymbol R_{\dH_0}^J
=
\limitcov_{J,\dH_0}
\bra{
\dtheta_0^J
}
\end{align}
as $N,T\rightarrow\infty$, where
\begin{align*}
\limitcov_J
\bra{
\dtheta_0^J
}
&=
-
\Ee\sbra{
\frac{
\partial
\boldsymbol s_t^J
\bra{
\dtheta_0^J
}
}{
\partial{\dtheta^J}\trans
}
}
=
\begin{pmatrix}
\limitcov_{\mu\mu}
\bra{
\dtheta_0^\mu
}
&
\boldsymbol 0
\\
\limitcov_{\sigma\mu}
\bra{
\dtheta_0^J
}
&
\limitcov_{\sigma\sigma}
\bra{
\dtheta_0^J
}
\end{pmatrix}.
\end{align*}
The upper-right block is zero because the first-stage criterion does not depend on $\dtheta^\sigma$, whereas the lower-left block accounts for the dependence of the second-stage score on $\dtheta^\mu$ through the residuals and the conditional variance filter. The diagonal blocks are nonsingular under Assumption \ref{ass:truePar}, hence, $\limitcov_J\bra{\dtheta_0^J}$ and $\limitcov_{J,\dH_0}\bra{\dtheta_0^J}$ are nonsingular.

By Proposition \ref{prop:perturb_full_model}$(d)$, Lemma \ref{lemma:rate_psit_first_derivs}$(d)$, and the moment conditions in Assumption \ref{ass:smooth_domination_score_Hessian}, the process $\cbra{\boldsymbol s_t^J\bra{\dtheta_0^J}}_{t\in\Zz}$ is near-epoch dependent of size $-1$ on $\cbra{\dvt}_{t\in\Zz}$. Therefore, the same Cramér--Wold argument used in Lemma \ref{lemma:CLT_NED_score_vector} implies that
\begin{align}
\label{eq:sequential_stacked_score_limit}
\frac{1}{\sqrt T}
\sum_{t=1}^{T}
\boldsymbol s_t^J
\bra{
\dtheta_0^J
}
\Rightarrow
\calN\bra{
\boldsymbol 0,
\dV_J
\bra{
\dtheta_0^J
}
}
\end{align}
as $T\rightarrow\infty$.

Then, by combining \eqref{eq:sequential_score_at_true_parameter} and \eqref{eq:sequential_QMLE_score_expansion} we can write
\begin{align}
\label{eq:sequential_QMLE_linear_representation}
&
\sqrt T
\bra{
\dthetamle^J
-
\boldsymbol r_{\widehat\dH}^J
\bra{
\dtheta_0^J
}
}
=
\widehat{\limitcov}_{J,T}^{-1}
{\boldsymbol R_{\widehat\dH}^J}\trans
\frac{1}{\sqrt T}
\sum_{t=1}^{T}
\boldsymbol s_t^J
\bra{
\dtheta_0^J
}
+
o_{\Pp}(1)
\end{align}
as $N,T\rightarrow\infty$. Finally, $\boldsymbol R_{\widehat\dH}^J \pconv \boldsymbol R_{\dH_0}^J$ as $N,T\rightarrow\infty$. Hence, \eqref{eq:sequential_stacked_Hessian_convergence}, \eqref{eq:sequential_stacked_score_limit}, and Slutsky's theorem imply that
\begin{align*}
\sqrt T
\bra{
\dthetamle^J
-
\boldsymbol r_{\widehat\dH}^J
\bra{
\dtheta_0^J
}
}
\Rightarrow
\calN\bra{
\boldsymbol 0,
\limitcov_{J,\dH_0}^{-1}
\bra{
\dtheta_0^J
}
\dV_{J,\dH_0}
\bra{
\dtheta_0^J
}
\bra{
\limitcov_{J,\dH_0}
\bra{
\dtheta_0^J
}\trans
}^{-1}
},
\end{align*}
which proves the result.
\qed

\section{Auxiliary lemmas}
\label{appendix_lemmas}

\begin{lemma}
\label{lemma:NED}
Under Assumptions \ref{ass:noise}--\ref{ass:contraction_cond}, the process $\cbra{ \bra{ \dbarFt\trans,\dbarpsitti(\dtheta_0)\trans
}\trans}_{t\in\Nn}$ is near-epoch dependent of size $-1$ on the geometrically $\beta$-mixing process $\cbra{\dvt}_{t\in\Zz}$. Consequently, both $\cbra{\dbarFt}_{t\in\Nn}$ and $\cbra{\dbarpsitti(\dtheta_0)}_{t\in\Nn}$ are near-epoch dependent
of size $-1$ on $\cbra{\dvt}_{t\in\Zz}$.
\end{lemma}
\begin{proof}
Let $k\geq1$ be as in Assumption \ref{ass:contraction_cond}. For any $\tau\geq1$, consider the $\tau$-truncated processes $\cbra{\dFt^{(\tau)}}_{t\in\Nn}$ and $\cbra{\dpsi_{t|t-1}^{(\tau)}(\dtheta_0)}_{t\in\Nn}$, obtained by initializing the score-driven recursion at $t-\tau$ with $\dpsi_{1|0}$ and subsequently using $\dvt,\ldots,\dv_{t-\tau}$. Therefore, $\bra{{\dFt^{(\tau)}}\trans,{\dpsi_{t|t-1}^{(\tau)}(\dtheta_0)}\trans
}\trans$ is a measurable function of $\dvt,\ldots,\dv_{t-\tau}$.

By repeated applications of the mean value theorem to the score-driven recursion for $t\leq \tau$, Assumption \ref{ass:contraction_cond} implies that
\begin{align}
\label{eq:ineq_NED2.2}
&\vvbra{
\dbarpsitpt(\dtheta_0)
-
\dpsi^{(\tau)}_{t+1|t}(\dtheta_0)
}_p
\nonumber\\
\leq{}&
\vbra{X_{t,k}(\dtheta_0)}
\vvbra{
\dbarpsi_{t-k+1|t-k}(\dtheta_0)
-
\dpsi^{(\tau)}_{t-k+1|t-k}(\dtheta_0)
}_p
+
\vbra{
\sum_{j=1}^{k}
\sup_{\dpsi\in\boldsymbol{\calP}}
\vvbra{
\boldsymbol{Y}_{t-j+1}(\dtheta_0)
}
\vvbra{
\dbarF_{t-j+1}
-
\dF_{t-j+1}^{(\tau)}
}
}_p
\nonumber\\
\leq{}&
\calX_k
\vvbra{
\dbarpsi_{t-k+1|t-k}(\dtheta_0)
-
\dpsi^{(\tau)}_{t-k+1|t-k}(\dtheta_0)
}_p
+
\vbra{
\sum_{j=1}^{k}
\sup_{\dpsi\in\boldsymbol{\calP}}
\vvbra{
\boldsymbol{Y}_{t-j+1}(\dtheta_0)
}
\vvbra{
\dbarF_{t-j+1}
-
\dF_{t-j+1}^{(\tau)}
}
}_p .
\end{align}
Thus, without loss of generality, assuming that $t\geq\tau+1$ and proceeding recursively, we get
\begin{align}
\label{eq:ineq_NED3}
&\vvbra{
\dbarpsitpt(\dtheta_0) - \dpsi^{(\tau)}_{t+1|t}(\dtheta_0)
}_p
\\
\leq{}&
\calX^{\lfloor t/k \rfloor}_{k}
\vvbra{
\dbarpsi_{1|0}(\dtheta_0) - \dpsi_{1|0}
}_p
+
\calX^{\lfloor \tau/k \rfloor}_{k}
\sum_{j=1}^{t-k}
\calX^{\lfloor (j-1)/k \rfloor}_{k}
\vvbra{
\sup_{\dpsi \in \boldsymbol{\calP}}
\vvbra{
\boldsymbol{Y}_{t-k-j+1}(\dtheta_0)
}_p
\vvbra{
\dbarF_{t-k-j+1} - \dF^{(\tau)}_{t-k-j+1}
}
}_p
\nonumber
,
\end{align}
and note that $\vvbra{\dbarpsi_{1|0}(\dtheta_0) - \dpsi_{1|0}}_p \leq \Delta_1 < \infty$, with $\dbarpsi_{1|0}(\dtheta_0) = \bra{\boldsymbol I_r - \dPhi_0}\inv\ddelta_0 + \sum_{k=0}^\infty \dPhi_0^k  \dK_0 \du_{-k}$ which converges almost surely, by Assumption \ref{ass:noise}, whereas \[
\vvbra{\sup_{\dpsi \in \boldsymbol{\calP}} \vvbra{\boldsymbol{Y}_{t-k-j+1}(\dtheta_0) }\vvbra{\dbarF_{t-k-j+1} - \dF^{(\tau)}_{t-k-j+1} }}_p \leq \Delta_2 < \infty,
\]
for some finite positive real $\Delta_2$ independent of $t$ and $\tau$, and $\calX_{k}< 1$, are ensured by Assumption \ref{ass:contraction_cond}.

Since $t\geq\tau+1$ and $0<\calX_k<1$, it follows that $\calX_k^{\lfloor t/k\rfloor}
\leq\calX_k^{\lfloor\tau/k\rfloor}$. Moreover, Assumption \ref{ass:contraction_cond}$(ii)$, Minkowski's inequality, and the geometric-sum bound imply that
\begin{align*}
&\sup_{t\in\Nn}
\vvbra{
\sum_{j=1}^{t-k}
\calX_k^{\lfloor(j-1)/k\rfloor}
\vbra{
\sup_{\dpsi\in\boldsymbol{\calP}}
\vvbra{
\boldsymbol{Y}_{t-k-j+1}(\dtheta_0)
}
\vvbra{
\dbarF_{t-k-j+1}
-
\dF_{t-k-j+1}^{(\tau)}
}
}
}_p
\\
&\qquad\leq
\Delta_2 \sum_{j=1}^{\infty}
\calX_k^{\lfloor(j-1)/k\rfloor}
=
{k \Delta_2}/\bra{1-\calX_k}
<\infty,
\end{align*}
Consequently, combining the cases $t\leq\tau$ and $t\geq\tau+1$, we obtain
\begin{align}
\label{eq:NED_psi_rate}
\sup_{t\in\Nn}
\vvbra{
\dbarpsitti(\dtheta_0)
-
\dpsi_{t|t-1}^{(\tau)}(\dtheta_0)
}_p
\leq
\calX^{\lfloor t/k \rfloor}_{k}
\cdot \Delta_1
+
\calX_k^{\lfloor\tau/k\rfloor}
\cdot
{k \Delta_2}/\bra{1-\calX_k}
=
O\bra{\calX_k^{\lfloor\tau/k\rfloor}}.
\end{align}
For $p>2$ and using H\"{o}lder's inequality, we get
\begin{align*}
\vvbra{
\dbarFt-\dFt^{(\tau)}
}_2
&=
\vvbra{
\boldsymbol q\bra{\dbarpsitti(\dtheta_0),\dvt}
-
\boldsymbol q\bra{\dpsi_{t|t-1}^{(\tau)}(\dtheta_0),\dvt}
}_2
\\
&\leq
\vvbra{
B_t
\bra{
\dbarpsitti(\dtheta_0),
\dpsi_{t|t-1}^{(\tau)}(\dtheta_0)
}
\vvbra{
\dbarpsitti(\dtheta_0)
-
\dpsi_{t|t-1}^{(\tau)}(\dtheta_0)
}
}_2.
\\
&\leq
\vvbra{
B_t\bra{
\dbarpsitti(\dtheta_0),
\dpsi_{t|t-1}^{(\tau)}(\dtheta_0)
}
}_{\frac{2p}{p-2}}
\vvbra{
\dbarpsitti(\dtheta_0)
-
\dpsi_{t|t-1}^{(\tau)}(\dtheta_0)
}_p
\\
&\leq
B\,
\vvbra{
\dbarpsitti(\dtheta_0)
-
\dpsi_{t|t-1}^{(\tau)}(\dtheta_0)
}_p.
\end{align*}
and hence, Assumptions \ref{ass:noise}--\ref{ass:contraction_cond}, and \eqref{eq:NED_psi_rate},
\begin{align}
\label{eq:NED_F_rate}
\sup_{t\in\Nn}
\vvbra{
\dbarFt-\dFt^{(\tau)}
}_2
\leq
B
\cdot\bra{
\calX^{\lfloor t/k \rfloor}_{k}
\cdot \Delta_1
+
\calX_k^{\lfloor\tau/k\rfloor}
\cdot
{k \Delta_2}/\bra{1-\calX_k}
}
.
\end{align}
Equations \eqref{eq:NED_psi_rate} and \eqref{eq:NED_F_rate} imply that
\begin{align}
\label{eq:NED_joint_rate}
&\sup_{t\in\Nn}
\vvbra{
\bra{
\dbarFt\trans,
\dbarpsitti(\dtheta_0)\trans
}\trans
-
\bra{
{\dFt^{(\tau)}}\trans,
{\dpsi_{t|t-1}^{(\tau)}(\dtheta_0)}\trans
}\trans
}_2
\nonumber
\\
\leq{}&
B^\prime
\cdot\bra{
\calX^{\lfloor t/k \rfloor}_{k}
\cdot \Delta_1
+
\calX_k^{\lfloor\tau/k\rfloor}
\cdot
{k \Delta_2}/\bra{1-\calX_k}
}
=
O\bra{\calX_k^{\lfloor\tau/k\rfloor}},
\end{align}
for some $B^\prime \geq B > 0.$ Since the truncated pair is a measurable function of $\dvt,\ldots,\dv_{t-\tau}$, it follows from \eqref{eq:NED_joint_rate} that
\begin{align*}
0\leq{}&
\sup_{t\in\Nn}
\vvbra{
\bra{\dbarFt\trans,\dbarpsitti(\dtheta_0)\trans}\trans
-
\Ee\sbra{
\left.
\bra{\dbarFt\trans,\dbarpsitti(\dtheta_0)\trans}\trans
\right|
\dvt,\ldots,\dv_{t-\tau}
}
}_2
\\
\leq{}&
\sup_{t\in\Nn}
\vvbra{
\bra{\dbarFt\trans,\dbarpsitti(\dtheta_0)\trans}\trans
-
\bra{
{\dFt^{(\tau)}}\trans,
{\dpsi_{t|t-1}^{(\tau)}(\dtheta_0)}\trans
}\trans
}_2
\\
=&\,
O\bra{\calX_k^{\lfloor\tau/k\rfloor}}.
\end{align*}
Because $0<\calX_k<1$, the approximation coefficients decrease geometrically and are therefore of size $-1$.  Hence, by an application of Theorem 6.12 (or, equivalently, Lemma D5 in Appendix D) of \cite{Potscher1997}, we obtain that the joint process
$
\cbra{
\bra{
\dbarFt\trans,
\dbarpsitti(\dtheta_0)\trans
}\trans
}_{t\in\Nn}
,$
is near-epoch dependent of size $-1$ on $\cbra{\dvt}_{t\in\Zz}$ when generated by the nonlinear system given by \eqref{eq:Model_xt0}--\eqref{eq:SD_Model_psit0}. The two marginal processes inherit the same property as coordinate projections of the joint process.
\end{proof}

\begin{lemma}
\label{lemma:rate_psit_first_derivs}
Consider the first derivative process associated with the score-driven
recursion,
\begin{align}
\label{eq:first_derivative_filter_recursion}
\nabla^{\dtheta}\dpsitpt(\dtheta)
=
\boldsymbol X_t(\dtheta)
\nabla^{\dtheta}\dpsitti(\dtheta)
+
\dDt(\dtheta),
\end{align}
where
\[
\boldsymbol X_t(\dtheta)
=
\dPhi
+
\dK
\frac{\partial\du_t(\dtheta)}
{\partial\dpsitti\trans}
\bigg|_{\dpsitti=\dpsitti(\dtheta)},
\quad\dDt(\dtheta) = \frac{\partial \dpsitpt(\dtheta) }{\partial \dtheta\trans},
\]
and the partial derivative defining $\dDt(\dtheta)$ is taken holding $\dpsitti$ fixed.

By differentiating \eqref{eq:psi_decomposition} with respect to $\dtheta$ we get
\begin{align}
\label{eq:psi_first_derivative_decomposition}
&
\nabla^{\dtheta}\dhathatpsithetatti
-
\nabla^{\dtheta}\dbarpsitti(\dtheta)
\\
={}&
\bra{
\nabla^{\dtheta}\dhathatpsithetatti
-
\nabla^{\dtheta}\dhatpsithetatti
}
+
\bra{
\nabla^{\dtheta}\dhatpsithetatti
-
\nabla^{\dtheta}\dtildepsithetatti
}
+
\bra{
\nabla^{\dtheta}\dtildepsithetatti
-
\nabla^{\dtheta}\dbarpsitti(\dtheta)
}.
\nonumber
\end{align}
Then, under Assumptions \ref{ass:noise}--\ref{ass:contraction_cond} and \ref{ass:smooth_domination_score_Hessian}, with $p>2$, the following results hold:
\begin{enumerate}

\item[(a)] For any $t\geq1$, as $N,T\rightarrow\infty$,
\begin{align*}
\sup_{\dtheta\in\dTheta}
\vvbra{
\nabla^{\dtheta}\dhathatpsithetatpt
-
\nabla^{\dtheta}\dhatpsithetatpt
}
=
O_{\Pp}\bra{
\frac{1}{\min\bra{\sqrt N,T}}
}
+
O_{a.s.}\bra{
\calX_k^{\lfloor t/(2k)\rfloor}
}.
\end{align*}

\item[(b)] For any $t\geq1$, as $N,T\rightarrow\infty$,
\begin{align*}
\sup_{\dtheta\in\dTheta}
\vvbra{
\nabla^{\dtheta}\dhatpsithetatti
-
\nabla^{\dtheta}\dtildepsithetatti
}
=
O_{\Pp}\bra{
\frac{1}{\min\bra{\sqrt N,\sqrt T}}
}
+
O_{a.s.}\bra{
\calX_k^{\lfloor t/(2k)\rfloor}
+
\varrho_k^{\lfloor t/(2k)\rfloor}
}.
\end{align*}

\item[(c)] For any fixed initialization $\dpsi_{1|0}\in\boldsymbol{\mathcal P}$, as $t\rightarrow\infty$,
\begin{align*}
\sup_{\dtheta\in\dTheta}
\vvbra{
\nabla^{\dtheta}\dtildepsithetatti
-
\nabla^{\dtheta}\dbarpsitti(\dtheta)
}
=
O_{a.s.}\bra{
\calX_k^{\lfloor t/(2k)\rfloor}
+
\varrho_k^{\lfloor t/(2k)\rfloor}
}.
\end{align*}

\item[(d)] The process $\cbra{ \nabla^{\dtheta}\dbarpsitti(\dtheta)}_{t\in\Zz}$ is strictly stationary and ergodic and satisfies
\[
\Ee\sbra{
\sup_{\dtheta\in\dTheta}
\vvbra{
\nabla^{\dtheta}\dbarpsitti(\dtheta)
}^{p}
}
<\infty.
\]
Moreover, $\cbra{ \nabla^{\dtheta}\dbarpsitti(\dtheta_0) }_{t\in\Zz}$ is near-epoch dependent of size $-1$ on the geometrically $\beta$-mixing process $\cbra{\dvt}_{t\in\Zz}$.

\item[(e)] For any fixed initialization $\dpsi_{1|0}\in\boldsymbol{\mathcal P}$ and any $t\geq1$, as $N,T\rightarrow\infty$,
\begin{align*}
\sup_{\dtheta\in\dTheta}
\vvbra{
\nabla^{\dtheta}\dhathatpsithetatpt
-
\nabla^{\dtheta}\dbarpsitpt(\dtheta)
}
=&
O_{\Pp}\bra{
\frac{1}{\min\bra{\sqrt N,T}}
}
+
O_{\Pp}\bra{
\frac{1}{\min\bra{\sqrt N,\sqrt T}}
}
\\
&+
O_{a.s.}\bra{
\calX_k^{\lfloor t/(2k)\rfloor}
+
\varrho_k^{\lfloor t/(2k)\rfloor}
}.
\end{align*}

\end{enumerate}
\end{lemma}

\begin{proof}
We first establish two preliminary results that will be used throughout the proof. Iterating \eqref{eq:first_derivative_filter_recursion} in blocks of length $k$ yields
\begin{align*}
\nabla^{\dtheta}\dbarpsitpt(\dtheta)
={}&
\bra{
\prod_{j=1}^{k}
\boldsymbol X_{t-j+1}(\dtheta)
}
\nabla^{\dtheta}
\dbarpsi_{t-k+1|t-k}(\dtheta)
+
\sum_{j=1}^{k}
\bra{
\prod_{h=1}^{j-1}
\boldsymbol X_{t-h+1}(\dtheta)
}
\dD_{t-j+1}(\dtheta),
\end{align*}
where the product corresponding to $j=1$ is understood to be $\boldsymbol I_r$. Therefore,
\begin{align}
\label{eq:bound_p_derivs_process}
&
\vvbra{
\sup_{\dtheta\in\dTheta}
\vvbra{
\nabla^{\dtheta}\dbarpsitpt(\dtheta)
}
}_p
\nonumber
\\
\leq{}&
\calX_k
\vvbra{
\sup_{\dtheta\in\dTheta}
\vvbra{
\nabla^{\dtheta}
\dbarpsi_{t-k+1|t-k}(\dtheta)
}
}_p
+
\vvbra{
\sup_{\dtheta\in\dTheta}
\vvbra{
\sum_{j=1}^{k}
\bra{
\prod_{h=1}^{j-1}
\boldsymbol X_{t-h+1}(\dtheta)
}
\dD_{t-j+1}(\dtheta)
}
}_p.
\end{align}
By the definition of $\dD_t(\dtheta)$, the matrix $\frac{\partial\vect{\dD_t(\dtheta)}}{\partial\dpsitti\trans}$ contains the derivative of the score entering
$\dK\,\frac{\partial\du_t(\dtheta)}{\partial\dpsitti\trans}$. Hence, the compactness of $\dTheta$ and Assumption \ref{ass:smooth_domination_score_Hessian}$(iv)$ imply that, for every $q>2$,
\begin{align}
\label{eq:first_derivative_X_moment}
\sup_{t\in\Zz}
\vvbra{
\sup_{\dtheta\in\dTheta}
\sup_{\dpsi\in\calP}
\vvbra{
\boldsymbol X_t(\dtheta)
}
}_q
<\infty.
\end{align}
Consequently, since $k$ is fixed, repeated applications of H\"older's inequality, together with \eqref{eq:first_derivative_X_moment} and Assumption \ref{ass:smooth_domination_score_Hessian}$(iv)$, give, for every $j=1,\ldots,k$,
\begin{align*}
&
\sup_{t\in\Zz}
\vvbra{
\sup_{\dtheta\in\dTheta}
\vvbra{
\bra{
\prod_{h=1}^{j-1}
\boldsymbol X_{t-h+1}(\dtheta)
}
\dD_{t-j+1}(\dtheta)
}
}_p
\\
\leq{}&
\bra{
\sup_{s\in\Zz}
\vvbra{
\sup_{\dtheta\in\dTheta}
\sup_{\dpsi\in\calP}
\vvbra{
\boldsymbol X_s(\dtheta)
}
}_{jp}
}^{j-1}
\sup_{s\in\Zz}
\vvbra{
\sup_{\dtheta\in\dTheta}
\vvbra{
\dD_s(\dtheta)
}
}_{jp}
<\infty.
\end{align*}
Since the sum in the second term on the right-hand side of \eqref{eq:bound_p_derivs_process} contains only $k$ terms, it follows that, for some finite positive real $K$ independent of $t$,
\begin{align}
\label{eq:K_bound}
\sup_{t\in\Zz}
\vvbra{
\sup_{\dtheta\in\dTheta}
\vvbra{
\sum_{j=1}^{k}
\bra{
\prod_{h=1}^{j-1}
\boldsymbol X_{t-h+1}(\dtheta)
}
\dD_{t-j+1}(\dtheta)
}
}_p
\leq K.
\end{align}
Now, for every $n\geq1$, let $\cbra{\nabla^{\dtheta}\dbarpsi^{[n]}_{s|s-1}(\dtheta)}_{s\in\Zz}$ denote the derivative recursion driven by the stationary processes $\cbra{\dH_0\dbarF_s}_{s\in\Zz}$ and $\cbra{\dbarpsi_{s|s-1}(\dtheta)}_{s\in\Zz}$ and initialized at time $t-nk$  according to $\nabla^{\dtheta} \dbarpsi^{[n]}_{t-nk+1|t-nk}(\dtheta)=\boldsymbol 0_{r\times d}$.

By an application of the $k$-step inequality in \eqref{eq:bound_p_derivs_process} recursively over the $n$ blocks between $t-nk$ and $t$, and using \eqref{eq:K_bound}, we get
\begin{align}
\label{eq:first_derivative_finite_past_moment}
\vvbra{
\sup_{\dtheta\in\dTheta}
\vvbra{
\nabla^{\dtheta}
\dbarpsi^{[n]}_{t+1|t}(\dtheta)
}
}_p
\leq
K
\sum_{j=0}^{n-1}
\calX_k^j
\leq
\frac{K}{1-\calX_k}.
\end{align}
The first inequality follows because the contribution generated within each block has $L_p$-norm bounded by $K$, while each preceding complete block contributes an additional factor $\calX_k$.

By the uniform $k$-step contraction,
\[
\sup_{\dtheta\in\dTheta}
\vvbra{
\nabla^{\dtheta}
\dbarpsi^{[n]}_{t+1|t}(\dtheta)
-
\nabla^{\dtheta}
\dbarpsitpt(\dtheta)
}
\easconv
0
,
\qquad n\rightarrow\infty.
\]
 Therefore, Fatou's lemma and
\eqref{eq:first_derivative_finite_past_moment} imply that
\begin{align*}
&
\Ee\sbra{
\sup_{\dtheta\in\dTheta}
\vvbra{
\nabla^{\dtheta}
\dbarpsitpt(\dtheta)
}^{p}
}
\\
\leq{}&
\liminf_{n\rightarrow\infty}
\Ee\sbra{
\sup_{\dtheta\in\dTheta}
\vvbra{
\nabla^{\dtheta}
\dbarpsi^{[n]}_{t+1|t}(\dtheta)
}^{p}
}
\leq
\bra{
\frac{K}{1-\calX_k}
}^{p}.
\end{align*}
Since the bound is independent of $t$, we take the $p$-th root in order to get
\begin{align}
\label{eq:first_derivative_uniform_moment}
\sup_{t\in\Zz}
\vvbra{
\sup_{\dtheta\in\dTheta}
\vvbra{
\nabla^{\dtheta}
\dbarpsitpt(\dtheta)
}
}_p
\leq
\frac{K}{1-\calX_k}
<\infty.
\end{align}
Moreover, the mean value theorem and Assumption \ref{ass:smooth_domination_score_Hessian}$(iii)$--$(iv)$ imply that, for any two filters and their respective factor inputs, the differences between the corresponding matrices $\boldsymbol X_t(\dtheta)$ and $\dDt(\dtheta)$ are bounded by the differences between the underlying filters and factor inputs. More precisely, let us denote with $\dpsi_{t|t-1}^{(1)}$ and $\dpsi_{t|t-1}^{(2)}$ two recursions being compared. Then, repeated applications of the mean value theorem, together with Assumption \ref{ass:smooth_domination_score_Hessian}$(iii)$--$(iv)$, 
\begin{align}
\label{eq:mvt_X_matrix}
&
\sup_{\dtheta\in\dTheta}
\vvbra{
\boldsymbol X_t^{(1)}(\dtheta)
-
\boldsymbol X_t^{(2)}(\dtheta)
}
\nonumber
\\
\leq{}&
\sup_{\dtheta\in\dTheta}
\sup_{\dpsi\in\calP}
\vvbra{
\frac{\partial\vect{\boldsymbol X_t(\dtheta)}}
{\partial\dpsi\trans}
}
\sup_{\dtheta\in\dTheta}
\vvbra{
\dpsi_{t|t-1}^{(1)}(\dtheta)
-
\dpsi_{t|t-1}^{(2)}(\dtheta)
}
+
\sup_{\dtheta\in\dTheta}
\vvbra{
\frac{\partial\vect{\boldsymbol X_t(\dtheta)}}
{\partial\dF_t\trans}
}
\vvbra{
\dF_t^{(1)}-\dF_t^{(2)}
}
,
\end{align}
almost surely, and
\begin{align}
\label{eq:mvt_D_matrix}
&
\sup_{\dtheta\in\dTheta}
\vvbra{
\dD_t^{(1)}(\dtheta)
-
\dD_t^{(2)}(\dtheta)
}
\nonumber
\\
\leq{}&
\sup_{\dtheta\in\dTheta}
\sup_{\dpsi\in\calP}
\vvbra{
\frac{\partial\vect{\dDt(\dtheta)}}
{\partial\dpsi\trans}
}
\sup_{\dtheta\in\dTheta}
\vvbra{
\dpsi_{t|t-1}^{(1)}(\dtheta)
-
\dpsi_{t|t-1}^{(2)}(\dtheta)
}
+
\sup_{\dtheta\in\dTheta}
\vvbra{
\frac{\partial\vect{\dDt(\dtheta)}}
{\partial\dF_t\trans}
}
\vvbra{
\dF_t^{(1)}-\dF_t^{(2)}
},
\end{align}
almost surely. Taking the $L_{p/2}$-norm on both sides and applying Minkowski's and H\"older's inequalities, Assumption \ref{ass:smooth_domination_score_Hessian}$(iii)$--$(iv)$ yields
\begin{align}
\label{eq:first_derivative_X_D_perturbation}
&
\vvbra{
\sup_{\dtheta\in\dTheta}
\vvbra{
\boldsymbol X_t^{(1)}(\dtheta)
-
\boldsymbol X_t^{(2)}(\dtheta)
}
+
\sup_{\dtheta\in\dTheta}
\vvbra{
\dD_t^{(1)}(\dtheta)
-
\dD_t^{(2)}(\dtheta)
}
}_{p/2}
\nonumber
\\
\leq{}&
\bra{K_X^\prime + K_D^{\prime\prime}}
\vvbra{
\sup_{\dtheta\in\dTheta}
\vvbra{
\dpsi_{t|t-1}^{(1)}(\dtheta)
-
\dpsi_{t|t-1}^{(2)}(\dtheta)
}
}_p
+
\bra{ K_X^{\prime\prime} + K_D^{\prime\prime\prime}}
\vvbra{
\dF_t^{(1)}-\dF_t^{(2)}
}_p.
\end{align}
We now apply this bound separately to the three terms on the right-hand side of \eqref{eq:psi_first_derivative_decomposition}.

\medskip

\noindent
$(a)$ By subtracting the derivative recursions associated with $\dhathatpsithetatpt$ and $\dhatpsithetatpt$ we have
\begin{align}
\label{eq:first_derivative_difference_a}
&
\nabla^{\dtheta}\dhathatpsithetatpt
-
\nabla^{\dtheta}\dhatpsithetatpt
\nonumber
\\
={}&
\doublehat{\boldsymbol X}_t(\dtheta)
\bra{
\nabla^{\dtheta}\dhathatpsithetatti
-
\nabla^{\dtheta}\dhatpsithetatti
}
+
\bra{
\doublehat{\boldsymbol X}_t(\dtheta)
-
\widehat{\boldsymbol X}_t(\dtheta)
}
\nabla^{\dtheta}\dhatpsithetatti
+
\doublehat{\dD}_t(\dtheta)
-
\widehat{\dD}_t(\dtheta).
\end{align}
Iterating \eqref{eq:first_derivative_difference_a} in blocks of length $k$, and using the expansions in \eqref{eq:mvt_X_matrix} and \eqref{eq:mvt_D_matrix}, together with Assumption \ref{ass:contraction_cond}, we obtain
\begin{align}
\label{eq:first_derivative_recursive_bound_a}
&
\sup_{\dtheta\in\dTheta}
\vvbra{
\nabla^{\dtheta}\dhathatpsithetatpt
-
\nabla^{\dtheta}\dhatpsithetatpt
}
\nonumber
\\
\leq{}&
\calX_k^{\lfloor t/k\rfloor}
\sup_{\dtheta\in\dTheta}
\vvbra{
\nabla^{\dtheta}\dpsi_{1|0}
-
\nabla^{\dtheta}\dpsi^\star_{1|0}
}
+
\sum_{s=1}^{t}
\calX_k^{\lfloor(t-s)/k\rfloor}
\calC^{\dpsi}_{s,NT}
\sup_{\dtheta\in\dTheta}
\vvbra{
\dhathatpsi_{s|s-1}(\dtheta)
-
\dhatpsi_{s|s-1}(\dtheta)
}
\nonumber
\\&+
\sum_{s=1}^{t}
\calX_k^{\lfloor(t-s)/k\rfloor}
\calC^{\dF}_{s,NT}
\vvbra{
\dhatF_s-\widehat\dH\dF_s
},
\end{align}
where
\begin{align*}
\calC^{\dpsi}_{s,NT}
=
\sup_{\dtheta\in\dTheta}
\sup_{\dpsi\in\calP}
\vvbra{
\frac{\partial\vect{\boldsymbol X_s(\dtheta)}}
{\partial\dpsi\trans}
}
\sup_{\dtheta\in\dTheta}
\vvbra{
\nabla^{\dtheta}\dhatpsi_{s|s-1}(\dtheta)
}
+
\sup_{\dtheta\in\dTheta}
\sup_{\dpsi\in\calP}
\vvbra{
\frac{\partial\vect{\dD_s(\dtheta)}}
{\partial\dpsi\trans}
},
\end{align*}
and
\begin{align*}
\calC^{\dF}_{s,NT}
=
\sup_{\dtheta\in\dTheta}
\vvbra{
\frac{\partial\vect{\boldsymbol X_s(\dtheta)}}
{\partial\dF_s\trans}
}
\sup_{\dtheta\in\dTheta}
\vvbra{
\nabla^{\dtheta}\dhatpsi_{s|s-1}(\dtheta)
}
+
\sup_{\dtheta\in\dTheta}
\vvbra{
\frac{\partial\vect{\dD_s(\dtheta)}}
{\partial\dF_s\trans}
}.
\end{align*}
Since both $\dpsi_{1|0}$ and $\dpsi^\star_{1|0}$ are fixed and do not depend on $\dtheta$, $\nabla^{\dtheta}\dpsi_{1|0}=\nabla^{\dtheta}\dpsi^\star_{1|0}=\boldsymbol 0_{r\times d}$. Hence, the first term on the right-hand side of \eqref{eq:first_derivative_recursive_bound_a} is identically zero. The initialization effect remaining in the derivative process is instead inherited from the difference between the initializations of the underlying filters through $\dhathatpsi_{s|s-1}(\dtheta)-\dhatpsi_{s|s-1}(\dtheta)$. 

By the same $k$-step moment argument used to establish \eqref{eq:first_derivative_uniform_moment}, applied to $\nabla^{\dtheta}\dhatpsi_{s|s-1}(\dtheta)$, together with Assumption \ref{ass:smooth_domination_score_Hessian}$(iii)$--$(iv)$, it holds that, for every fixed $t$ and every $M>0$,
\begin{align}
\label{eq:first_derivative_psi_coefficient_tightness}
\lim_{M\rightarrow\infty}
\limsup_{N,T\rightarrow\infty}
\Pp\bra{
\max_{1\leq s\leq t}
\calC^{\dpsi}_{s,NT}
>
M
}
&=
0,
\\
\label{eq:first_derivative_F_coefficient_tightness}
\lim_{M\rightarrow\infty}
\limsup_{N,T\rightarrow\infty}
\Pp\bra{
\max_{1\leq s\leq t}
\calC^{\dF}_{s,NT}
>
M
}
&=
0
.
\end{align}
To control the contribution of $\dhathatpsi_{s|s-1}(\dtheta)-\dhatpsi_{s|s-1}(\dtheta)$, recall from \eqref{eq:ineq_prop_inv} that, almost surely,
\begin{align}
\label{eq:first_derivative_underlying_filter_bound}
&
\sup_{\dtheta\in\dTheta}
\vvbra{
\dhathatpsi_{s|s-1}(\dtheta)
-
\dhatpsi_{s|s-1}(\dtheta)
}
\nonumber\\
\leq{}&
\calX_k^{\lfloor(s-1)/k\rfloor}\Delta_1
+
\sum_{j=1}^{s-1}
\calX_k^{\lfloor(j-1)/k\rfloor}
\vbra{
\sup_{\dtheta\in\dTheta}
\vbra{
Y_{s-j}(\dtheta)
}
\vvbra{
\dhatF_{s-j}
-
\widehat\dH\dF_{s-j}
}
},
\end{align}
where
$
\Delta_1
=
\vvbra{
\dpsi_{1|0}
-
\dpsi^\star_{1|0}
}
<\infty.
$ Substituting \eqref{eq:first_derivative_underlying_filter_bound} into the corresponding term of \eqref{eq:first_derivative_recursive_bound_a}
\begin{align}
&
\sum_{s=1}^{t}
\calX_k^{\lfloor(t-s)/k\rfloor}
\calC^{\dpsi}_{s,NT}
\sup_{\dtheta\in\dTheta}
\vvbra{
\dhathatpsi_{s|s-1}(\dtheta)
-
\dhatpsi_{s|s-1}(\dtheta)
}
\nonumber
\\
\leq{}&
\Delta_1
\sum_{s=1}^{t}
\calX_k^{
\lfloor(t-s)/k\rfloor
+
\lfloor(s-1)/k\rfloor
}
\calC^{\dpsi}_{s,NT}
\nonumber
\\
&+
\sum_{s=1}^{t}
\sum_{j=1}^{s-1}
\calX_k^{
\lfloor(t-s)/k\rfloor
+
\lfloor(j-1)/k\rfloor
}
\calC^{\dpsi}_{s,NT}
\vbra{
\sup_{\dtheta\in\dTheta}
\vbra{
Y_{s-j}(\dtheta)
}
\vvbra{
\dhatF_{s-j}
-
\widehat\dH\dF_{s-j}
}}.
\label{eq:first_derivative_filter_contribution}
\end{align}
For the first term on the right-hand side of \eqref{eq:first_derivative_filter_contribution}, the same $k$-step argument used to obtain \eqref{eq:first_derivative_uniform_moment}, together with Assumption \ref{ass:smooth_domination_score_Hessian}$(iii)$--$(iv)$, implies
\[
\sup_{N,T}
\sup_{1\leq s\leq T}
\vvbra{
\sup_{\dtheta\in\dTheta}
\vvbra{
\nabla^{\dtheta}\dhatpsi_{s|s-1}(\dtheta)
}
}_p
\leq
\frac{K}{1-\calX_k}
<\infty.
\]
Consequently, H\"older's inequality and Assumption \ref{ass:smooth_domination_score_Hessian}$(iii)$--$(iv)$ imply that,
with $q=p/2>1$,
\begin{align}
\label{eq:first_derivative_coefficient_moment}
&
\max\Bigg\{
\sup_{N,T}\sup_{1\leq s\leq T}
\vvbra{
\calC^{\dpsi}_{s,NT}
}_q,
\sup_{N,T}\sup_{1\leq s\leq T}
\vvbra{
\calC^{\dF}_{s,NT}
}_q
\Bigg\}
\nonumber
\\
\leq{}&
\frac{K}{1-\calX_k}
\max\cbra{
K_X^\prime,
K_X^{\prime\prime}
}
+
\max\cbra{
K_D^{\prime\prime},
K_D^{\prime\prime\prime}
}
<\infty.
\end{align}
Thus, the required uniform moment bound holds with $q=p/2>1$. Therefore, for every $\varepsilon>0$, the Markov's inequality implies
\begin{align*}
&\sum_{s=1}^{\infty}
\Pp\bra{
\calC^{\dpsi}_{s,NT}
>
s^{1/q+\varepsilon}
}
\leq
\sup_{N,T}\sup_{1\leq s\leq T}
\Ee\sbra{
\bra{\calC^{\dpsi}_{s,NT}}^{q}
}
\sum_{s=1}^{\infty}
s^{-1-q\varepsilon}
<\infty,
\end{align*}
and thus, the Borel–Cantelli lemma gives $\calC^{\dpsi}_{s,NT} = O_{a.s.}\bra{s^{1/q+\varepsilon}}$. Moreover, by using similar arguments as in \eqref{eq:O_sum_calXk} in the proof of Proposition \ref{prop:rate_psit_decomp}$(a)$, 
\begin{align}
\label{eq:first_derivative_filter_initialization_component}
&
\Delta_1
\sum_{s=1}^{t}
\calX_k^{
\lfloor(t-s)/k\rfloor
+
\lfloor(s-1)/k\rfloor
}
\calC^{\dpsi}_{s,NT}
\nonumber
\\
={}&
O_{a.s.}\bra{
t^{1+1/q+\varepsilon}
\calX_k^{\lfloor(t-1)/k\rfloor-1}
}
=
O_{a.s.}\bra{
\calX_k^{\lfloor t/(2k)\rfloor}
}.
\end{align}
Now, for the second term on the right-hand side of \eqref{eq:first_derivative_filter_contribution}, we have
\begin{align*}
&
\sum_{s=1}^{t}
\sum_{j=1}^{s-1}
\calX_k^{
\lfloor(t-s)/k\rfloor
+
\lfloor(j-1)/k\rfloor
}
\calC^{\dpsi}_{s,NT}
\vbra{
\sup_{\dtheta\in\dTheta}
\vbra{
Y_{s-j}(\dtheta)
}
\vvbra{
\dhatF_{s-j}
-
\widehat\dH\dF_{s-j}
}
}
\\
\leq{}&
\bra{
\frac{k}{1-\calX_k}
}^{2}
\max_{1\leq s\leq t}
\calC^{\dpsi}_{s,NT}
\max_{1\leq u\leq t}
\vbra{
\sup_{\dtheta\in\dTheta}
\vbra{
Y_u(\dtheta)
}
\max_{1\leq u\leq t}
\vvbra{
\dhatF_u-\widehat\dH\dF_u
}}.
\end{align*}
The first two maxima are tight by Assumptions \ref{ass:contraction_cond} and \ref{ass:smooth_domination_score_Hessian}, while Proposition \ref{prop:rate_yt_decomp}(a) and the finite-maximum argument give
$
\max_{1\leq u\leq t}
\vvbra{
\dhatF_u-\widehat\dH\dF_u
}
=
O_{\Pp}\bra{
\frac{1}{\min\bra{\sqrt N,T}}
}$. Consequently,
\begin{align}
\label{eq:first_derivative_filter_PC_component}
&
\sum_{s=1}^{t}
\sum_{j=1}^{s-1}
\calX_k^{
\lfloor(t-s)/k\rfloor
+
\lfloor(j-1)/k\rfloor
}
\calC^{\dpsi}_{s,NT}
\vbra{
\sup_{\dtheta\in\dTheta}
\vbra{
Y_{s-j}(\dtheta)
}
\vvbra{
\dhatF_{s-j}
-
\widehat\dH\dF_{s-j}
}}
\nonumber
\\
={}&
O_{\Pp}\bra{
\frac{1}{\min\bra{\sqrt N,T}}
}.
\end{align}
It remains to control the third term in the right-hand side of \eqref{eq:first_derivative_recursive_bound_a}. The finite-maximum argument used in the proof of Proposition \ref{prop:rate_psit_decomp}$(a)$ gives, for every $L>0$,
\begin{align}
\label{eq:fixed_t_factor_error_tightness}
\lim_{L\rightarrow\infty}
\limsup_{N,T\rightarrow\infty}
\Pp\bra{
\min\bra{\sqrt N,T}
\max_{1\leq s\leq t}
\vvbra{
\dhatF_s-\widehat\dH\dF_s
}
>
L
}
=
0.
\end{align}
Moreover, almost surely, 
\begin{align*}
\sum_{s=1}^{t}
\calX_k^{\lfloor(t-s)/k\rfloor}
\calC^{\dF}_{s,NT}
\vvbra{
\dhatF_s-\widehat\dH\dF_s
}
\leq
\frac{k}{1-\calX_k}
\max_{1\leq s\leq t}
\calC^{\dF}_{s,NT}
\max_{1\leq s\leq t}
\vvbra{
\dhatF_s-\widehat\dH\dF_s
}.
\end{align*}
Therefore, for every $M,L>0$,
\begin{align*}
&
\Pp\bra{
\sum_{s=1}^{t}
\calX_k^{\lfloor(t-s)/k\rfloor}
\calC^{\dF}_{s,NT}
\vvbra{
\dhatF_s-\widehat\dH\dF_s
}
>
\frac{kML}
{\bra{1-\calX_k}\min\bra{\sqrt N,T}}
}
\\
\leq{}&
\Pp\bra{
\max_{1\leq s\leq t}
\calC^{\dF}_{s,NT}>M
}
+
\Pp\bra{
\min\bra{\sqrt N,T}
\max_{1\leq s\leq t}
\vvbra{
\dhatF_s-\widehat\dH\dF_s
}
>L
}.
\end{align*}
Letting $N,T\rightarrow\infty$, followed by $M,L\rightarrow\infty$, and using \eqref{eq:first_derivative_F_coefficient_tightness}–\eqref{eq:fixed_t_factor_error_tightness}, 
\begin{align}
\label{eq:first_derivative_filter_factor_error_component}
&
\sum_{s=1}^{t}
\calX_k^{\lfloor(t-s)/k\rfloor}
\calC^{\dF}_{s,NT}
\vvbra{
\dhatF_s-\widehat\dH\dF_s
}
=
O_{\Pp}\bra{
\frac{1}{\min\bra{\sqrt N,T}}
}.
\end{align}
Combining
\eqref{eq:first_derivative_filter_initialization_component}, \eqref{eq:first_derivative_filter_PC_component}, and \eqref{eq:first_derivative_filter_factor_error_component}, and substituting these bounds into \eqref{eq:first_derivative_recursive_bound_a}, 
\[
\sup_{\dtheta\in\dTheta}
\vvbra{
\nabla^{\dtheta}\dhathatpsithetatpt
-
\nabla^{\dtheta}\dhatpsithetatpt
}
=
O_{\Pp}\bra{
\frac{1}{\min\bra{\sqrt N,T}}
}
+
O_{a.s.}\bra{
\calX_k^{\lfloor t/(2k)\rfloor}
},
\]
which proves part $(a)$.

\medskip

\noindent
$(b)$ The filters $\dhatpsithetatti$ and $\dtildepsithetatti$ have the same fixed initialization $\dpsi^\star_{1|0}$ and are driven, respectively, by $\widehat\dH\dF_t$ and $\dH_0\dbarF_t$. Subtracting their derivative recursions 
\begin{align}
\label{eq:first_derivative_difference_b}
&
\nabla^{\dtheta}\dhatpsithetatpt
-
\nabla^{\dtheta}\dtildepsithetatpt
\nonumber
\\
={}&
\widehat{\boldsymbol X}_t(\dtheta)
\bra{
\nabla^{\dtheta}\dhatpsithetatti
-
\nabla^{\dtheta}\dtildepsithetatti
}
+
\bra{
\widehat{\boldsymbol X}_t(\dtheta)
-
\widetilde{\boldsymbol X}_t(\dtheta)
}
\nabla^{\dtheta}\dtildepsithetatti
+
\widehat{\dD}_t(\dtheta)
-
\widetilde{\dD}_t(\dtheta).
\end{align}
Therefore, as in point $(a)$, iterating \eqref{eq:first_derivative_difference_b} in blocks of length $k$, and using the expansions in \eqref{eq:mvt_X_matrix} and \eqref{eq:mvt_D_matrix}, together with Assumption \ref{ass:contraction_cond}, 
\begin{align}
\label{eq:first_derivative_recursive_bound_b}
&
\sup_{\dtheta\in\dTheta}
\vvbra{
\nabla^{\dtheta}\dhatpsithetatpt
-
\nabla^{\dtheta}\dtildepsithetatpt
}
\nonumber
\\
\leq{}&
\sum_{s=1}^{t}
\calX_k^{\lfloor(t-s)/k\rfloor}
\calC^{\dpsi}_{s,NT}
\sup_{\dtheta\in\dTheta}
\vvbra{
\dhatpsi_{s|s-1}(\dtheta)
-
\dtildepsi_{s|s-1}(\dtheta)
}
+
\sum_{s=1}^{t}
\calX_k^{\lfloor(t-s)/k\rfloor}
\calC^{\dF}_{s,NT}
\vvbra{
\widehat\dH\dF_s-\dH_0\dbarF_s
},
\end{align}
where $\calC^{\dpsi}_{s,NT}$ and $\calC^{\dF}_{s,NT}$ are defined as in part $(a)$, with $\nabla^{\dtheta}\dhatpsi_{s|s-1}(\dtheta)$ replaced by $\nabla^{\dtheta}\dtildepsi_{s|s-1}(\dtheta)$. By the same argument used in \eqref{eq:first_derivative_coefficient_moment}, with $q=p/2>1$,
\begin{align}
\label{eq:first_derivative_coefficients_b}
\max\cbra{
\sup_{N,T}
\sup_{1\leq s\leq T}
\vvbra{\calC^{\dpsi}_{s,NT}}_q,
\sup_{N,T}
\sup_{1\leq s\leq T}
\vvbra{\calC^{\dF}_{s,NT}}_q
}
<\infty.
\end{align}
Consequently, for every $\varepsilon>0$, Markov's inequality and the Borel--Cantelli lemma imply that
\begin{align}
\label{eq:first_derivative_coefficients_as_b}
\max\cbra{
\calC^{\dpsi}_{s,NT},
\calC^{\dF}_{s,NT}
}
=
O_{a.s.}\bra{
s^{1/q+\varepsilon}
}.
\end{align}
We first consider the first term on the right-hand side of \eqref{eq:first_derivative_recursive_bound_b}. From the recursive bound used in the proof of Proposition \ref{prop:rate_psit_decomp}$(b)$ we get, almost surely,
\begin{align}
\label{eq:first_derivative_underlying_filter_bound_b}
&
\sup_{\dtheta\in\dTheta}
\vvbra{
\dhatpsi_{s|s-1}(\dtheta)
-
\dtildepsi_{s|s-1}(\dtheta)
}
\nonumber
\\
\leq{}&
\sum_{j=1}^{s-1}
\calX_k^{\lfloor(j-1)/k\rfloor}
\vbra{
\sup_{\dtheta\in\dTheta}
\vbra{
Y_{s-j}(\dtheta)
}
\vvbra{
\bra{
\widehat\dH-\dH_0
}
\dF_{s-j}
}
}
\nonumber
\\
&+
\sum_{j=1}^{s-1}
\calX_k^{\lfloor(j-1)/k\rfloor}
\vbra{
\sup_{\dtheta\in\dTheta}
\vbra{
Y_{s-j}(\dtheta)
}
\vvbra{
\dH_0
\bra{
\dF_{s-j}-\dbarF_{s-j}
}
}
}.
\end{align}
Substituting \eqref{eq:first_derivative_underlying_filter_bound_b} into the first term on the right-hand side of \eqref{eq:first_derivative_recursive_bound_b} gives the sum of the following two terms:
\begin{align}
\label{eq:first_derivative_filter_rotation_b}
I_t
&=
\sum_{s=1}^{t}
\sum_{j=1}^{s-1}
\calX_k^{
\lfloor(t-s)/k\rfloor+\lfloor(j-1)/k\rfloor
}
\calC^{\dpsi}_{s,NT}
\vbra{
\sup_{\dtheta\in\dTheta}
\vbra{
Y_{s-j}(\dtheta)
}
\vvbra{
\bra{
\widehat\dH-\dH_0
}
\dF_{s-j}
}},
\\
\label{eq:first_derivative_filter_stationary_b}
II_t
&=
\sum_{s=1}^{t}
\calX_k^{\lfloor(t-s)/k\rfloor}
\calC^{\dpsi}_{s,NT}
\sum_{j=1}^{s-1}
\calX_k^{\lfloor(j-1)/k\rfloor}
\vbra{
\sup_{\dtheta\in\dTheta}
\vbra{
Y_{s-j}(\dtheta)
}
\vvbra{
\dH_0
\bra{
\dF_{s-j}-\dbarF_{s-j}
}
}
}.
\end{align}
For \eqref{eq:first_derivative_filter_rotation_b}, geometric summability implies, almost surely,
\begin{align*}
I_t
\leq
\bra{
\frac{k}{1-\calX_k}
}^{2}
\max_{1\leq s\leq t}
\calC^{\dpsi}_{s,NT}
\vbra{
\max_{1\leq u\leq t}
\sup_{\dtheta\in\dTheta}
\vbra{
Y_u(\dtheta)
}
\max_{1\leq u\leq t}
\vvbra{
\bra{
\widehat\dH-\dH_0
}
\dF_u
}
}.
\end{align*}
Proposition \ref{prop:rate_yt_decomp} and the finite-maximum argument imply
\begin{align}
\label{eq:first_derivative_rotation_factor_rate_b}
\max_{1\leq u\leq t}
\vvbra{
\bra{
\widehat\dH-\dH_0
}
\dF_u
}
=
O_{\Pp}\bra{
\frac{1}{\min\bra{\sqrt N,\sqrt T}}
}.
\end{align}
The other two maxima are tight by \eqref{eq:first_derivative_coefficients_b} and Assumption \ref{ass:contraction_cond}. Hence, by applying the probability argument
used in \eqref{eq:first_derivative_F_coefficient_tightness}--\eqref{eq:first_derivative_filter_factor_error_component} 
\begin{align}
\label{eq:first_derivative_filter_rotation_rate_b}
I_t
=
O_{\Pp}\bra{
\frac{1}{\min\bra{\sqrt N,\sqrt T}}
}.
\end{align}
For \eqref{eq:first_derivative_filter_stationary_b}, the bound established in \eqref{eq:O_sum_calXk2}, applied with $t=s-1$, implies that
\begin{align}
\label{eq:first_derivative_underlying_filter_stationary_b}
&
\sum_{j=1}^{s-1}
\calX_k^{\lfloor(j-1)/k\rfloor}
\vbra{
\sup_{\dtheta\in\dTheta}
\vbra{
Y_{s-j}(\dtheta)
}
\vvbra{
\dH_0
\bra{
\dF_{s-j}-\dbarF_{s-j}
}
}}
=
O_{a.s.}\bra{
\varrho_k^{\lfloor(s-1)/k\rfloor}
}.
\end{align}
Combining \eqref{eq:first_derivative_coefficients_as_b} and \eqref{eq:first_derivative_underlying_filter_stationary_b} in \eqref{eq:first_derivative_filter_stationary_b} 
\begin{align}
\label{eq:first_derivative_filter_stationary_rate_b}
II_t
=
O_{a.s.}\bra{
\sum_{s=1}^{t}
s^{1/q+\varepsilon}
\calX_k^{\lfloor(t-s)/k\rfloor}
\varrho_k^{\lfloor(s-1)/k\rfloor}
}
=
O_{a.s.}\bra{
\calX_k^{\lfloor t/(2k)\rfloor}
+
\varrho_k^{\lfloor t/(2k)\rfloor}
}.
\end{align}
It follows from \eqref{eq:first_derivative_filter_rotation_rate_b} and \eqref{eq:first_derivative_filter_stationary_rate_b} that the first term on the right-hand side of \eqref{eq:first_derivative_recursive_bound_b} is
\begin{align}
\label{eq:first_derivative_filter_component_b}
&
\sum_{s=1}^{t}
\calX_k^{\lfloor(t-s)/k\rfloor}
\calC^{\dpsi}_{s,NT}
\sup_{\dtheta\in\dTheta}
\vvbra{
\dhatpsi_{s|s-1}(\dtheta)
-
\dtildepsi_{s|s-1}(\dtheta)
}
\nonumber
\\
={}&
O_{\Pp}\bra{
\frac{1}{\min\bra{\sqrt N,\sqrt T}}
}
+
O_{a.s.}\bra{
\calX_k^{\lfloor t/(2k)\rfloor}
+
\varrho_k^{\lfloor t/(2k)\rfloor}
}.
\end{align}
We now consider the second term on the right-hand side of \eqref{eq:first_derivative_recursive_bound_b}. Since
\[
\widehat\dH\dF_s-\dH_0\dbarF_s
=
\bra{
\widehat\dH-\dH_0
}
\dF_s
+
\dH_0
\bra{
\dF_s-\dbarF_s
},
\]
we have, almost surely,
\begin{align}
\label{eq:first_derivative_factor_contribution_b}
&
\sum_{s=1}^{t}
\calX_k^{\lfloor(t-s)/k\rfloor}
\calC^{\dF}_{s,NT}
\vvbra{
\widehat\dH\dF_s-\dH_0\dbarF_s
}
\nonumber
\\
\leq{}&
\sum_{s=1}^{t}
\calX_k^{\lfloor(t-s)/k\rfloor}
\calC^{\dF}_{s,NT}
\vvbra{
\bra{
\widehat\dH-\dH_0
}
\dF_s
}
+
\sum_{s=1}^{t}
\calX_k^{\lfloor(t-s)/k\rfloor}
\calC^{\dF}_{s,NT}
\vvbra{
\dH_0
\bra{
\dF_s-\dbarF_s
}
}.
\end{align}
For the first term on the right-hand side of \eqref{eq:first_derivative_factor_contribution_b}, by the geometric summability, we obtain
\begin{align*}
&
\sum_{s=1}^{t}
\calX_k^{\lfloor(t-s)/k\rfloor}
\calC^{\dF}_{s,NT}
\vvbra{
\bra{
\widehat\dH-\dH_0
}
\dF_s
}
\leq
\frac{k}{1-\calX_k}
\max_{1\leq s\leq t}
\calC^{\dF}_{s,NT}
\max_{1\leq s\leq t}
\vvbra{
\bra{
\widehat\dH-\dH_0
}
\dF_s
}.
\end{align*}
Therefore, by \eqref{eq:first_derivative_coefficients_b} and \eqref{eq:first_derivative_rotation_factor_rate_b},
\begin{align}
\label{eq:first_derivative_factor_rotation_rate_b}
&
\sum_{s=1}^{t}
\calX_k^{\lfloor(t-s)/k\rfloor}
\calC^{\dF}_{s,NT}
\vvbra{
\bra{
\widehat\dH-\dH_0
}
\dF_s
}
=
O_{\Pp}\bra{
\frac{1}{\min\bra{\sqrt N,\sqrt T}}
}.
\end{align}
For the second term on the right-hand side of \eqref{eq:first_derivative_factor_contribution_b}, using the geometric convergence established in the proof of Proposition \ref{prop:rate_psit_decomp}$(b)$ we get
\[
\vvbra{
\dH_0
\bra{
\dF_s-\dbarF_s
}
}
=
O_{a.s.}\bra{
\calX_k^{\lfloor s/k\rfloor}
}.
\]
Hence, by \eqref{eq:first_derivative_coefficients_as_b},
\begin{align}
\label{eq:first_derivative_factor_stationary_rate_b}
&
\sum_{s=1}^{t}
\calX_k^{\lfloor(t-s)/k\rfloor}
\calC^{\dF}_{s,NT}
\vvbra{
\dH_0
\bra{
\dF_s-\dbarF_s
}
}
=
O_{a.s.}\bra{
\sum_{s=1}^{t}
s^{1/q+\varepsilon}
\calX_k^{
\lfloor(t-s)/k\rfloor+\lfloor s/k\rfloor
}
}
=
O_{a.s.}\bra{
\calX_k^{\lfloor t/(2k)\rfloor}
}.
\end{align}
Combining \eqref{eq:first_derivative_factor_rotation_rate_b} and \eqref{eq:first_derivative_factor_stationary_rate_b} in
\eqref{eq:first_derivative_factor_contribution_b} 
\begin{align}
\label{eq:first_derivative_factor_component_b}
&
\sum_{s=1}^{t}
\calX_k^{\lfloor(t-s)/k\rfloor}
\calC^{\dF}_{s,NT}
\vvbra{
\widehat\dH\dF_s-\dH_0\dbarF_s
}
=
O_{\Pp}\bra{
\frac{1}{\min\bra{\sqrt N,\sqrt T}}
}
+
O_{a.s.}\bra{
\calX_k^{\lfloor t/(2k)\rfloor}
}.
\end{align}
Finally, substituting \eqref{eq:first_derivative_filter_component_b} and \eqref{eq:first_derivative_factor_component_b} into \eqref{eq:first_derivative_recursive_bound_b} yields
\[
\sup_{\dtheta\in\dTheta}
\vvbra{
\nabla^{\dtheta}\dhatpsithetatpt
-
\nabla^{\dtheta}\dtildepsithetatpt
}
=
O_{\Pp}\bra{
\frac{1}{\min\bra{\sqrt N,\sqrt T}}
}
+
O_{a.s.}\bra{
\calX_k^{\lfloor t/(2k)\rfloor}
+
\varrho_k^{\lfloor t/(2k)\rfloor}
},
\]
which proves part $(b)$.

\medskip

\noindent
$(c)$ The processes $\dtildepsithetatti$ and $\dbarpsitti(\dtheta)$ are driven by the same stationary factor process $\dH_0\dbarF_t$. By subtracting their derivative recursions we get
\begin{align}
\label{eq:first_derivative_difference_c}
&
\nabla^{\dtheta}\dtildepsithetatpt
-
\nabla^{\dtheta}\dbarpsitpt(\dtheta)
\nonumber
\\
={}&
\widetilde{\boldsymbol X}_t(\dtheta)
\bra{
\nabla^{\dtheta}\dtildepsithetatti
-
\nabla^{\dtheta}\dbarpsitti(\dtheta)
}
+
\bra{
\widetilde{\boldsymbol X}_t(\dtheta)
-
\overline{\boldsymbol X}_t(\dtheta)
}
\nabla^{\dtheta}\dbarpsitti(\dtheta)
+
\widetilde{\dD}_t(\dtheta)
-
\overline{\dD}_t(\dtheta).
\end{align}
Since the factor input is the same in the two recursions, a straightforward application of the mean-value theorem shows that
\begin{align}
\label{eq:first_derivative_recursive_bound_c}
&
\sup_{\dtheta\in\dTheta}
\vvbra{
\nabla^{\dtheta}\dtildepsithetatpt
-
\nabla^{\dtheta}\dbarpsitpt(\dtheta)
}
\nonumber
\\
\leq{}&
\calX_k^{\lfloor t/k\rfloor}
\sup_{\dtheta\in\dTheta}
\vvbra{
\nabla^{\dtheta}\dpsi^\star_{1|0}
-
\nabla^{\dtheta}\dbarpsi_{1|0}(\dtheta)
}
+
\sum_{s=1}^{t}
\calX_k^{\lfloor(t-s)/k\rfloor}
\calC^{\dpsi}_{s}
\sup_{\dtheta\in\dTheta}
\vvbra{
\dtildepsi_{s|s-1}(\dtheta)
-
\dbarpsi_{s|s-1}(\dtheta)
},
\end{align}
where $\calC^{\dpsi}_{s}$ is defined as in part $(a)$ with $\nabla^{\dtheta}\dhatpsi_{s|s-1}(\dtheta)$ replaced by $\nabla^{\dtheta}\dbarpsi_{s|s-1}(\dtheta)$. By \eqref{eq:first_derivative_uniform_moment}, H\"older's inequality, and Assumption \ref{ass:smooth_domination_score_Hessian}$(iii)$--$(iv)$, with
$q=p/2>1$,
\begin{align}
\label{eq:first_derivative_coefficient_moment_c}
\sup_{s\in\Zz}
\vvbra{
\calC^{\dpsi}_{s}
}_q
&\leq
K_X^\prime
\sup_{s\in\Zz}
\vvbra{
\sup_{\dtheta\in\dTheta}
\vvbra{
\nabla^{\dtheta}\dbarpsi_{s|s-1}(\dtheta)
}
}_p
+
K_D^{\prime\prime}
<\infty.
\end{align}
Consequently, for every $\varepsilon>0$,
\begin{align*}
\sum_{s=1}^{\infty}
\Pp\bra{
\calC^{\dpsi}_{s}
>
s^{1/q+\varepsilon}
}
&\leq
\sup_{s\in\Zz}
\Ee\sbra{
\bra{
\calC^{\dpsi}_{s}
}^{q}
}
\sum_{s=1}^{\infty}
s^{-1-q\varepsilon}
<\infty.
\end{align*}
The Borel--Cantelli lemma therefore implies
\begin{align}
\label{eq:first_derivative_coefficient_as_c}
\calC^{\dpsi}_{s}
=
O_{a.s.}\bra{
s^{1/q+\varepsilon}
}.
\end{align}
Again, since $\dpsi^\star_{1|0}$ is fixed, $\nabla^{\dtheta}\dpsi^\star_{1|0} = \boldsymbol 0_{r\times d}$, and \eqref{eq:first_derivative_uniform_moment} implies that
\[
\sup_{\dtheta\in\dTheta}
\vvbra{
\nabla^{\dtheta}\dbarpsi_{1|0}(\dtheta)
}
<\infty
\qquad
\text{almost surely}.
\]
Hence, the first term on the right-hand side of \eqref{eq:first_derivative_recursive_bound_c} satisfies
\[
\calX_k^{\lfloor t/k\rfloor}
\sup_{\dtheta\in\dTheta}
\vvbra{
\nabla^{\dtheta}\dpsi^\star_{1|0}
-
\nabla^{\dtheta}\dbarpsi_{1|0}(\dtheta)
}
=
O_{a.s.}\bra{
\calX_k^{\lfloor t/k\rfloor}
}.
\]
Moreover, Proposition \ref{prop:rate_psit_decomp}$(c)$ further implies that
\[
\sup_{\dtheta\in\dTheta}
\vvbra{
\dtildepsi_{s|s-1}(\dtheta)
-
\dbarpsi_{s|s-1}(\dtheta)
}
=
O_{a.s.}\bra{
\varrho_k^{\lfloor(s-1)/k\rfloor}
}.
\]
Combining this result with \eqref{eq:first_derivative_coefficient_as_c}, the second term on the right-hand side of \eqref{eq:first_derivative_recursive_bound_c}
satisfies
\begin{align*}
&
\sum_{s=1}^{t}
\calX_k^{\lfloor(t-s)/k\rfloor}
\calC^{\dpsi}_{s}
\sup_{\dtheta\in\dTheta}
\vvbra{
\dtildepsi_{s|s-1}(\dtheta)
-
\dbarpsi_{s|s-1}(\dtheta)
}
\\
={}&
O_{a.s.}\bra{
\sum_{s=1}^{t}
s^{1/q+\varepsilon}
\calX_k^{\lfloor(t-s)/k\rfloor}
\varrho_k^{\lfloor(s-1)/k\rfloor}
}
=
O_{a.s.}\bra{
\calX_k^{\lfloor t/(2k)\rfloor}
+
\varrho_k^{\lfloor t/(2k)\rfloor}
},
\end{align*}
so that, by substituting the preceding two bounds into \eqref{eq:first_derivative_recursive_bound_c} we clearly get
\[
\sup_{\dtheta\in\dTheta}
\vvbra{
\nabla^{\dtheta}\dtildepsithetatpt
-
\nabla^{\dtheta}\dbarpsitpt(\dtheta)
}
=
O_{a.s.}\bra{
\calX_k^{\lfloor t/(2k)\rfloor}
+
\varrho_k^{\lfloor t/(2k)\rfloor}
},
\]
which proves part $(c)$.

\medskip

\noindent
$(d)$ When evaluated using the stationary and ergodic processes $\cbra{\dH_0\dbarF_t}_{t\in\Zz}$ and $\cbra{\dbarpsitti(\dtheta)}_{t\in\Zz}$, the sequences
$\cbra{\boldsymbol X_t(\dtheta)}_{t\in\Zz}$ and $\cbra{\dD_t(\dtheta)}_{t\in\Zz}$ are themselves strictly stationary and ergodic. Moreover, Assumption
\ref{ass:smooth_domination_score_Hessian}$(iv)$ implies that
\[
\Ee\sbra{
\log^+
\sup_{\dtheta\in\dTheta}
\vvbra{
\dD_t(\dtheta)
}
}
<\infty,
\]
whereas Assumption \ref{ass:contraction_cond}$(i)$ implies that the $k$-step Lipschitz coefficient of the derivative recursion is bounded almost surely by $\calX_k<1$. Therefore, Theorem 3.1 of \citet{Bougerol1993}, applied to \eqref{eq:first_derivative_filter_recursion}, implies that the derivative recursion admits a unique causal strictly stationary and ergodic solution given by
\begin{align}
\label{eq:first_derivative_stationary_solution}
\nabla^{\dtheta}\dbarpsitpt(\dtheta)
=
\dD_t(\dtheta)
+
\sum_{j=1}^{\infty}
\bra{
\prod_{h=0}^{j-1}
\boldsymbol X_{t-h}(\dtheta)
}
\dD_{t-j}(\dtheta).
\end{align}
The series in \eqref{eq:first_derivative_stationary_solution} converges almost surely. Furthermore, by \eqref{eq:first_derivative_uniform_moment} 
\[
\sup_{t\in\Zz}
\Ee\sbra{
\sup_{\dtheta\in\dTheta}
\vvbra{
\nabla^{\dtheta}\dbarpsitti(\dtheta)
}^{p}
}
<\infty.
\]
It remains to establish near-epoch dependence. For any $\tau\geq1$, consider the $\tau$-truncated factor and filter processes used in the proof of Lemma \ref{lemma:NED}, and construct the corresponding derivative recursion by initializing it at $t-\tau$ with the fixed value $\boldsymbol 0_{r\times d}$. The truncated derivative is a measurable function of $\dvt,\ldots,\dv_{t-\tau}$. An application of the derivative perturbation bound \eqref{eq:first_derivative_X_D_perturbation}, the contraction condition, and the geometric approximation obtained in Lemma \ref{lemma:NED}, reveals that
\[
\vvbra{
\nabla^{\dtheta}\dbarpsitti(\dtheta_0)
-
\nabla^{\dtheta}\dpsi_{t|t-1}^{(\tau)}(\dtheta_0)
}_2
=
O\bra{
\calX_k^{\lfloor\tau/(2k)\rfloor}
}.
\]
Consequently, by the contraction property of conditional expectations,
\begin{align*}
&
\vvbra{
\nabla^{\dtheta}\dbarpsitti(\dtheta_0)
-
\Ee\sbra{
\left.
\nabla^{\dtheta}\dbarpsitti(\dtheta_0)
\right|
\dvt,\ldots,\dv_{t-\tau}
}
}_2
\\
\leq{}&
2
\vvbra{
\nabla^{\dtheta}\dbarpsitti(\dtheta_0)
-
\nabla^{\dtheta}\dpsi_{t|t-1}^{(\tau)}(\dtheta_0)
}_2
=
O\bra{
\calX_k^{\lfloor\tau/(2k)\rfloor}
}.
\end{align*}
The approximation coefficients therefore decrease geometrically and are of size $-1$. Hence, the process $\cbra{ \nabla^{\dtheta}\dbarpsitti(\dtheta_0)}_{t\in\Zz}$ is near-epoch dependent of size $-1$ on $\cbra{\dvt}_{t\in\Zz}$, which is geometrically $\beta$-mixing by Assumption \ref{ass:noise}$(i)$.

\medskip

\noindent
$(e)$ Finally, using the decomposition in \eqref{eq:psi_first_derivative_decomposition} it is straightforward to see that the result follows immediately by combining parts $(a)$--$(c)$ and the triangle inequality.
\end{proof}

\begin{lemma}
\label{lemma:rate_psit_second_derivs}
Consider the second derivative process associated with the score-driven recursion,
\begin{align}
\label{eq:second_derivative_filter_recursion}
\nabla^{\dtheta\dtheta}\dpsitpt(\dtheta)
={}&
\bra{
\boldsymbol I_d
\otimes
\boldsymbol X_t(\dtheta)
}
\nabla^{\dtheta\dtheta}\dpsitti(\dtheta)
\nonumber
\\
&+
\bra{
\nabla^{\dtheta}\dpsitti(\dtheta)\trans
\otimes
\boldsymbol I_r
}
\frac{
\partial\vect{\boldsymbol X_t(\dtheta)}
}{
\partial\dtheta\trans
}
+
\frac{
\partial\vect{\dD_t(\dtheta)}
}{
\partial\dtheta\trans
}.
\end{align}
By differentiating \eqref{eq:psi_first_derivative_decomposition} with respect to $\dtheta$ we get
\begin{align}
\label{eq:psi_second_derivative_decomposition}
&
\nabla^{\dtheta\dtheta}\dhathatpsithetatti
-
\nabla^{\dtheta\dtheta}\dbarpsitti(\dtheta)
\\
={}&
\bra{
\nabla^{\dtheta\dtheta}\dhathatpsithetatti
-
\nabla^{\dtheta\dtheta}\dhatpsithetatti
}
+
\bra{
\nabla^{\dtheta\dtheta}\dhatpsithetatti
-
\nabla^{\dtheta\dtheta}\dtildepsithetatti
}
\nonumber
\\
&+
\bra{
\nabla^{\dtheta\dtheta}\dtildepsithetatti
-
\nabla^{\dtheta\dtheta}\dbarpsitti(\dtheta)
}.
\nonumber
\end{align}
Then, under Assumptions \ref{ass:noise}--\ref{ass:contraction_cond}
and \ref{ass:smooth_domination_score_Hessian}, with $p>2$, the
following results hold:
\begin{enumerate}

\item[(a)] For any $t\geq1$, as $N,T\rightarrow\infty$,
\begin{align*}
&
\sup_{\dtheta\in\dTheta}
\vvbra{
\nabla^{\dtheta\dtheta}\dhathatpsithetatpt
-
\nabla^{\dtheta\dtheta}\dhatpsithetatpt
}
=
O_{\Pp}\bra{
\frac{1}{\min\bra{\sqrt N,T}}
}
+
O_{a.s.}\bra{
\calX_k^{\lfloor t/(4k)\rfloor}
}.
\end{align*}

\item[(b)] For any $t\geq1$, as $N,T\rightarrow\infty$,
\begin{align*}
&
\sup_{\dtheta\in\dTheta}
\vvbra{
\nabla^{\dtheta\dtheta}\dhatpsithetatti
-
\nabla^{\dtheta\dtheta}\dtildepsithetatti
}
=
O_{\Pp}\bra{
\frac{1}{\min\bra{\sqrt N,\sqrt T}}
}
+
O_{a.s.}\bra{
\calX_k^{\lfloor t/(4k)\rfloor}
+
\varrho_k^{\lfloor t/(4k)\rfloor}
}.
\end{align*}

\item[(c)] For any fixed initialization
$\dpsi^\star_{1|0}\in\boldsymbol{\mathcal P}$, as
$t\rightarrow\infty$,
\begin{align*}
&
\sup_{\dtheta\in\dTheta}
\vvbra{
\nabla^{\dtheta\dtheta}\dtildepsithetatti
-
\nabla^{\dtheta\dtheta}\dbarpsitti(\dtheta)
}
=
O_{a.s.}\bra{
\calX_k^{\lfloor t/(4k)\rfloor}
+
\varrho_k^{\lfloor t/(4k)\rfloor}
}.
\end{align*}

\item[(d)] The process $\cbra{ \nabla^{\dtheta\dtheta}\dbarpsitti(\dtheta) }_{t\in\Zz}$ is strictly stationary and ergodic and satisfies
\[
\Ee\sbra{
\sup_{\dtheta\in\dTheta}
\vvbra{
\nabla^{\dtheta\dtheta}\dbarpsitti(\dtheta)
}^{p}
}
<\infty.
\]
Moreover, $\cbra{\nabla^{\dtheta\dtheta}\dbarpsitti(\dtheta_0) }_{t\in\Zz}$ is near-epoch dependent of size $-1$ on the geometrically $\beta$-mixing process $\cbra{\dvt}_{t\in\Zz}$.

\item[(e)] For any fixed initialization $\dpsi_{1|0}\in\boldsymbol{\mathcal P}$ and any $t\geq1$, as $N,T\rightarrow\infty$,
\begin{align*}
\sup_{\dtheta\in\dTheta}
\vvbra{
\nabla^{\dtheta\dtheta}\dhathatpsithetatpt
-
\nabla^{\dtheta\dtheta}\dbarpsitpt(\dtheta)
}
=&
O_{\Pp}\bra{
\frac{1}{\min\bra{\sqrt N,T}}
}
+
O_{\Pp}\bra{
\frac{1}{\min\bra{\sqrt N,\sqrt T}}
}
\nonumber
\\
&+
O_{a.s.}\bra{
\calX_k^{\lfloor t/(4k)\rfloor}
+
\varrho_k^{\lfloor t/(4k)\rfloor}
}.
\end{align*}

\end{enumerate}
\end{lemma}

\begin{proof}
The coefficient multiplying the lagged second derivative in \eqref{eq:second_derivative_filter_recursion} is $\boldsymbol I_d\otimes\boldsymbol X_t(\dtheta)$. Therefore,
\begin{align*}
&
\sup_{\dtheta\in\dTheta}
\vvbra{
\prod_{j=1}^{k}
\bra{
\boldsymbol I_d
\otimes
\boldsymbol X_{t-j+1}(\dtheta)
}
}
=
\sup_{\dtheta\in\dTheta}
\vvbra{
\boldsymbol I_d
\otimes
\prod_{j=1}^{k}
\boldsymbol X_{t-j+1}(\dtheta)
}
\leq
\calX_k
\end{align*}
almost surely. Hence, the second derivative recursion inherits the same $k$-step contraction as the first derivative recursion.

We first establish the required moment bound. By the equality of mixed partial derivatives, the matrices
\[
\frac{\partial\vect{\boldsymbol X_t(\dtheta)}}
{\partial\dtheta\trans}
\qquad\text{and}\qquad
\frac{\partial\vect{\dD_t(\dtheta)}}
{\partial\dpsitti\trans}
\]
contain the same mixed derivatives, up to a fixed rearrangement of their elements. The additional terms arising from evaluation along the filtered process are controlled by $\nabla^{\dtheta}\dpsitti(\dtheta)$ and the derivative envelopes in Assumption \ref{ass:smooth_domination_score_Hessian}$(iii)$--$(vi)$.
Consequently, Lemma \ref{lemma:rate_psit_first_derivs}$(d)$, H\"older's inequality, and Assumption \ref{ass:smooth_domination_score_Hessian}$(iii)$--$(vi)$ imply that
\begin{align*}
&
\vvbra{
\sup_{\dtheta\in\dTheta}
\vvbra{
\bra{
\nabla^{\dtheta}\dbarpsitti(\dtheta)\trans
\otimes
\boldsymbol I_r
}
\frac{
\partial\vect{\boldsymbol X_t(\dtheta)}
}{
\partial\dtheta\trans
}
}
}_p
\\
\leq{}&
\vvbra{
\sup_{\dtheta\in\dTheta}
\vvbra{
\nabla^{\dtheta}\dbarpsitti(\dtheta)
}
}_{2p}
\vvbra{
\sup_{\dtheta\in\dTheta}
\vvbra{
\frac{
\partial\vect{\boldsymbol X_t(\dtheta)}
}{
\partial\dtheta\trans
}
}
}_{2p}
<\infty,
\end{align*}
The same bound holds uniformly for the corresponding nonstationary and feasible recursions.

Applying the $k$-step moment argument used in \eqref{eq:first_derivative_X_moment}--\eqref{eq:first_derivative_uniform_moment} to \eqref{eq:second_derivative_filter_recursion}, and using the fact that the finite partial products of $\boldsymbol X_t(\dtheta)$ have
bounded moments, yields
\begin{align}
\label{eq:second_derivative_uniform_moment}
\sup_{t\in\Zz}
\vvbra{
\sup_{\dtheta\in\dTheta}
\vvbra{
\nabla^{\dtheta\dtheta}
\dbarpsitti(\dtheta)
}
}_p
<\infty.
\end{align}
The same argument gives the corresponding uniform moment bounds for the second derivative processes associated with $\dhathatpsithetatti$, $\dhatpsithetatti$, and $\dtildepsithetatti$.

We now compare any two of these derivative recursions. Subtracting their respective versions of \eqref{eq:second_derivative_filter_recursion}, applying the mean value theorem, and iterating the resulting inequality in blocks of length $k$ 
\begin{align}
\label{eq:second_derivative_generic_bound}
&
\sup_{\dtheta\in\dTheta}
\vvbra{
\nabla^{\dtheta\dtheta}
\dpsi^{(1)}_{t+1|t}(\dtheta)
-
\nabla^{\dtheta\dtheta}
\dpsi^{(2)}_{t+1|t}(\dtheta)
}
\nonumber
\\
\leq{}&
\calX_k^{\lfloor t/k\rfloor}
\sup_{\dtheta\in\dTheta}
\vvbra{
\nabla^{\dtheta\dtheta}
\dpsi^{(1)}_{1|0}(\dtheta)
-
\nabla^{\dtheta\dtheta}
\dpsi^{(2)}_{1|0}(\dtheta)
}
\nonumber
\\
&+
\sum_{s=1}^{t}
\calX_k^{\lfloor(t-s)/k\rfloor}
\calC^{(2)}_{s,NT}
\Bigg(
\sup_{\dtheta\in\dTheta}
\vvbra{
\dpsi^{(1)}_{s|s-1}(\dtheta)
-
\dpsi^{(2)}_{s|s-1}(\dtheta)
}
\nonumber
\\
&\hspace{10em}
+
\vvbra{
\dF_s^{(1)}-\dF_s^{(2)}
}
+
\sup_{\dtheta\in\dTheta}
\vvbra{
\nabla^{\dtheta}\dpsi^{(1)}_{s|s-1}(\dtheta)
-
\nabla^{\dtheta}\dpsi^{(2)}_{s|s-1}(\dtheta)
}
\Bigg).
\end{align}
Here, $\calC^{(2)}_{s,NT}$ collects the derivative envelopes and the norms of the first- and second-derivative processes arising from the mean-value expansion. By \eqref{eq:first_derivative_uniform_moment}, \eqref{eq:second_derivative_uniform_moment}, H\"older's inequality, and Assumption \ref{ass:smooth_domination_score_Hessian}$(iii)$--$(vi)$, applied at sufficiently high moment orders, there exists $q>1$ such that
\[
\sup_{N,T}
\sup_{1\leq s\leq T}
\vvbra{
\calC^{(2)}_{s,NT}
}_q
<\infty.
\]
Thus, the coefficients are tight for every fixed $t$, while the same Markov and Borel--Cantelli argument used in the proof of Lemma \ref{lemma:rate_psit_first_derivs} controls the almost-sure components.

For part $(a)$, apply \eqref{eq:second_derivative_generic_bound} using Proposition \ref{prop:rate_yt_decomp}$(a)$, Proposition
\ref{prop:rate_psit_decomp}$(a)$, and Lemma \ref{lemma:rate_psit_first_derivs}$(a)$. For part $(b)$, use the corresponding results in Proposition
\ref{prop:rate_yt_decomp}, Proposition \ref{prop:rate_psit_decomp}$(b)$, and Lemma \ref{lemma:rate_psit_first_derivs}$(b)$. Finally, for part $(c)$, the
factor inputs coincide, and therefore Proposition \ref{prop:rate_psit_decomp}$(c)$ and Lemma \ref{lemma:rate_psit_first_derivs}$(c)$ apply directly.

The stochastic components retain the rates stated in parts $(a)$ and $(b)$ because the corresponding coefficients are tight for every fixed $t$. For the almost-sure components, the geometric-convolution bounds
\begin{align*}
&
\sum_{s=1}^{t}
s^{1/q+\varepsilon}
\calX_k^{\lfloor(t-s)/k\rfloor}
\calX_k^{\lfloor s/(2k)\rfloor}
=
O\bra{
\calX_k^{\lfloor t/(4k)\rfloor}
},
\\
&
\sum_{s=1}^{t}
s^{1/q+\varepsilon}
\calX_k^{\lfloor(t-s)/k\rfloor}
\varrho_k^{\lfloor s/(2k)\rfloor}
=
O\bra{
\calX_k^{\lfloor t/(4k)\rfloor}
+
\varrho_k^{\lfloor t/(4k)\rfloor}
}
\end{align*}
give the rates stated in parts $(a)$--$(c)$.

When evaluated using the stationary processes $\cbra{\dH_0\dbarF_t}_{t\in\Zz}$ and $\cbra{\dbarpsitti(\dtheta)}_{t\in\Zz}$, the second derivative
recursion has stationary and ergodic coefficients. The $k$-step contraction established above and the moment bound in \eqref{eq:second_derivative_uniform_moment} therefore imply that it admits a unique causal strictly stationary and ergodic solution.

For any $\tau\geq1$, construct the corresponding truncated second derivative recursion using the truncated factor, filter, and first
derivative processes. Applying \eqref{eq:second_derivative_generic_bound}, Lemma \ref{lemma:NED}, and Lemma \ref{lemma:rate_psit_first_derivs}$(d)$ shows that its $L_2$-approximation error decreases geometrically in $\tau$. By construction, the truncated second derivative is a measurable
function of $\dvt,\ldots,\dv_{t-\tau}$. Consequently,
\begin{align*}
&
\vvbra{
\nabla^{\dtheta\dtheta}\dbarpsitti(\dtheta_0)
-
\Ee\sbra{
\left.
\nabla^{\dtheta\dtheta}\dbarpsitti(\dtheta_0)
\right|
\dvt,\ldots,\dv_{t-\tau}
}
}_2
\\
\leq{}&
2
\vvbra{
\nabla^{\dtheta\dtheta}\dbarpsitti(\dtheta_0)
-
\nabla^{\dtheta\dtheta}
\dpsi_{t|t-1}^{(\tau)}(\dtheta_0)
}_2,
\end{align*}
which also decreases geometrically in $\tau$. Hence, $\cbra{\nabla^{\dtheta\dtheta}\dbarpsitti(\dtheta_0)}_{t\in\Zz}$ is near-epoch dependent of size $-1$ on
$\cbra{\dvt}_{t\in\Zz}$.

Finally, part $(e)$ follows from \eqref{eq:psi_second_derivative_decomposition}, parts $(a)$--$(c)$, and the triangle inequality.
\end{proof}


\begin{lemma}
\label{lemma:U_convergence_score_vector}
Consider the score vector
\begin{equation}
\label{eq:score_vector}
\nabla^{\dtheta}
\ell_{\dF}
\bra{
\dFt;\dtheta
}
=
\nabla^{\dtheta}
\dpsithetatti\trans
\nabla^{\dpsi}
\ell_{\dF}
\bra{
\dFt;\dtheta
}
+
\frac{
\partial\ell_{\dF}
\bra{
\dFt;\dtheta
}
}{
\partial\dtheta
},
\end{equation}
where the partial derivative in the second term is computed holding $\dpsitti$ and $\dFt$ fixed and is subsequently evaluated at $\dpsitti=\dpsitti\bra{\dtheta}$.

Under Assumptions \ref{ass:noise}--\ref{ass:truePar}, with $p>2$, for every fixed initial condition $\dpsi_{1|0}\in\calP$, as $N,T\rightarrow\infty$,
\begin{align}
\label{eq:uniform_score_population_rate}
&
\sup_{\dtheta\in\dTheta}
\vvbra{
\frac{1}{T}
\sum_{t=1}^T
\nabla^{\dtheta}
\doublehat{\ell}_{\dF}
\bra{
\dhatFt;
\boldsymbol r_{\widehat\dH}\bra{\dtheta}
}
-
\Ee\sbra{
\nabla^{\dtheta}
\ell_{\dF}
\bra{
\dbarFt;
\dtheta
}
}
}
=
O_{\Pp}\bra{\frac{1}{\min\bra{N,\sqrt T}}},
\end{align}
where the derivative of the first term is taken with respect to $\dtheta$ through the mapping $\boldsymbol r_{\widehat\dH}\bra{\dtheta}$.

Moreover,
\begin{align}
\label{eq:uniform_score_sample_approximation}
&
\frac{1}{\sqrt T}
\sup_{\dtheta\in\dTheta}
\vvbra{
\sum_{t=1}^T
\nabla^{\dtheta}
\doublehat{\ell}_{\dF}
\bra{
\dhatFt;
\boldsymbol r_{\widehat\dH}\bra{\dtheta}
}
-
\sum_{t=1}^T
\nabla^{\dtheta}
\ell_{\dF}
\bra{
\dbarFt;
\dtheta
}
}
=
O_{\Pp}\bra{
\frac{1}{\sqrt N}
+
\frac{\sqrt T}{N}
+
\frac{1}{\sqrt T}
}.
\end{align}
Consequently, if $\frac{\sqrt T}{N}\longrightarrow0$, then, as $N,T\rightarrow\infty$,
\begin{align}
\label{eq:uniform_score_root_T_equivalence}
&
\frac{1}{\sqrt T}
\sup_{\dtheta\in\dTheta}
\vvbra{
\sum_{t=1}^T
\nabla^{\dtheta}
\doublehat{\ell}_{\dF}
\bra{
\dhatFt;
\boldsymbol r_{\widehat\dH}\bra{\dtheta}
}
-
\sum_{t=1}^T
\nabla^{\dtheta}
\ell_{\dF}
\bra{
\dbarFt;
\dtheta
}
}
=
o_{\Pp}(1).
\end{align}
Finally, the stationary score process $\cbra{ \nabla^{\dtheta}\ell_{\dF}\bra{\dbarF_t;\dtheta_0}}_{t\in\Zz}$ has mean zero and is near-epoch dependent of size $-1$ on the geometrically $\beta$-mixing process $\cbra{\dvt}_{t\in\Zz}$.
\end{lemma}

\begin{proof}
Similarly to the proof of Proposition \ref{lemma:ULLN}, an application of the triangle inequality shows that 
\begin{align}
\label{eq:score_decomposition}
&
\sup_{\dtheta\in\dTheta}
\vvbra{
\frac{1}{T}
\sum_{t=1}^T
\nabla^{\dtheta}
\doublehat{\ell}_{\dF}
\bra{
\dhatF;
\boldsymbol r_{\widehat\dH}\bra{\dtheta}
}
-
\Ee\sbra{
\nabla^{\dtheta}
\ell_{\dF}
\bra{
\dbarFt;
\dtheta
}
}
}
\nonumber
\\
\leq{}&
\sup_{\dtheta\in\dTheta}
\vvbra{
\frac{1}{T}
\sum_{t=1}^T
\nabla^{\dtheta}
\doublehat{\ell}_{\dF}
\bra{
\dhatFt;
\boldsymbol r_{\widehat\dH}\bra{\dtheta}
}
-
\frac{1}{T}
\sum_{t=1}^T
\nabla^{\dtheta}
\widehat{\ell}_{\dF}
\bra{
\dFt\widehat\dH\trans;
\boldsymbol r_{\widehat\dH}\bra{\dtheta}
}
}
\nonumber
\\
&+
\sup_{\dtheta\in\dTheta}
\vvbra{
\frac{1}{T}
\sum_{t=1}^T
\nabla^{\dtheta}
\widehat{\ell}_{\dF}
\bra{
\dFt\widehat\dH\trans;
\boldsymbol r_{\widehat\dH}\bra{\dtheta}
}
-
\frac{1}{T}
\sum_{t=1}^T
\nabla^{\dtheta}
\ell_{\dF}
\bra{
\dFt;
\dtheta
}
}
\nonumber
\\
&+
\sup_{\dtheta\in\dTheta}
\vvbra{
\frac{1}{T}
\sum_{t=1}^T
\nabla^{\dtheta}
\ell_{\dF}
\bra{
\dFt;
\dtheta
}
-
\frac{1}{T}
\sum_{t=1}^T
\nabla^{\dtheta}
\ell_{\dF}
\bra{
\dbarFt;
\dtheta
}
}
\nonumber
\\
&+
\sup_{\dtheta\in\dTheta}
\vvbra{
\frac{1}{T}
\sum_{t=1}^T
\nabla^{\dtheta}
\ell_{\dF}
\bra{
\dbarFt;
\dtheta
}
-
\Ee\sbra{
\nabla^{\dtheta}
\ell_{\dF}
\bra{
\dbarFt;
\dtheta
}
}
}.
\end{align}
For the likelihoods evaluated at $\boldsymbol r_{\widehat\dH}\bra{\dtheta}$, the derivatives in \eqref{eq:score_decomposition} are taken with respect to $\dtheta$ through the  mapping $\boldsymbol r_{\dH}(\cdot)$. The four terms on the right-hand side of \eqref{eq:score_decomposition} account, respectively, for factor estimation and its recursive propagation through the filter, the effect of the possibly different filter initializations under the alignment mapping, convergence to the stationary factor and filter processes, and the sampling variation of the stationary score.


\bigskip
\noindent
\textbf{Controlling the first term in \eqref{eq:score_decomposition}.}\\
In the following, we use the notation
\begin{align*}
\nabla^{\dtheta\dpsi^\prime}
\ell_{\dF}
\bra{
\dFt;\dtheta
}
&=
\frac{
\partial
\nabla^{\dtheta}
\ell_{\dF}
\bra{
\dFt;\dtheta
}
}{
\partial
\vect{
\nabla^{\dtheta}\dpsitti\bra{\dtheta}
}^{\trans}
},
\\
\nabla^{\dtheta\dpsi}
\ell_{\dF}
\bra{
\dFt;\dtheta
}
&=
\frac{
\partial
\nabla^{\dtheta}
\ell_{\dF}
\bra{
\dFt;\dtheta
}
}{
\partial
\dpsitti\bra{\dtheta}^{\trans}
},
\\
\nabla^{\dtheta\dF}
\ell_{\dF}
\bra{
\dFt;\dtheta
}
&=
\frac{
\partial
\nabla^{\dtheta}
\ell_{\dF}
\bra{
\dFt;\dtheta
}
}{
\partial
\dFt\trans
}.
\end{align*}
Regarding the full score contribution as a function of $\nabla^{\dtheta}\dpsitti$, $\dpsitti$, and $\dFt$, a first-order expansion around
$
\bra{
\nabla^{\dtheta}
\dhatpsitti
\bra{
\boldsymbol r_{\widehat\dH}\bra{\dtheta}
},
\dhatpsitti
\bra{
\boldsymbol r_{\widehat\dH}\bra{\dtheta}
},
\widehat\dH\dFt
}
$
shows that, uniformly over $\dtheta\in\dTheta$,
\begin{align}
\label{eq:score_signed_expansion}
&
\frac{1}{T}
\bra{
\sum_{t=1}^T
\nabla^{\dtheta}
\doublehat{\ell}_{\dF}
\bra{
\dhatFt;
\boldsymbol r_{\widehat\dH}\bra{\dtheta}
}
-
\sum_{t=1}^T
\nabla^{\dtheta}
\widehat{\ell}_{\dF}
\bra{
\dFt\widehat\dH\trans;
\boldsymbol r_{\widehat\dH}\bra{\dtheta}
}
}
\nonumber
\\
={}&
\frac{1}{T}
\sum_{t=1}^{T}
\bra{
\nabla^{\dtheta}\doublehat{\ell}_{\dF}
\bra{
\dhatFt;
\boldsymbol r_{\widehat\dH}(\dtheta)
}
-
\nabla^{\dtheta}\widehat{\ell}_{\dF}
\bra{
\widehat\dH\dFt;
\boldsymbol r_{\widehat\dH}(\dtheta)
}
}
\nonumber
\\
={}&
\frac{1}{T}
\sum_{t=1}^{T}
\nabla^{\dtheta\dpsi^\prime}
\widehat{\ell}_{\dF}
\bra{
\widehat\dH\dFt;
\boldsymbol r_{\widehat\dH}(\dtheta)
}
\vect\bra{
\nabla^{\dtheta}\dhathatpsitti
\bra{
\boldsymbol r_{\widehat\dH}(\dtheta)
}
-
\nabla^{\dtheta}\dhatpsitti
\bra{
\boldsymbol r_{\widehat\dH}(\dtheta)
}
}
\nonumber
\\
&+
\frac{1}{T}
\sum_{t=1}^{T}
\nabla^{\dtheta\dpsi}
\widehat{\ell}_{\dF}
\bra{
\widehat\dH\dFt;
\boldsymbol r_{\widehat\dH}(\dtheta)
}
\bra{
\dhathatpsitti
\bra{
\boldsymbol r_{\widehat\dH}(\dtheta)
}
-
\dhatpsitti
\bra{
\boldsymbol r_{\widehat\dH}(\dtheta)
}
}
\nonumber
\\
&+
\frac{1}{T}
\sum_{t=1}^{T}
\nabla^{\dtheta\dF}
\widehat{\ell}_{\dF}
\bra{
\widehat\dH\dFt;
\boldsymbol r_{\widehat\dH}(\dtheta)
}
\bra{
\dhatFt-\widehat\dH\dFt
}
\nonumber
\\
&+
\frac{1}{T}
\sum_{t=1}^{T}
\boldsymbol R^\prime_{t,NT}(\dtheta).
\end{align}
By the uniform second-order smoothness in Assumption \ref{ass:smooth_domination_score_Hessian}, the multivariate remainder $\boldsymbol R^\prime_{t,NT}(\dtheta)$ is
quadratic in the factor, filter, and first-derivative-filter perturbations. Moreover, the argument used in the proof of Lemma \ref{lemma:rate_psit_first_derivs}$(a)$ shows that the first derivative filter perturbation inherits the average squared rate of the factor and filter perturbations. Therefore,
\begin{align}
\label{eq:quadratic_first_derivative_filter_error}
&
\frac{1}{T}
\sum_{t=1}^{T}
\sup_{\dtheta\in\dTheta}
\vvbra{
\nabla^{\dtheta}
\dhathatpsitti
\bra{
\boldsymbol r_{\widehat\dH}(\dtheta)
}
-
\nabla^{\dtheta}
\dhatpsitti
\bra{
\boldsymbol r_{\widehat\dH}(\dtheta)
}
}^{2}
=
O_{\Pp}\bra{
\frac{1}{N}
+
\frac{1}{T}
}.
\end{align}
Indeed, the stochastic component in Lemma \ref{lemma:rate_psit_first_derivs}$(a)$ is generated by the factor and filter perturbations, whose average squared rates are given in \eqref{eq:quadratic_factor_error} and \eqref{eq:quadratic_filter_error}, whereas its geometrically decreasing initialization component has average squared order $O\bra{T^{-1}}$. Hence, from the Cauchy--Schwarz inequality, \eqref{eq:quadratic_factor_error}, \eqref{eq:quadratic_filter_error}, and
\eqref{eq:quadratic_first_derivative_filter_error} we get
\begin{align*}
&
\sup_{\dtheta\in\dTheta}
\vvbra{
\frac{1}{T}
\sum_{t=1}^{T}
\boldsymbol R^\prime_{t,NT}(\dtheta)
}
\leq
\frac{1}{T}
\sum_{t=1}^{T}
\sup_{\dtheta\in\dTheta}
\vvbra{
\boldsymbol R^\prime_{t,NT}(\dtheta)
}
=
O_{\Pp}\bra{
\frac{1}{N}
+
\frac{1}{T}
}.
\end{align*}
Given this result, we first expand the two recursive differences appearing in the first two terms of \eqref{eq:score_signed_expansion}.  From \eqref{eq:filter_iterated_expansion}, we know that for every $t\geq2$,
\begin{align}
\label{eq:score_filter_iterated_expansion}
&
\dhathatpsitti
\bra{
\boldsymbol r_{\widehat\dH}\bra{\dtheta}
}
-
\dhatpsitti
\bra{
\boldsymbol r_{\widehat\dH}\bra{\dtheta}
}
\nonumber
\\
={}&
\bra{
\prod_{h=1}^{t-1}
\boldsymbol X_{t-h}
\bra{
\boldsymbol r_{\widehat\dH}\bra{\dtheta}
}
}
\bra{
\dpsi_{1|0}
-
\dpsi^\star_{1|0}
}
+
\sum_{s=1}^{t-1}
\bra{
\prod_{h=s+1}^{t-1}
\boldsymbol X_h
\bra{
\boldsymbol r_{\widehat\dH}\bra{\dtheta}
}
}
\boldsymbol Y_s
\bra{
\boldsymbol r_{\widehat\dH}\bra{\dtheta}
}
\bra{
\dhatF_s-\widehat\dH\dF_s
}
\nonumber
\\
&+
\text{second-order terms}.
\end{align}
Analogously, from \eqref{eq:first_derivative_filter_recursion}, another first-order expansion around 
$
\bra{
\nabla^{\dtheta}
\dhatpsitti
\bra{
\boldsymbol r_{\widehat\dH}\bra{\dtheta}
},
\dhatpsitti
\bra{
\boldsymbol r_{\widehat\dH}\bra{\dtheta}
},
\widehat\dH\dFt
}
$
reveals that
\begin{align}
\label{eq:first_derivative_filter_signed_expansion}
&
\vect{\bra{
\nabla^{\dtheta}
\dhathatpsi_{t+1|t}
\bra{
\boldsymbol r_{\widehat\dH}\bra{\dtheta}
}
-
\nabla^{\dtheta}
\dhatpsi_{t+1|t}
\bra{
\boldsymbol r_{\widehat\dH}\bra{\dtheta}
}
}
}
\nonumber
\\
={}&
\bra{
\boldsymbol I_d
\otimes
\boldsymbol X_t
\bra{
\boldsymbol r_{\widehat\dH}\bra{\dtheta}
}
}
\vect{\bra{
\nabla^{\dtheta}
\dhathatpsitti
\bra{
\boldsymbol r_{\widehat\dH}\bra{\dtheta}
}
-
\nabla^{\dtheta}
\dhatpsitti
\bra{
\boldsymbol r_{\widehat\dH}\bra{\dtheta}
}
}
}
\nonumber
\\
&+
\Bigg\{
\bra{
\nabla^{\dtheta}
\dhatpsitti
\bra{
\boldsymbol r_{\widehat\dH}\bra{\dtheta}
}^{\trans}
\otimes
\boldsymbol I_r
}
\frac{
\partial
\vect{
\boldsymbol X_t
\bra{
\boldsymbol r_{\widehat\dH}\bra{\dtheta}
}
}
}{
\partial
\dhatpsitti
\bra{
\boldsymbol r_{\widehat\dH}\bra{\dtheta}
}^{\trans}
}
\nonumber
\\
&
\qquad\qquad \qquad\qquad \qquad\qquad \qquad
+
\frac{
\partial
\vect{
\dD_t
\bra{
\boldsymbol r_{\widehat\dH}\bra{\dtheta}
}
}
}{
\partial
\dhatpsitti
\bra{
\boldsymbol r_{\widehat\dH}\bra{\dtheta}
}^{\trans}
}
\Bigg\}
\bra{
\dhathatpsitti
\bra{
\boldsymbol r_{\widehat\dH}\bra{\dtheta}
}
-
\dhatpsitti
\bra{
\boldsymbol r_{\widehat\dH}\bra{\dtheta}
}
}
\nonumber
\\
&+
\Bigg\{
\bra{
\nabla^{\dtheta}
\dhatpsitti
\bra{
\boldsymbol r_{\widehat\dH}\bra{\dtheta}
}^{\trans}
\otimes
\boldsymbol I_r
}
\frac{
\partial
\vect{
\boldsymbol X_t
\bra{
\boldsymbol r_{\widehat\dH}\bra{\dtheta}
}
}
}{
\partial
\bra{
\widehat\dH\dFt
}^{\trans}
}
\nonumber
\\
&\qquad\qquad \qquad\qquad \qquad\qquad \qquad
+
\frac{
\partial
\vect{
\dD_t
\bra{
\boldsymbol r_{\widehat\dH}\bra{\dtheta}
}
}
}{
\partial
\bra{
\widehat\dH\dFt
}^{\trans}
}
\Bigg\}
\bra{
\dhatFt-\widehat\dH\dFt
}
\nonumber
\\
&+
\text{second-order terms}.
\end{align}
Since the two initial conditions are fixed and do not depend on $\dtheta$, $\nabla^{\dtheta}\dpsi_{1|0}=\nabla^{\dtheta}\dpsi^\star_{1|0}=\boldsymbol 0_{r\times d}$. Consequently, iterating \eqref{eq:first_derivative_filter_signed_expansion} gives, for every $t\geq2$,
\begin{align}
\label{eq:first_derivative_filter_iterated_expansion}
&
\vect{
\nabla^{\dtheta}
\dhathatpsitti
\bra{
\boldsymbol r_{\widehat\dH}\bra{\dtheta}
}
-
\nabla^{\dtheta}
\dhatpsitti
\bra{
\boldsymbol r_{\widehat\dH}\bra{\dtheta}
}
}
\nonumber
\\
={}&
\sum_{s=1}^{t-1}
\bra{
\prod_{h=s+1}^{t-1}
\bra{
\boldsymbol I_d
\otimes
\boldsymbol X_h
\bra{
\boldsymbol r_{\widehat\dH}\bra{\dtheta}
}
}
}
\nonumber
\\
&\quad\times
\Bigg\{
\bra{
\nabla^{\dtheta}
\dhatpsi_{s|s-1}
\bra{
\boldsymbol r_{\widehat\dH}\bra{\dtheta}
}^{\trans}
\otimes
\boldsymbol I_r
}
\frac{
\partial
\vect{
\boldsymbol X_s
\bra{
\boldsymbol r_{\widehat\dH}\bra{\dtheta}
}
}
}{
\partial
\dhatpsi_{s|s-1}
\bra{
\boldsymbol r_{\widehat\dH}\bra{\dtheta}
}^{\trans}
}
\nonumber
\\
&
\qquad\qquad \qquad\qquad \qquad\qquad \qquad
+
\frac{
\partial
\vect{
\dD_s
\bra{
\boldsymbol r_{\widehat\dH}\bra{\dtheta}
}
}
}{
\partial
\dhatpsi_{s|s-1}
\bra{
\boldsymbol r_{\widehat\dH}\bra{\dtheta}
}^{\trans}
}
\Bigg\}
\bra{
\dhathatpsi_{s|s-1}
\bra{
\boldsymbol r_{\widehat\dH}\bra{\dtheta}
}
-
\dhatpsi_{s|s-1}
\bra{
\boldsymbol r_{\widehat\dH}\bra{\dtheta}
}
}
\nonumber
\\
&+
\sum_{s=1}^{t-1}
\bra{
\prod_{h=s+1}^{t-1}
\bra{
\boldsymbol I_d
\otimes
\boldsymbol X_h
\bra{
\boldsymbol r_{\widehat\dH}\bra{\dtheta}
}
}
}
\nonumber
\\
&\quad\times
\Bigg\{
\bra{
\nabla^{\dtheta}
\dhatpsi_{s|s-1}
\bra{
\boldsymbol r_{\widehat\dH}\bra{\dtheta}
}^{\trans}
\otimes
\boldsymbol I_r
}
\frac{
\partial
\vect{
\boldsymbol X_s
\bra{
\boldsymbol r_{\widehat\dH}\bra{\dtheta}
}
}
}{
\partial
\bra{
\widehat\dH\dF_s
}^{\trans}
}
\nonumber
\\
&
\qquad\qquad \qquad\qquad \qquad\qquad \qquad+
\frac{
\partial
\vect{
\dD_s
\bra{
\boldsymbol r_{\widehat\dH}\bra{\dtheta}
}
}
}{
\partial
\bra{
\widehat\dH\dF_s
}^{\trans}
}
\Bigg\}
\bra{
\dhatF_s-\widehat\dH\dF_s
}
\nonumber
\\
&+
\text{second-order terms}.
\end{align}
All products in \eqref{eq:score_filter_iterated_expansion} and \eqref{eq:first_derivative_filter_iterated_expansion} are ordered with the largest time index on the left. An empty product in \eqref{eq:score_filter_iterated_expansion} is equal to $\boldsymbol I_r$, whereas an empty product in \eqref{eq:first_derivative_filter_iterated_expansion} is equal to $\boldsymbol I_{rd}$.


Since $\dhatF_s-\widehat\dH\dF_s$ does not depend on $\dtheta$, the linear terms obtained by substituting \eqref{eq:score_filter_iterated_expansion} and
\eqref{eq:first_derivative_filter_iterated_expansion} into \eqref{eq:score_signed_expansion}, and then interchanging the resulting time sums, coincide with the derivative with respect to $\dtheta$ of the effective likelihood weight defined in \eqref{eq:effective_likelihood_weight}. Accordingly, for every
$\dH\in\calH$, define
\begin{align}
\label{eq:effective_score_weight}
\boldsymbol{\cal S}_{s,T}
\bra{
\dtheta,\dH
}
:={}&
\nabla^{\dtheta}
\boldsymbol{\cal G}_{s,T}
\bra{
\dtheta,\dH
}
\nonumber
\\
={}&
\nabla^{\dtheta}
\Bigg[
\nabla^{\dF}
\widehat{\ell}_{\dF}
\bra{
\dH\dF_s;
\boldsymbol r_{\dH}\bra{\dtheta}
}
\nonumber
\\
&\quad+
\boldsymbol Y_s
\bra{
\boldsymbol r_{\dH}\bra{\dtheta}
}^{\trans}
\sum_{t=s+1}^{T}
\bra{
\prod_{h=s+1}^{t-1}
\boldsymbol X_h
\bra{
\boldsymbol r_{\dH}\bra{\dtheta}
}
}^{\trans}
\nabla^{\dpsi}
\widehat{\ell}_{\dF}
\bra{
\dH\dF_t;
\boldsymbol r_{\dH}\bra{\dtheta}
}
\Bigg].
\end{align}
Here, the derivative in \eqref{eq:effective_score_weight} acts on every occurrence of $\boldsymbol r_{\dH}\bra{\dtheta}$, including its occurrence through the filtered process, $\boldsymbol X_h$, $\boldsymbol Y_s$, and the derivatives of the likelihood. Thus, the product-rule expansion of $\boldsymbol{\cal S}_{s,T}\bra{\dtheta,\dH}$ contains exactly the linear terms generated by the filter and first-derivative-filter expansions above.

It follows that
\begin{align}
\label{eq:weighted_score_factor_error}
&
\frac{1}{T}
\bra{
\sum_{t=1}^T
\nabla^{\dtheta}
\doublehat{\ell}_{\dF}
\bra{
\dhatFt;
\boldsymbol r_{\widehat\dH}\bra{\dtheta}
}
-
\sum_{t=1}^T
\nabla^{\dtheta}
\widehat{\ell}_{\dF}
\bra{
\dFt\widehat\dH\trans;
\boldsymbol r_{\widehat\dH}\bra{\dtheta}
}
}
\nonumber
\\
={}&
\frac{1}{T}
\sum_{s=1}^{T}
\boldsymbol{\cal S}_{s,T}
\bra{
\dtheta,\widehat\dH
}^{\trans}
\bra{
\dhatF_s-\widehat\dH\dF_s
}
+
O_{\Pp}\bra{
\frac{1}{N}
+
\frac{1}{T}
},
\end{align}
uniformly over $\dtheta\in\dTheta$. By substituting the PC representation in \eqref{eq:evecXX3} into \eqref{eq:weighted_score_factor_error} we therefore obtain
\begin{align}
\label{eq:weighted_score_PC_expansion}
&
\frac{1}{T}
\bra{
\sum_{t=1}^T
\nabla^{\dtheta}
\doublehat{\ell}_{\dF}
\bra{
\dhatFt;
\boldsymbol r_{\widehat\dH}\bra{\dtheta}
}
-
\sum_{t=1}^T
\nabla^{\dtheta}
\widehat{\ell}_{\dF}
\bra{
\dFt\widehat\dH\trans;
\boldsymbol r_{\widehat\dH}\bra{\dtheta}
}
}
\nonumber
\\
={}&
\frac{1}{T}
\sum_{s=1}^{T}
\boldsymbol{\cal S}_{s,T}
\bra{
\dtheta,\widehat\dH
}^{\trans}
\bra{
\frac{\widehat{\boldsymbol M}}{N}
}^{-1}
\bra{
A_s+B_s+C_s+D_s
}
+
O_{\Pp}\bra{
\frac{1}{N}
+
\frac{1}{T}
}.
\end{align}
Now, By Assumption \ref{ass:smooth_domination_score_Hessian}, Lemmas \ref{lemma:rate_psit_first_derivs}$(d)$ and \ref{lemma:rate_psit_second_derivs}$(d)$, and the geometric summability of the products of $\boldsymbol X_t$, it holds that
\begin{align}
\label{eq:effective_score_weight_moment}
\sup_{T\geq1}
\sup_{1\leq s\leq T}
\Ee\sbra{
\sup_{\dH\in\calH}
\sup_{\dtheta\in\dTheta}
\vvbra{
\boldsymbol{\cal S}_{s,T}
\bra{
\dtheta,\dH
}
}^{2}
}
<
\infty.
\end{align}
The same conditions, together with the compactness of $\dTheta$ and $\calH$, imply that $\boldsymbol{\cal S}_{s,T}\bra{\dtheta,\dH}$ is locally Lipschitz in
$\dtheta$ and $\dH$, uniformly over $\dTheta\times\calH$, with a random Lipschitz coefficient having a bounded second moment. Consequently, the arguments used in \eqref{eq:weighted_A_contribution_rate_lemma}, \eqref{eq:weighted_B_contribution_rate_lemma}, \eqref{eq:weighted_C_contribution_rate_lemma}, and \eqref{eq:weighted_D_contribution_rate_lemma} apply with $\boldsymbol{\cal G}_{s,T}$ replaced by $\boldsymbol{\cal S}_{s,T}$, since $d$ is fixed. By combining those bounds with \eqref{eq:weighted_score_PC_expansion} we get
\begin{align}
\label{eq:weighted_score_rate}
&
\sup_{\dtheta\in\dTheta}
\vvbra{
\frac{1}{T}
\sum_{t=1}^T
\nabla^{\dtheta}
\doublehat{\ell}_{\dF}
\bra{
\dhatFt;
\boldsymbol r_{\widehat\dH}\bra{\dtheta}
}
-
\frac{1}{T}
\sum_{t=1}^T
\nabla^{\dtheta}
\widehat{\ell}_{\dF}
\bra{
\dFt\widehat\dH\trans;
\boldsymbol r_{\widehat\dH}\bra{\dtheta}
}
}
\nonumber
\\
={}&
O_{\Pp}\bra{
\frac{1}{\sqrt{NT}}
+
\frac{1}{N}
+
\frac{1}{T}
}.
\end{align}

\bigskip
\noindent
\textbf{Controlling the second term in \eqref{eq:score_decomposition}.}\\
Since $\log\vbra{\dH}$ does not depend on $\dtheta$, differentiating \eqref{eq:log_likelihood_equivariance_total} with respect to $\dtheta$ yields
\begin{align}
\label{eq:score_equivariance_total}
\nabla^{\dtheta}
\ell_{\dF}
\bra{
\dFt\dH\trans;
\boldsymbol r_{\dH}\bra{\dtheta}
}
=
\nabla^{\dtheta}
\ell_{\dF}
\bra{
\dFt;
\dtheta
},
\end{align}
whenever the two filters are initialized equivariantly.

It therefore remains only to control the possibly different filter initializations. Using the same block iteration used in \eqref{eq:rotated_filter_initialization_rate}, applied also to \eqref{eq:first_derivative_filter_recursion}, we obtain, for some
$0<\varrho_k<1$,
\begin{align}
\label{eq:rotated_score_filter_initialization_rate}
&
\sup_{\dH\in\calH}
\sup_{\dtheta\in\dTheta}
\Bigg\{
\vvbra{
\dhatpsi_{t|t-1}
\bra{
\boldsymbol r_{\dH}\bra{\dtheta}
}
-
\dpsi_{t|t-1}
\bra{
\boldsymbol r_{\dH}\bra{\dtheta}
}
}
\nonumber
\\
&\hspace{7em}
+
\vvbra{
\nabla^{\dtheta}
\dhatpsi_{t|t-1}
\bra{
\boldsymbol r_{\dH}\bra{\dtheta}
}
-
\nabla^{\dtheta}
\dpsi_{t|t-1}
\bra{
\boldsymbol r_{\dH}\bra{\dtheta}
}
}
\Bigg\}
=
O_{a.s.}\bra{
\varrho_k^{\lfloor(t-1)/(2k)\rfloor}
}.
\end{align}
Using the score representation in \eqref{eq:score_vector}, the mean value theorem, Assumptions \ref{ass:moment_score} and \ref{ass:smooth_domination_score_Hessian}, and \eqref{eq:rotated_score_filter_initialization_rate}, we obtain
\begin{align*}
&
\sup_{\dH\in\calH}
\sup_{\dtheta\in\dTheta}
\vvbra{
\nabla^{\dtheta}
\widehat{\ell}_{\dF}
\bra{
\dH\dF_t;
\boldsymbol r_{\dH}\bra{\dtheta}
}
-
\nabla^{\dtheta}
\ell_{\dF}
\bra{
\dH\dF_t;
\boldsymbol r_{\dH}\bra{\dtheta}
}
}
=
O_{a.s.}\bra{
\varrho_k^{\lfloor(t-1)/(2k)\rfloor}
},
\end{align*}
after enlarging $\varrho_k$, if necessary, exactly as in \eqref{eq:stationary_likelihood_contribution_geometric_rate}. Consequently,
\begin{align}
\label{eq:rotated_score_initialization_rate}
&
\sup_{\dH\in\calH}
\sup_{\dtheta\in\dTheta}
\vvbra{
\frac{1}{T}
\sum_{t=1}^T
\nabla^{\dtheta}
\widehat{\ell}_{\dF}
\bra{
\dFt\dH\trans;
\boldsymbol r_{\dH}\bra{\dtheta}
}
-
\frac{1}{T}
\sum_{t=1}^T
\nabla^{\dtheta}
\ell_{\dF}
\bra{
\dFt\dH\trans;
\boldsymbol r_{\dH}\bra{\dtheta}
}
}
=
O_{\Pp}\bra{\frac{1}{T}}.
\end{align}
Since $\Pp\bra{\widehat\dH\in\calH}\rightarrow1$ as $N,T\rightarrow\infty$, combining \eqref{eq:score_equivariance_total} and \eqref{eq:rotated_score_initialization_rate}, evaluated at $\dH=\widehat\dH$, yields
\begin{align}
\label{eq:rotated_score_equivariance_rate}
&
\sup_{\dtheta\in\dTheta}
\vvbra{
\frac{1}{T}
\sum_{t=1}^T
\nabla^{\dtheta}
\widehat{\ell}_{\dF}
\bra{
\dFt\widehat\dH\trans;
\boldsymbol r_{\widehat\dH}\bra{\dtheta}
}
-
\frac{1}{T}
\sum_{t=1}^T
\nabla^{\dtheta}
\ell_{\dF}
\bra{
\dFt;
\dtheta
}
}
=
O_{\Pp}\bra{\frac{1}{T}}.
\end{align}

\bigskip
\noindent
\textbf{Controlling the third term in \eqref{eq:score_decomposition}.}\\
By \eqref{eq:stationary_factor_geometric_rate}, \eqref{eq:stationary_filter_geometric_rate}, and Lemma \ref{lemma:rate_psit_first_derivs}$(c)$, after enlarging
$\varrho_k$, if necessary,
\begin{align}
\label{eq:stationary_score_arguments_rate}
&
\vvbra{
\dFt-\dbarFt
}
+
\sup_{\dtheta\in\dTheta}
\vvbra{
\dpsithetatti-\dbarpsitti\bra{\dtheta}
}
\nonumber
\\
&\quad+
\sup_{\dtheta\in\dTheta}
\vvbra{
\nabla^{\dtheta}\dpsithetatti
-
\nabla^{\dtheta}\dbarpsitti\bra{\dtheta}
}
=
O_{a.s.}\bra{
\varrho_k^{\lfloor(t-1)/(2k)\rfloor}
}.
\end{align}
The score representation in \eqref{eq:score_vector}, the mean value theorem, and the derivative envelopes in Assumptions \ref{ass:moment_score} and
\ref{ass:smooth_domination_score_Hessian} therefore imply
\begin{align}
\label{eq:stationary_score_contribution_geometric_rate}
&
\sup_{\dtheta\in\dTheta}
\vvbra{
\nabla^{\dtheta}
\ell_{\dF}
\bra{
\dFt;
\dtheta
}
-
\nabla^{\dtheta}
\ell_{\dF}
\bra{
\dbarFt;
\dtheta
}
}
=
O_{a.s.}\bra{
\varrho_k^{\lfloor(t-1)/(2k)\rfloor}
},
\end{align}
after one further enlargement of $\varrho_k$, using the same Markov and Borel--Cantelli arguments as in \eqref{eq:derivative_envelope_summability}--\eqref{eq:derivative_envelope_geometric_growth}. Since the right-hand side of \eqref{eq:stationary_score_contribution_geometric_rate} is summable,
\begin{align}
\label{eq:stationary_score_rate}
&
\sup_{\dtheta\in\dTheta}
\vvbra{
\frac{1}{T}
\sum_{t=1}^T
\nabla^{\dtheta}
\ell_{\dF}
\bra{
\dFt;
\dtheta
}
-
\frac{1}{T}
\sum_{t=1}^T
\nabla^{\dtheta}
\ell_{\dF}
\bra{
\dbarFt;
\dtheta
}
}
=
O_{\Pp}\bra{\frac{1}{T}}.
\end{align}

\bigskip
\noindent
\textbf{Controlling the fourth term in \eqref{eq:score_decomposition}.}\\
Consider the centered stationary score family
$
\bra{
\nabla^{\dtheta}
\ell_{\dF}
\bra{
\dbarF_t;
\dtheta
}
-
\Ee\sbra{
\nabla^{\dtheta}
\ell_{\dF}
\bra{
\dbarF_t;
\dtheta
}}}$. The uniform contraction argument in Lemma \ref{lemma:NED}, together with the corresponding argument for the first derivative recursion in Lemma \ref{lemma:rate_psit_first_derivs}$(d)$, gives geometrically decreasing approximation coefficients for the jointly truncated process
$
\cbra{
\bra{
\dbarF_t^{(\tau)\trans},
\dbarpsitti^{(\tau)}\bra{\dtheta}^{\trans},
\vect
\nabla^{\dtheta}
\dbarpsitti^{(\tau)}\bra{\dtheta}^{\trans}
}^{\trans}
}_{t\in\Zz},
$
uniformly over $\dtheta\in\dTheta$. The score representation in \eqref{eq:score_vector}, the mean value theorem, and the moment conditions in Assumptions \ref{ass:moment_score} and \ref{ass:smooth_domination_score_Hessian} therefore imply that the centered stationary score family is uniformly near-epoch dependent
with geometrically decreasing approximation coefficients.

Moreover, Assumption \ref{ass:smooth_domination_score_Hessian} and Lemma \ref{lemma:rate_psit_first_derivs}$(d)$ give the required uniform moment and Lipschitz bounds. Hence, the finite-window decomposition and the $\beta$-mixing empirical-process argument used in \eqref{eq:centered_stationary_likelihood_Lipschitz_bound}--\eqref{eq:stationary_likelihood_root_T_maximal_bound}
applies componentwise to the centered stationary score family. Consequently,
\begin{align}
\label{eq:stationary_uniform_score_rate}
&
\sup_{\dtheta\in\dTheta}
\vvbra{
\frac{1}{T}
\sum_{t=1}^T
\nabla^{\dtheta}
\ell_{\dF}
\bra{
\dbarFt;
\dtheta
}
-
\Ee\sbra{
\nabla^{\dtheta}
\ell_{\dF}
\bra{
\dbarFt;
\dtheta
}
}
}
=
O_{\Pp}\bra{\frac{1}{\sqrt T}}.
\end{align}
The combination of \eqref{eq:weighted_score_rate}, \eqref{eq:rotated_score_equivariance_rate}, \eqref{eq:stationary_score_rate}, and \eqref{eq:stationary_uniform_score_rate} shows that
\begin{align*}
&
\sup_{\dtheta\in\dTheta}
\vvbra{
\frac{1}{T}
\sum_{t=1}^T
\nabla^{\dtheta}
\doublehat{\ell}_{\dF}
\bra{
\dhatFt;
\boldsymbol r_{\widehat\dH}\bra{\dtheta}
}
-
\Ee\sbra{
\nabla^{\dtheta}
\ell_{\dF}
\bra{
\dbarFt;
\dtheta
}
}
}
\\
={}&
O_{\Pp}\bra{
\frac{1}{\sqrt T}
+
\frac{1}{\sqrt{NT}}
+
\frac{1}{N}
+
\frac{1}{T}
}
=
O_{\Pp}\bra{
\frac{1}{\sqrt T}
+
\frac{1}{N}
}.
\end{align*}
This proves \eqref{eq:uniform_score_population_rate}.

Combining only \eqref{eq:weighted_score_rate}, \eqref{eq:rotated_score_equivariance_rate}, and \eqref{eq:stationary_score_rate} we get
\begin{align*}
&
\frac{1}{\sqrt T}
\sup_{\dtheta\in\dTheta}
\vvbra{
\sum_{t=1}^T
\nabla^{\dtheta}
\doublehat{\ell}_{\dF}
\bra{
\dhatFt;
\boldsymbol r_{\widehat\dH}\bra{\dtheta}
}
-
\sum_{t=1}^T
\nabla^{\dtheta}
\ell_{\dF}
\bra{
\dbarFt;
\dtheta
}
}
\\
={}&
O_{\Pp}\bra{
\frac{1}{\sqrt N}
+
\frac{\sqrt T}{N}
+
\frac{1}{\sqrt T}
},
\end{align*}
which proves \eqref{eq:uniform_score_sample_approximation}. If $\sqrt T/N\rightarrow0$, the right-hand side is $o_{\Pp}(1)$ as $N,T\rightarrow\infty$, proving
\eqref{eq:uniform_score_root_T_equivalence}.

Finally, evaluating the preceding truncated process argument at $\dtheta=\dtheta_0$ shows that the true score process
$
\cbra{
\nabla^{\dtheta}
\ell_{\dF}
\bra{
\dbarF_t;
\dtheta_0
}
}_{t\in\Zz}$
is near-epoch dependent of size $-1$ on the $\beta$-mixing process $\cbra{\dvt}_{t\in\Zz}$. Since $\dtheta_0$ is an interior maximizer of the stationary population criterion, differentiation under the expectation implies that
$
\Ee\sbra{
\nabla^{\dtheta}
\ell_{\dF}
\bra{
\dbarF_t;
\dtheta_0
}
}
=
\boldsymbol 0_d.
$
Thus, the stationary score process is zero mean and near-epoch dependent of size $-1$ on the geometrically $\beta$-mixing process $\cbra{\dvt}_{t\in\Zz}$.
\end{proof}


\begin{lemma}
\label{lemma:ULLN_Hessian_matrix}
Consider the Hessian matrix 
\begin{align}
\label{eq:Hessian_matrix}
\nabla^{\dtheta\dtheta}
\ell_{\dF}
\bra{
\dFt;\dtheta
}
={}&
\nabla^{\dtheta}
\dpsithetatti\trans
\nabla^{\dpsi\dpsi}
\ell_{\dF}
\bra{
\dFt;\dtheta
}
\nabla^{\dtheta}
\dpsithetatti
+
\bra{
\boldsymbol I_d
\otimes 
\nabla^{\dpsi}
\ell_{\dF}
\bra{
\dFt;\dtheta
}^{\trans}
}
\nabla^{\dtheta\dtheta}
\dpsithetatti
\nonumber
\\
&+
\nabla^{\dtheta}
\dpsithetatti\trans
\frac{
\partial^2
\ell_{\dF}
\bra{
\dFt;\dtheta
}
}{
\partial\dpsitti
\partial\dtheta\trans
}
+
\frac{
\partial^2
\ell_{\dF}
\bra{
\dFt;\dtheta
}
}{
\partial\dtheta
\partial\dpsitti\trans
}
\nabla^{\dtheta}
\dpsithetatti
+
\frac{
\partial^2
\ell_{\dF}
\bra{
\dFt;\dtheta
}
}{
\partial\dtheta
\partial\dtheta\trans
},
\end{align}
where all partial derivatives in \eqref{eq:Hessian_matrix} are computed holding $\dpsitti$ and $\dFt$ fixed and is subsequently evaluated at $\dpsitti=\dpsitti\bra{\dtheta}$.

Under Assumptions \ref{ass:noise}--\ref{ass:truePar}, as $T\rightarrow\infty$,
\begin{align}
\label{eq:uniform_Hessian_ULLN}
\sup_{\dtheta\in\dTheta}
\vvbra{
\frac{1}{T}
\sum_{t=1}^T
\nabla^{\dtheta\dtheta}
\ell_{\dF}
\bra{
\dFt;\dtheta
}
-
\Ee\sbra{
\nabla^{\dtheta\dtheta}
\ell_{\dF}
\bra{
\dbarFt;\dtheta
}
}
}
\asconv
0.
\end{align}
Moreover, as $T\rightarrow\infty$,
\begin{align}
\label{eq:information_matrix_limit}
-
\Ee\sbra{
\nabla^{\dtheta\dtheta}
\ell_{\dF}
\bra{
\dFt;\dtheta_0
}
}
\longrightarrow
\limitcov\bra{\dtheta_0}
=
-
\Ee\sbra{
\nabla^{\dtheta\dtheta}
\ell_{\dF}
\bra{
\dbarF_t;\dtheta_0
}
},
\end{align}
where $\limitcov\bra{\dtheta_0}$ is positive definite and therefore invertible.
\end{lemma}

\begin{proof}
By \eqref{eq:stationary_factor_geometric_rate},
\eqref{eq:stationary_filter_geometric_rate}, Lemma
\ref{lemma:rate_psit_first_derivs}$(c)$, and Lemma
\ref{lemma:rate_psit_second_derivs}$(c)$, after enlarging
$\varrho_k$, if necessary,
\begin{align}
\label{eq:stationary_Hessian_arguments_rate}
&
\vvbra{
\dFt-\dbarFt
}
+
\sup_{\dtheta\in\dTheta}
\vvbra{
\dpsithetatti-\dbarpsitti\bra{\dtheta}
}
\nonumber
\\
&\quad+
\sup_{\dtheta\in\dTheta}
\vvbra{
\nabla^{\dtheta}\dpsithetatti
-
\nabla^{\dtheta}\dbarpsitti\bra{\dtheta}
}
\nonumber
\\
&\quad+
\sup_{\dtheta\in\dTheta}
\vvbra{
\nabla^{\dtheta\dtheta}\dpsithetatti
-
\nabla^{\dtheta\dtheta}\dbarpsitti\bra{\dtheta}
}
=
O_{a.s.}\bra{
\varrho_k^{\lfloor(t-1)/(4k)\rfloor}
}.
\end{align}
By Assumption \ref{ass:smooth_domination_score_Hessian}, the mean value theorem, and \eqref{eq:stationary_Hessian_arguments_rate} we therefore get
\begin{align}
\label{eq:stationary_Hessian_contribution_geometric_rate}
&
\sup_{\dtheta\in\dTheta}
\vvbra{
\nabla^{\dtheta\dtheta}
\ell_{\dF}
\bra{
\dFt;\dtheta
}
-
\nabla^{\dtheta\dtheta}
\ell_{\dF}
\bra{
\dbarF_t;\dtheta
}
}
=
O_{a.s.}\bra{
\varrho_k^{\lfloor(t-1)/(4k)\rfloor}
},
\end{align}
after one further enlargement of $\varrho_k$, if necessary, using the same Markov and Borel--Cantelli arguments employed in \eqref{eq:derivative_envelope_summability}--\eqref{eq:derivative_envelope_geometric_growth}.

Since the right-hand side of \eqref{eq:stationary_Hessian_contribution_geometric_rate} is summable, we obtain
\begin{align}
\label{eq:stationary_Hessian_initialization_rate}
&
\sup_{\dtheta\in\dTheta}
\vvbra{
\frac{1}{T}
\sum_{t=1}^T
\nabla^{\dtheta\dtheta}
\ell_{\dF}
\bra{
\dFt;\dtheta
}
-
\frac{1}{T}
\sum_{t=1}^T
\nabla^{\dtheta\dtheta}
\ell_{\dF}
\bra{
\dbarFt;\dtheta
}
}
=
O_{a.s.}\bra{
\frac{1}{T}
}.
\end{align}
By Lemmas \ref{lemma:NED}, \ref{lemma:rate_psit_first_derivs}$(d)$, and \ref{lemma:rate_psit_second_derivs}$(d)$, the process
$\cbra{
\nabla^{\dtheta\dtheta}
\ell_{\dF}\bra{\dbarF_t;\dtheta}
}_{t\in\Zz}$
is strictly stationary and ergodic for every $\dtheta\in\dTheta$. Moreover, Assumption \ref{ass:smooth_domination_score_Hessian} implies that it is almost
surely continuous in $\dtheta$ and satisfies
\begin{align*}
\Ee\sbra{
\sup_{\dtheta\in\dTheta}
\vvbra{
\nabla^{\dtheta\dtheta}
\ell_{\dF}
\bra{
\dbarF_t;\dtheta
}
}
}
<\infty.
\end{align*}
Since $\dTheta$ is compact, the uniform ergodic law of large numbers implies that
\begin{align}
\label{eq:stationary_Hessian_ULLN}
&
\sup_{\dtheta\in\dTheta}
\vvbra{
\frac{1}{T}
\sum_{t=1}^T
\nabla^{\dtheta\dtheta}
\ell_{\dF}
\bra{
\dbarFt;\dtheta
}
-
\Ee\sbra{
\nabla^{\dtheta\dtheta}
\ell_{\dF}
\bra{
\dbarFt;\dtheta
}
}
}
\asconv
0
\end{align}
as $T\rightarrow\infty$.

An application of the triangle inequality then shows
\begin{align*}
&
\sup_{\dtheta\in\dTheta}
\vvbra{
\frac{1}{T}
\sum_{t=1}^T
\nabla^{\dtheta\dtheta}
\ell_{\dF}
\bra{
\dFt;\dtheta
}
-
\Ee\sbra{
\nabla^{\dtheta\dtheta}
\ell_{\dF}
\bra{
\dbarFt;\dtheta
}
}
}
\\
\leq{}&
\sup_{\dtheta\in\dTheta}
\vvbra{
\frac{1}{T}
\sum_{t=1}^T
\nabla^{\dtheta\dtheta}
\ell_{\dF}
\bra{
\dFt;\dtheta
}
-
\frac{1}{T}
\sum_{t=1}^T
\nabla^{\dtheta\dtheta}
\ell_{\dF}
\bra{
\dbarFt;\dtheta
}
}
\\
&+
\sup_{\dtheta\in\dTheta}
\vvbra{
\frac{1}{T}
\sum_{t=1}^T
\nabla^{\dtheta\dtheta}
\ell_{\dF}
\bra{
\dbarFt;\dtheta
}
-
\Ee\sbra{
\nabla^{\dtheta\dtheta}
\ell_{\dF}
\bra{
\dbarFt;\dtheta
}
}
}.
\end{align*}
Equations \eqref{eq:stationary_Hessian_initialization_rate} and \eqref{eq:stationary_Hessian_ULLN} prove \eqref{eq:uniform_Hessian_ULLN} as $T\rightarrow\infty$.

Finally, the corresponding $L_1$ version of \eqref{eq:stationary_Hessian_contribution_geometric_rate} is given by
\begin{align*}
&
\vvbra{
\Ee\sbra{
\nabla^{\dtheta\dtheta}
\ell_{\dF}
\bra{
\dFt;\dtheta_0
}
-
\nabla^{\dtheta\dtheta}
\ell_{\dF}
\bra{
\dbarFt;\dtheta_0
}
}
}
=
O\bra{
\frac{1}{T}
}.
\end{align*}
By stationarity,
\begin{align*}
-
\Ee\sbra{
\nabla^{\dtheta\dtheta}
\ell_{\dF}
\bra{
\dbarFt;\dtheta_0
}
}
=
-
\Ee\sbra{
\nabla^{\dtheta\dtheta}
\ell_{\dF}
\bra{
\dbarF_t;\dtheta_0
}
}
=
\limitcov\bra{\dtheta_0}.
\end{align*}
These two results prove \eqref{eq:information_matrix_limit}. Finally, $\limitcov\bra{\dtheta_0}$ is positive definite by Assumption \ref{ass:uniformly_pos_def} and is therefore invertible.\end{proof}


\begin{lemma}
\label{lemma:CLT_NED_score_vector}
Under Assumptions \ref{ass:noise}--\ref{ass:truePar}, with $p>2$, for every fixed initial condition $\dpsi_{1|0}\in\calP$, define
\begin{align}
\label{eq:stationary_score_variance}
\dV_T\bra{\dtheta_0}
=
\sum_{t=1}^{T}
\sum_{s=1}^{T}
\Ee\sbra{
\nabla^{\dtheta}
\ell_{\dF}
\bra{
\dbarF_t;\dtheta_0
}
\nabla^{\dtheta}
\ell_{\dF}
\bra{
\dbarF_s;\dtheta_0
}^{\trans}
}.
\end{align}
Then, as $T\rightarrow\infty$,
\begin{align}
\label{eq:long_run_score_variance}
\frac{1}{T}
\dV_T\bra{\dtheta_0}
\longrightarrow
\dV\bra{\dtheta_0},
\end{align}
where
\begin{align}
\label{eq:long_run_score_covariance_sum}
\dV\bra{\dtheta_0}
=
\sum_{h=-\infty}^{\infty}
\Ee\sbra{
\nabla^{\dtheta}
\ell_{\dF}
\bra{
\dbarF_t;\dtheta_0
}
\nabla^{\dtheta}
\ell_{\dF}
\bra{
\dbarF_{t-h};\dtheta_0
}^{\trans}
}
\end{align}
exists and is finite. Moreover, as $T\rightarrow\infty$,
\begin{align}
\label{eq:initialized_score_CLT}
\frac{1}{\sqrt T}
\sum_{t=1}^T
\nabla^{\dtheta}
\ell_{\dF}
\bra{
\dFt;\dtheta_0
}
\Rightarrow
\calN
\bra{
\boldsymbol 0_d,
\dV\bra{\dtheta_0}
}.
\end{align}
\end{lemma}

\begin{proof}
By Lemma \ref{lemma:U_convergence_score_vector}, the stationary  score process $\cbra{ \nabla^{\dtheta}\ell_{\dF}\bra{\dbarF_t;\dtheta_0
}}_{t\in\Zz}$ has mean zero and is near-epoch dependent of size $-1$ on the geometrically $\beta$-mixing process $\cbra{\dvt}_{t\in\Zz}$. Moreover, Assumption \ref{ass:smooth_domination_score_Hessian} and Lemma \ref{lemma:rate_psit_first_derivs}$(d)$ imply that
\begin{align}
\label{eq:stationary_score_pth_moment}
\Ee\sbra{
\vvbra{
\nabla^{\dtheta}
\ell_{\dF}
\bra{
\dbarF_t;\dtheta_0
}
}^{p}
}
<\infty.
\end{align}
To establish the absolute summability of the score autocovariances, let
\[
\boldsymbol s_t
=
\nabla^{\dtheta}
\ell_{\dF}
\bra{
\dbarF_t;\dtheta_0
},
\]
and let $\boldsymbol s_t^{(\tau)}$ denote the corresponding score contribution computed from the $\tau$-truncated factor, filter, and first-derivative filter processes. By construction, $\boldsymbol s_t^{(\tau)}$ is measurable with respect to $\sigma\bra{\dvt,\ldots,\dv_{t-\tau}}$. The truncated-process argument
used in the proof of Lemma \ref{lemma:U_convergence_score_vector}, together with Assumption \ref{ass:smooth_domination_score_Hessian}, gives, for some finite
positive real $C$ and some $0<\varrho_k<1$,
\begin{align}
\label{eq:stationary_score_truncation_bound}
\sup_{t\in\Zz}
\vvbra{
\boldsymbol s_t-\boldsymbol s_t^{(\tau)}
}_2
&\leq
C\varrho_k^{\lfloor\tau/k\rfloor},
\\
\label{eq:truncated_stationary_score_moment}
\sup_{\tau\geq1}
\sup_{t\in\Zz}
\vvbra{
\boldsymbol s_t^{(\tau)}
}_p
&\leq C.
\end{align}
For all sufficiently large $h$, set $\tau=\lfloor h/3\rfloor$. The
truncated variables $\boldsymbol s_t^{(\tau)}$ and
$\boldsymbol s_{t-h}^{(\tau)}$ are measurable with respect to $\sigma\bra{\dvt,\ldots,\dv_{t-\tau}}$ and $\sigma\bra{\dv_{t-h},\ldots,\dv_{t-h-\tau}}$, respectively. These two innovation blocks are separated by $h-\tau$ periods, and
$
h-\tau
=
h-\lfloor h/3\rfloor
\geq
\lfloor h/3\rfloor.
$
Thus, the choice $\tau=\lfloor h/3\rfloor$ ensures that both the truncation depth and the separation between the resulting innovation
blocks increase proportionally with $h$.

Using the covariance decomposition
\begin{align*}
\operatorname{Cov}
\bra{
\boldsymbol s_t,\boldsymbol s_{t-h}
}
=
\operatorname{Cov}
\bra{
\boldsymbol s_t-\boldsymbol s_t^{(\tau)},
\boldsymbol s_{t-h}
}
+
\operatorname{Cov}
\bra{
\boldsymbol s_t^{(\tau)},
\boldsymbol s_{t-h}-\boldsymbol s_{t-h}^{(\tau)}
}
+
\operatorname{Cov}
\bra{
\boldsymbol s_t^{(\tau)},
\boldsymbol s_{t-h}^{(\tau)}
},
\end{align*}
the Cauchy--Schwarz inequality, \eqref{eq:stationary_score_pth_moment}, and \eqref{eq:stationary_score_truncation_bound} give
\begin{align}
\label{eq:stationary_score_covariance_truncation_terms}
&
\vvbra{
\operatorname{Cov}
\bra{
\boldsymbol s_t-\boldsymbol s_t^{(\tau)},
\boldsymbol s_{t-h}
}
}
+
\vvbra{
\operatorname{Cov}
\bra{
\boldsymbol s_t^{(\tau)},
\boldsymbol s_{t-h}-\boldsymbol s_{t-h}^{(\tau)}
}
}
\leq
C\varrho_k^{\lfloor h/(3k)\rfloor}.
\end{align}
Moreover, every $\beta$-mixing process is $\alpha$-mixing, with $\alpha_{\dv}(h)\leq\beta_{\dv}(h)$. Hence, by using the covariance inequality
for $\alpha$-mixing random variables of \cite{Davydov1968}, with $q=p$ and $1/r=(p-2)/p$, coordinatewise to the truncated score vectors, whose underlying
innovation blocks are separated by $h-\tau$ periods, and using $\alpha_{\dv}(h)\leq\beta_{\dv}(h)$ together with \eqref{eq:truncated_stationary_score_moment}, we obtain 
\begin{align*}
\vvbra{
\operatorname{Cov}
\bra{
\boldsymbol s_t^{(\tau)},
\boldsymbol s_{t-h}^{(\tau)}
}
}
\leq
C
\beta_{\dv}\bra{h-\tau}^{\frac{p-2}{p}}
\vvbra{
\boldsymbol s_t^{(\tau)}
}_p
\vvbra{
\boldsymbol s_{t-h}^{(\tau)}
}_p.
\end{align*}
By \eqref{eq:truncated_stationary_score_moment}, the two $L_p$ norms are uniformly bounded. Since $\tau=\lfloor h/3\rfloor$ and the mixing coefficients are nonincreasing, $\beta_{\dv}\bra{h-\tau}\leq\beta_{\dv}\bra{\lfloor h/3\rfloor}$, and hence
\begin{align}
\label{eq:stationary_score_covariance_mixing_term}
\vvbra{
\operatorname{Cov}
\bra{
\boldsymbol s_t^{(\tau)},
\boldsymbol s_{t-h}^{(\tau)}
}
}
\leq
C
\beta_{\dv}
\bra{
\lfloor h/3\rfloor
}^{\frac{p-2}{p}}.
\end{align}
Since $\Ee\sbra{\boldsymbol s_t}=\boldsymbol 0_d$, $\operatorname{Cov}\bra{\boldsymbol s_t,\boldsymbol s_{t-h}}=\Ee\sbra{\boldsymbol s_t\boldsymbol s_{t-h}^{\trans}}$ and therefore, combining \eqref{eq:stationary_score_covariance_truncation_terms} and \eqref{eq:stationary_score_covariance_mixing_term} yields
\begin{align}
\label{eq:stationary_score_covariance_bound}
&
\vvbra{
\Ee\sbra{
\nabla^{\dtheta}
\ell_{\dF}
\bra{
\dbarF_t;\dtheta_0
}
\nabla^{\dtheta}
\ell_{\dF}
\bra{
\dbarF_{t-h};\dtheta_0
}^{\trans}
}
}
\nonumber
\\
\leq{}&
C
\bra{
\varrho_k^{\lfloor h/(3k)\rfloor}
+
\beta_{\dv}
\bra{
\lfloor h/3\rfloor
}^{\frac{p-2}{p}}
}.
\end{align}
By Assumption \ref{ass:noise}$(i)$, $\beta_{\dv}(h)$ decreases geometrically, and thus, it clearly holds that $\sum_{h=1}^{\infty}\varrho_k^{\lfloor h/(3k)\rfloor}<\infty,$ and $\sum_{h=1}^{\infty}\beta_{\dv}\bra{\lfloor h/3\rfloor}^{\frac{p-2}{p}}<\infty.$ The covariance at $h=0$ is finite by
\eqref{eq:stationary_score_pth_moment}, while stationarity gives the same bound for negative lags. Consequently,
\begin{align}
\label{eq:stationary_score_absolute_covariance_summability}
\sum_{h=-\infty}^{\infty}
\vvbra{
\Ee\sbra{
\nabla^{\dtheta}
\ell_{\dF}
\bra{
\dbarF_t;\dtheta_0
}
\nabla^{\dtheta}
\ell_{\dF}
\bra{
\dbarF_{t-h};\dtheta_0
}^{\trans}
}
}
<\infty.
\end{align}
By stationarity and the zero-mean property of the stationary score we further obtain
\begin{align*}
\frac{1}{T}
\dV_T\bra{\dtheta_0}
={}&
\sum_{h=-(T-1)}^{T-1}
\bra{
1-\frac{\vbra{h}}{T}
}
\Ee\sbra{
\nabla^{\dtheta}
\ell_{\dF}
\bra{
\dbarF_t;\dtheta_0
}
\nabla^{\dtheta}
\ell_{\dF}
\bra{
\dbarF_{t-h};\dtheta_0
}^{\trans}
}.
\end{align*}
Therefore, \eqref{eq:stationary_score_absolute_covariance_summability} and the dominated convergence theorem imply \eqref{eq:long_run_score_variance} as $T\rightarrow\infty$, with $\dV\bra{\dtheta_0}$ given by \eqref{eq:long_run_score_covariance_sum}.

Now fix any $\boldsymbol a\in\Rr^d$. The scalar process
$
\cbra{
\boldsymbol a\trans
\nabla^{\dtheta}
\ell_{\dF}
\bra{
\dbarF_t;\dtheta_0
}
}_{t\in\Zz}
$ has mean zero, has a bounded $p$-th moment, and is near-epoch dependent of size $-1$ on the geometrically $\beta$-mixing process $\cbra{\dvt}_{t\in\Zz}$. Since every $\beta$-mixing process is also $\alpha$-mixing and its $\alpha$-mixing coefficients are bounded by its $\beta$-mixing coefficients, Assumption \ref{ass:noise}$(i)$ implies that $\cbra{\dvt}_{t\in\Zz}$ is $\alpha$-mixing of size $-2p/(p-2)$. Hence, by Theorem 10.2 of
\citet{Potscher1997} we obtain
\begin{align*}
\frac{1}{\sqrt T}
\sum_{t=1}^T
\boldsymbol a\trans
\nabla^{\dtheta}
\ell_{\dF}
\bra{
\dbarFt;\dtheta_0
}
\Rightarrow
\calN
\bra{
0,
\boldsymbol a\trans
\dV\bra{\dtheta_0}
\boldsymbol a
}
\end{align*}
as $T\rightarrow\infty$. If $\boldsymbol a\trans\dV\bra{\dtheta_0}\boldsymbol a=0$, then, since the stationary score has mean zero,
\begin{align*}
\Vv\sbra{
\frac{1}{\sqrt T}
\sum_{t=1}^T
\boldsymbol a\trans
\nabla^{\dtheta}
\ell_{\dF}
\bra{
\dbarFt;\dtheta_0
}
}
&=
\boldsymbol a\trans
\frac{\dV_T\bra{\dtheta_0}}{T}
\boldsymbol a
\longrightarrow
\boldsymbol a\trans
\dV\bra{\dtheta_0}
\boldsymbol a
=
0
\end{align*}
as $T\rightarrow\infty$, by \eqref{eq:long_run_score_variance}. Therefore, for every $\varepsilon>0$, the Chebyshev's inequality implies
\begin{align*}
&
\Pp\bra{
\left|
\frac{1}{\sqrt T}
\boldsymbol a\trans
\nabla^{\dtheta}
\ell_{\dF}
\bra{
\dbarF;\dtheta_0
}
\right|
>
\varepsilon
}
\leq
\frac{1}{\varepsilon^2}
\boldsymbol a\trans
\frac{\dV_T\bra{\dtheta_0}}{T}
\boldsymbol a
\longrightarrow
0
\end{align*}
as $T\rightarrow\infty$. Hence, $\frac{1}{\sqrt T}
\sum_{t=1}^T
\boldsymbol a\trans
\nabla^{\dtheta}
\ell_{\dF}
\bra{
\dbarFt;\dtheta_0
}
\pconv
0$, and therefore it converges in distribution to the degenerate normal distribution $\calN\bra{0,0}$. The Cramér--Wold device consequently yields
\begin{align}
\label{eq:stationary_score_CLT}
\frac{1}{\sqrt T}
\sum_{t=1}^T
\nabla^{\dtheta}
\ell_{\dF}
\bra{
\dbarFt;\dtheta_0
}
\Rightarrow
\calN
\bra{
\boldsymbol 0_d,
\dV\bra{\dtheta_0}
}
\end{align}
as $T\rightarrow\infty$.

Finally, by evaluating \eqref{eq:stationary_score_rate} at $\dtheta=\dtheta_0$ we get
\begin{align*}
&
\frac{1}{\sqrt T}
\vvbra{
\sum_{t=1}^T
\nabla^{\dtheta}
\ell_{\dF}
\bra{
\dFt;\dtheta_0
}
-
\sum_{t=1}^T
\nabla^{\dtheta}
\ell_{\dF}
\bra{
\dbarFt;\dtheta_0
}
}
\\
={}&
\sqrt T
\vvbra{
\frac{1}{T}
\sum_{t=1}^T
\nabla^{\dtheta}
\ell_{\dF}
\bra{
\dFt;\dtheta_0
}
-
\frac{1}{T}
\sum_{t=1}^T
\nabla^{\dtheta}
\ell_{\dF}
\bra{
\dbarFt;\dtheta_0
}
}
=
O_{\Pp}\bra{
\frac{1}{\sqrt T}
}
,
\end{align*}
and thus, combining this result with \eqref{eq:stationary_score_CLT} and applying Slutsky's theorem proves \eqref{eq:initialized_score_CLT}.
\end{proof}





\newpage

\section{Additional Monte Carlo studies}

\subsection{Trace statistics}\label{appendix_MC}
We measure the performance of the estimated unobservable process $\big\{\dhathatpsithetatti\big\}_{t\in\Nn}$ by means of the following trace statistics, which is not affected by the identification indeterminacy of the factors:
\begin{equation*}
TR_{MLE}
=
\frac{\trace \bra{\dpsi\trans\dhathatpsi(\dthetamle)\bra{\dhathatpsi(\dthetamle)\trans\dhathatpsi(\dthetamle)}\inv\dhathatpsi(\dthetamle)\trans\dpsi}}
{\trace \bra{\dpsi\trans\dpsi}}
,
\end{equation*}
where $\dpsi = \big(\dpsi_{1|0}, \dots, \dpsi_{T|T-1}\big)\trans$ and $\dhathatpsi(\dthetamle) = \big(\dhathatpsi_{1|0}(\dthetamle),\dots, \dhathatpsi_{T|T-1}(\dthetamle)\big)\trans$ denote the true and filtered (which is also a function of the MLE) time-varying parameter processes, respectively. The closer $TR_{MLE}$ is to one, the closer the space spanned  by $\dhathatpsi(\dthetamle)$ is to the space spanned by $\dpsi$. Results are in Table \ref{tab:sim_cond_loc_var_pca}.

\begin{table}[!ht]
	\centering
	\caption{Monte Carlo simulation results for the filtered conditional
		locations and variances}
	\label{tab:sim_cond_loc_var_pca}
	\scriptsize
	\begingroup
	\setlength{\tabcolsep}{5pt}
	\begin{tabular}{@{}lccrrrr@{}}
		\toprule
		& & &
		\multicolumn{2}{c}{Conditional locations} &
		\multicolumn{2}{c}{Conditional variances} \\
		\cmidrule(lr){4-5}\cmidrule(lr){6-7}
		Distribution & $N$ & $T$
		& One factor & Three factors
		& One factor & Three factors \\
		\midrule
		Gaussian
		& 100 & 200
		& 0.9334 & 0.9248
		& 0.9778 & 0.9515 \\
		& 100 & 500
		& 0.9361 & 0.9359
		& 0.9859 & 0.9730 \\
		& 200 & 200
		& 0.9245 & 0.9346
		& 0.9758 & 0.9578 \\
		& 200 & 500
		& 0.9475 & 0.9552
		& 0.9932 & 0.9794 \\
		& 500 & 200
		& 0.9320 & 0.9343
		& 0.9785 & 0.9608 \\
		& 500 & 500
		& 0.9426 & 0.9531
		& 0.9886 & 0.9815 \\
		& 500 & 1000
		& 0.9559 & 0.9672
		& 0.9946 & 0.9915 \\
		\addlinespace
		
		Student's $t_4$
		& 100 & 200
		& 0.9087 & 0.8412
		& 0.9628 & 0.9368 \\
		& 100 & 500
		& 0.9190 & 0.8410
		& 0.9819 & 0.9563 \\
		& 200 & 200
		& 0.9118 & 0.9243
		& 0.9626 & 0.9439 \\
		& 200 & 500
		& 0.9229 & 0.9242
		& 0.9799 & 0.9661 \\
		& 500 & 200
		& 0.9152 & 0.9254
		& 0.9625 & 0.9454 \\
		& 500 & 500
		& 0.9342 & 0.9418
		& 0.9770 & 0.9660 \\
		& 500 & 1000
		& 0.9331 & 0.9417
		& 0.9868 & 0.9828 \\
		\addlinespace
		
		Student's $t_5$
		& 100 & 200
		& 0.9068 & 0.9161
		& 0.9653 & 0.9377 \\
		& 100 & 500
		& 0.9171 & 0.9185
		& 0.9782 & 0.9579 \\
		& 200 & 200
		& 0.9225 & 0.9195
		& 0.9652 & 0.9441 \\
		& 200 & 500
		& 0.9203 & 0.9304
		& 0.9799 & 0.9671 \\
		& 500 & 200
		& 0.9265 & 0.9355
		& 0.9740 & 0.9458 \\
		& 500 & 500
		& 0.9284 & 0.9318
		& 0.9817 & 0.9710 \\
		& 500 & 1000
		& 0.9434 & 0.9380
		& 0.9913 & 0.9848 \\
		\bottomrule
	\end{tabular}
	\endgroup
	
	\vspace{0.4em}
	\parbox{0.96\linewidth}{\footnotesize
		Monte Carlo averages of $TR_{\mathrm{MLE}}$ for the filtered vectors
		of conditional locations and conditional variances. Results are based
		on 100 replications.}
\end{table}

\clearpage
\subsection{Results when using spatial Kendall's PCs as in \citet{he2020elliptic}}\label{app:kendall}


\begin{table}[!ht]
	\centering
	\caption{Monte Carlo simulation results for the QMLE of the conditional
		location and variance models using spatial Kendall's tau factor estimates}
	\label{tab:sim_qmle_kendall_t4}
	\scriptsize
	\begingroup
	\setlength{\tabcolsep}{3pt}
	\begin{tabular}{@{}lccrrrrrrrr@{}}
		\toprule
		& & &
		\multicolumn{4}{c}{Conditional location model} &
		\multicolumn{4}{c}{Conditional variance model} \\
		\cmidrule(lr){4-7}\cmidrule(lr){8-11}
		& & &
		\multicolumn{2}{c}{One factor} &
		\multicolumn{2}{c}{Three factors} &
		\multicolumn{2}{c}{One factor} &
		\multicolumn{2}{c}{Three factors} \\
		\cmidrule(lr){4-5}\cmidrule(lr){6-7}
		\cmidrule(lr){8-9}\cmidrule(lr){10-11}
		Distribution & $N$ & $T$
		& Bias$^2$ & Var.
		& Bias$^2$ & Var.
		& Bias$^2$ & Var.
		& Bias$^2$ & Var. \\
		\midrule
		Student's $t_4$
		& 100 & 200
		& 0.00040 & 0.00333 & 0.00107 & 0.00403
		& 0.00246 & 0.01545 & 0.01336 & 0.03907 \\
		& 100 & 500
		& 0.00008 & 0.00237 & 0.00018 & 0.00332
		& 0.00048 & 0.00893 & 0.00197 & 0.01296 \\
		& 200 & 200
		& 0.00035 & 0.00327 & 0.00117 & 0.00605
		& 0.00040 & 0.01174 & 0.01232 & 0.03887 \\
		& 200 & 500
		& 0.00007 & 0.00247 & 0.00047 & 0.00338
		& 0.00052 & 0.01073 & 0.00225 & 0.01375 \\
		& 500 & 200
		& 0.00024 & 0.00365 & 0.00151 & 0.02293
		& 0.00092 & 0.01525 & 0.00983 & 0.03512 \\
		& 500 & 500
		& 0.00004 & 0.00212 & 0.00077 & 0.01423
		& 0.00037 & 0.00847 & 0.00148 & 0.01076 \\
		& 500 & 1000
		& 0.00001 & 0.00184 & 0.00013 & 0.00508
		& 0.00040 & 0.00450 & 0.00045 & 0.00331 \\
		\bottomrule
	\end{tabular}
	\endgroup
	
	\vspace{0.4em}
	\parbox{0.98\linewidth}{\footnotesize
		Monte Carlo averages of the squared bias and variance of the QMLE under
		Student's $t_4$ innovations, using spatial Kendall's tau factor estimates in the
		first stage. For the conditional location models, entries are arithmetic
		means of the coordinate-specific values for $\operatorname{vec}(\dPhi)$ and
		$\operatorname{vec}(\dK)$. For the conditional variance models, the averages
		additionally include the intercept parameters $\ddelta$. Results are based on
		100 replications.}
\end{table}

\begin{table}[!ht]
	\centering
	\caption{Monte Carlo simulation results for the filtered conditional
		locations and variances using spatial Kendall's tau factor estimates}
	\label{tab:sim_filter_kendall_t4}
	\scriptsize
	\begingroup
	\setlength{\tabcolsep}{5pt}
	\begin{tabular}{@{}lccrrrr@{}}
		\toprule
		& & &
		\multicolumn{2}{c}{Conditional locations} &
		\multicolumn{2}{c}{Conditional variances} \\
		\cmidrule(lr){4-5}\cmidrule(lr){6-7}
		Distribution & $N$ & $T$
		& One factor & Three factors
		& One factor & Three factors \\
		\midrule
		Student's $t_4$
		& 100 & 200 & 0.9688 & 0.9128 & 0.9760 & 0.9389 \\
		& 100 & 500 & 0.9694 & 0.9147 & 0.9821 & 0.9576 \\
		& 200 & 200 & 0.9730 & 0.9246 & 0.9780 & 0.9434 \\
		& 200 & 500 & 0.9753 & 0.9199 & 0.9830 & 0.9647 \\
		& 500 & 200 & 0.9736 & 0.9252 & 0.9785 & 0.9447 \\
		& 500 & 500 & 0.9805 & 0.9368 & 0.9841 & 0.9732 \\
		& 500 & 1000 & 0.9811 & 0.9412 & 0.9906 & 0.9856 \\
		\bottomrule
	\end{tabular}
	\endgroup
	
	\vspace{0.4em}
	\parbox{0.96\linewidth}{\footnotesize
		Monte Carlo averages of $TR_{\mathrm{MLE}}$ for the filtered vectors of
		conditional locations and conditional variances under Student's $t_4$
		innovations, using spatial Kendall's tau factor estimates in the first stage.
		Results are based on 100 replications.}
\end{table}

\end{document}